\documentclass[aps,prx,twocolumn,showpacs,amsfonts,amssymb,amsmath]{revtex4-2}

\usepackage{graphicx}
\usepackage{natbib}
\usepackage{bm}
\usepackage{bbm}
\usepackage{physics}
\usepackage{amsmath}
\usepackage{amsthm}
\usepackage{dsfont}
\usepackage{dcolumn}
\usepackage[colorlinks,citecolor=blue,linkcolor=blue,urlcolor=blue,bookmarks=false,hypertexnames=true]{hyperref}
\usepackage[mathscr]{eucal}

\newtheorem{definition}{Definition}

\newtheorem{theorem}{Theorem}

\begin{document}

\title{Modeling of decoherence and fidelity enhancement during transport of entangled qubits}
\author{Aleksandr S. Mokeev}
\author{Yu-Ning Zhang}
\author{Viatcheslav V. Dobrovitski}
 \email{V.V.Dobrovitski@tudelft.nl}
 \affiliation{QuTech and Kavli Institute of Nanoscience, Delft University of Technology,\\
 PO Box 5046, 2600 GA Delft, The Netherlands}

\date{\today}

\begin{abstract}
Entangled qubits transported through space is a key element in many prospective quantum information systems, from long-distance quantum communication to large modular quantum processors. 
The moving qubits are decohered by time- and space-dependent noises with finite extent of correlations. 
%
%
Since the qubit paths are interrelated, 
the phase fluctuations experienced by the qubits exhibit peculiar correlations, which are difficult to analyze with the conventional theory of random processes. 
We propose an approach to this problem based on the concept of trajectories on random sheets.
We establish its efficacy with a specific example of the electron spins shuttled in semiconductor structures, and derive explicit solutions, revealing the role of unusual noise correlations. 
We also demonstrate how the analysis of noise correlations can enhance the shuttling fidelity, by studying transport of two entangled spins encoding a logical qubit. 
%
The proposed approach enabled us to specify the favorable conditions for this encoding, and find a particularly beneficial shuttling regime. We also describe application of this approach to other shuttling scenarios, and explain its utility for analysis of other types of systems, such as atomic qubits and photons 
in quantum networks.
\end{abstract}

\maketitle

\section*{Introduction \label{sec:intro}}

Transport of qubits through space is a central component of many quantum information processing platforms; one can mention
propagation of 
photonic qubits in quantum communication networks \cite{KimbleQuInternet08,AwschEtalQuInterconnects21,WehnerEtalQuInternet18,TutorialEntanglProtocols24HansonBorregaardEtal}, transport of atoms between different locations in large reconfigurable atomic arrays \cite{BeugnonGrangierEtalOptTweezers07,ResizeAtomicArraysAhn2016,%
BernienEtalTwoSpeciesAtomArray22,
KaufmanOptTweezerReview21,AtomicArraysLukin2022}, transport of ions between different regions in prospective trapped-ion quantum processors \cite{KielpMonroeWineland02,PinoNeyenhuisEtal21}, or shuttling of electrons between different quantum dots in scalable semiconductor architectures \cite{FaultTolQuDotTaylorLukin05,BenjaminBaughEtalTopoQCSi19,VdSEtalSingleSpinCCD16,VdSBluhmVeldhorstEtalVision17,langrock_blueprint_2023,KunneEtalSpinBusArch23,BauerleTakadaEtalAcousticTransportElectrons22,mills_shuttling_2019,BoterVdsEtalSpiderWebArray22}.
%
%
In all these examples, the physical carrier (a photon in the optical fiber, an atom in 
an atomic array, or an electron in a semiconductor structure) encodes the qubit 
state in some internal degree of freedom (photon polarization, spin of an electron, angular momentum of an atom, etc); the carrier propagates through a medium in the form of a spatially localized wavepacket, transporting the qubit. 

Going beyond single qubits, and transporting batches of entangled qubits, in parallel or sequentially, is a crucial step towards combining the qubit transport with quantum error suppression and correction in large fault-tolerant platforms. For instance, transport of multiple entangled atoms \cite{AtomicArraysLukin2022} enables quantum computation with error correction in large-scale atomic arrays
\cite{%
ChongBernienEtalCircuitsAtomArrays23,
XuEtalQECforAtomArrays2023,AtomicProcessorWithQECLukin2023,SiegelEtalFaultTolShuttl24}. 
Transmission of large photonic cluster states can be used to overcome the problem of the photon loss in quantum communication and greatly reduce the classical overhead \cite{BorregaardQuRepeater20,EconomouPhotonClusters17,AzumaPhotonicQuRepeaters17,CoganPhotonClusterState23}. 
Shuttling multiple electrons with entangled spins in semiconductor quantum processors would enable quantum error suppression and correction, making an important step towards fault-tolerance \cite{FaultTolQuDotTaylorLukin05,BenjaminBaughEtalTopoQCSi19,BenjaminEtalMulticoreQComp22,BoterVdsEtalSpiderWebArray22}. 

%

Random inhomogeneities in the medium produce the noise that decoheres the moving qubits; thus, reliable modeling of decoherence and analysis of the fidelity become important and timely issues, needed for realistic assessment and optimization of the multi-qubit transport.
%
%
%
For instance, random variations of the optical path length along optical fiber, caused by fluctuations of its refractive index in space and time, scramble the phase and the polarization of photonic qubits; this dephasing, together with the photon loss, constitute most significant problems for long-distance quantum communication \cite{NemannEntanglDistr248km,LiuGuoDistribQuComp23,PompiliHansonQuNetwork22,BersinLukinDixonMetroAreaQuNetwork23,PanEtalMetroAreaQuNetwork21}.  
Similarly, when electrons are shuttled in a semiconducting structure, the 
time- and space-dependent magnetic noise dephases the electron spins \cite{seidler_conveyor-mode_2022,yoneda_coherent_2021,zwerver_shuttling_2023,van_riggelen-doelman_coherent_2023,BurkardEtalTwoDotShuttl20,noiri_shuttling-based_2022,StruckSchreiberEtalSpinPairShuttling23,SmetVandersypenEtal24SpinShuttlSilicon,JadotMeunierEtalTwoSpinShuttling21,BoterJoyntVdSNoiseCorrBellStates20,MortemousqueMeunierEtalShuttl2DArray21}.
This noise can be produced e.g.\ by the nuclear spins, randomly located in the 
shuttling channel and randomly flipping in time. It can also be caused by 
fluctuating charge defects, which randomly displace and distort the electron's 
wavepacket, and, in the presence of the magnetic field gradient or 
inhomogeneity of the $g$-factor, create the time- and space-varying random magnetic field $B({\vec r},t)$ \cite{BoscoZouLossHighFidShuttlingSOI23,HuangHuSpinRelaxShuttl13,StruckCywinskiSchreiber20QubitNoise,KepaCywinskiKrzywdaSpinNoise23,zou_spatially_2023,ShalakDelerueNiquetChargeNoiseSiHole23,SpenceNiquetMeunierEtalChargeNoise22,ShehataVanDorpeEtalChargeNoiseQuDots23,BurkardLaddPanNicholReviewQuDots23}.
%
%
The noise field $B({\vec r},t)$ (or, in the case of photonic qubits in optical fibers, the electric permittivity $\epsilon({\vec r},t)$  \cite{CorneyDrummond01NoiseInOptFibersI,CorneyDrummond01NoiseInOptFibersII,DongHuangLiLiuNoiseOptFibers,JeunhommeSingleModeFiberOpt,WanserFluctOptFibers92})
is correlated both in time
and space, while the standard stochastic description of decoherence is based on the notion of random processes depending only on time  \cite{kubo_statistical_1985,GardinerZoller,Feynman2000,AndersonWeiss53,Kubo1954,KlauderAnderson62,Kubo1963}.

This makes the correlated dephasing of entangled qubits an intricate problem. 
%
If approached in a standard manner, it gives rise to bizarre random processes, e.g.\ whose correlations increase with time, or whose statistics at a given instant is most strongly correlated with the future values of other processes. 
%
%
In this work we present an approach that can treat a broad range of time- and space-correlated realistic noise models for $B({\vec r},t)$ using the notion of a random sheet \cite{Chentsov_1956,Kitagava_1951}, which generalizes the idea of a random process. By analyzing trajectories on random sheets, we show that this approach allows for explicit analytical solution for nontrivially correlated noises in many cases, and provides mathematically justified numerical modeling methods for more complex situations. 

We demonstrate application of this approach in a specific setttings, considering dephasing of the electron spins shuttled along a semiconducting channel (e.g.\ in a Si-, Si/SiGe- or GaAs-based structure), and study two problems, not investigated before, where the effects of nontrivial  noise are prominent. 
First, we analyze the forth-and-back shuttling of a single qubit, inspired by recent experiments \cite{StruckSchreiberEtalSpinPairShuttling23,VolmerStruckEtalValleySplit23,SmetVandersypenEtal24SpinShuttlSilicon}. We obtain the analytical solution and compare it with the numerical results, confirming good computational performance of the proposed modeling methods. Next, we establish practical relevance and usefulness of our approach by studying protection of the qubits from dephasing during transport. We consider 
encoding of a logical qubit
in a decoherence-free subspace \cite{LidarChuangWhaleyDFS98,ZanardiRasettiDFS97,ViolaCoryEtalDFSExp01} formed by the singlet and triplet states of two electron spins, which are shuttled with some delay one after another. We provide explicit analytical solution and numerical simulation results, which demonstrate that such an encoding can greatly improve the fidelity of the qubit state transfer, and identify the range of the system parameters where such an encoding is beneficial. In particular, we  establish that dephasing accumulates slowly even at very low shuttling velocity, such that high-fidelity low-speed shuttling is possible for arbitrarily long distances, as long as the delay time is small enough. Major benefits of this particular shuttling regime are explained in detail.

Finally, in Sec.~\ref{sec:summary} we give brief overview of the approach, and demonstrate its extensions to more general shuttling scenarios and other semiconductor platforms. We also explain value and utility of this approach in analyzing dephasing of other types of systems, such as qubits in atomic arrays and photonic qubits propagating in complex quantum communication networks. 
Thus, our approach is expected to be valuable for realistic assessment, benchmarking and optimization of multi-qubit transport in a wide range of relevant physical systems.

\section{Description of the system and formulation of the problem}
\label{sec:formulation}

Many approaches can be used to fight decoherence, from decoherence suppression via dynamical decoupling, to encoding of the logical qubit in a state of several physical qubits, using e.g.\ decoherence-free subspaces of multiple physical qubits, or employing a range of quantum error-correcting codes. However, in order to effectively apply these approaches in practice, it is vital to have a detailed understanding of the decoherence process, and to quantitatively model and describe different decoherence channels. 
In this work, we analyze dephasing of traveling qubits using shuttling of spins in semiconductors as an instructive and transparent model case. 

\begin{figure}[tbp]
\centering
\includegraphics[width=\columnwidth]{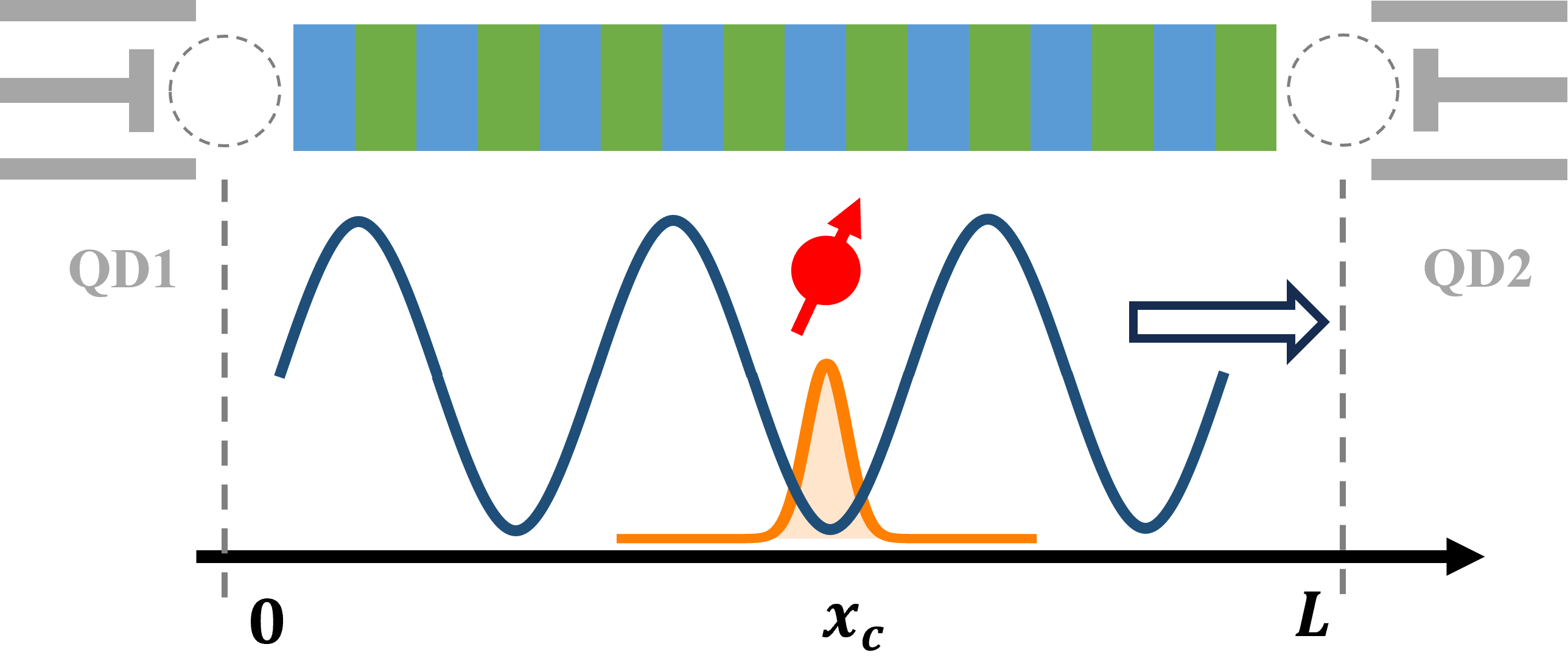}
\caption{Schematic representation of a single spin shuttling between quantum dots. The electron, whose spin (shown as a red arrow) encodes the qubit, is confined within a minimum of the sine-like moving potential (solid dark blue line), forming a tightly localized wavepacked (shown with orange) centered at $x=x_c$. The moving potential transports the electron from left (quantum dot QD1, located at $x=0$) to right (quantum dot QD2, located at $x=L$). Both quantum dots shown as grey circles at the top of the figure, along with the clavier gates, generating the moving potential and represented as interspersed blue and green rectangles.}
\label{fig:shuttling}
\end{figure}

For conceptual simplicity, we consider the conveyor-belt type shuttling, when a long (length $L\sim 1$--10~$\mu$m) 
one-dimensional channel is formed between two quantum dots, as shown in Fig.~\ref{fig:shuttling}. Initially, at $t=0$, an electron is loaded into the channel from the quantum dot at $x=0$, and, being confined in a moving potential well, is transported to the other quantum dot located at $x=L$, reaching it at the time moment $t=t_0$. The moving confining potential can be created in Si/SiGe structures 
using a set of clavier gates deposited on top of the shuttling channel, by applying to these gates the sinusoidally varying voltage whose phase changes  along the channel \cite{seidler_conveyor-mode_2022,langrock_blueprint_2023,JeonBenjaminFisher24RobustChargeShuttl}. Alternatively, e.g.\ in GaAs-based structures, one can employ the surface acoustic wave  \cite{ShuttlingSAWRitchie11,ShuttlingSAWMeunier11,HuangHuSpinRelaxShuttl13,SchuetzLukinVandersypenEtal17SAW} 
propagating along the shuttling channel, which creates the moving electric confining potential via piezoelectric effect.

The shuttled electron forms a wavepacket, tightly localized along all three dimensions (such that its size is much smaller than the shuttling length $L$), with the center at the point $x=x_c$; we denote the corresponding spatial distribution of the electron density as 
$\rho(x,y,z;x_c)\equiv\rho({\vec r};x_c)$. The center of the wavepacket moves along the shuttling channel, such that $x_c$ changes with time: $x_c=0$ at $t=0$ and $x_c=L$ at $t=t_0$. An important requirement for the shuttling process is its (at least, approximate) adiabaticity, that is, the wavepacket is assumed to be confined within the minimum of the shuttling potential without leaking to the adjacent minima, and moves without being excited to higher orbital states within a given minimum and/or to the excited valley states.

In most experiments, a nominally uniform quantizing magnetic field $B_Q\sim 0.1$--1~T is applied to the system, but the effective $g$-factor of the electron slightly varies in space due to random variations in the material properties, 
such that the Zeeman energy of the electron spin varies in space. Besides, if randomly located nuclear spins are present in the system, then the position-dependent random hyperfine field $B_{\rm hf}({\vec r})$ should be taken into account. 
As a result, for an electron with the wavepacket density profile $\rho({\vec r};x_c)$, the effective Zeeman energy splitting experienced by the electron spin $S=1/2$ is described by the Hamiltonian
\begin{equation}
\label{eq:hz}
{\tilde H}_Z (x_c) = S_z \int d^3 r\, \mu_B\, [g({\vec r})\,B_Q + g_s\,B_{\rm hf}({\vec r})]\,\rho({\vec r};x_c),
\end{equation}
where $S_z$ is the operator of the $z$-component of the electron spin, $g({\vec r})$ is the position-dependent effective $g$-factor of the electron, $\mu_B$ is the Bohr magneton, and $g_s\approx 2.002$ is the electron spin $g$-factor.

Here and below we assume that the {\em direction} of the effective magnetic field acting on the shuttled spin is almost uniform in space and time, and that the longitudinal relaxation can be neglected; this is a good approximation for Si- and Si/SiGe-based structures. In GaAs-based systems, the spin-orbit coupling creates a fluctuating transverse component, such that the longitudinal relaxation also must  be taken into account \cite{HuangHuSpinRelaxShuttl13}, along with the dephasing studied here. Shuttling in Ge-based systems, where the spin-orbit effects are more pronounced, would require to take into account significant spatial variation of the direction of the effective magnetic field acting on the shuttled spin \cite{van_riggelen-doelman_coherent_2023, WangVeldhorstEtalGeQDs24,BoscoZouLossHighFidShuttlingSOI23,BoscoZouLossHighFidShuttlingSOI23}, and the approach presented here needs minor modifications, see Sec.~\ref{sec:summary}.

Next, we need to take into account the fluctuations of the shape of the wavepacket in space and time. The shape of the wavepacket in an ideal system would stay constant during the shuttling process, i.e.\ $\rho({\vec r};x_c)=\rho_0(x-x_c,y,z)$, but in reality the confining potential, which determines the shape of $\rho({\vec r};x_c)$, fluctuates in time and space 
\footnote{The confining potential created by the clavier gates slightly differs from an ideal traveling wave due to the finite gate pitch: the potential periodically changes its shape, and so does the wavepacket of the shuttled electron. If the phase shift between the adjacent pairs of the gates is small enough then these changes are small~\cite*{langrock_blueprint_2023,JeonBenjaminFisher24RobustChargeShuttl} and can be neglected. Otherwise, the periodic changes in the shape of the wavepacket, and the associated time variation of $B_0$ in Eq.~\ref{eq:tildeB} can be taken into account within the outlined approach.}. 
Firstly, due to random variation of the material properties along the shuttling channel, the shape of the confining potential changes, such that the the density distribution $\rho(x,y,z;x_c)$ slightly varies as the wavepacket center $x_c$ moves.
Secondly, there is a large number of charged defects present in real systems, such as the charge traps, which can randomly trap and release electrons, or where the trapped charge can randomly jump between two nearby positions. These events affect the electrostatic potential felt by the shuttled electron, and lead to small random displacements and distortions of the wavepacket, such that the density $\rho(x,y,z;x_c)$ becomes explicitly dependent not only on $x_c$ but also on time:
\begin{equation}
\label{eq:wavepacket}
\rho({\vec r},t;x_c)=\rho_0(x-x_c,y,z) + \Delta\rho(x,y,z,t;x_c).
\end{equation}
As seen from Eq.~\ref{eq:hz}, such variations, in the presence of the space-dependent field $B({\vec r})$ and the $g$-factor $g({\vec r})$, make the Zeeman energy depend on both $x_c$ and $t$:
\begin{equation}
\label{eq:fullhz}
\begin{split}
{\tilde H}_Z (x_c,t) &= \\
\mu_B S_z\! 
&\int\!d^3 r\, [g({\vec r}) B_Q + g_s B_{\rm hf}({\vec r},t)]\,\rho({\vec r},t;x_c),
\end{split}
\end{equation}
where we also took into account that the flip-flops in the nuclear spin bath make the field $B_{\rm hf}({\vec r},t)$ time dependent.
As a result, the Zeeman Hamiltonian describing the shuttling electron can be written in a familiar form
\begin{equation}
\label{eq:tildeB}
{\tilde H}_Z (x_c,t) = g_0 \mu_B \left[B_0(x_c) + {\tilde B}(x_c,t)\right] S_z 
\end{equation}
where $g_0$ is a nominal $g$-factor of the electron in the semiconducting channel (the appropriately averaged value of $g({\vec r})$, see below), while $B_0(x_c,t)$ and ${\tilde B}(x_c,t)$ represent, respectively, the deterministic and the random part of the effective magnetic field; more precise meaning of these terms will be discussed at the end of this Section.

Our goal is to describe dephasing of such an electron spin (or several such spins) moving along certain trajectory $x_c(t)$ under the action of the fluctuating magnetic field ${\tilde B}(x_c,t)$.
In principle, it is possible to approach the problem directly using Eqs.~\ref{eq:hz} and \ref{eq:wavepacket}, calculating random fluctuations of the wavepacket $\rho({\vec r},t; x_c(t))$ and analyzing their impact on the electron spin, similarly to the approach traditionally used in analyzing the optical waves or pulses propagating through random media \cite{RytovEtalStatRad,SobczykStochWave}. However, such an analysis is very difficult, and usually requires other radical simplifications \cite{SmithRaynerTwoPhotonWaveMechanics06,PatersonOrbAngMomPhotons05,
SemenovVogelEntanglPhotonsTurbAtmosph10,VasylyevSemenovVogelAtmosphTransmittance16,PerinaTeichEtalPhotonDetectStatistics73,Tatarskii72}, such that the treatment based on Eq.~\ref{eq:tildeB}, as described below, is more suitable for analysis of multiple qubits shuttled along non-trivial trajectories (see also Sec.~\ref{sec:summary} about possible combination of the two methods).

A powerful approach, widely used for studying dephasing of {\em stationary} qubits, is based on the concept of a random process. Originally developed for magnetic resonance experiments \cite{BloembPurcellPound48,AndersonWeiss53,Kubo1954,KlauderAnderson62,Kubo1963}, this approach replaces the complex problem of the dynamics of macroscopically large number of interacting quantum spins by a much simpler analysis of a relevant spin subjected to a random time-varying magnetic field ${\tilde B}(t)$, which causes fluctuations of the qubit's Larmor frequency $\omega(t)= \mu [B_Q+{\tilde B}(t)]/\hbar$, where $\mu$ is the magnetic moment of the relevant spin. In the course of a single experiment of duration $t_0$ the qubit acquires, along with the deterministic phase $\alpha=\mu B_Q t_0$, the random phase 
\begin{equation}
\phi(t_0) = \mu \int_0^{t_0} {\tilde B}(t)\,dt,
\end{equation}
which changes every time the experiment is performed, such that the overall signal after many experimental shots is an average over all possible realizations of the random process ${\tilde B}(t)$. The averaging leads to reduction of the coherences (off-diagonal elements of the qubit density matrix) by the dephasing factor 
\begin{equation}
\label{eq:wRP}
W(t_0)={\mathbb E}\left(\exp{-i\phi(t_0)}\right),
\end{equation}
where ${\mathbb E}(\dots)$ denotes the expected value of the corresponding random quantity, i.e.\ an average over ${\tilde B}(t)$, such that the dephasing factor is determined by the characteristic functional of the random process ${\tilde B}(t)$ \cite{Kubo1954,KlauderAnderson62,UhlenbeckOrnsteinRandomProc,ChandrasekharRandomProc,WangUhlenbeckRandomProc}. 
The noises encountered in experiments are often well approximated as random processes with favorable properties (such as Markov
\cite{MarkovNote}%
%
%
, Gaussian, stationary, etc.), which make them efficiently tractable:  
the well-developed mathematical theory of random processes \cite{UhlenbeckOrnsteinRandomProc,ChandrasekharRandomProc,WangUhlenbeckRandomProc} enables analytical solutions in many simpler cases, and provides a number of rigorously justified, efficient numerical methods for more complex situations.

\begin{figure}[tbp]
\includegraphics[width=\linewidth]{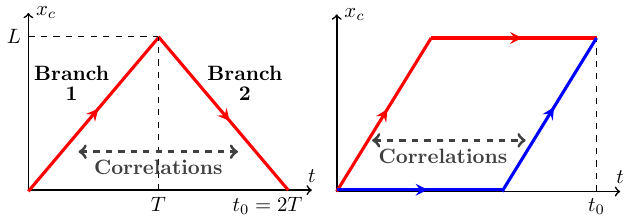}
\caption{Examples of space-time trajectories leading to non-trivial correlated random fields $\mathrm{B}(t)$. {\bf (a)} Forth-and-back shuttling of a single spin qubit. The trajectory $x_c(t)$ consists of two branches: branch 1 corresponds to shuttling of a qubit from its initial location $x_c=0$ to the destination at $x_c=L$, and branch 2 corresponds to shuttling back to $x_c=0$ with the same velocity. {\bf (b)} Sequential shuttling of two qubits. Trajectory $x_{c1}(t)$ of the first qubit is shown in red, and $x_{c2}(t)$ of the second qubit is in blue.}
\label{fig:traj1and2}
\end{figure}

Dephasing of the shuttled qubit can be analyzed in the same manner: for a given spacetime trajectory, i.e.\ for a given dependence $x_c(t)$, the random phase acquired by the shuttled qubit during a single experiment is
\begin{equation}
\label{eq:phit0onespin}
\phi(t_0) = g_0 \mu_B \int_0^{t_0} {\tilde B}(x_c(t),t)\,dt,
\end{equation}
such that the averaging over many experimental shots leads to the dephasing factor 
\begin{equation}
\label{eq:wRS}
W(t_0)={\mathbb E}\left(\exp{-i\phi(t_0)}\right),
\end{equation}
where the expectation ${\mathbb E}(\dots)$ now corresponds to averaging over ${\tilde B}(x_c(t),t)$.
For a single qubit with a reasonably simple trajectory $x_c(t)$ (for instance, $x_c(t)=v t$ with constant velocity $v$), one can introduce a random process ${\mathrm B}(t)\equiv {\tilde B}(x_c(t),t)$, and analyze the problem making standard approximations about ${\mathrm B}(t)$ (e.g.\ that it is Markov, stationary, etc). This approach can successfully treat a number of special models for ${\tilde B}(x_c,t)$, such as spatially uncorrelated noise, static noise, or the noise produced by a few point sources \cite{langrock_blueprint_2023,BoscoZouLossHighFidShuttlingSOI23,HuangHuSpinRelaxShuttl13}
(similar simplifications were also used for photonic qubits, see e.g.\ \cite{SmithRaynerTwoPhotonWaveMechanics06,PatersonOrbAngMomPhotons05,SemenovVogelEntanglPhotonsTurbAtmosph10}), or in the limit of zero correlation time via Bloch--Redfield theory \cite{JeskeColeSpatiallyCorrDecoh13,JeskeVogtColeSpatiallyCorrNoise13}.

However, if ${\tilde B}(x,t)$ has finite correlations in time and space, this approach encounters serious difficulties when applied to nontrivial trajectories $x_c(t)$ or to the entangled qubits.

For example, consider the forth-and-back shuttling of an electron spin, studied in recent experiments \cite{StruckSchreiberEtalSpinPairShuttling23,VolmerStruckEtalValleySplit23,SmetVandersypenEtal24SpinShuttlSilicon}.
The spacetime trajectory (see Fig.~\ref{fig:traj1and2}a) consists of two branches, with the electron starting at $x_c=0$ at $t=0$ and moving forth to $x_c=L$ with the constant velocity $v$ (first branch), and then immediately shuttled back to $x_c=0$ with the same velocity (second branch). Within each branch, the noise process ${\mathrm B}(t)$ can be modeled as Gaussian, stationary and Markovian, i.e.\ Ornstein--Uhlenbeck (OU) process with the effective correlation time $t_c=[\tau_c^{-1} + v/\lambda_c]^{-1}$, where $\lambda_c$ and $\tau_c$ are the correlation length and time, respectively, of the original noise field ${\tilde B}(x_c,t)$. However, calculations of the total dephasing factor should take into account that the phases accumulated during motion along each branch are correlated, which is trivial only for $\tau_c\to 0$ (uncorrelated branches) or for $\tau_c\to\infty$, when the underlying noise ${\tilde B}(x_c(t),t)$ is static and the phases accumulated along each branch are the same. 

\begin{figure}[tbp!]
\includegraphics[width=\linewidth]{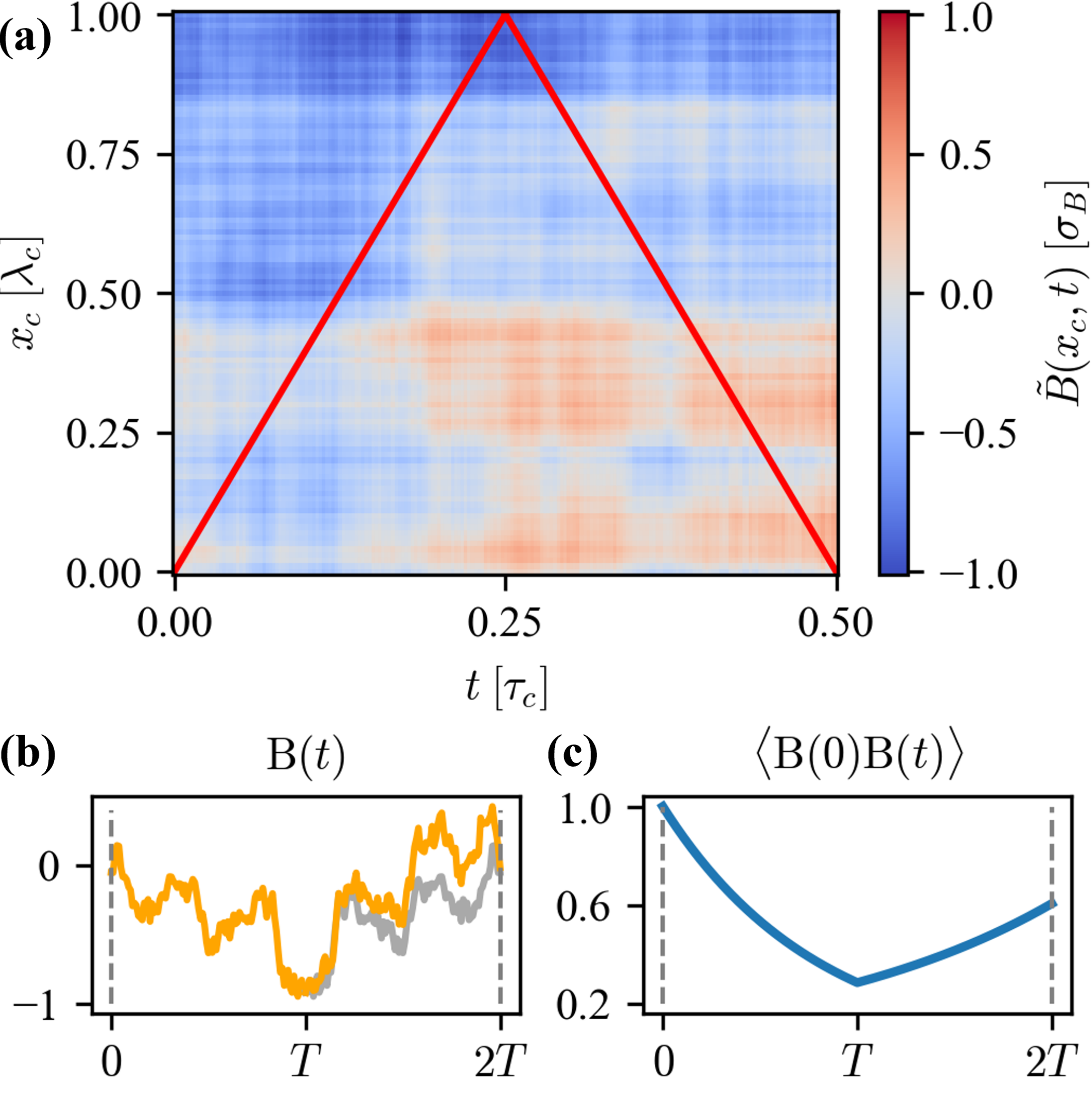}
\caption{During forth-and-back shuttling, the random magnetic field $\mathrm{B}(t)$ exhibits re-emerging correlations, occuring because the qubit passes the same spatial location $x_c$ twice, during the first and the second branch of the shuttling process. 
{\bf (a)} One realization of the noise $\tilde{B}(x_c,t)$ (in units of $\sigma_\mathrm{B}$, see Eq.~\ref{eq:sigmaBdef}), modeled as an OU random sheet (see Definition \ref{def:OU}); the trajectory of a qubit shuttled forth and back (shown also in Fig.~\ref{fig:traj1and2}a), is displayed as a solid red line. Time $t$ and coordinate $x$ are measured in units of the correlation time $\tau_c$ and length $\lambda_c$, respectively.
{\bf (b)} The corresponding realization of the random field $\mathrm{B}(t)$ (in units of $\sigma_\mathrm{B}$), exhibiting re-emerging correlations. Within branch 1 of the trajectory $x_c(t)$ (at $0<t<T$), the field $\mathrm{B}(t)$ (solid orange line) quickly becomes uncorrelated with its earlier values. In the second branch, the correlations are regained as the qubit passes the same locations $x_c$ again; as a result, the function $\mathrm{B}(t)$ at $T<t<2T$ (branch 2) is close to its own mirror image (shown as a grey line) in the region $0<t<T$ (branch 1).
{\bf (c)} Correlation function $\langle\mathrm{B}(0)\mathrm{B}(t)\rangle$ (in units of $\sigma^2_\mathrm{B}$) of the noise acting on the qubit during the forth-and-back shuttling, demonstrating the re-emerging correlations in the branch 2 of the shuttling process.}
\label{fig:corrnoise}
\end{figure}

For general values 
of $\tau_c$ and $\lambda_c$, the correlations are non-trivial, and can have very unusual form, e.g.\ when $\lambda_c$ is small while $\tau_c$ is large, see Fig.~\ref{fig:corrnoise}
\footnote{Data and codes used in this work could be accessed at \url{https://data.4tu.nl/datasets/d0d1007f-c27d-491d-b7e1-cc60e38047b4}, DOI 10.4121/d0d1007f-c27d-491d-b7e1-cc60e38047b4. The codes for modeling of the multi-qubit shuttling are available on Github repository \url{https://github.com/EigenSolver/SpinShuttling.jl} and \url{https://eigensolver.github.io/SpinShuttling.jl/dev/}.}.
In that case $t_c\ll\tau_c$, and the correlations along the first branch decay fast, such that ${\mathrm B}(T)$ is practically uncorrelated with ${\mathrm B}(0)$. But when the electron is shuttled back, the correlations are restored due to long correlation time $\tau_c$, and if $T\ll\tau_c$ then ${\mathrm B}(t_0=2T)$ (at the end of the shuttling) is close to its initial value ${\mathrm B}(0)$. In other words, the correlations of the process ${\mathrm B}(t)$ {\em increase with time}, from almost zero (the correlator $\langle{\mathrm B}(0){\mathrm B}(T)\rangle\approx 0$) to almost maximal value (the correlator 
$\langle{\mathrm B}(0){\mathrm B}(t_0)\rangle\approx\langle{\mathrm B}(0)^2\rangle$).
Equally subtle is the Markov property: the process ${\mathrm B}(t)$ is Markovian within each branch separately, but is non-Markov when both branches are considered together. Below we show that such a behavior is typical for the random processes ${\mathrm B}(t)$ generated by the noise ${\tilde B}(x_c(t),t)$.
For more general trajectories, which may include several forth-and-back moves with different velocities and possible spin flips between the branches (if we analyze the spin echo or other decoupling protocols during shuttling), the properties of the random process ${\mathrm B}(t)$ can be even more extraordinary, rendering many standard approximations and methods of the theory of random processes \cite{GardinerRandomProc,VanKampenRandomProc} unusable.

These problems become even more prominent when we consider two electrons with entangled spins, shuttled one after another with some delay, as it was done e.g.\ in Ref.~\onlinecite{JadotMeunierEtalTwoSpinShuttling21}, where the shuttling of a singlet was investigated.
In this work, we establish that such a two-qubit transport is a relatively simple way to protect the qubit state against dephasing during shuttling. Specifically, we consider encoding the state of the qubit, which is to be shuttled, in a decoherence-free subspace \cite{LidarChuangWhaleyDFS98,ZanardiRasettiDFS97,ViolaCoryEtalDFSExp01} formed by the singlet and triplet states of two electron spins.
These two spins are shuttled with some delay one after another. 
The element $\langle\uparrow\downarrow|\rho(t_0)|\downarrow\uparrow\rangle$
of the two-spin density matrix $\rho(t_0)$, which determines the mutual phase between the spins and distinguishes singlet from triplet, is diminished by the noise, with the dephasing factor 
\begin{equation}
\label{eq:twospinW}
W(t_0)={\mathbb E}\left(\exp{-i[\phi_1(t_0)-\phi_2(t_0)]}\right),
\end{equation}
where $\phi_1(t_0)$ and $\phi_2(t_0)$ are the phases acquired by the first and the second spins along their corresponding space-time trajectories $x_{c1}(t)$ and $x_{c2}(t)$ (shown in Fig.~\ref{fig:traj1and2}b) under the action of the random processes ${\mathrm B}_1(t)={\tilde B}(x_{c1}(t),t)$ and ${\mathrm B}_2(t)={\tilde B}(x_{c2}(t),t)$. 

If the underlying noise ${\tilde B}(x_c,t)$ is quasi-static, and does not vary in time on a timescale of a single shuttling event, then the random phases $\phi_1(t_0)$ and $\phi_2(t_0)$ are approximately the same, provided that both qubits are shuttled along the same channel with approximately the same velocity and the delay between them is not too long. In such a case, the difference between $\phi_1(t_0)$ and $\phi_2(t_0)$ would be almost zero, even if each phase by itself were large, and the dephasing factor $W(t_0)$ would be almost one, ensuring high fidelity of the shuttling process.  

However, this proposal would be unreliable and vacuous without a detailed study of the dephasing under realistic circumstances. E.g., it is crucial to know, how the finite speed of time fluctuations of the noise ${\tilde B}(x_c,t)$ affects the dephasing, how large can the delay time be to ensure satisfactory fidelity, etc. Such an analysis presents the same type of difficulties as the forth-and-back shuttling. The processes ${\mathrm B}_1(t)$ and ${\mathrm B}_2(t)$ are correlated, being produced by the same underlying noise ${\tilde B}(x_c,t)$, and their correlations can be quite unusual: e.g., depending on the values of $\tau_c$ and $\lambda_c$, the value of ${\mathrm B}_1(t)$ at the end of the trajectory $x_{c1}(t)$ can be strongly correlated with ${\mathrm B}_2(t)$ either at the beginning or at the end of the trajectory $x_{c2}(t)$. In this case, even the physical and mathematical properties of the processes ${\mathrm B}_{1,2}(t)$ are somewhat strange: while each of them separately can be modeled as a well-behaved process (e.g.\ an OU process), the mutual correlations can increase with time and/or with distance, and, as we go along the trajectories  $x_{c1}(t)$ and $x_{c2}(t)$, both mutual and self-correlations of ${\mathrm B}_{1,2}(t)$ can decrease, increase, or behave non-monotonically. 

Such problems become complex in a general case, when many entangled qubits are shuttled along several neighboring shuttling channels, experiencing multiple random magnetic fields with non-trivial correlations, produced by the space- and time-correlated noise ${\tilde B}(x_c,t)$.
In order to be useful, stochastic description of the dephasing process should provide a range of sufficiently realistic but solvable mathematical models.
For that, we first need to represent ${\tilde B}(x,t)$ as a mathematical object that can express the expected properties of the noise, e.g.\ the finite correlation time $\tau_c$ and length $\lambda_c$, or the specific noise spectral density $S(\omega)$ (e.g.\ $1/f$ noise or Nyquist--Johnson noise) at a given point in space, etc. 
Second, we need to describe and characterize the random processes ${\mathrm B}(t)$ produced by restricting ${\tilde B}(x,t)$ to the trajectories $x_c(t)$ of the qubits, and ensure that the resulting process ${\mathrm B}(t)$ is physically meaningful (e.g., free of singularities often encountered for random fields in several dimensions \cite{DynkinMarkovProcAndRandFields,Walsh1986,WernerPowellLectNotesGFF21}). Finally, we should be able to calculate the characteristic functional associated with the dephasing, in analytic form for simpler cases, or employing mathematically justified numerical methods for more complex situations.
For the latter, it is crucial to understand how the random processes ${\mathrm B}(t)$, behave under discretization, when the distance (in time and space) between the numerical mesh points is small. The associated subtleties (such as the difference between the It\^{o} and the Stratonovich conventions, singularities arising in the limit of delta-correlated processes, etc.) are well known for the ``usual'' random processes \cite{GardinerRandomProc,VanKampenRandomProc}, and the necessary adjustments in numerical methods have been studied in detail \cite{MilsteinTretyakovBook}, but it is not a priori clear what should be expected from the processes generated from ${\tilde B}(x_c,t)$ on the trajectories $x_c(t)$. 

In this work we employ the concept of the Gaussian random sheet to model the noise ${\tilde B}(x,t)$ with experimentally relevant properties. 
In this way we can satisfy the requirements formulated above, obtaining physically meaningful and mathematically tractable processes ${\mathrm B}(t)$ for rather general qubits trajectories $x_c(t)$, even in the situations where ${\tilde B}(x,t)$ or ${\mathrm B}(t)$ have the singularities which represent the point-like dephasing hotspots \cite{langrock_blueprint_2023,DodsonFriesenValleyAndInterface22,LosertFriesenValleySplit23} or the instant electron spin flips performed in the spin echo and dynamical decoupling experiments. We demonstrate performance of this concept by studying the two cases discussed above, the forth-and-back shuttling of a spin and the shuttling of two entangled spins. 

It is important to explicate a subtle feature of the Hamiltonian ${\tilde H}_Z (x_c,t)$ in Eq.~\ref{eq:tildeB}, which contains both the random part ${\tilde B}(x_c,t)$ and the deterministic part $B_0(x_c)$ of the effective magnetic field. The part ${\tilde B}(x_c,t)$ randomly varies from one experiment to another, such that its value cannot be predicted in advance, while the value of $B_0(x_c)$ can be predicted, at least in principle. Correspondingly, the delineation between these parts depends on the specific experimental setting. For instance, if we consider a device where the quenched disorder in $g(\vec{r})$ leads to appearance of an effective field $B_{\rm q.d.}({\vec r})=g(\vec{r})\mu_B B_Q$ that does not vary with time, then $B_{\rm q.d.}$ should be viewed as deterministic: while the profile $B_{\rm q.d.}({\vec r})$ is not known in advance, but it can be measured once (at least, in principle) and used later to reliably predict its influence on the shuttled qubits. But if we are studying a large batch of similar devices with different profiles $B_{\rm q.d.}({\vec r})$, or the same device where $B_{\rm q.d.}({\vec r})$ fluctuates slowly but randomly (such that the previous measurements cannot be re-used later) then this field should be considered as random, a part of ${\tilde B}(x_c,t)$. 
In a similar way, if the trajectories $x_c(t)$ of the qubits slightly fluctuate from one experimental shot to another, then the deterministic quantity $B_0(x_c)$ gives rise to a random contribution to the magnetic field acting on the spin. 
In this work, we consider all qubits trajectories $x_c(t)$ and the field $B_0(x_c)$ as deterministic, we also assume that the ideal-case wavepacket $\rho_0(x-x_c,y,z)$ does not explicitly depend on time and that the average hyperfine field is zero, such that the condition
\begin{equation}
g_0 B_0(x_c) = \int\!d^3 r\, g({\vec r}) B_Q\,\rho_0(x-x_c,y,z),
\end{equation}
see Eqs.~\ref{eq:fullhz} and \ref{eq:tildeB}, determines the quantities $g_0$ and $B_0$.

The approach outlined in this section, Eqs.~\ref{eq:tildeB}--\ref{eq:twospinW}, is most useful in the situations where the total shuttling distance is noticeably larger than the characteristic width of the envelope $\rho({\vec r},t;x_c)$. Otherwise, it may be more advantageous to 
explicitly determine the shape of $\rho({\vec r},t;x_c)$ and work directly with Eq.~\ref{eq:fullhz}; the representation of the noise ${\tilde B}(x,t)$ as a random Gaussian sheet would still be helpful.
The approach presented here can also be extended to include other important features of the actual devices, e.g.\ to take into account that the the effective magnetic field acting on the shuttled spin is a vector ${\vec B}(x_c,t)$, such that the spin decoherence includes random rotations around the $x$ and $y$ axes. It may be also important to include into consideration the spin-orbit interaction and excitations to the excited valley states and higher orbitals of the shuttling potential \cite{langrock_blueprint_2023,BoscoZouLossHighFidShuttlingSOI23,StruckSchreiberEtalSpinPairShuttling23,VolmerStruckEtalValleySplit23,JeonBenjaminFisher24RobustChargeShuttl}. Such extensions, discussed in more detail in Sec.~\ref{sec:summary}, are conceptually straightforward but laborious, and fall out of the scope of this work.

\section{Stochastic sheet as a noise model \label{sec:mathGen}}

In this section we introduce the notion of a Gaussian random sheet (and, more generally, random process indexed by multiple parameters) and discuss two important examples, the Brownian sheet and the Ornstein--Uhlenbeck sheet, which lay the mathematical foundation for the rest of the paper. We prove that the restriction of the Gaussian sheet to a rather general set of the qubits trajectories $x_c(t)$ produces well-defined and physically meaningful Gaussian random processes ${\mathrm B}(t)$. We give explicit expressions for the dephasing factors, and propose two methods for their numerical evaluation, providing mathematical justification for correctness of these methods and the estimates for the associated numerical errors.

\subsection{Gaussian random sheets}

Let us begin with the standard setting of the probability theory, considering a probability space $(\Omega, \mathcal{B}, \mathbb{P})$, where $\Omega$ is a sample space, $\mathcal{B}\subseteq 2^{\Omega}$ is a $\sigma$-algebra of events, and $\mathbb{P}$ is a probability measure. We can introduce an arbitrary set $\mathcal{X}$, called the index set, and consider a collection $Z$ of real-valued random variables indexed by the elements $\theta\in \mathcal{X}$, represented as $Z:\Omega \times \mathcal{X} \rightarrow \mathbb{R}$, i.e.\ a collection of $\mathbb{P}$-measurable functions $Z(\cdot,\theta): \Omega \rightarrow \mathbb{R}$ at a given point $\theta\in \mathcal{X}$. 
The term ``random process'' is commonly used for the conventional case \cite{VanKampenRandomProc,GardinerRandomProc,kubo_statistical_1985}, where the index set is $\mathcal{X} = \mathbb{R}$, or some proper subset (such as the positive half-line $\mathbb{R}_{+}$ 
\footnote{We assume the convention that $\mathbb{R}_+$ includes zero.}%
), or an interval $[a,b] \subset \mathbb{R}$, etc. 
It is equally possible to consider random processes indexed by multiple  parameters, such as the noise ${\tilde B}(x_c,t)$ indexed by two real parameters, such that $\theta=(x_c,t)\in \mathcal{X} = \mathbb{R}^{2}$. The processes indexed by two parameters are often referred to as stochastic or random sheets, while the processes with arbitrary (e.g., higher-dimensional) index sets are sometimes called random or stochastic fields 
\footnote{Note that the term ``random field'' is used in literature to describe a number of similar but different objects with different properties, such that its concrete meaning depends on the context.
In particular, it is important not to confuse an arbitrary Gaussian field, considered in this section, with one specific case, the standard object in statistical physics known as Gaussian Free Fields (GFF)~\cite*{friedli2017statistical,sheffield2007gaussian}. GFF is a specific type of Gaussian field on a graph (discrete case) or on a Sobolev space of functions on some continuous domain $D \subset \mathbb{R}^{n}$ (continuous case). The discrete GFF is a random function $h(x)$ on the graph with the probability density 
$
\exp\left(-\sum_{x \sim y} p_{x,y} (h(x) - h(y))^2 \right),
$
where the summation goes over all pairs of adjacent vertices, and $p_{x,y}$ are the Markov chain transition probabilities. In the continuum limit, when the domain $D$ is two- or higher-dimensional, the variance of a GFF becomes infinite, and therefore $h(x)$ has to be understood as a random element (random variable with non-numerical values), whose values belong to the space of bounded linear functionals on the Sobolev space $H^1(D)$, i.e., it is a distribution over distributions or a stochastic field indexed by functions. In contrast, the Gaussian fields considered in this paper have no such singularities, and admit simpler mathematical description.
}.

The processes indexed by multiple parameters exhibit a number of novel features, so that some fundamental notions have to be reconsidered in the case of multiple parameters: for instance, the standard way to define the transition probabilities and the Markov property is via the natural linear time ordering, which is clearly defined only for a single-parameter index; extending this definition to the multi-parameter case is non-trivial, see Appendix \ref{app:markov}.
Below we introduce and discuss the most relevant properties of the processes indexed by multiple parameters.

Let us fix the sample $\omega\in\Omega$; the function $Z(\omega,\cdot):\mathcal{X} \rightarrow \mathbb{R}$ is called a realization, or sample, of the random field $Z$. 
%
%
Realizations of the random field belong to the linear space $\mathfrak{L}(\mathcal{X})$ of all functions on $\mathcal{X}$, and one can ask a question about the probability for a given realization to be in the subset $\mathfrak{Z} \subset \mathfrak{L}(\mathcal{X})$. This construction turns $\mathfrak{L}(\mathcal{X})$ into the probability space by the pushforward of the initial probability measure, producing the following construction
\begin{equation}
\mathbb{P}_{\rm law} \left( \mathfrak{Z} \right)= \mathbb{P}\left(\{\omega \in \Omega : \  Z(\omega,\cdot)\in \mathfrak{Z} \} \right),\  \mathfrak{Z} \subset \mathfrak{L}(\mathcal{X}).
\end{equation}
The new $\sigma$-algebra of events is the smallest by inclusion $\sigma$-algebra containing all cylindrical sets of functions
\begin{equation}
\begin{split}
&\mathfrak{U}\left(\theta_1, A_1; \ldots ;\theta_n, A_n \right)=\\
& \{f(\theta) \in \mathfrak{L}(\mathcal{X}) : \  f(\theta_1)\in A_1, \ldots,f(\theta_n)\in A_n \},
\end{split}
\end{equation}
where $n\in \mathbb{N}$, $\theta_k \in \mathcal{X}$, and $A_k \subset \mathbb{R}$ is the Borel subset $\forall k\in \mathbb{N}$, $0<k \le n$. Informally speaking, this is a set of functions ``passing through'' the sets $A_k$ at the ``time moments'' $\theta_k$ indexed by the multi-dimensional ``time'' $\theta$.

The random variables in the new probability space are $\mathbb{P}_{\rm law}$-measurable functionals on $\mathfrak{L}(\mathcal{X})$, and the stochastic fields are sets of functionals. 
Below, for brevity of notations, we will omit the first argument in the functions denoting the random fields, writing $Z(\theta)$, or even $Z$, instead of $Z(\omega,\theta)$ everywhere.

If we take the set of all functionals returning value of a function at a given point, i.e.\
\begin{equation}
f_{\theta^*}(h(\cdot)) = h(\theta^{*}),\quad h\in \mathfrak{L}(\mathcal{X}), \theta^{*}\in \mathcal{X},
\end{equation}
then we obtain the version of the initial stochastic field. 
\begin{definition}[Version equivalence] \label{version_def}
Two stochastic fields $Z_1$ and $Z$ 
are the versions of each other if the distributions of $Z_1(\theta)$ and $Z(\theta)$ 
are the same for all $\theta$.
\end{definition}
The version of the random field (in the sense of Definition \ref{version_def} above) that is given by the law of the process could be considered as canonical.

We are mainly interested in stochastic fields indexed by topological spaces, in fact, by metric spaces. It is possible to speak about the continuity of realizations of such fields. The field $Y$ is a field with continuous realizations if almost surely $Y(\omega,\cdot)$ is continuous. In that case, we can restrict our discussion to the space $C(\mathcal{X})$ of continuous functions on $\mathcal{X}$ instead of studying general functions in $\mathfrak{L}\left(\mathcal{X}\right)$.
\begin{definition}[Gaussian field] \label{def:Gaussianfield}
A field $Z$ is said to be Gaussian if, for any finite set of indices ${\theta_1, \theta_2, \ldots, \theta_n} \subseteq \mathcal{X}$, the random vector $\left[Z(\theta_1), Z(\theta_2), \ldots, Z(\theta_n)\right]$ has a multivariate Gaussian distribution.
\end{definition}
Statistical properties of a Gaussian field are fully characterized by its first two moments, the mean $m(\theta)$ and the covariance $K(\theta_1,\theta_2)$ functions:
\begin{eqnarray}
&m(\theta) = \mathbb{E} [Z(\theta)], \\
&K(\theta,\theta') = \operatorname{cov}(Z(\theta), Z(\theta')), \quad \theta,\theta'\in \mathcal{X}.    
\end{eqnarray}
In this article, we are considering noise with zero average; this is not a very restrictive assumption. A zero-mean (centered) process can be modified to have a desired mean by adding a deterministic function.

In this work we are dealing mainly with the random sheets whose index sets $\mathcal{X}$ are two-dimensional, such as $\mathbb{R}^2$ or $\mathbb{R}_+^2$. Two basic examples of the Gaussian sheets are Brownian (sometimes also called Wiener) sheet \cite{Kitagava_1951,Chentsov_1956} and Ornstein--Uhlenbeck (OU) sheet.
\begin{definition}[Brownian sheet] \label{wsheetdef}
A zero-mean Gaussian sheet ${\mathscr W}: \Omega\times \mathbb{R}_+^{2} \rightarrow \mathbb{R}$ is called Brownian sheet if its covariance function is
\begin{equation}
K_{\mathscr W} (\theta,\theta') =\min \{x,x'\}\cdot \min \{t,t'\},
\end{equation}
where $\theta = (x,t)$ and $\theta' = (x',t')$ are points in $\mathbb{R}_+^2$.
\end{definition}
\begin{definition}[Ornstein--Uhlenbeck sheet] \label{def:OU}
A zero-mean Gaussian sheet $U:\ \Omega\times \mathbb{R}_+^{2} \rightarrow \mathbb{R}$ is called Ornstein--Uhlenbeck (OU) sheet \cite{Walsh1986,Baran2003, Baran2012} if its covariance function is
\begin{equation}\label{eq:random_sheet_corr}
\begin{split}
&K_{\mathrm{OU}}(\theta,\theta') = \\
&\frac{\sigma^2}{4 \kappa_x \kappa_t} \exp(-\kappa_x |x-x'| - \kappa_t |t-t'|),
\end{split}
\end{equation}
where $\theta = (x,t)$ and $\theta' = (x',t')$ are points in $\mathbb{R}_+^2$, and $\sigma, \kappa_x, \kappa_t \in \mathbb{R}_{+}$.
\end{definition}

Brownian sheet and OU sheet are related to each other via Lamperti transform
\begin{equation}\label{lamperti_transform}
U(x,t) ={\rm e}^{-\kappa_x x} \ {\rm e}^{-\kappa_t t} \cdot {\mathscr W}\left(\frac{\sigma}{2\kappa_x} {\rm e}^{2\kappa_x x}, \frac{\sigma}{2\kappa_t} {\rm e}^{2\kappa_t t}\right).   
\end{equation}
Both sheets have versions (in the sense of Definition \ref{version_def}) whose realizations are continuous functions on $\mathbb{R}_{+}^2$ \cite{Chentsov_1956}; we consider only these versions throughout the article.

An important property of OU sheet is stationarity (also called strict homogeneity, e.g.\ in \cite{AdlerBookGeomRandF}). For the case considered here, when the index set $\mathcal X$ coincides with $\mathbb{R}^{n}_{+}$, 
stationarity of a random field is defined in a rather straightforward manner:
\begin{definition}[Stationarity] \label{stat}
Let $Z: \Omega \times \mathbb{R}^{n}_{+} \rightarrow \mathbb{R}$ be a stochastic field with the index set $\mathbb{R}^{n}_{+}$. It is called stationary if for any $k\in \mathbb{N}$, and for any $(\theta_1, \theta_2, \ldots, \theta_k) \subset \mathbb{R}^{n}_{+}$ and any $\vartheta \in \mathbb{R}^{n}_{+}$, the random vectors
$\left(Z(\theta_1), Z(\theta_2), \ldots, Z(\theta_n)\right)$ and $\left(Z(\theta_1+\vartheta), Z(\theta_2 + \vartheta), \ldots, Z(\theta_n + \vartheta)\right)$ are equal in distribution.
\end{definition}
The finite-dimensional distributions of the OU sheet are invariant under $n$-dimensional ``time'' shifts since its covariance (\ref{eq:random_sheet_corr}) depends only on the difference of coordinates, therefore OU sheet is stationary.

\subsection{Gaussian process defined by kernel}

As it was previously mentioned, a Gaussian field is characterized by its first two moments: the mean and the covariance. In this duet, the covariance may be seen as playing a pivotal role. The mean acts as a backbone of the process: the process fluctuates around the mean function, and all processes considered here are centered. Meanwhile, the covariance defines the stochastic behavior, which is a primary feature distinguishing the Gaussian processes from regular functions. 

Consider an arbitrary zero-mean Gaussian field $Z(\theta)$. The covariance function $K_Z$ determines the stochastic field up to equivalence of versions, i.e.\ all zero-mean Gaussian fields with the same covariance function share the same distribution at each point. Covariance function of any Gaussian field is a kernel function.
\begin{definition}[Kernel function]
A function
\begin{equation}
K(\theta,\vartheta): \mathcal{X}\times\mathcal{X} \rightarrow \mathbb{R}
\end{equation}
is called a kernel function if it is symmetric, i.e.\ $K(\theta,\vartheta) = K(\vartheta,\theta)$, and positively semi-definite. That is, for all $N \in \mathbb{N}$, and arbitrary set of real numbers $b_1, \ldots, b_N$ ($b_j\in\mathbb{R}$, $j=1,\dots N$) and for any collection of points $\theta_1,\ldots,\theta_N$ ($\theta_j\in\mathcal{X}$, $j=1,\dots N$), the linear form
\begin{equation}
\sum_{k} \sum_{j} b_k b_j K(\theta_k,\theta_j) \ge 0.
\end{equation}
\end{definition}
The kernels can be combined to produce new ones. For example, if $K_1$ and $K_2$ are kernels, then $K'=a K_1+b K_2$ with non-negative real numbers $a$ and $b$ is also a kernel, as well as a product of the two kernels $K''=K_1 K_2$. 
Moreover, for any non-negative function $f(\theta):\mathcal{X} \rightarrow \mathbb{R}_{+}$, the product $K'''=f(\theta) f(\vartheta) K(\theta,\vartheta)$ is also a kernel. We also notice that $K = 1$ is a valid kernel on a compact space. We use the kernel modification to address the problem of the dephasing hotspots, see Appendix~\ref{app:hotspot}.

Kernel functions, also known as reproducing kernels, help establishing the connection between the theory of Gaussian processes and the theory of reproducing kernel Hilbert spaces. 
Namely, the kernel $K(\theta,\vartheta)$ defines an integral operator $L_K$, which acts on the space $L^2_{\mu} \left(\mathcal{X} \right)$ of the square integrable functions, and whose action on a function $f(\theta)$ from this space is described as
\begin{equation}
\label{eq:mercerintop}
(L_K f)(\theta) = \int_{\mathcal{X}} K(\theta,\vartheta) f(\vartheta) d \mu(\vartheta).
\end{equation}
The celebrated Mercer's theorem \cite{Sun2005MercerThm,AdlerBookGeomRandF} 
establishes important and extremely useful properties of the operator $L_K$.
\begin{theorem}[Mercer's theorem]\label{thm:mercer}
Let $\mathcal{X}$ be a complete separable metric space with Borel measure $\mu$. 
Let the kernel function $K(\theta,\vartheta)$ be square integrable with respect to the measure $\mu$ in the variable $\theta$, i.e.\ 
\begin{equation}
K(\theta,\cdot) \in L^2_{\mu} \left(\mathcal{X} \right) ,\quad \forall \theta\in \mathcal{X},
\end{equation}
and, due to the symmetry, the same holds for $\vartheta$. Let $K(\theta,\vartheta)$ be also square integrable in both variables $\theta$ and $\vartheta$ simultaneously, i.e.\ square integrable on $\mathcal{X} \times \mathcal{X}$ with respect to the measure $\mu \times \mu$.

Under these conditions, $L_K$ is a self-adjoint compact operator with a bounded non-negative eigenvalues $\lambda_1 \ge \lambda_2 \ge \ldots \ge 0$. Eigenvectors $e_k(\theta)$ of $L_k$ form an orthonormal basis in $L^2_\mu(\mathcal X)$, and the kernel $K$ can be represented as a series
\begin{equation}
\label{eq:mercerexp}
K(\theta,\vartheta) = \sum_{k} \lambda_k\,e_k(\theta)\,e_k(\vartheta),
\end{equation}
which converges in $L^2_{\mu \times \mu} \left(\mathcal{X} \times \mathcal{X} \right)$ metric.
\end{theorem}
Note that convergence in $L^2_{\mu \times \mu} \left(\mathcal{X} \times \mathcal{X} \right)$ metric may be insufficient for applications, guaranteeing only that the integral of the square distance between $K(\theta,\vartheta)$ and the sum in Eq.~\ref{eq:mercerexp} approaches zero, while the difference between the two functions can be large, e.g.\ on any set of measure zero, which may include the qubit trajectories.   

Fortunately, on a compact space $\mathcal{X}$, any continuous kernel satisfies the condition of the Mercer's theorem, and convergence in this case is uniform and absolute. This opens a convenient way to define and sample any zero-mean Gaussian field with prescribed continuous covariance function $K(\theta,\vartheta)$. This is achieved by employing the Kosambi--Karhunen--Lo{\`e}ve expansion: 
\begin{theorem}[Kosambi--Karhunen--Lo{\`e}ve expansion]
Consider the set of functions $e_k(\theta)$ from Theorem \ref{thm:mercer}, which form an orthonormal basis in $L^2_\mu(\mathcal X)$ with compact $\mathcal X$. Let $(w_1, w_2, \ldots)$ be a countable family of independent standard normal random variables, i.e.\ Gaussian random variables with zero mean and unit variance. 
Then there exists a random field $X(\theta)$ defined by the series
\begin{equation}\label{eq:KKLE}
X(\theta) = \sum_{k} w_k\,\sqrt{\lambda_k}\,e_k(\theta),
\end{equation}
that is, the series converges, in the mean-square sense and uniformly with respect to $\theta$, to the Gaussian random field $X(\theta)$ which has zero mean and the covariance function $K(\theta,\vartheta)$ given by Eq.~\ref{eq:mercerexp}. 
Inverse is also true, i.e.\ a centered Gaussian random field $X(\theta)$ with the covariance $K(\theta,\vartheta)$ can be represented in such form.
\end{theorem}

The Kosambi--Karhunen--Lo{\`e}ve expansion is a very important result, providing a constructive way to generate and sample a Gaussian process with an arbitrary covariance function that represents the specific statistical features of our model. Moreover, uniform convergence of the series allows truncation of the sum in (\ref{eq:KKLE}) at a suitable level of numerical accuracy.

\subsection{Restrictions of Gaussian random sheets on trajectories}
\label{sec:RSontraj}

The random magnetic field ${\mathrm B}(t)$ acting on a shuttled qubit is determined by restricting the underlying noise ${\tilde B}(x_c,t)$ to the qubit trajectory $x_c(t)$, as ${\mathrm B}(t)={\tilde B}(x_c(t),t)$. If we describe ${\tilde B}(x_c,t)$ as a Gaussian random sheet, then Gaussian property is inherited by the process ${\mathrm B}(t)$.

Indeed, let us take a centered random sheet $J(\theta)$ with $\theta=(x,t)\in\mathbb{R}_+^2$ and covariance $K_J(\theta,\theta')$, and consider a (continuous) path ${\Gamma}(s): [0,S] \rightarrow \mathbb{R}_+^2$, 
parametrized by an arbitrary suitable variable $s$, with $0\le s\le S<\infty$. We emphasize that the path $\Gamma(s)$ does not necessarily have to take the form $(x(t),t)$; for example, it could be a closed loop. The restriction of the sheet $J$ on the path $\Gamma(s)$ defines the stochastic process ${\mathrm V}(s) = J(\Gamma(s))$. As follows from the definition \ref{def:Gaussianfield}, this process is Gaussian: for any finite set of indices $(\theta_1, \theta_2, \ldots, \theta_n)$ belonging to the path $\Gamma(s)$, one can define the random vector $\left[{\mathrm V}(s_1), {\mathrm V}(s_2), \ldots, {\mathrm V}(s_n)\right]$, where ${\mathrm V}(s_j)=J(\theta_j)$ with $\theta_j=\Gamma(s_j)$, $j=1,\dots n$. The vector $\left[J(\theta_1),\dots J(\theta_n)\right]$ follows multivariate Gaussian distribution with zero mean and covariance $K_J(\theta_j,\theta_k)$
so that $\mathrm{V}(s)$ is a Gaussian random process (of a conventional kind, indexed by a single parameter $s$) with the covariance function
\begin{equation}
\label{eq:covarV}
K_{\mathrm V}(s,s') = K_J(\Gamma(s),\Gamma(s')).
\end{equation}

In the case of several qubits, when $M$ electronic spins are shuttled along each own trajectory $x_{c\,k}(t)$ ($k=1,\dots M$), there are $M$ random processes ${\mathrm B}_k(t)={\tilde B}(x_{c\,k}(t),t)$. The dynamics of decoherence of such a multi-qubit system is determined by linear combinations of the processes ${\mathrm B}_k(t)$, see Eqs.~\ref{eq:twospinW} above and \ref{eq:manyspinB} below. 

The key issue here is that the processes ${\mathrm B}_k(t)$ are not statistically independent, so that even if each process ${\mathrm B}_k(t)$ is Gaussian, their linear combination is not necessarily Gaussian \cite{FujitaYoshidaNonGaussian23}. 
Fortunately, since they are derived from the same Gaussian random sheet ${\tilde B}(x_c,t)$, their linear combination does retain the Gaussian property.

To establish that, let us take the same centered random sheet $J(\theta)$ with $\theta=(x,t)\in\mathbb{R}_+^2$, and, similar to the single-qubit case above, consider a set of (continuous) paths ${\Gamma}_k(s)$, $k=1,\dots M$, all parametrized by the same variable $s\in[0,S]$. Each path defines a centered Gaussian process $\mathrm{V}_k(s)=J(\Gamma_k(s))$, and we are studying the process
\begin{equation}
\mathrm{Y}(s) = \sum_{k=1}^M a_k \mathrm{V}_k(s),
\end{equation}
where $a_k$ are arbitrary real numbers. As before, for any $n\in\mathbb N$ we take an arbitrary set of values $s_1, s_2,\dots s_n$; for each $s_j$
\begin{equation}
\mathrm{Y}(s_j) = \sum_{k=1}^M a_k \mathrm{V}_k(s_j) = \sum_{k=1}^M a_k J(\Gamma_k(s_j)).
\end{equation}
For each set of points $\theta_{k,j}=\Gamma_k(s_j)$, the random vector $\left[J(\theta_{1,1}),\dots J(\theta_{M,1}), J(\theta_{1,2}),\dots J(\theta_{M,2}),\dots\right]$, constructed from the entries $J(\theta_{k,j})$, follows multivariate Gaussian distribution. It is easy to calculate (e.g.\ by calculating the characteristic function) that the vector $\left[Y_1,\dots Y_n\right]$, constructed from the sums 
$Y_j=\sum_{k=1}^M a_k J(\theta_{k,j})$, also follows multivariate Gaussian distribution, thus establishing that $\mathrm{Y}(s)$ is a centered Gaussian random process with the covariance function
\begin{equation}
K_\mathrm{Y}(s,s') = \sum_{k,p=1}^M a_k a_p K_J(\Gamma_k(s),\Gamma_p(s')).
\end{equation}
This argument could be easily generalized for Gaussian fields with multi-parameter index, or even for Gaussian fields on some abstract sets. 
For our work, it is important that the Gaussian property of the underlying random sheet is reliably inherited after restriction to a set of paths $\Gamma_k(s)$. It is also important that continuity of realizations of $J$ and of the path ${\Gamma}(s)$ implies continuity of realizations for the process $\mathrm{Y}(s)$, which also implies continuity of the 
covariance function $K_\mathrm{Y}$.

\subsection{Characteristic functionals}

As we mentioned in the previous Section (see Eqs.~\ref{eq:wRP} and \ref{eq:wRS} and the related discussion), the central object in the standard stochastic description of dephasing of the stationary qubits is the dephasing factor $W(t_0)$, which is determined by the characteristic functional of the corresponding random process. In a similar way, we can define a characteristic functional for an abstract random field $Z:\Omega \times \mathcal{X} \rightarrow \mathbb{R}$. 
\begin{definition}[Characteristic functional]
Let $\mathfrak{L}(\mathcal X)$ denote the space containing all realizations of $Z$. Let us take a linear functional $\mathbf{f}$ acting on the space $\mathfrak{L}(\mathcal{X})$, i.e.\ $\mathbf{f}$ belongs to the dual space of $\mathfrak{L}(\mathcal{X})$, $\mathbf{f} \in \mathfrak{L}^*(\mathcal{X})$. We assume that $Z$ is a correctly defined random variable, in a sense that each functional $\mathbf{f}$ restricted to the set of all realizations of $Z$ defines the  $\mathbb{P}$-measurable function $\mathbf{f}[Z(\omega, \cdot)]$. Characteristic functional ${\mathscr F}_Z(\mathbf{f})$ of the random field $Z$ is a bounded nonlinear functional acting on $\mathbf{f}$ as follows:
\begin{equation}
{\mathscr F}_Z(\mathbf{f}) =  \mathbb{E}[\exp{i \mathbf{f}[Z(\omega, \cdot)]}], 
\end{equation}
\end{definition}
An excellent review of characteristic functionals is given in \cite{Skorokhod1965}.

Using the definition above, it is possible to develop a theory for calculating the dephasing factors by working directly with the Gaussian sheet ${\tilde B}(x,t)$, using Eqs.~\ref{eq:fullhz} and/or \ref{eq:tildeB}, but it is much easier to work with the Gaussian process $\mathrm{B}(t)={\tilde B}(x_c(t),t)$, whose properties have been discussed above.

The Riesz--Markov--Kakutani theorem states that each linear functional $\mathbf{f}$ acting on $C([a,b])$ could be represented as a Stieltjes integral with the corresponding bounded variation integrator $F(s),\ s\in [a,b]$, i.e.\ 
\begin{equation}
\mathbf{f}[g(\cdot)]=\int_a^b g(s)\, dF(s).
\end{equation}
Correspondingly, if $Z$ is a stochastic process $Z(s), s\in [a,b]$ with continuous realizations, the characteristic functional should be applied to $\mathbf{f}$ as
\begin{equation}
\label{char_fun}
{\mathscr F}_Z(\mathbf{f}) = \mathbb{E}\left[\exp\left(i \int_{a}^{b} Z(s)\, dF(s)\right)\right].
\end{equation}

If the process $Z(s)$ is Gaussian, then the characteristic functional can be expressed in explicit form via the covariance function $K_Z(\cdot,\cdot)$ \cite{Skorokhod1965}:
\begin{equation}\label{char_fun_gauss}
{\mathscr F}_{Z}(\mathbf{f}) = \exp\left(-\frac{1}{2}\, \int\limits_{a}^{b}\!\int\limits_{a}^{b}  K_Z (s,u)\,dF(s)\,dF(u) \right).
\end{equation}

In our case, the role of the general Gaussian process $Z$ is played by the specific process $\mathrm{V}(s)$, which is obtained by restricting 
a centered Gaussian sheet $J(\theta),\ \theta=(x,t)\in\mathbb{R}_+^2$ with the covariance $K_J(\theta,\theta')$ to the continuous path ${\Gamma}(s)$. 
Let $F(s)$ be a piecewise-continuous function of bounded variation on $[0,S]$
\footnote{Although in almost all real physical situations the function $F(s)$ is continuous, we include the case of discontinuities mostly for generality of treatment.},
and denote the discontinuity points of $F(s)$ as ${s_1, \ldots, s_d}$, in ascending order. 
%
%
%
At every discontinuity, the function $F(s)$ has a jump of the height
\begin{equation}
\zeta_{k} = \lim\limits_{\tau \rightarrow 0+}[F(s_k+\tau) - F(s_k-\tau)],
\end{equation}
while at all other points it has derivative $\xi(s)=dF(s)/ds$. Then, according to the definition, the Stieltjes integral in the exponent (\ref{char_fun}) for the process $\mathrm{V}$ can be represented as
\begin{equation}\label{char_fun_integr}
%
%
\int\limits_0^S \!\mathrm{V}(s)\, dF(s) = \sum\limits_{k=1}^{d}\!\zeta_k \mathrm{V}(s_k) + \sum\limits_{k=0}^{d}\!\int\limits_{s_k}^{s_{k+1}}\!\! \mathrm{V}(s)\,\xi(s)\,ds,
\end{equation}
where, for convenience of notations, we introduced two additional points, $s_0=0$ and $s_{d+1}=S$ (note that $s_1$ may coincide with $s_0$, and $s_d$ may coincide with $s_{d+1}$).
Thus, the characteristic functional of the process $\mathrm{V}(s)$ with the integrator $F(s)$ can be expressed as ${\mathscr F}_{\mathrm{V}}= \exp{-\chi(S)}$, where
\begin{widetext}
\begin{equation}\label{char_fun_dbl_integr}
\chi(S) = \frac{1}{2} \sum\limits_{j,k=0}^{d} \int\limits_{s_k}^{s_{k+1}} \int\limits_{s_j}^{s_{j+1}}\! K_\mathrm{V}(u,v)\, \xi(u) \xi (v)\, du dv + \frac{1}{2} \sum\limits_{j,k=1}^{d} \zeta_k \zeta_j K_\mathrm{V}(s_k,s_j) + 
\sum\limits_{j=1}^{d} \sum\limits_{k=0}^{d} \zeta_j\!\!\int\limits_{s_k}^{s_{k+1}}\!\! K_\mathrm{V}(u,s_j)\xi(u)\,du.
\end{equation}
\end{widetext}
For many processes $\mathrm{V}$, the expression above can be evaluated directly in analytical form. In more complex cases, the integration can be performed numerically.

\subsection{Numerical evaluation of the characteristic functional\label{sec:method}}

\subsubsection{Numerical integration via quadratures}

When analytical evaluation of Eq.~\ref{char_fun_dbl_integr} is impossible or too difficult, the required integrals can be calculated numerically, using the well-known quadrature formulas \cite{suli2003introduction,NumRecipes}. This could also be a method of choice in the situations when the covariance function of the underlying random sheet is not known analytically, but has been extracted from experimental measurements \cite{yoneda_noise-correlation_2023,StruckCywinskiSchreiber20QubitNoise,rojas-arias_spatial_2023,BoterJoyntVdSNoiseCorrBellStates20,SpenceNiquetMeunierEtalChargeNoise22,ShehataVanDorpeEtalChargeNoiseQuDots23,ViolaSpaceTimeNoise22} or from numerical simulations \cite{KepaFockeCywinskiKrzywdaChargeNoise23,KepaCywinskiKrzywdaSpinNoise23,ShalakDelerueNiquetChargeNoiseSiHole23}.

\subsubsection{Integration using the Monte Carlo sampling}
\label{subsubsec:MC}

In some situations, it could be advantageous, instead of numerically calculating the integrals in Eq.~\ref{char_fun_dbl_integr}, to calculate the characteristic functional by sampling the values of the integral 
\begin{equation} 
\label{int_basic}
\int\limits_{0}^{S} \mathrm{V}(s)\,dF(s),
\end{equation}
where $\mathrm{V}(s) = J(\Gamma(s))$ is obtained by restricting a Gaussian sheet $J(\theta) = J(x,t)$ on the path $\Gamma(s)$, and calculating the average in   (\ref{char_fun}) via the Monte Carlo method \cite{KalosWhitlock}. 

We treat this problem in two steps. Firstly, we discretize the integral, turning it into a sum, and show that such discretization is a valid approximation by providing an estimate of the error. Secondly, we describe the way to sample the Gaussian process $\mathrm{V}(s)$ with an arbitrary covariance function at the nodes of our discretization mesh.

The standard error analysis of the discretized integral \cite{suli2003introduction,NumRecipes} is not readily applicable, because  realizations of many Gaussian processes are not smooth, e.g.\ for OU process they are continuous but nowhere differentiable with probability 1.
However, it can be shown that realizations of the OU process are almost surely $\alpha$-H{\"o}lder continuous, i.e.\
\begin{equation}
|X(r,\omega) - X(s,\omega)| \le C^{(X)}_{\alpha}(\omega) |r-s|^{\alpha} ,
\end{equation}
for each $0<\alpha <1/2$ with some $\alpha$- and $\omega$-dependent H{\"o}lder constant $C^{(X)}_{\alpha}(\omega)$. This constant is bounded from above with an exponentially high probability, see \cite{Azmoodeh2014}. More specifically,  Theorem~1 of Ref.~\onlinecite{Azmoodeh2014} makes the following statement:
\begin{theorem}
Consider the Gaussian process $Z(s,\omega)$ indexed by real parameter $s$ and defined on the probability space $(\Omega, \mathcal{B}, \mathbb{P})$. Realizations of the process $Z$ are $\alpha$-H{\"o}lder continuous with probability 1, i.e.
\begin{equation}\label{hol}
|Z(r,\omega) - Z(s,\omega)| \le C_{\alpha}(\omega) |r-s|^{\alpha} ,
\end{equation}
for any $\alpha<\alpha_0$, where $\alpha_0$ is a positive number, 
if and only if for each $\alpha<\alpha_0$ there exists $c_{\alpha}$ such that
\begin{equation}
\label{hol_nes_suf}
\sqrt{\mathbb{E}\left[\left(Z(r,\omega) - Z(s,\omega)\right)^2\right]} \le c_{\alpha} |r-s|^{\alpha}.
\end{equation}
In that case, the H{\"o}lder constant $C_{\alpha}(\omega)$ is bounded with exponentially high probability, in the sense that
\begin{equation}
\label{hol_est}
\mathbb{E}\left[ \exp{m \left(C_{\alpha}(\omega)\right)^{\kappa}} \right] < \infty
\end{equation}
for any $m\in \mathbb{R}$ and $\kappa < 2$. For $\kappa = 2$ the statement is true for small enough positive $m$.
\label{thm:MCalpha}
\end{theorem}

In the case of the OU process $X(s)$, this theorem gives the value of the constant $\alpha_0=1/2$ because
\begin{equation}
\begin{split}
&\sqrt{\mathbb{E}\left[\left(X(r,\omega) - X(s,\omega)\right)^2\right]} \propto \\
&\sqrt{1 - \exp|r-s|} = O(\sqrt{|r-s|}).
\end{split}
\end{equation}
In the case when the relevant process is $\mathrm{V}(s)$, which was obtained by restriction of a OU sheet $U(x,t)$ on a path $\Gamma$ parametrized by specifying $x(s)$ and $t(s)$, we have 
\begin{equation}
\begin{split}
&\sqrt{\mathbb{E}\left[\left(\mathrm{V}(r,\omega) - \mathrm{V}(s,\omega)\right)^2\right]} \propto \\
&\qquad\sqrt{1-\exp(-\kappa_t |t(r) - t(s)|-\kappa_x |x(r) - x(s)|)},
\end{split}
\end{equation}
if we suppose that both $x(s)$ and $t(s)$ are at least $\beta$-H{\"o}lder continuous ($\beta>0$), then the estimate is $\alpha< \alpha_0 = \min \{1/2,  \beta/2 \}$.

We are interested in the case of Lipschitz integrator $F(s)$, thus we have the option to use quadrature formulae for Stieltjes integral \cite{Alomari2014, Dragomir2015}. For example
\begin{equation}
\label{quad_fl}
\int\limits_{c}^{d} f(s) d F(s) \approx \left(F(d) - F(c)\right)\,f(x), x\in [c,d],
\end{equation}
with the error $Q(\omega)(d-c)^{\alpha+1}$, where 
\begin{equation}
\label{eq:q_quad}
Q(\omega)=L_{F} C_{\alpha}(\omega)/(1+\alpha),
\end{equation}
and $L_{F}$ is the Lipschitz constant of $F$. 

To assure validity of this approximation for a general Gaussian process $Z(s,\omega)$, we need to estimate $C_{\alpha}(\omega)$; we can use the condition (\ref{hol_est}) for that. 
To estimate the integral in (\ref{char_fun}) using the quadrature formula (\ref{quad_fl}), we, as usual, divide the interval $[a,b]$ into $n$ equal subintervals. Denoting the integration nodes as $s_j$ ($j=0,\dots n$), with $s_0 = a$ and $s_n = b$, we approximate the integral as
\begin{equation}
\Sigma_n(\omega) =\sum_{j=1}^{n} \left(F(s_j) - F(s_{j-1}) \right) Z(s_j,\omega).
\end{equation}
The corresponding error is bounded as 
\begin{equation}
\left| \int\limits_{a}^{b} Z(s,\omega) dF(s) - \Sigma_n(\omega) \right| \le 
n \left(\frac{b-a}{n}\right)^{\alpha + 1} \!\!\!  Q(\omega),
\end{equation}
so that the approximation is valid as long as $C_{\alpha}(\omega) = {\bar{o}}(n^{\alpha})$. 

In order to show that, we consider $\varepsilon$ such that $\alpha > \varepsilon > 0$, and consider $0< \kappa < 2$. We use the exponential Chebyshev's inequality 
\begin{equation}
\mathbb{P}(z\ge a) \le \exp{-q a}\,\mathbb{E}\left[\exp{q z}\right],
\end{equation}
which is valid for a random variable $z$ and for any positive bound $a$ and arbitrary positive parameter $q$, and apply it to the random variable $[C_{\alpha}(\omega)]^\kappa$ and the bound $n^{\kappa(\alpha -\varepsilon)}$, setting $q=1$. Since $\kappa$ is positive, this is equivalent to the statement that 
\begin{equation}
\begin{split}
&\mathbb{P}\left( C_{\alpha}(\omega) > n^{\alpha -\varepsilon} \right) = 
\mathbb{P}\left( C^\kappa_{\alpha}(\omega) > n^{\kappa(\alpha -\varepsilon)} \right) \leq \\
&\qquad \mathbb{E}\left[ \exp{\left(C_{\alpha}(\omega)\right)^{\kappa}} \right]\, \exp{-n^{\kappa(\alpha -\varepsilon)}},
\end{split}
\end{equation}
that is,
\begin{equation} 
\label{eq:rvar_esteem}
\mathbb{P}\left( C_{\alpha}(\omega) \ge n^{\alpha -\varepsilon} \right) \leq \Delta_M = M_0 \exp(-n^{\kappa(\alpha -\varepsilon)}),
\end{equation}
with $M_0=\mathbb{E}\left[ \exp{\left(C_{\alpha}(\omega)\right)^{\kappa}} \right]$, which is bounded due to the statement (\ref{hol_est}) with $m=1$. 

Thus, we can separate the whole probability space $\Omega$ into the unfavorable part $\Omega_u$, where $C_{\alpha}(\omega) > n^{\alpha -\varepsilon}$, and its  complement, the favorable part $\Omega_f$. The integral (\ref{char_fun}) over the unfavorable part is 
\begin{equation}
\int\limits_{\Omega_u} \exp\left( i\int_{a}^{b} Z(s,\omega) dF(s) \right)\,
d\mathbb{P}(\omega),
\end{equation}
which is the expectation value of an imaginary exponential over the subset $\Omega_u$ of the probability space. Since the modulus of the imaginary exponential is limited by 1, the absolute value of this integral does not exceed the measure of 
$\Omega_u$, which is the probability
$\mathbb{P}\left( C_{\alpha}(\omega) > n^{\alpha -\varepsilon} \right)$ given in Eq.~\ref{eq:rvar_esteem}. Thus, 
\begin{eqnarray}
&&\mathbb{E}\left[ \exp\left( i\int_{a}^{b} Z(s,\omega) dF(s) \right) \right] = \Delta_M + \\ \nonumber
&&\int_{\Omega_f}\exp\left(i\int_{a}^{b} Z(s,\omega) dF(s) \right) d\mathbb{P}(\omega),
\end{eqnarray}
For every point $\omega$ in the favorable subset $\Omega_f$ the approximation 
\begin{equation}
\int\limits_{a}^{b} Z(t,\omega) d \xi(t) = \Sigma_n(\omega) + O(n^{-\varepsilon}),
\end{equation}
is valid with the overall error $O(n^{-\varepsilon})$, and $\varepsilon <\alpha < \alpha_0$. Thus, the accuracy of the approximation 
\begin{equation}
\label{mcest}
\mathbb{E}\left[ \exp\left( i\int\limits_{a}^{b} Z(s,\omega) dF(s) \right) \right] \approx 
\mathbb{E} \left[ \exp\left( i \Sigma_n(\omega) \right) \right]
\end{equation}
is a sum of $\Delta_M$ in Eq.~\ref{eq:rvar_esteem} and the discretization error $O(n^{-\varepsilon})$. The overall error is minimized by judiciously choosing the parameter $\varepsilon=\alpha_0-\delta$, then the overall accuracy of the approximation is $O(n^{-\alpha_0}\log{n})$, see Appendix~\ref{app:mceval} for details.

In particular, $\alpha_0=1/2$ for the OU process and also for the random processes discussed below in this work. However, for more complex types of noise, the value of $\alpha_0$ can be different.

In particular, if the integrator $F(s) = s$ then the right hand side in (\ref{mcest}) simplifies to
\begin{equation}
\label{eq:discr1}
\mathbb{E} \left[ \exp\left( i\,\,\frac{b-a}{n} \sum_{j=1}^{n} Z(s_j,\omega) \right) \right], 
\end{equation}
More generally, if $d^2 F(s)/ds^2$ is continuous on $[a,b]$, then the right-hand side of (\ref{mcest}) can be approximated as
\begin{equation}
\label{eq:discr2}
\mathbb{E} \left[ \exp\left( i\,\,\frac{b-a}{n} \sum_{j=1}^{n} \xi\left(\frac{s_j + s_{j-1}}{2} \right) Z(s_j,\omega) \right) \right],
\end{equation}
where $\xi(s)=dF(s)/ds$.

These conclusions do not change in the case of many qubits or in the case of piecewise-continuous $F(s)$: we are still dealing with the sum of the integrals of the type (\ref{int_basic}), and from the triangle inequality we obtain the value of $\alpha_0$ which is the minimal among such values of all qubits. 
The factors similar to those appearing in Eq.~\ref{eq:rvar_esteem} could be estimated uniformly, since the number of the qubits is finite and not very large. 

Having justified the discretization of the integral, we now present an algorithm for sampling the values of the random process $Z$ at the nodes $s_j$ of the discretization mesh, i.e.\ sampling of the random vector $\left(Z(s_j,\omega)\right)_{j=1}^{n}$. One possibility is to use the Kosambi--Karhunen--Lo{\`e}ve representation. However, this approach may present serious numerical difficulties. Consider, for instance, the OU sheet $U(x,t), 0 \le x \le X, 0 \le t \le T$, for which the truncated Kosambi--Karhunen--Lo{\`e}ve expansion has the form \cite{Deheuvels2006, Baran2012}
\begin{widetext}
\begin{equation}
\begin{split} 
&U(x,t) \approx \sum \limits_{j=1}^{D} \sum \limits_{k=1}^{D} w_{j,k} \frac{4\sigma \exp(\kappa_x \left(X - x \right) + \kappa_t \left(T - t \right))}{\pi^2 \sqrt{\kappa_x\kappa_t} (2k-1)(2j-1)} \sin\left(\frac{\pi (2j-1) \exp\left(-\kappa_x \left(X - x \right) \right)}{2} \right)\\
&\sin\left(\frac{\pi (2k-1) \exp\left(-\kappa_t \left(T - t \right) \right)}{2} \right),
\end{split}
\end{equation}
\end{widetext}
where the first $D$ terms were retained; $w_{j,k}$ are independent standard normal random variables, now indexed with the double index ${j,k}$. As $T$ and/or $X$ become large, the variance of the double sum, although being of the order of $O(1/D^2)$, grows due to the exponential factors $\exp{\kappa_x \left(X - x \right)}$ and $\exp{\kappa_t \left(T - t \right)}$, such that the number $D$ of terms to be retained should also increase exponentially, which is a typical case of the infamous sign problem \cite{KalosWhitlock}. The sign problem arises due to the adopted approach of obtaining the OU sheet using the Lamperti transform~(\ref{lamperti_transform}) from the Wiener sheet.

Therefore, here we employ another approach, used before for sampling OU sheet  \cite{Skordos2008}. Having discretized the integral (\ref{int_basic}) by choosing the set of nodes $\{ s_1, s_2, \ldots , s_n \}$, we calculate the covariance matrix $\mathbf{C}$ with the elements
\begin{equation}
\label{eq:discrcov}
\mathbf{C}_{j,p} = K_{\mathrm{V}}(s_j,s_p), 
\end{equation}
and apply Cholesky decomposition, expressing it as $\mathbf{C} = L L^{T}$. If $w = (w_1,w_2,\ldots, w_n)$ is, as before, a vector of i.i.d.\ normal variables with zero mean and unit variance, then $L w$ is the required centered Gaussian random vector with the covariance matrix $\mathbf{C}$.

\section{Stochastic modeling of dephasing during qubit shuttling \label{sec:model}}

In this Section we demonstrate, how the dynamics of the qubits dephasing under nontrivially correlated noise can be analyzed using the notion of the Gaussian random sheet. We assume that the noise ${\tilde B}(x_c,t)$ can be modeled as an OU sheet (see Definition \ref{def:OU}) with the correlation length $\lambda_c=1/\kappa_x$ and the correlation time $\tau_c=1/\kappa_t$. This model is, on one hand, sufficiently realistic to capture the basic physics of the real noise, and on the other hand, makes it possible to analyze the dephasing dynamics for various  shuttling scenarios, and obtain explicit solutions. 

Note that we do not consider the errors associated with loading the electrons from the quantum dots to the shuttling channel and back. While significant in some cases, these errors are unrelated to the main goal of this work, and also largely depend on specific experimental details.

Our approach is based on the Hamiltonian ${\tilde H}_Z (x_c,t)$ specified in Eq.~\ref{eq:tildeB}. If a single electron is shuttled in an arbitrary way, specified by the trajectory $x_c(t)$, then its spin experiences the time-dependent magnetic field $B_{\rm tot}(t)\equiv B_0(x_c(t)) + {\tilde B}(x_c(t),t)$, which is a sum of the deterministic contribution $B_0(t)\equiv B_0(x_c(t))$ and the random field ${\mathrm B}(t)\equiv {\tilde B}(x_c(t),t)$, such that the spin evolution is determined by the Hamiltonian
\begin{equation}
{\tilde H}_Z(t) = g_0 \mu_B \left[B_0(t) + {\mathrm B}(t)\right] S_z.
\end{equation}
Below, for simplicity of notations, we set $g_0\mu_B=1$.
For a single shuttling experiment of duration $t_0$, the corresponding evolution operator is
\begin{equation}
{\tilde U}(t_0) = \exp{-i \left[\alpha(t_0) + \phi(t_0)\right] S_z},
\end{equation}
with the deterministic $\alpha(t_0)$ and the random $\phi(t_0)$ contributions to the phase
\begin{equation}
\label{eq:phases1spin}
\alpha(t_0)=\int_0^{t_0}\!\!B_0(t)\,dt,\quad \phi(t_0)=\int_0^{t_0}\!\!{\mathrm B}(t)\,dt.
\end{equation}
If the initial state of the qubit was described by the density matrix $\rho(0)$, then the density matrix at the end of the shuttling experiment 
$\rho(t_0)={\tilde U}(t_0) \rho(0) {\tilde U}^\dag(t_0)$ has a simple form in the basis of the states $|\uparrow\rangle$ and $|\downarrow\rangle$: each matrix element acquires the corresponding phase
\begin{equation}
\rho_{p,p'}(t_0) = \rho_{p,p'}(0)\, 
{\mathrm e}^{-i (m-m') \left[\alpha(t_0) + \phi(t_0)\right]},
\end{equation}
where $\rho_{p,p'}\equiv \langle p|\rho|p'\rangle$, the indices $p$ and $p'$ each take the values $\uparrow$ or $\downarrow$, and where $m$ and $m'$ are the shorthand notations for $m(p)$ and $m(p')$ respectively, where $m(\uparrow)=+1/2$ and $m(\downarrow)=-1/2$. Thus, the diagonal elements of the density matrix are unaltered, and the off-diagonal elements are multiplied by the phase factor $\exp{i\left[\alpha(t_0) + \phi(t_0)\right]}$ or its conjugate. Since the phase $\alpha(t_0)$ is deterministic, and can be corrected or taken into account in advance (e.g.\ by using an appropriate rotating frame), below we set $\alpha(t_0)=0$ for simplicity. Averaging over realizations of the field ${\mathrm B}(t)$ leads to the density matrix ${\bar\rho}(t_0)$ with the elements
\begin{equation}
\label{eq:barrho1}
{\bar\rho}_{p,p'}(t_0) = \rho_{p,p'}(0)\,{\mathbb E}\left[{\mathrm e}^{-i (m-m') \phi(t_0)}\right],
\end{equation}
whose diagonal elements are the same as in $\rho(0)$, while the off-diagonal elements ${\bar\rho}_{\uparrow,\downarrow}(t_0)$ and ${\bar\rho}_{\downarrow,\uparrow}(t_0)$ are multiplied, respectively, by the decoherence factor (cf.\ Eq.~\ref{eq:wRS} and recall that we set $g_0\mu_B=1$ here)
\begin{equation}
\label{eq:WSingleSpin}
W(t_0) = {\mathbb E}\left[{\mathrm e}^{-i\phi(t_0)}\right] = 
{\mathbb E}\left[\exp{-i \int\limits_0^{t_0}\!\!{\mathrm B}(t) dt}\right],
\end{equation}
and its complex conjugate $W^*(t_0)$.

These calculations are easily extended to a system of $M$ non-interacting entangled qubits, shuttled along the trajectories $x_{c1}(t)$, $x_{c2}(t)$, $\dots, x_{c\,M}(t)$, under the action of the time-dependent magnetic fields comprised of the deterministic parts $B_{01}(t)=B_0(x_{c1}(t))$, $B_{02}(t)=B_0(x_{c2}(t))$, $\dots, B_{0\,n}(t)=B_0(x_{c\,n}(t))$ and the random fields ${\mathrm B}_1(t)={\tilde B}(x_{c1}(t),t)$, ${\mathrm B}_2(t)={\tilde B}(x_{c2}(t),t)$, $\dots, {\mathrm B}_n(t)={\tilde B}(x_{c\,n}(t),t)$, 
Note that the qubit trajectories $x_{c\,k}(t)$ ($k=1\dots M$) differ from each other even if all qubits are shuttled through the same channel in the same manner: the finite delay $\Delta t$ between shuttling of the first and the second qubit already makes $x_{c1}(t)$ different from $x_{c2}(t)$, since in that case $x_{c2}(t)=x_{c1}(t+\Delta t)$, and the random fields ${\mathrm B}_1(t)$ and ${\mathrm B}_2(t)$ are different due to the fluctuations of ${\tilde B}(x,t)$ in time. 

Since the qubits do not interact during shuttling, their dynamics is described by the Hamiltonian
\begin{equation}
{\tilde H}_Z(t) = \sum_{k=1}^M g_0 \mu_B \left[ B_{0k}(t) + {\mathrm B}_k(t)\right] S_k^z.
\end{equation}
In general, if the qubits are shuttled in a different manner and $\rho_0(x-x_c,y,z)$ varies from one qubit to another, the values of $g_0$ can differ for different spins, but here we neglect such situations, and set, as before, $g_0\mu_B=1$. 

We use the qubit basis states $|\uparrow\rangle$ and $|\downarrow\rangle$, and specify the many-qubit product basis states via the multi-index ${\mathbf p}=(p_1,p_2,\dots p_M)$, such that $|{\mathbf p}\rangle=|p_1\rangle\otimes\dots |p_M\rangle$; as before, each index $p_k$ can take the values $\uparrow$ or $\downarrow$. In this basis, the elements of the multi-qubit density matrix after shuttling are modified by acquiring the phase factors
\begin{eqnarray}
\rho_{{\mathbf p},{\mathbf p'}}(t_0) &=& \rho_{{\mathbf p},{\mathbf p'}}(0)\, {\mathrm e}^{-i \Theta_{{\mathbf p},{\mathbf p'}}},\\ \nonumber
\Theta_{{\mathbf p},{\mathbf p'}} &=& \sum_{k=1}^M (m_k-m'_k) \left[\alpha_k(t_0) + \phi_k(t_0)\right],
\end{eqnarray}
where $\rho_{{\mathbf p},{\mathbf p'}}=\langle {\mathbf p}|\rho|{\mathbf p'}\rangle$, while $m_k$ and $m'_k$ are the shorthand notations for $m(p_k)$ and $m(p'_k)$, with $m(\uparrow)=+1/2$ and $m(\downarrow)=-1/2$. The deterministic and the random phases are
\begin{equation}
\alpha_k(t_0)= \int_0^{t_0}\!\!B_{0k}(t)\,dt,\quad  \phi_k(t_0)= \int_0^{t_0}\!\!{\mathrm B}_k(t)\,dt,
\end{equation}
respectively. By setting $\alpha_k(t_0)=0$ as before, and averaging over the random fields ${\mathrm B}_k(t)$, we obtain the result similar to Eq.~\ref{eq:barrho1}: the diagonal elements of the averaged $n$-qubit density matrix ${\bar\rho}(t_0)$ remain unaltered after shuttling, while the off-diagonal elements are multiplied by the decoherence factors: 
\begin{equation}
{\bar\rho}_{{\mathbf p},{\mathbf p'}}(t_0) = \rho_{{\mathbf p},{\mathbf p'}}(0)\, W_{{\mathbf p},{\mathbf p'}}(t_0),
\end{equation}
and the decoherence factors $W_{{\mathbf p},{\mathbf p'}}(t_0)$ are
\begin{eqnarray}
\label{eq:wManyQ}
W_{{\mathbf p},{\mathbf p'}}(t_0) &=& {\mathbb E}\left({\mathrm e}^{-i \Theta_{{\mathbf p},{\mathbf p'}}}\right),\\ \nonumber
\Theta_{{\mathbf p},{\mathbf p'}} &=& \int_0^{t_0} \sum_{k=1}^M (m_k-m'_k)\, {\mathrm B}_k(t)\,dt.
\end{eqnarray}
This expression demonstrates that each decoherence factor $W_{{\mathbf p},{\mathbf p'}}(t_0)$ is determined by the characteristic functional of the random process
\begin{equation}
\label{eq:manyspinB}
{\mathrm B}_{{\mathbf p},{\mathbf p'}}(t)=\sum_{k=1}^M (m_k-m'_k)\, {\mathrm B}_k(t).
\end{equation}
By modeling the noise ${\tilde B}(x_c,t)$ as a Gaussian random sheet, we can calculate all decoherence factors $W_{{\mathbf p},{\mathbf p'}}(t_0)$. 
Indeed, the zero-mean Gaussian sheet is specified by its covariance function $K_{\tilde B}(\theta,\theta')=K_{\tilde B}(x,t;x',t')$, 
see Definition \ref{def:Gaussianfield}; for the case of the OU sheet, adopted in this work, it is given in Eq.~\ref{eq:random_sheet_corr}. The qubit trajectories specify a set of paths $\Gamma_k$ on the OU sheet; for the situations considered below, all paths are continuous, and, additionally, can be parametrized by the time variable $s=t$, such that $\Gamma_k(s)$ are just the pairs $(x_{c\,k}(t),t)$ on the $(x_c,t)$ plane. The processes ${\mathrm B}_k(t)$ are Gaussian, and the process ${\mathrm B}_{{\mathbf p},{\mathbf p'}}(t)$ is also Gaussian, as we have established in Sec.~\ref{sec:RSontraj}; the covariance of ${\mathrm B}_{{\mathbf p},{\mathbf p'}}(t)$ is
\begin{equation}
K_{{\mathbf p},{\mathbf p'}}(t_1,t_2) = \sum_{j,k=1}^M \Delta_j \Delta_k K_{j,k}(t_1,t_2),
\end{equation}
where $K_j(t_1,t_2)$ is the correlation function of ${\mathrm B}_j(t)$ and  ${\mathrm B}_k(t)$, and $\Delta_j= m_j-m'_j$ (and, similarly, $\Delta_k$). 
Furthermore, below we consider only continuous trajectories $x_{c\,k}(t)$ and the free coherence decay of the shuttled spins (without the dynamical decoupling pulses). Thus, in the characteristic functional that corresponds to the decoherence factor $W_{{\mathbf p},{\mathbf p'}}$ in Eq.~\ref{eq:wManyQ}, the integrator function $F(t)$ is simply $F(t)=t$. 
Therefore, applying Eq.~\ref{char_fun_dbl_integr}, we obtain simply
\begin{eqnarray}
\label{eq:wdoubleint}
W_{{\mathbf p},{\mathbf p'}}(t_0) &=& \exp{-\chi_{{\mathbf p},{\mathbf p'}}(t_0)}\\ \nonumber
\chi_{{\mathbf p},{\mathbf p'}}(t_0) &=& \frac{1}{2} \int\limits_{0}^{t_0} \int\limits_{0}^{t_0}\! K_{{\mathbf p},{\mathbf p'}}(t_1,t_2)\, dt_1 dt_2.
\end{eqnarray}
Thus, the problem is reduced to finding the appropriate kernel for each shuttling scenario and integrating it analytically or numerically, as discussed in Sec.~\ref{sec:method}.

In more complex scenarios involving several entangled qubits,
it may happen that the trajectories of the qubits $x_{c\,k}(t)$ define the path $\Gamma$ (see Sec.~\ref{sec:RSontraj}) which is not continuous. In that case, one cannot guarantee continuity of realizations of the relevant random process, which is required by the mathematical theory developed in Sec.~\ref{sec:mathGen}. In that case, the integrator $F(s)$ should be adjusted: e.g.\ we can connect the continuous components of the path $\Gamma$ by line segments and set the integrator $F(t)$ equal to a constant there, such that the control function $\xi(t)=dF/dt$ in Eq.~\ref{char_fun_dbl_integr} is zero on these line segments. 
The action of the dynamical decoupling pulses applied to the shuttled electron spins, can also be taken into account using appropriate function $\xi(t)=dF/dt$ \cite{KlauderAnderson62,CywinskiLutchynDasSarmaDD08,ViolaLloydDD}. Approximating the pulses as instant, we set the control function $\xi(t)$ to change sign, flipping from +1 to -1 or back after every pulse, thus producing a piecewise-constant $\xi(t)$ and continuous pile-type integrator function $F(t)$.
The corresponding modifications in Eq.~\ref{eq:wdoubleint}, although not explicitly considered in this work, are rather straightforward.

\subsection{Shuttling of a single qubit}
\label{sec:singlequbitshuttling}

Let us start with analyzing the simplest scenario, where a single electron spin is transported one way, from the initial position $x_c=0$ at $t=0$ to $x_c=L$ at the time moment $t=T$ with a constant velocity $v=L/T$, such that the total shuttling time $t_0=T$. The shuttling trajectory in a two-dimensional plane $\theta=(x_c,t)$ is a straight line segment $x_c(t)=v\,t$ connecting the points $\theta_0=(0,0)$ and $\theta_1=(L,T)$. Dephasing of the electron spin is completely specified by the decoherence factor $W(t_0)$ given in  Eq.~\ref{eq:WSingleSpin}, which now takes the form
\begin{equation}
W(T) = {\mathbb E}\left[\exp{-i \int_0^{T}\!\!{\mathrm B}(t) dt}\right],
\end{equation}
with ${\mathrm B}(t)={\tilde B}(x_c(t),t)={\tilde B}(v\,t,t)$ and $t_0=T$. Since  ${\tilde B}(x_c,t)$ is represented as an OU sheet, characterized by the mean and the covariance function given in Eq.~\ref{eq:random_sheet_corr}, the field ${\mathrm B}(t)$ is also Gaussian, with zero mean and the covariance function
\begin{eqnarray}
\label{eq:sigmaBdef}
K_{\mathrm B}(t_1,t_2) &=& K_{OU}(x_1=v t_1,t_1; x_2=v t_2,t_2)  \\ \nonumber
&=& \frac{\sigma^2}{4 \kappa_x \kappa_t} \exp{-\kappa_x v |t_1 - t_2|-\kappa_t |t_1 - t_2|}\\ \nonumber
&=& \sigma_{\mathrm B}^2 \exp{-(\kappa_x v + \kappa_t) |t_1 - t_2|},
\end{eqnarray}
where $\sigma_{\mathrm B}=\sigma/\sqrt{4\kappa_x \kappa_t}$, and 
the parameters $\kappa_x=1/\lambda_c$ and $\kappa_t=1/\tau_c$, where $\lambda_c$ is the correlation length and $\tau_c$ is the correlation time, characterize, respectively,  spatial and temporal extent of the correlations.
That is, the process ${\mathrm B}(t)$ in that case is a standard one-parameter OU process with the effective correlation decay rate $\kappa = \kappa_x v + \kappa_t$ and the variance $\sigma_{\mathrm B}^2$. This is a very natural result. Firstly, note that the shuttling trajectory is a segment of a straight line, invariant with respect to translations along itself, and the underlying OU sheet ${\tilde B}(x_c,t)$ is stationary, so that the resulting process ${\mathrm B}(t)$ is stationary. Secondly, since $x_c(t)$ is a monotonically increasing function of $t$, the shuttling trajectory defines the flow (in the sense of Definition~\ref{def:flow}, i.e.\ a set of nested rectangles in the $(x_c,t)$ plane), such that ${\mathrm B}(t)$ is a Markov process, as discussed in Appendix~\ref{app:markov}. Thirdly, ${\mathrm B}(t)$ inherits the Gaussian property from the OU sheet ${\tilde B}(x_c,t)$. Therefore, the process ${\mathrm B}(t)$ is Gaussian, stationary, and Markovian, and therefore, according to the Doob's theorem \cite{GardinerRandomProc,VanKampenRandomProc}, is an OU process.
The decoherence factor in the case of OU noise is \cite{kubo_statistical_1985,KlauderAnderson62},
\begin{eqnarray}
\label{eq:wT}
W(T) &=& \exp{-\chi(T)},\\ \nonumber
\chi(T)&=&(\sigma_{\mathrm B}/\kappa)^2 \,[\kappa T+{\mathrm e}^{-\kappa T}-1].
\end{eqnarray}
In the case of slow strong noise, when $\sigma_{\mathrm B}\gg\kappa$, most of the coherence decay happens at short times $T\ll 1/\kappa$, and in this region   $W(T)\approx \exp{-\sigma_{\mathrm B}^2 T^2 /2}$ (Gaussian decay, the regime of quasi-static noise). For fast weak noise with $\sigma_{\mathrm B}\ll\kappa$, the decay happens mostly at long times $T\gg 1/\kappa$, and $W(T)\approx \exp{-\sigma_{\mathrm B}^2 T/\kappa}$ in this region (exponential decay, the motional narrowing regime).

The simplest scenario, where the non-trivial correlations of the random sheet ${\tilde B}(x,t)$ start playing an important role, is the forth-and-back shuttling, where a single electron spin is first shuttled forth, from $x_c=0$ at $t=0$ to $x_c=L$ at $t=T$ with a constant velocity $v=L/T$, and then is moved back with the same velocity, from $x_c=L$ at $t=T$ to $x_c=0$ at $t=2T$, so the total shuttling time $t_0=2T$; this shuttling scenario was studied in recent experiments \cite{StruckSchreiberEtalSpinPairShuttling23,VolmerStruckEtalValleySplit23,SmetVandersypenEtal24SpinShuttlSilicon}.
The corresponding shuttling trajectory in the plane $(x_c,t)$ consists of two branches: the first branch is a segment $x_c(t)=v\,t$ connecting the points $\theta_0=(0,0)$ and $\theta_1=(L,T)$, and the second branch is a segment $x_c(t)=2L-v\,t$ connecting the points $\theta_1$ and $\theta_2=(0,2T)$, see Fig.~\ref{fig:traj1and2}. The random magnetic field ${\mathrm B}(t)$ generated by such a trajectory is subtle. Taken separately within each branch, ${\mathrm B}(t)$ is just the usual OU process: the arguments of the previous paragraph show that it is Gaussian, stationary and Markovian. However, when taken over the whole trajectory, it becomes neither stationary nor Markovian, and for every realization of this random process, the part of the function ${\mathrm B}(t)$ in the region $t\in [T,2T]$ is close to the mirror image of itself in the region $t\in [0,T]$, see Fig.~\ref{fig:corrnoise}. The only useful property reliably inherited by ${\mathrm B}(t)$ is Gaussianity, which is enough to analytically calculate the decoherence factor $W(t_0=2T) = \exp{-\chi(t_0)}$ (\ref{eq:WSingleSpin}) via the double integral (cf.\ Eq.~\ref{eq:wdoubleint})
\begin{equation}
\label{eq:specific1}
\chi(t_0=2T) = \frac{1}{2} \int\limits_{0}^{t_0} \int\limits_{0}^{t_0}\! K_\mathrm{B}(t_1,t_2)\, dt_1 dt_2,
\end{equation}
see Appendix \ref{app:forthback} for details, giving the answer 
\begin{equation}
\label{eq:forthbackMain}
\chi(t_0) = (\sigma_{\mathrm B}/\kappa_t)^2 \Lambda(\beta,\gamma),\quad \Lambda(\beta,\gamma)=\Lambda_1+\Lambda_2 
\end{equation}
where 
\begin{eqnarray}
\Lambda_1 &=& 2\frac{(\beta+\gamma+{\mathrm e}^{-\beta-\gamma}-1)} {(1+\gamma/\beta)^2},\\ \nonumber
\Lambda_2 &=& \frac{\left({\mathrm e}^{-2\beta}-1\right) (\gamma /\beta) - 2 {\mathrm e}^{-(\beta+\gamma)}+{\mathrm e}^{-2\beta}+1}{1-\gamma ^2/\beta ^2},
\end{eqnarray}
and where we introduced the dimensionless shuttling length $\gamma=\kappa_x L=L/\lambda_c$ and time $\beta=\kappa_t T=T/\tau_c$, measured in units of the correlation length $\lambda_c$ and the correlation time $\tau_c$ of the underlying OU sheet.
Note that $\Lambda_1$ has the form similar to the result for dephasing by a standard OU process, cf.\ Eq.~\ref{eq:wT}, since this contribution comes from integration over the regions where $t_1$ and $t_2$ both belong to either branch 1 or branch 2, and where $\mathrm{B}(t)$ behaves as a regular OU process. The  influence of correlations on dephasing is mostly encoded in $\Lambda_2$.

In order to explicitly see the role of correlations during forth-and-back shuttling, let us compare it with the equivalent one-way shuttling, when the electron is shuttled in one direction over the total distance $\ell=2L$ during the total time $t_0=2T$. In that case, the decoherence factor is $W_0(t_0)=\exp{-\chi_0(t_0)}$, where 
\begin{eqnarray}
\label{eq:forthOnly}
\chi_0(t_0) &=& (\sigma_{\mathrm B}/\kappa_t)^2 \Lambda_0(\beta,\gamma),\\ \nonumber 
\Lambda_0(\beta,\gamma) &=& \frac{2(\beta+\gamma)+{\mathrm e}^{-2(\beta+\gamma)}-1}{(1+\gamma/\beta)^2},
\end{eqnarray}
see Eq.~\ref{eq:wT} with $v=\ell/t_0=L/T$.

\begin{figure}[tbp!]
\includegraphics[width=\columnwidth]{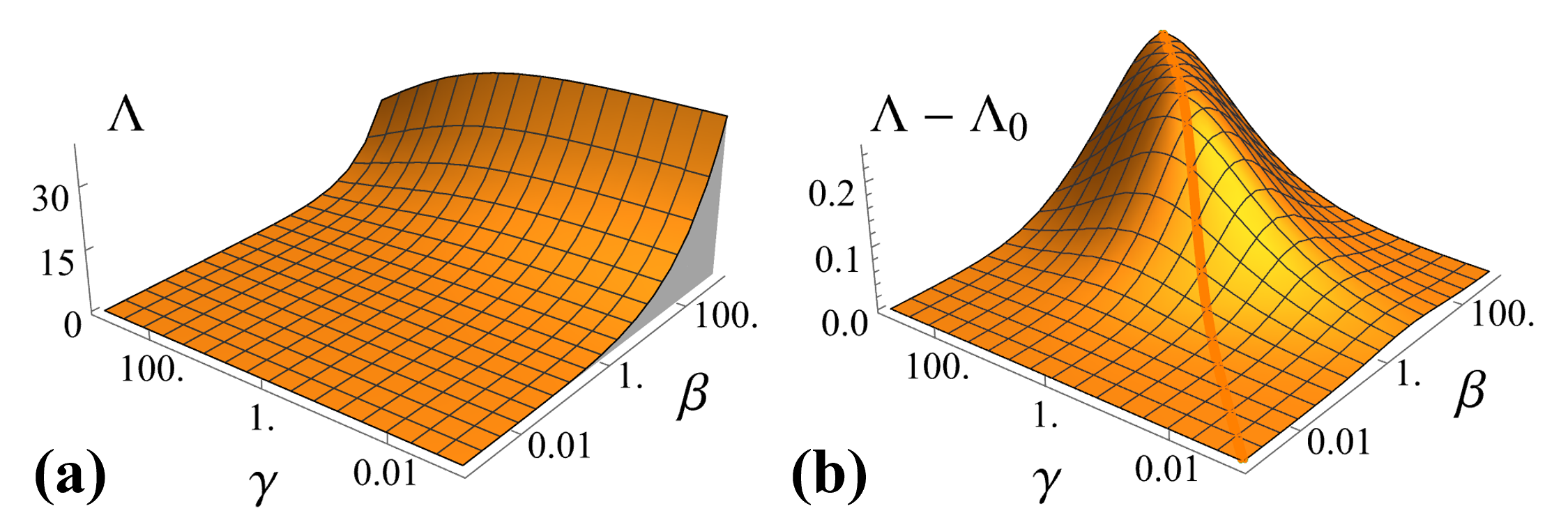}
\caption{ {\bf (a)}: Dependence of the function $\Lambda(\beta,\gamma)$, which determines the decoherence factor $W(t_0)$ in the case of the forth-and-back shuttling, on the dimensionless time $\beta=\kappa_t T$ and length $\gamma=\kappa_x L$. The function $\Lambda_0(\beta,\gamma)$, determining the decoherence factor $W_0(t_0)$ in the case of the equivalent one-way shuttling, is qualitatively similar, but quantitatively different. 
{\bf (b)}: Dependence of the difference $\Lambda-\Lambda_0$ on $\beta$ and $\gamma$. The difference is positive and is bounded by the value of $1/4$, which is achieved at $\beta=\gamma$ in the limit $\beta\to\infty$}
\label{fig:Lambdas}
\end{figure}

Two-dimensional plot of $\Lambda$ as a function of $\beta$ and $\gamma$ is given in Fig.~\ref{fig:Lambdas}a. The function $\Lambda_0(\beta,\gamma)$ looks qualitatively similar, and the difference between the two is shown in Fig.~\ref{fig:Lambdas}b. The quantity $\Lambda-\Lambda_0$ is always positive, and is bounded by the value of $1/4$, which is achieved at $\beta=\gamma$ in the limit $\beta\to\infty$ (apparent singularity of $\Lambda$ at $\beta=\gamma$ is removable). Thus, correlations of the noise in space and time during the forth-and-back shuttling are always detrimental, leading to stronger decoherence, in comparison with the equivalent one-way shuttling.

\begin{figure}[tbp]
\includegraphics[width=\columnwidth]{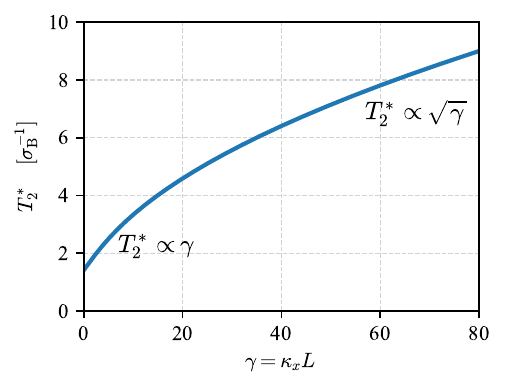}
\caption{The dephasing time $T_2^*$ as a function of the shuttling length $\gamma=\kappa_x L = L/\lambda_c$ in the case of the forth-and-back shuttling. The decoherence factor $W(t_0)$ at short times $t_0\ll\tau_c$ exhibits Gaussian decay as a function of time $t_0$, but the coherence time $T_2^*$ as a function of the shuttling length $\gamma$ changes its character from ballistic ($T_2^*\propto \gamma$) to diffusive ($T_2^*\propto \sqrt{\gamma}$) decay.}
\label{fig:T2forthback}
\end{figure}

It is instructive to examine the behavior of $\Lambda(\beta,\gamma)$ at short times, i.e.\ at $\beta\ll 1$. In this regime 
\begin{equation}
\Lambda\approx 4 \beta^2\,\frac{\gamma-1+\mathrm{e}^{-\gamma}}{\gamma^2},
\end{equation}
so that the coherence decay has Gaussian form, with the decoherence factor
\begin{equation}
\label{eq:shorttime}
W(t_0)=\mathrm{e}^{-\left(t_0/T_2^*\right)^2},\ \ T_2^*=\frac{\gamma}{\sigma_{\mathrm B}}\,\left(\gamma-1+\mathrm{e}^{-\gamma}\right)^{-1/2}.
\end{equation}
The coherence decay time $T_2^*$ as a function of the dimensionless shuttling length $\gamma=\kappa_x L$ is shown in Fig.~\ref{fig:T2forthback}. The decay time $T_2^*$ has the finite value $\sqrt{2}/\sigma_\mathrm{B}$ at $\gamma=0$, and grows linearly with the shuttling length. At large $\gamma$, the time $T_2^*$ still grows, but at a slower rate, as $\sqrt{\gamma}/\sigma_\mathrm{B}$. 
That is, the character of decoherence changes from ballistic ($T_2^*$ growing linearly with $L$) to diffusive ($T_2^*\propto\sqrt{L}$), in spite of showing no signs of motional narrowing in time decay, with $W(t_0)$ keeping its Gaussian form.

It is interesting to point out that Ref.~\onlinecite{StruckSchreiberEtalSpinPairShuttling23}, where the forth-and-back shuttling has been experimentally implemented, reports the measured dependence of the oscillations decay time (also denoted $T_2^*$ in that work) on the shuttling distance. Their experimental curve (Fig.~2g in \cite{StruckSchreiberEtalSpinPairShuttling23}) has a rather similar overall form. Unfortunately, quantitative comparison requires knowledge of some experimental details, which we found difficult to extract from the manuscript.

\begin{figure}[bp!]
\includegraphics[width=\columnwidth]{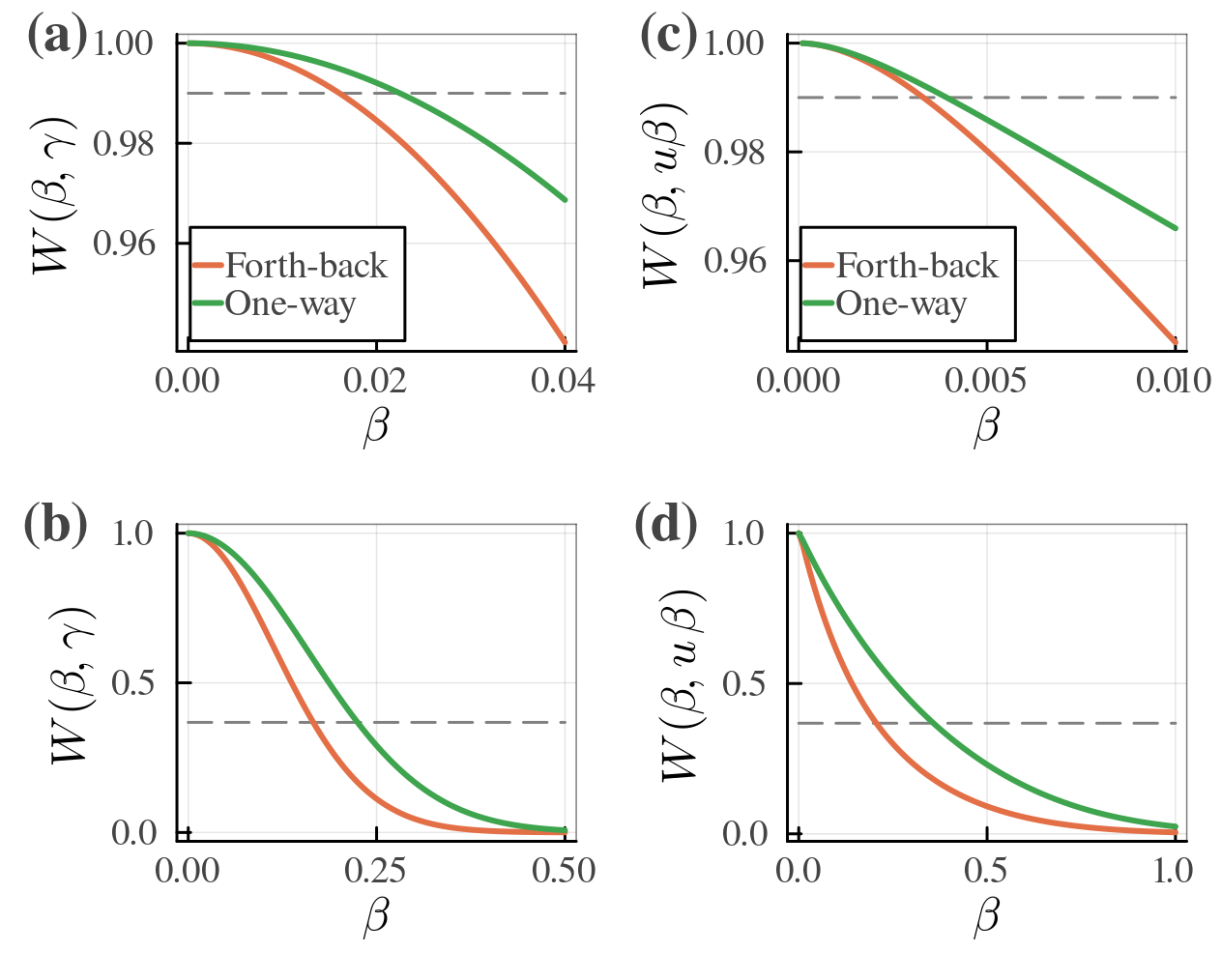}
\caption{Comparison of the dephasing dynamics for the forth-and-back shuttling and the corresponding one-way shuttling. {\bf (a)}, {\bf (b)}: The regime of short times $\beta=T/\tau_c$ but long shuttling length $\gamma=L/\lambda_c$, when both $W(t_0)$ and $W_0(t_0)$ exhibit Gaussian decay vs.\ time $t_0=2T$.  
The graphs correspond to $\gamma=40$ and $\sigma_{\mathrm B}/\kappa_t=20$. 
{\bf (a)}: Initial stage of the dephasing, when $W(t_0)$ and $W_0(t_0)$ are close to 1, the horizontal grey line shows the level $W=0.99$; {\bf (b)}: overall shape of $W(t_0)$ and $W_0(t_0)$, the horizontal grey line shows the level $W=1/e$, which determines the times $T_2^*$ and $T_{2,0}^*$; these times differ by a factor close to $\sqrt{2}$.
{\bf (c)}, {\bf (d)}: The regime of long times $\beta=T/\tau_c$ and high shuttling velocity $u=\gamma/\beta$, when both $W(t_0)$ and $W_0(t_0)$ exhibit exponential decay vs.\ time $t_0=2T$. 
The graphs correspond to fixed $u=300$ and $\sigma_{\mathrm B}/\kappa_t=25$.  
{\bf (c)}: Initial stage of the dephasing, when $W(t_0)$ and $W_0(t_0)$ are close to 1, the grey horizontal line shows the level $W=0.99$; {\bf (b)}: overall shape of the curves $W(t_0)$ and $W_0(t_0)$, the grey horizontal line shows the level $W=1/e$, which determines the times $T_2^*$ and $T_{2,0}^*$; these times differ by a factor close to 2. }
\label{fig:FABvsOne}
\end{figure}

The function $\Lambda_0(\beta,\gamma)$, which characterizes dephasing during the equivalent one-way shuttling, also depends quadratically on $\beta$ at short times $\beta\ll 1$, such that the decoherence is also Gaussian: 
\begin{eqnarray}
\label{eq:shorttime0}
\Lambda_0 &\approx & \beta^2\,\frac{2\gamma-1+\mathrm{e}^{-2\gamma}}{\gamma^2}, \\ \nonumber
W_0(t_0)&=&\mathrm{e}^{-\left(t_0/T_{2,0}^*\right)^2},\ \ T_{2,0}^*=\frac{2\gamma}{\sigma_{\mathrm B}}\,\left(2\gamma-1+\mathrm{e}^{-2\gamma}\right)^{-1/2}.
\end{eqnarray}
For the experimentally interesting regime of high-velocity shuttling, when the shuttling time is short, $\beta\ll 1$, but the length is large, $\gamma\gg 1$, we obtain
\begin{equation}
\Lambda\approx 4\,\beta^2/\gamma,\quad \Lambda_0\approx 2\,\beta^2/\gamma,
\end{equation}
such that the dephasing times $T_2^*$ differ by a factor of $\sqrt{2}$ for the two shuttling scenarios, see Fig~\ref{fig:FABvsOne}a and b.
More generally, for arbitrary shuttling times $\beta$ in the high-velocity regime, when the dimensionless velocity $u=\gamma/\beta= v\,(\kappa_x/\kappa_t)$ is large,
\begin{equation}
\Lambda\approx \frac{2\beta+1-\mathrm{e}^{-2\beta}}{u},\quad
\Lambda_0\approx \frac{2\beta}{u}.
\end{equation}
Interestingly, in this parameter region, the decay is always exponential in time  for the one-way shuttling case (the motional narrowing regime), but in the case of the forth-and-back shuttling, the decay changes from Gaussian to exponential as the shuttling time $\beta$ grows, exhibiting transition from the quasi-static to the motional narrowing regime. But even in the motional narrowing regime, when $\beta\gg 1$, the difference between the two shuttling scenarios remains noticeable:
\begin{equation}
W=\exp{-2 t_0 \frac{\sigma_\mathrm{B}^2}{v \kappa_x}},\ \  
W_0=\exp{-t_0 \frac{\sigma_\mathrm{B}^2}{v \kappa_x}},
\end{equation}
where we used dimensional velocity $v$ and time $t_0$, so that decoherence in the case of the forth-and-back shuttling is twice faster than in the one-way case, see Fig.~\ref{fig:FABvsOne}c and d.

Note that the high-velocity regime $u=\beta/\gamma\gg 1$ is determined by the ratio of the dimensional velocity $v=L/T$ to the characteristic velocity $c=\kappa_t/\kappa_x$, which is dictated by the properties of the random sheet ${\tilde B}(x_c,t)$, i.e.\ by the material properties of the sample.

\subsection{Protection from dephasing during transport: Shuttling of two entangled qubits}

Decoherence during a single-qubit shuttling considered above is finite even if the noise ${\tilde B}(x_c,t)$ is quasi-static in time, with $\kappa_t\to 0$, as seen e.g.\ from Eqs.~\ref{eq:shorttime} and \ref{eq:shorttime0}. This happens because the random field $\mathrm{B}(t)$ acquires nontrivial dependence on time due to variation of the noise in space, through the time variation of $x_c(t)$. 
For a static qubit subjected to a quasi-static noise, dephasing is known to be easily eliminated by the spin echo that reverses the dephasing dynamics and eliminates the random phase acquired by the qubit. However, it may be difficult to do the same for a qubit which is being transported through space.

A possibly simpler alternative, analyzed below, is to encode the state of the qubit in a decoherence-free subspace of a logical qubit \cite{LidarChuangWhaleyDFS98,ZanardiRasettiDFS97,ViolaCoryEtalDFSExp01}, formed by the singlet and triplet states of two electron spins. These spins are shuttled with some delay one after another (similar to e.g.\  Ref.~\onlinecite{JadotMeunierEtalTwoSpinShuttling21}, where the shuttling of the singlet state was studied). In order to substantiate this proposal, it is crucial to analyze its performance under realistic circumstances, and identify the range of parameters where such an encoding would be beneficial.

\begin{figure}[tbp]
\includegraphics[width=\linewidth]{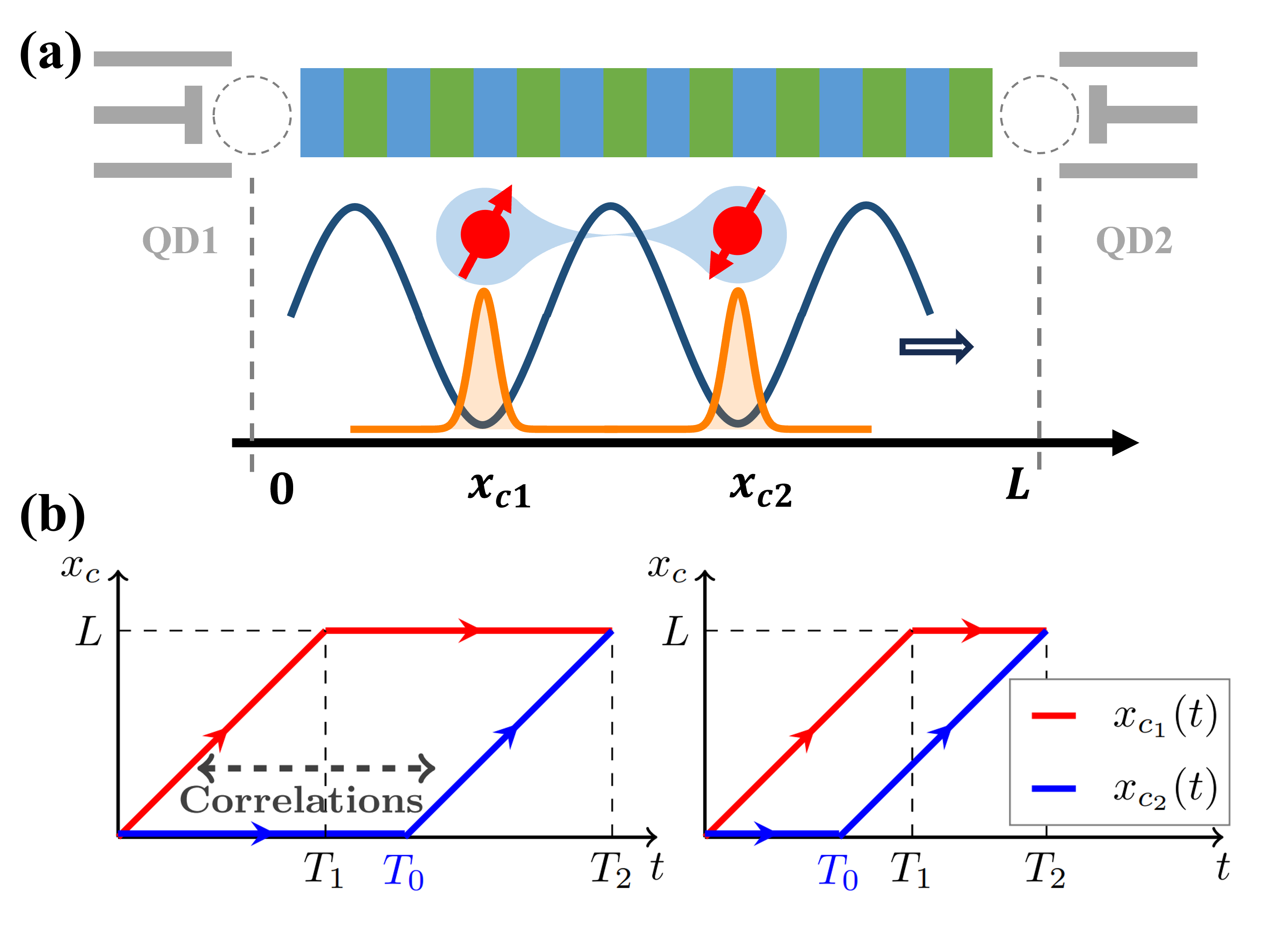}
\caption{{\bf (a)}: Schematic representation of a sequential two-spin shuttling. The first electron is loaded into the shuttling channel at $t=0$, while the second electron is waiting in the original quantum dot, at $x=0$. After the delay $T_0$, the second electron starts shuttling. During the shuttling, both electrons, whose spins (shown as red arrows) encode the logical qubit, are confined in the minima of the moving potential (solid dark blue line), forming localized wavepackets (shown with orange) with centers at $x=x_{c1}$ and $x=x_{c2}$.
The moving potential transports the electrons from left to right; the minimal delay is determined by the period of the confining potential. After the first electron reaches at time $t=T_1$ the destination quantum dot ($x=L$), it stays there while the second electron finishes the shuttling.
{\bf (b)}: Space-time trajectories of the two sequentially shuttled spins. Trajectory $x_{c1}(t)$ of the first qubit is shown in red, and $x_{c2}(t)$ of the second qubit is in blue. Depending on the travel time $T_1$ and the delay $T_0$, two cases are possible, $T_0>T_1$ and $T_0<T_1$, shown in the left and the right panel, respectively.}
\label{fig:two_spin_sequence}
\end{figure}

Assume that we have a state of a spin qubit $a|\!\uparrow\rangle + b|\!\downarrow\rangle$ that is to be shuttled from one quantum dot to another. We use an ancilla spin, located e.g.\ in an adjacent dot and prepared in a state $|\!\uparrow\rangle$, 
such that the two-spin wavefunction is a product $\left[a|\!\uparrow_1\rangle + b|\!\downarrow_1\rangle\right]\,|\!\uparrow_2\rangle$, where the subscripts 1 and 2 denote the states of the qubit and the ancilla, respectively. Then we apply a CNOT gate with the qubit as a control and ancilla as a target, turning the two-spin state into 
\begin{eqnarray}
|\psi(0)\rangle &=& a\,|\!\uparrow_1 \downarrow_2\rangle + b\,|\!\downarrow_1\uparrow_2\rangle \\ \nonumber
&=&  \frac{a+b}{\sqrt{2}}\,|T_0\rangle + \frac{a-b}{\sqrt{2}}\,|S\rangle,
\end{eqnarray}
encoding the original qubit state into a linear combination of the singlet 
$|S\rangle = {\textstyle\frac{1}{\sqrt{2}}} \left[|\!\uparrow_1\downarrow_2\rangle - |\!\downarrow_1\uparrow_2\rangle\right]$ and the triplet 
$|T_0\rangle = {\textstyle\frac{1}{\sqrt{2}}} \left[|\!\uparrow_1\downarrow_2\rangle + |\!\downarrow_1\uparrow_2\rangle\right]$. After the encoding, the two spins are sequentially shuttled to the destination, where the decoding is done with another CNOT gate, turning the states $|\!\uparrow_1 \downarrow_2\rangle$ and $|\!\downarrow_1\uparrow_2\rangle$ back into the qubit states $|\!\uparrow\rangle$ and $|\!\downarrow\rangle$.

Dephasing during the shuttling erases the mutual phase between the qubit and the ancilla, turning both $|S\rangle$ and $|T_0\rangle$ into the same incoherent mixture of the states $|\mathbf p\rangle=|\!\uparrow_1\rangle|\!\downarrow_2\rangle$ and $|\mathbf p'\rangle=|\!\downarrow_1\rangle|\!\uparrow_2\rangle$, so that the fidelity of shuttling is controlled by the dephasing factor $W_{\mathbf{p},\mathbf{p'}}$ given in Eq.~\ref{eq:wManyQ}, whose explicit form is
\begin{eqnarray}
\label{eq:wtwospins}
W(t_0) &=& \mathbb{E}\left[{\mathrm{e}}^{i\Theta(t_0)}\right],\\ \nonumber
\Theta(t_0) &=& \int_0^{t_0} \left({\mathrm B}_1(t)-{\mathrm B}_2(t)\right)\,dt = \int_0^{t_0} \delta{\mathrm B}(t)\,dt,
\end{eqnarray}
where we omitted the indices $\mathbf{p}$ and $\mathbf{p}'$ because no other states are involved. As before, we do not consider the errors associated with loading the electrons from the quantum dots to the shuttling channel and back.

Shuttling of the two-spin state $|\psi(0)\rangle$ instead of the original single spin qubit can greatly improve the transfer fidelity: 
if the noise ${\tilde B}(x_c,t)$ varies slowly on the timescale $t_0$ then the random processes ${\mathrm B}_1(t)$ and ${\mathrm B}_2(t)$ are close to each other during each shuttling event, so that the phase $\Theta(t_0)$ is always close to zero, and $W(t_0)$ is close to 1. We need to calculate $W(t_0)$ in order to quantify the possible advantage, and identify the favorable region of parameters.

We consider the situation where shuttling of the first spin starts at $t=0$ from the initial point $x_c=0$; the shuttling takes place with the constant velocity $v$ until the final point $x_c=L$ is reached at $t=T_1=L/v$. While the first electron is loaded into the shuttling channel and starts moving, the second electron is waiting in the original quantum dot, and starts shuttling with some delay, at $t=T_0$. Shuttling of the second electron takes place with the same velocity and the same initial and final points as the first electron, such that the second electron arrives to $x_c=L$ at $t=T_2=T_1+T_0$, and the total duration of the shuttling is $t_0=T_2$.
Thus, the trajectories of the spins are
\begin{eqnarray}
&& x_{c1}(t)=
\begin{cases}
v t,\quad 0<t\le T_1\\
L,\quad T_1<t<T_2
\end{cases}
\\ 
&& x_{c2}(t)=
\begin{cases}
0, \quad 0<t\le T_0\\
v(t-T_0), \quad T_0<t<T_2
\end{cases}
\end{eqnarray}
Depending on the shuttling velocity $v$ and the delay $T_0$, the second spin starts shuttling before or after the first spin reaches the final point, i.e.\ $T_0\le T_1$ and $T_0>T_1$, respectively, shown in Fig.~\ref{fig:two_spin_sequence}.

As discussed above, the fields ${\mathrm B}_1(t)$ and ${\mathrm B}_2(t)$, defined by the trajectories $x_{c1}(t)$ and $x_{c2}(t)$, and their difference $\delta{\mathrm B}(t)$, are Gaussian processes, such that $W(t_0)$ in Eq.~\ref{eq:wtwospins} can be calculated (see Eq.~\ref{eq:wdoubleint}) as
\begin{eqnarray}
\label{eq:chi12}
W(t_0) &=& \exp{-\chi(t_0)}\\ \nonumber
\chi(t_0) &=& \frac{1}{2} \int\limits_{0}^{t_0} \int\limits_{0}^{t_0}\! K_{\delta\mathrm{B}}(t_1,t_2)\, dt_1 dt_2,
\end{eqnarray}
where the covariance of the process $\delta{\mathrm B}(t)$ is
\begin{eqnarray}
\label{eq:cov2}
K_{\delta \mathrm{B}}(t_1, t_2) &=& \mathbb{E}\left[\left(\mathrm{B}_1(t_1)-\mathrm{B}_2(t_1)\right)\left(\mathrm{B}_1(t_2)-\mathrm{B}_1(t_2)\right)\right]\nonumber \\
&=&\mathbb{E}\left[\mathrm{B}_1(t_1) \mathrm{B}_1(t_2)\right] + \mathbb{E}\left[\mathrm{B}_2(t_1) \mathrm{B}_2(t_2)\right]
\nonumber\\
&-& \mathbb{E}\left[\mathrm{B}_2(t_1) \mathrm{B}_1(t_2)\right] - \mathbb{E}\left[\mathrm{B}_1(t_1) \mathrm{B}_2(t_2)\right].
\end{eqnarray}
The first two terms represent the autocorrelations of $\mathrm{B}_1$ and $\mathrm{B}_2$, while the last two terms represent the cross-correlation between them, which are non-trivial because both processes are derived from the same underlying random sheet $\tilde{B}(x_c,t)$.

The double integral in Eq.~\ref{eq:chi12} can be calculated analytically using the covariance functions of $\mathrm{B}_1(t)$ and $\mathrm{B}_2(t)$ as given by Eq.~\ref{eq:cov2}. Details of the calculation are described in Appendix \ref{apdx:derivation}. 
By introducing the dimensionless quantities 
\begin{equation}
\eta =\kappa_t T_1,\;\; \tau = \kappa_t T_0,\;\; \gamma= \kappa_x L,
\end{equation}
which denote the dimensionless shuttling time, shuttling length, and the delay, respectively, measured in the units of the correlation time $\tau_c$ and the correlation length $\lambda_c$, we can express the result in the following form:
\begin{equation}
\label{eq:epr_pair_solution}
\chi(t_0) \equiv \chi(\eta,\gamma,\tau) = (\sigma_{\mathrm B}/\kappa_t)^2 \left[\sum_{j=1}^4 P_j - \sum_{j=1}^4 C_j\right],
\end{equation}
where the quantities
\begin{eqnarray}
P_1&=&2\, \eta^2 \frac{e^{-\eta-\gamma}+\eta+\gamma-1}{(\eta+\gamma)^2},
\label{eq:noncorr_terms_1}
\\ \nonumber
P_2&=&2 (\tau+e^{-\tau}-1),
\\ \nonumber
P_3&=&P_4=\eta \frac{\left(1-e^{-\tau}\right) \left(1-e^{-\eta-\gamma}\right)}{\eta+\gamma},
\label{eq:noncorr_terms_2}
\end{eqnarray}
come from the autocorrelations of $\mathrm{B}_1(t)$ and $\mathrm{B}_2(t)$, while $C_j$ describe the cross-correlation terms, and have different form for $T_0>T_1$ and $T_0\le T_1$. In the former case ($T_0>T_1$), $C_j=C'_j$, where
\begin{widetext}
\begin{eqnarray}
C_1' &=& {\mathrm e}^{-\eta -\gamma -\tau} - 2 {\mathrm e}^{-\gamma} (\eta -\tau) + {\mathrm e}^{\eta-\gamma-\tau} - 2 {\mathrm e}^{-\eta-\gamma} 
\\ \nonumber
C_3' &=& \eta^2 {\mathrm e}^{-\tau} \left[\frac{{\mathrm e}^{-\eta-\gamma}+\eta+\gamma-1}{(\eta+\gamma)^2} 
+\frac{{\mathrm e}^{-(\gamma-\eta)}+\gamma-\eta-1}{(\gamma-\eta)^2}\right] 
\\ \nonumber
C_2' &=& C_4' = \eta\left[\frac{2(1-{\mathrm e}^{-\gamma})}{\gamma} - \frac{{\mathrm e}^{-\tau}(1-{\mathrm e}^{\eta-\gamma})}{\gamma-\eta} - \frac{1-{\mathrm e}^{-\eta-\gamma}}{\eta+\gamma}\right],
\label{eq:corr_terms_1}
\end{eqnarray}
while in the latter case, $T_0\le T_1$, the quantities $C_j=C''_j$, where
\begin{eqnarray}
C_1'' &=& \left({\mathrm e}^{\tau}-1\right)^2 {\mathrm e}^{-\gamma-\eta-\tau}
\\ \nonumber
C_3'' &=& \eta^2 
\biggl\{\frac{{\mathrm e}^{-\tau\gamma/\eta} \left[\eta-\gamma - \tau(1-\gamma/\eta)+1\right] - {\mathrm e}^{-\tau} \left(\eta-\gamma+1\right)}{(\eta-\gamma)^2}
\\ \nonumber
&&\quad+\frac{{\mathrm e}^{-\tau} \left({\mathrm e}^{-\gamma-\eta}+\gamma+\eta-1\right) + {\mathrm e}^{\tau-\eta-\gamma} + {\mathrm e}^{-\frac{\tau \gamma}{\eta}} \left[\eta+\gamma-\tau(1+ \gamma/\eta)-1\right]}{(\eta+\gamma)^2}
\biggr\}
\\ \nonumber
C_2'' &=& C_4'' = \eta \biggl\{\frac{\eta}{\gamma}
\left(\frac{1}{\eta+\gamma}+\frac{1}{\eta-\gamma}\right)\left(1 - {\mathrm e}^{-\frac{\gamma\tau}{\eta}}\right) - (1-{\mathrm e}^{-\tau})\left[
\frac{1}{\eta -\gamma }+\frac{{\mathrm e}^{-\eta -\gamma +\tau }}{\eta +\gamma}
\right]
\biggr\}.
\label{eq:corr_terms_2}
\end{eqnarray}
\end{widetext}
The term $P_1$ characterizes the phase accumulation of the first spin during its travel, while $P_2$ represents the effect of the delay $T_0$, and has the same form as dephasing under the action of an OU noise. The terms $P_3$ and $P_4$ correspond to the autocorrelations of the noise during the delay ($0<t<T_0$) and during travel of the second spin ($T_0<t<T_1+T_0$). The cross-correlation terms $C_j$ have similar meaning, but more complicated form.

\begin{figure}[tbp!]
\centering
\includegraphics[width=1\linewidth]{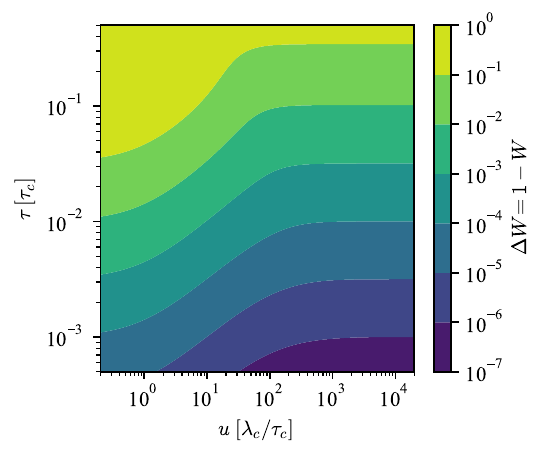}
\caption{Fidelity drop $\Delta W$ during shuttling of two entangled qubits as a function of the dimensionless delay time $\tau=\kappa_t T_0=T_0/\tau_c$ and dimensionless velocity $u=v\,\tau_c/\lambda_c=v\,\kappa_x/\kappa_t$. The qubits are assumed to be shuttled over the distance $L=100\,\lambda_c$, the noise amplitude is $\sigma_{\mathrm B}=\sqrt{2}\,\kappa_t$. With the experimental parameters given in Table~\ref{tab:experimental_params}, this corresponds to $L = 10\,\mu\textrm{m}$, $\tau_c=1/\kappa_t=20\,\mu\textrm{s}$, $\lambda_c=1/\kappa_x=0.1\,\mu \textrm{m}$, and the dephasing time of a static qubit $T_{2,s}^*=\sqrt{2}/\sigma_{\mathrm B} = 20\,\mu\textrm{s}$; the delay time (vertical axis) would be measured in units of $\tau_c$, while the dimensional velocity $v$ would be measured in units of $\lambda_c/\tau_c=5\,\mathrm{mm}/{\mathrm s}$.}
\label{fig:F_surface_scan}
\end{figure}

Sufficiently full analytical investigation of the result given above is rather difficult due to its complexity. 
Therefore, we study its properties in several limiting cases, which seem most interesting and relevant.

First, in order to get a general idea about the dephasing factor in the experimentally relevant region of parameters, in Fig.~\ref{fig:F_surface_scan} we plot the fidelity drop during the shuttling, quantified as 
\begin{equation}
\Delta W=1-W(t_0),
\end{equation}
as a function of the velocity $v$ and the delay $T_0$, using the typical parameters specified in Table~\ref{tab:experimental_params}.
As intuitively expected, high velocity and small delay greatly improve the fidelity of the logical qubit shuttling. In order to connect these results to the experiment, we use the typical expected parameters of Si/SiGe semiconductor shuttling systems given in Table~\ref{tab:experimental_params}. With these parameters, Fig.~\ref{fig:F_surface_scan} graph suggests that with shuttling velocity of the order of 1--10~$\mathrm{m}/\mathrm{s}$ and the delay in the range 10--100~ns it is possible to achieve the error around $10^{-5}$ when shuttling the logical qubit over 10~${\mu}$m.
However, these estimates do not take into account other sources of dephasing and errors, and cannot be taken as definitive. 

In order to gain better qualitative understanding of the regimes favoring the two-qubit shuttling in comparison with shuttling of a single qubit, let us consider two intuitively clear limits, $\kappa_x=0$ and $\kappa_x\to\infty$. When $\kappa_x=0$, in the limit of infinitely large correlation length, the random fields $\mathrm{B}_1(t)$ and $\mathrm{B}_2(t)$ are equal, and the two-spin system is subjected to a collective noise, where the concept of the decoherence-free subspace works the best. In that case, $W(t_0)=1$, and the two-qubit system is completely protected from decoherence. In that limit, obviously, the two-qubit shuttling is advantageous.

\begin{table}[tbp!]
\caption{Typical parameters of a realistic experimental device.}
\label{tab:experimental_params}
\begin{ruledtabular}
\begin{tabular}{lll}
Symbol  & Range  & Description
\\ 
\hline
$v$  & $10^{-3}$--$10^2\,\textrm{m/s}$  & Shuttling velocity \\ 
$L$     & $10\,\mu \textrm{m}$  & Shuttling length   \\ 
$\tau_{c}$  & $20\,\mu \textrm{s}$ & Correlation time of $\tilde{B}(x,t)$ \\
$\lambda_{c}$  & $100\, \textrm{nm}$ & Correlation length of $\tilde{B}(x,t)$ \\
$T_{2,s}^{*}$ & $20\,\mu \textrm{s}$ & Dephasing time of a static spin qubit
\end{tabular}
\end{ruledtabular}
\end{table}

In the opposite limit of $\kappa_x\to\infty$, the quantity $\chi=-\ln{W(t_0)}$ in Eq.~\ref{eq:epr_pair_solution} is
\begin{equation}
\chi(\eta, \gamma, \tau)\approx 2 \sigma_\mathrm{B}^2 \kappa_t^{-2} \left(\tau +{\mathrm e}^{-\tau }-1\right),
\label{eq:uncorrelated_noise_soluation}
\end{equation}
such that dephasing is controlled solely by the delay $\tau$.
Indeed, for $\kappa_x\to\infty$, the random magnetic fields at different spatial locations are uncorrelated, and the two spins are dephased independently of each other; besides, their dephasing during shuttling happens in the regime of extreme motional narrowing (see Eq.~\ref{eq:wT} in the limit $v\to\infty$), when the dephasing rate is very small. Thus, the only time interval where dephasing takes place is the delay time $T_0$, when the dephasing happens at the fixed spatial locations ($x_c=0$ and $x_c=L$) under the action of an OU noise with the correlation constant equal to $\kappa_t$. In that case, shuttling of a single qubit is better, while the two-qubit system suffers decoherence during the waiting time.

\begin{figure}[tbph!]
\centering
\includegraphics[width=\linewidth]{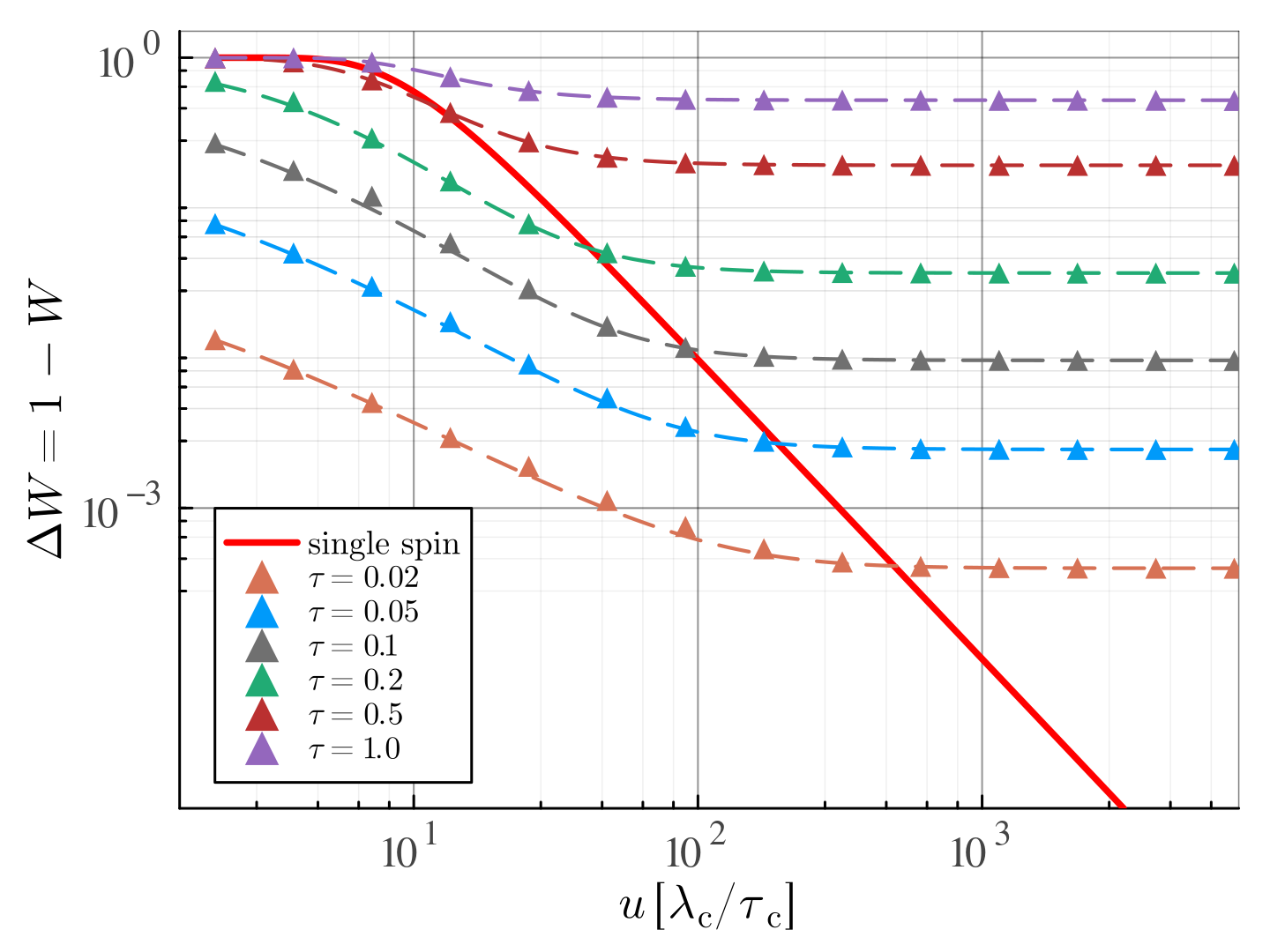}
\caption{Fidelity drop $\Delta W$ during shuttling of two entangled qubits (dashed lines with symbols) as a function of the dimensionless velocity $u=v\,\kappa_x/\kappa_t$ for different values of the delay $\tau=\kappa_t T_0$, compared with the fidelity drop of a single qubit (solid red line). Calculations have been performed for $\gamma=100$ and $\sigma_{\mathrm B}=\sqrt{2}\,\kappa_t$, same as in Fig.~\ref{fig:F_surface_scan}.
With the experimental parameters given in Table~\ref{tab:experimental_params}, the delay times corresponds to the delay times $T_0=0.4$, $1.0$, $2.0$, $4.0$, $10.0$ and $20\ \mu$s.}
\label{fig:F_single_vs_entangle}
\end{figure}

Extending the comparison between the single- and the two-qubit shuttling, we analyze the dependence of the fidelity drop on the shuttling velocity. 
In the limit of large velocities, $u\to\infty$, keeping the delay $\tau$ and the shuttling length $\gamma$ fixed, we have
\begin{equation}
\label{eq:stlargeu}
\chi(\eta=\gamma/u,\gamma,\tau)\approx 2\, \frac{\sigma_\mathrm{B}^2}{\kappa_t^2}\,(1-\mathrm{e}^{-\gamma})\left(\tau + \mathrm{e}^{-\tau }-1\right),
\end{equation}
where we used Eqs.~\ref{eq:corr_terms_1} corresponding to $\tau>\eta$.
In the opposite limit $u\to 0$, with fixed $\tau$ and $\gamma$, when the travel time $\eta =\gamma/u\to\infty$, we obtain 
\begin{equation}
\label{eq:stsmallu}
\chi(\eta=\gamma/u, \gamma,\tau)\approx 2 \left(\sigma_\mathrm{B}/\kappa_t\right)^2 \gamma \left(\tau +e^{-\tau}-1\right).
\end{equation}
It is important to note that in both limits the dephasing factor saturates at the finite values, determined by the shuttling length $\gamma$ and the delay $\tau$. In contrast, when a single unprotected qubit is shuttled over the length $\gamma$ with the velocity $u$, the corresponding decoherence factor $W_0(t_0)$ quickly goes to zero for slow shuttling,
\begin{equation}
\label{eq:w0smallu}
W_0(t_0)\approx \exp{-\frac{\sigma_\mathrm{B}^2}{\kappa_t^2}\, \frac{\gamma}{u}}\ \ \mathrm{at}\ \ u\to 0,
\end{equation}
but approaches 1 for fast shuttling,
\begin{equation}
\label{eq:w0largeu}
W_0(t_0)\approx \exp{-\frac{\sigma_\mathrm{B}^2}{\kappa_t^2}\, \frac{\gamma+\mathrm{e}^{-\gamma}-1}{u^2}}\ \ \mathrm{at}\ \ u\to\infty.
\end{equation}
The two modes of shuttling are compared in Fig.~\ref{fig:F_single_vs_entangle}. 

As follows from the analysis above, fidelity of the shuttling of a single physical qubit is always better if the shuttling speed can be increased indefinitely. However, at large shuttling speed the qubit becomes susceptible to  other sources of errors, such as the loss of wavepacket's adiabaticity (the wavepacket is no longer confined to a single minimum of the shuttling potential), excitation to higher orbital states within the minimum of the shuttling potential and to excited valley states; also, the influence of the spin-orbit coupling and the errors caused by the interface roughness (such as atomistic steps at the interface) become devastating at high shuttling velocities 
\cite{seidler_conveyor-mode_2022,langrock_blueprint_2023,StruckSchreiberEtalSpinPairShuttling23,HuangHuSpinRelaxShuttl13,BoscoZouLossHighFidShuttlingSOI23}. Thus, shuttling at very high speed may be not only difficult, but also extremely counter-productive.

In contrast, our analysis shows that even very sluggish shuttling of two entangled qubits can achieve high fidelity as long as the delay time is kept short enough. In this regime, high fidelity of the shuttling can be maintained over large distances (see Eq.~\ref{eq:stsmallu} for $\gamma\gg 1$) provided that $\tau$ is sufficiently small. Slow shuttling nullifies various errors associated with non-adiabaticity and suppresses the loss of fidelity caused by the interface roughness and spin-orbit coupling. Moreover, sufficiently slow shuttling may become insensitive to the spatial variations of the quantization axis in Ge-based semiconductor structures \cite{WangVeldhorstEtalGeQDs24} if the shuttling is slow enough to be adiabatic with respect to such variations, thus eliminating yet another possible problem.

More generally, considering shuttling over large distances in the limit of short delays, we find in the limit $\tau\to 0$
\begin{eqnarray}
\chi(\eta=\gamma/u,\gamma,\tau) &\approx & \frac{\sigma_\mathrm{B}^2}{\kappa_t^2}\, \tau^2\,\left[\frac{\gamma}{1+u} \right. 
\\ \nonumber
&+& \left. \left(1-\mathrm{e}^{-\gamma\frac{1+u}{u}}\right)\frac{u^2}{(1+u)^2} \right],
\end{eqnarray}
which for the experimentally interesting case of large-distance shuttling, $\gamma\gg 1$, becomes 
\begin{equation}
\label{eq:stsmalltau}
\chi(\eta=\gamma/u,\gamma,\tau)\approx \left(\sigma_\mathrm{B}/\kappa_t\right)^2 \tau^2 \frac{\gamma}{(1+u)}.
\end{equation}
The behavior described by Eqs.~\ref{eq:stlargeu}, \ref{eq:stsmallu} and \ref{eq:stsmalltau} is seen in Fig.~\ref{fig:F_single_vs_entangle}: for small $\tau$, the curves describing dephasing of two entangled spins demonstrate the approximate dependence $\propto 1/(u+1)$ before saturating at the finite value given by Eq.~\ref{eq:stlargeu}, and the onset of saturation begins at larger and larger values of $u$ as the delay $\tau$ gets smaller.
In comparison, the fidelity drop for a single qubit decreases faster, as $1/u^2$ for large $u$ (as given by Eq.~\ref{eq:w0largeu}), but at small velocities has a very flat top, corresponding to extremely fast decoherence predicted by Eq.~\ref{eq:w0smallu}.

Thus, our analysis shows that slow shuttling of two entangled qubits that encode a single logical qubit is a very attractive possibility. It eliminates a plethora of problems associated with the loss of shuttling adiabaticity, suppresses several important decoherence channels, but allows shuttling over large distances with very high fidelity, simply by making the delay $T_0$ sufficiently small. It is also worthwhile to point out a very favorable scaling of the fidelity as a function of $\tau=T_0/\tau_c$: Eqs.~\ref{eq:stsmallu} and \ref{eq:stsmalltau} predict that the fidelity drop $\Delta W$ and the achievable shuttling distance $L=\gamma \lambda_c$ both improve quadratically as $\tau$ becomes smaller.

\section{Numerical simulations: Benchmarks and accuracy 
\label{sec:results}}

In Sec.~\ref{sec:method}, we discussed two numerical methods for calculating general dephasing factors originating from the noise produced by a random sheet: one method is based on numerical integration (NI) of the double integral in Eq.~\ref{char_fun_dbl_integr} and the other is based on the Monte Carlo (MC) sampling of the integral (\ref{int_basic}). In that section we focused on mathematical justification and evaluating the accuracy of the MC sampling.
The three examples of different shuttling regimes, which we have considered in the previous Section, constitute excellent benchmark examples for the numerical methods proposed in Sec.~\ref{sec:method}: they involve sufficiently complex noise correlations, but at the same time admit explicit analytical solutions.

Correspondingly, we performed both NI and MC numerical calculations of the dephasing factors for all three regimes (one-way shuttling of a single qubit, forth-and-back shuttling of a single qubit, and shuttling of two entangled qubits), and compared them with the analytical solutions given in the previous Section. In all three cases, with sufficiently large number of mesh points, the numerical results practically coincided with the analytical predictions, thus producing a useful cross-check of the two approaches. Thus, the main focus on this Section is the analysis of accuracy of the NI and MC numerical methods.

We point out right away that for sampling of the noise produced by the OU sheet, we did not use the popular method based on the Kosambi--Karhunen--Lo{\`e}ve expansion, since it requires exponential number of the mesh points, as discussed in Sec.~\ref{sec:method}. Instead, we used the approach based on Cholesky decomposition of the correlation matrix \cite{Skordos2008} (see the end of Sec.~\ref{sec:method}), which does not suffer from such a problem.

Within both numerical approaches, we discretized the continuous trajectories $x_c(t)$ (or $x_{c1}(t)$ and $x_{c2}(t)$ respectively) using equidistant set of time points $t^{(1)}\dots t^{(n+1)}$, thus producing a set of mesh points $(t^{(1)},x^{(1)}), (t^{(2)},x^{(2)}) \dots (t^{(n+1)},x^{(n+1)})$. The number $n$ of the mesh intervals is chosen large enough to ensure that the distances between the adjacent points in time, $\Delta t$, and in  space, $\Delta x$, are small in comparison with the corresponding correlation scales, $\Delta t\ll \tau_c$ and $\Delta x\ll \lambda_c$; also, for all examples considered here, $\Delta x$ does not exceed $v\,\Delta t$, where $v$ is the shuttling velocity.

\begin{figure}[h!]
\centering
\includegraphics[width=1\linewidth]{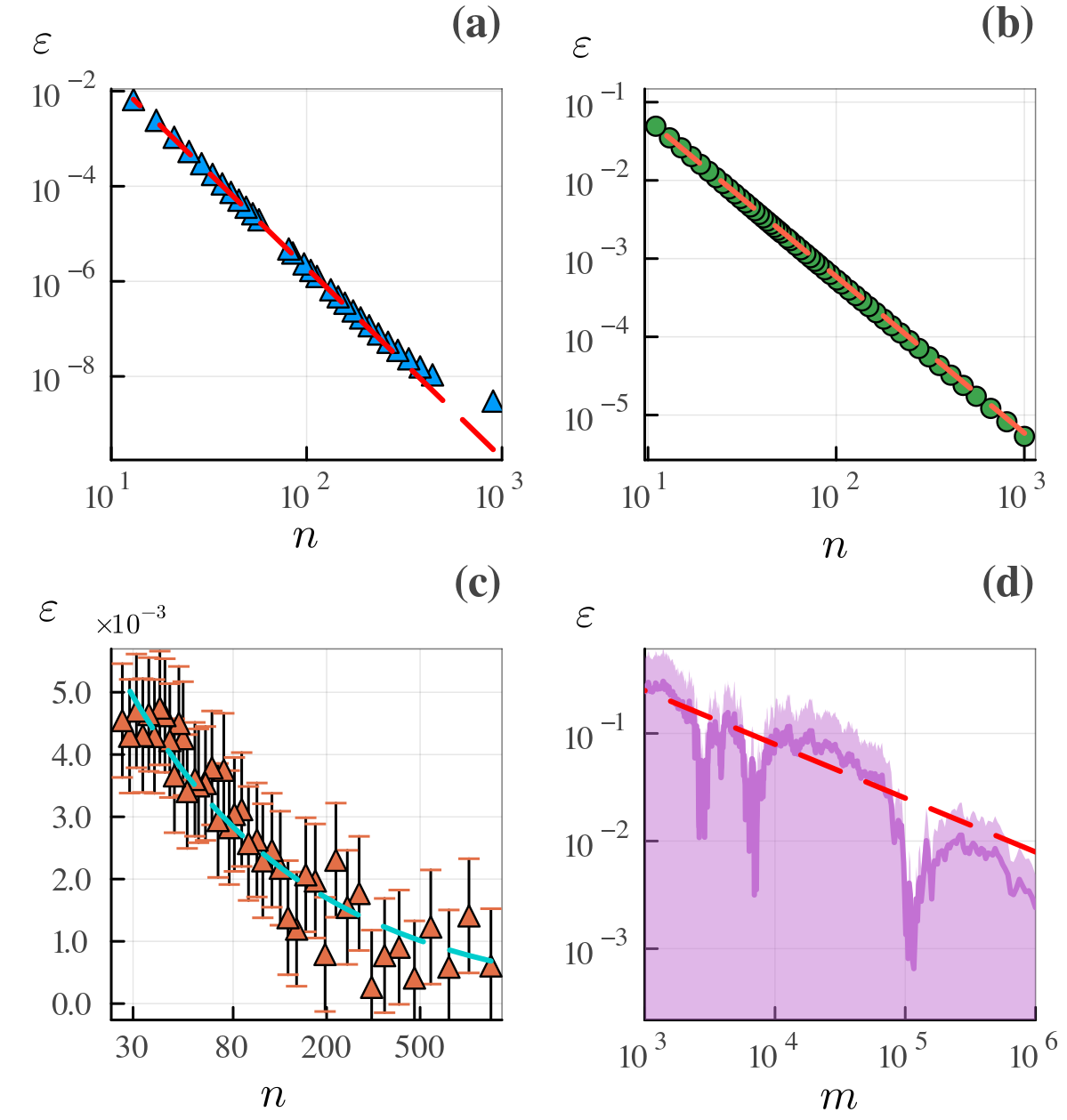}
\caption{Accuracy of the numerical methods, NI and MC sampling, for one-way shuttling of a single spin. In all panels, the error is defined as $\varepsilon=|W(T)-W_\mathrm{num}|/W(T)$, i.e.\ the relative difference between the exact analytical answer (\ref{eq:wT}) for the decoherence factor $W(T)$ and the numerically calculated value $W_\mathrm{num}$ of this quantity.
All graphs correspond to $\beta=T/\tau_c=1$, $\gamma=L/\lambda_c=5$ and $\sigma_\mathrm{B}=\kappa_t=1/\tau_c$.
{\bf (a)}: Error of the numerical integration taking into account the corner singularity of the correlation function $K_\mathrm{B}(t_1,t_2)$ as a function of the number $n$ of the mesh intervals. The dashed red line shows the results of fitting to the function $\varepsilon=A/n^4$, demonstrating very good agreement. 
{\bf (b)}: Error of the numerical integration when the corner singularity is ignored. The dashed red line shows the results of fitting to the function $\varepsilon=A/n^2$, showing very good agreement. {\bf (c)}: Error of the MC evaluation of the dephasing factor as a function of $n$. The numerical results were fitted with the function $\varepsilon=A\,n^{-\varkappa}$ (blue dashed line), the least-square fitting procedure gives  $\varkappa=0.563$, $A=3.346\cdot 10^{-2}$, close to the expected estimate $O(\log{n}/\sqrt{n})$ obtained in Sec.~\ref{subsubsec:MC} and Appedix~\ref{app:mceval}. The total number of the MC samples is $m=10^6$. 
{\bf (d)}: Statistical MC sampling error as a function of the total number $m$ of the MC samples for $n=300$ mesh intervals. The numerical results were fitted with the function $\varepsilon=A\,\sqrt{m}$ (dashed red line), showing the expected agreement.}
\label{fig:num_singlespin}
\end{figure}

For NI method, the specific forms of the general double integral (\ref{char_fun_dbl_integr}), given by Eqs.~\ref{eq:specific1} and \ref{eq:chi12}, were calculated as repeated one-dimensional integrals, subsequently over $t_1$ and then over $t_2$. Each one-dimensional integral was calculated using the Simpson's $1/3$ rule \cite{suli2003introduction,NumRecipes}
\begin{equation}
\int\limits_{t+\Delta t}^{t-\Delta t}f(s)\,ds \approx \frac {\Delta t}{3}
\left[f(t-\Delta t) + 4 f(t) + f(t+\Delta t)\right].
\end{equation}
When the integration is performed over the whole time interval $[0,t_0]$ of the shuttling process, the numerical error is equal to 
\begin{equation}
{\frac {{\Delta t}^4}{180}}\, t_0\, \max_{t\in [0,t_0]} \left|f^{(\mathrm{iv})}(t)\right|,
\label{eq:simpson_err}
\end{equation}
scaling as $1/n^4$ with increasing the number of intervals $n$, provided that the function $f(t)$ has finite derivative of the fourth order $f^{(\mathrm{iv})}(t)$.

However, the correlation function $K_{\mathrm{OU}}((x,t),(x',t'))$ of the OU sheet (\ref{eq:random_sheet_corr}) has a corner-type singularity along the lines $t=t'$ and $x=x'$. As a result, already for a simplest case of one-way shuttling of a single spin, the corresponding correlation function $K_\mathrm{B}(t_1,t_2)$ has the same singularity at $t_1=t_2$, with formally infinite fourth derivative, such that the estimate (\ref{eq:simpson_err}) above cannot be used.
For a simple case of one-way shuttling of a single spin, this difficulty is easily circumvented: the corresponding correlation function $K_\mathrm{B}(t_1,t_2)$, given by Eq.~\ref{eq:sigmaBdef}, can be integrated separately over the regions $t_1>t_2$ and $t_1<t_2$ 
\footnote{Due to the symmetry of $K_\mathrm{B}(t_1,t_2)$ in that case, it is sufficient to perform integration over only one of those regions.}, 
avoiding singularity at $t_1=t_2$. In that case, the accuracy of the order $1/n^4$ is achieved. This is confirmed by the numerical results shown in Fig.~\ref{fig:num_singlespin}(a). The accuracy of the numerical calculations is quantified using the relative error $\varepsilon=|W(T)-W_\mathrm{num}|/W(T)$, i.e.\ the relative difference between the exact decoherence factor $W(T)$ given by Eq.~\ref{eq:wT} and the numerically calculated value $W_\mathrm{num}$ of this quantity.

On the other hand, shuttling of two spins (logical singlet-triplet qubit) already involves a much more complex pattern of the corner-type singularities; such patterns may become rather difficult to handle in the case of more spins traveling along more complex trajectories. Moreover, in more complex situations with more complex noise, possible singularities of the correlation function can be difficult to predict and locate. Thus, it is useful to test the NI method when the corner-type singularities are ignored, as if we are not aware of their existence.
The corresponding results for a single-spin one-way shuttling are shown in Fig.~\ref{fig:num_singlespin}(b), showing that the NI error scales as $1/n^2$, which is expected when a corner-type singularity is approximated by a parabola. Thus, numerical integration could be a viable and efficient way of calculating the decoherence factors even if the singularities of the correlation function are not known.

\begin{figure}[h!]
\centering
\includegraphics[width=1\linewidth]{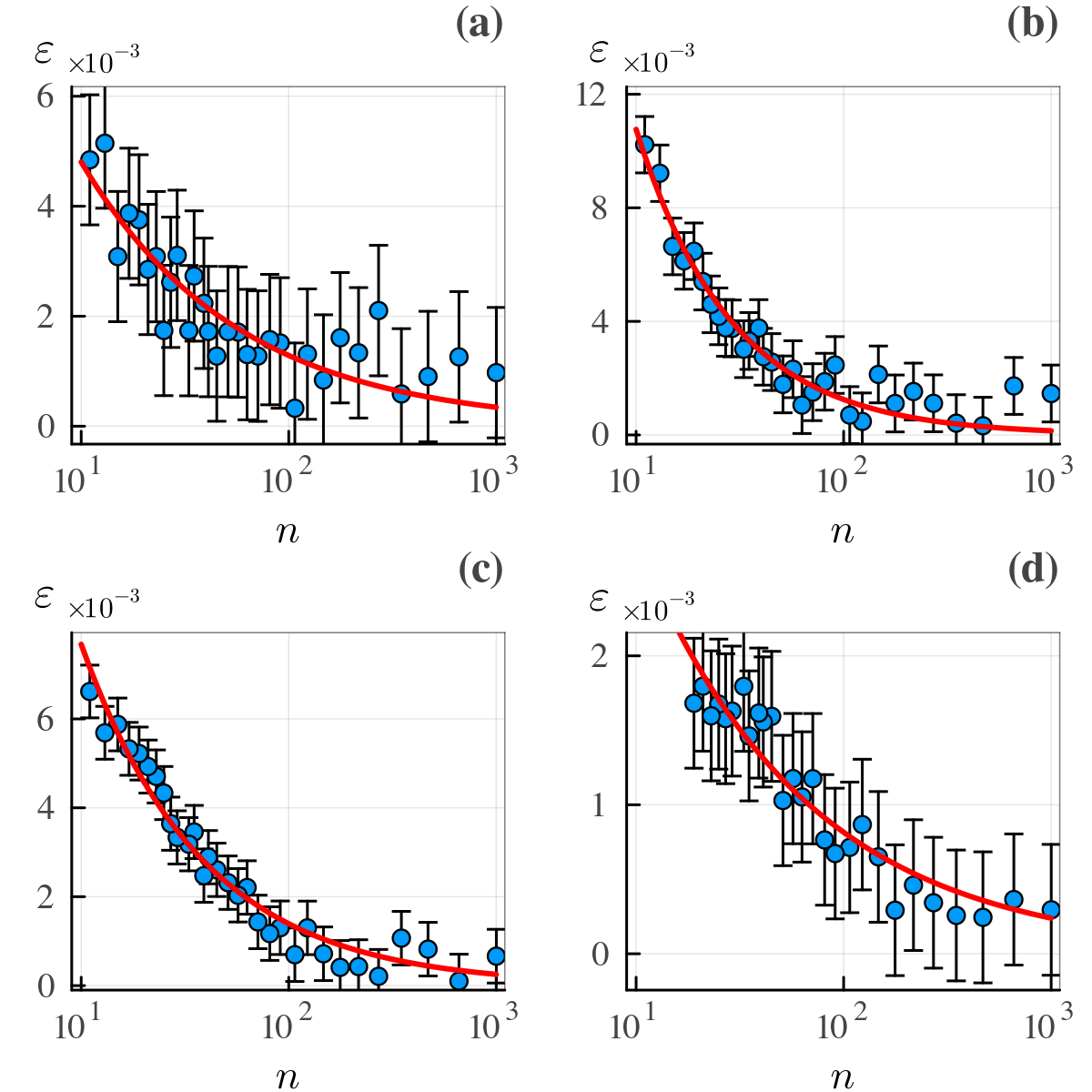}
\caption{Numerical error $\varepsilon$ of the MC sampling as a function of the number $n$ of the mesh intervals. 
All panels present numerical results for the single-spin one-way shuttling, with the shuttling time $\beta=T/\tau_c=0.1$ and $\sigma_\mathrm{B}=\kappa_t=1/\tau_c$, with varying shuttling length, {\bf (a)}: $\gamma=0.1$, {\bf (b)}: $\gamma=0.5$, {\bf (c)}: $\gamma=2.5$, {\bf (d)}: $\gamma=5$.
The error $\varepsilon$ is defined in the same way as in Fig.~\ref{fig:num_singlespin}. 
The calculated values of $\varepsilon$ were fitted with a power-law function 
$\varepsilon=A n^{-\varkappa}$ (solid red lines), the least-square fitting parameters are 
$\varkappa=0.572$, $A=1.794\cdot 10^{-2}$ in (a); $\varkappa=0.9355$, $A=9.265\cdot 10^{-2}$ in (b); 
$\varkappa=0.741$, $A=4.232\cdot 10^{-2}$ in (c); $\varkappa=0.533$, $A=0.949\cdot 10^{-2}$ in (d). 
In all cases, the error decreases in reasonable agreement with (or faster than) the expected scaling $\varepsilon=C\,\log{n}/\sqrt{n}$.}
\label{fig:num_1spin_det}
\end{figure}

Along with the NI results, Fig.~\ref{fig:num_singlespin} presents the results of the method based on MC sampling. The accuracy of the MC method for calculation of the integral (\ref{int_basic}) is determined by two factors: (i) the accuracy of discretization of the integral, i.e.\ representing it as a sum, as in Eqs.~\ref{eq:discr1} or \ref{eq:discr2}, and (ii) accuracy of approximating the expectation value via average over a finite number $m$ of the samples of the corresponding random process. These errors are assessed, respectively, in panels (c) and (d) of Fig.~\ref{fig:num_singlespin}. The results show that the statistical error, associated with finite number $m$ of the MC samples used for averaging, decreases in accordance with the expected scaling $1/\sqrt{m}$. The discretization error also decreases in accordance with (or faster than) the estimate $O(\log{n}/\sqrt{n})$ suggested by the analysis in Sec.~\ref{subsubsec:MC} and Appendix~\ref{app:mceval}. 

The latter aspect is illustrated in a more detailed manner in Fig.~\ref{fig:num_1spin_det}, where the discretization error of the MC method is plotted for different values of the shuttling length $\gamma$ with fixed shuttling time $\beta=0.1$, i.e.\ for different shuttling velocities. These simulations were performed with $m=10^6$ (to decrease the statistical sampling error), and the number of the discretization intervals was varied from $n=10$ to $n=10^3$. Again, the results indicate that the estimate $\varepsilon\sim \log{n}/\sqrt{n}$ describes the actual numerical simulations reasonably well (or could be even somewhat pessimistic for some regimes).

\begin{figure}[hp!]
\centering
\includegraphics[width=1\linewidth]{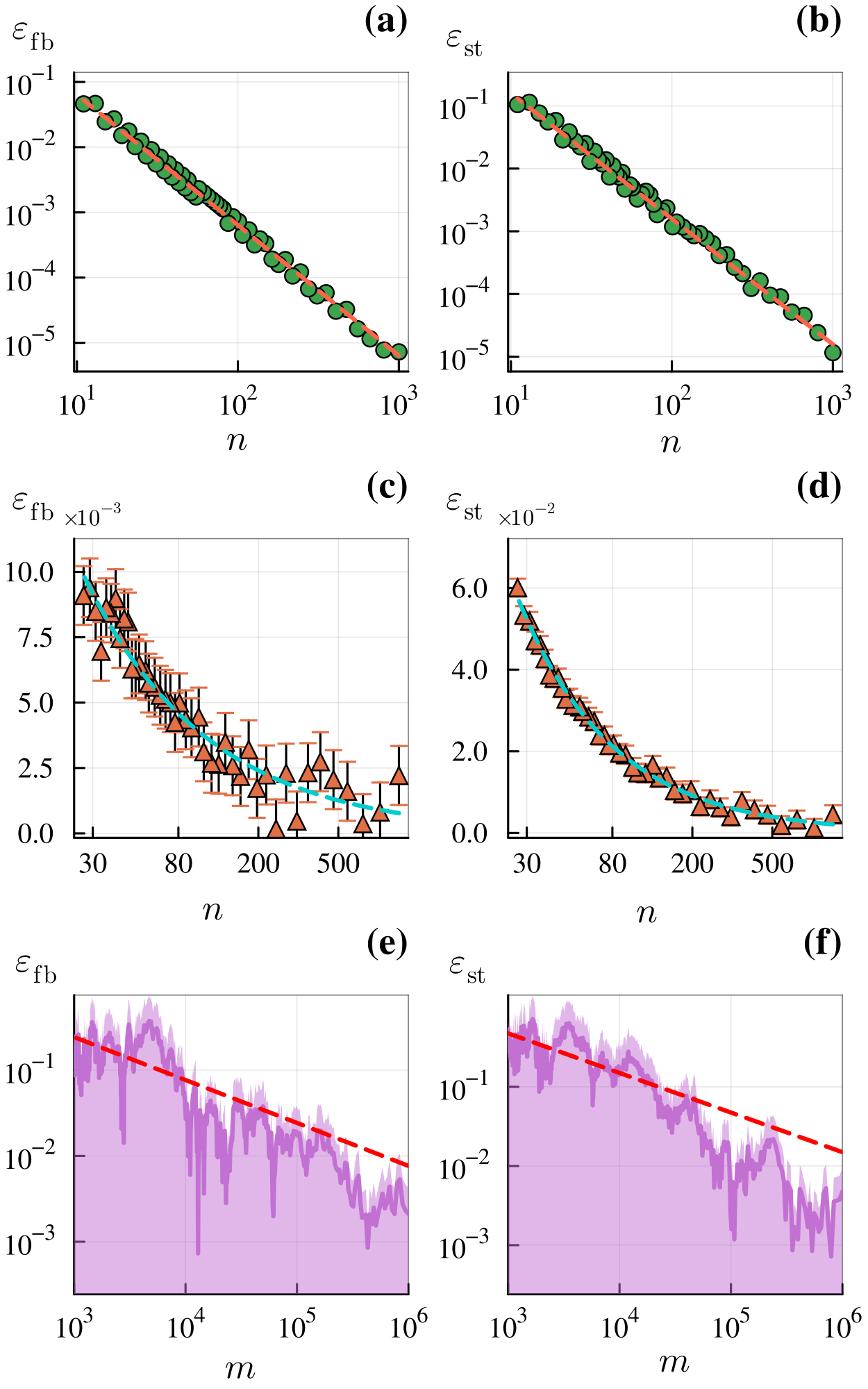}
\caption{
Accuracy of the numerical calculations of the dephasing factor for the forth-and-back shuttling of a single qubit (panels {\bf a}, {\bf c} and {\bf e}), and for the shuttling of two entangled qubits encoding the logical singlet-triplet qubit (panels {\bf b}, {\bf d} and {\bf f}).
Calculations for the forth-and-back shuttling were performed for $\beta=1$, $\gamma=5$ and $\sigma_\mathrm{B}/\kappa_t=1$; for the singlet-triplet qubit, the parameters $\tau=1$, $\eta=1.5$, $\gamma=10$ and $\sigma_\mathrm{B}/\kappa_t=1$ were used. The error $\varepsilon$ is defined in the same way as in Fig.~\ref{fig:num_singlespin}. 
{\bf (a), (b)}: Error of the NI method as a function of the number $n$ of the mesh intervals, calculated with ignoring the corner singularities of the covariance function. The dashed red lines show the fitting to the function $\varepsilon=A/n^2$, demonstrating good agreement with the expected scaling $1/n^2$.
{\bf (c), (d)}. Error $\varepsilon$ of the MC sampling as a function of the number $n$ of the mesh intervals. The number of the MC samples is $m=10^6$.
The calculated values of $\varepsilon$ were fitted with a power-law function 
$\varepsilon=A n^{-\varkappa}$ (dashed blue lines), the least-square fitting parameters are 
$\varkappa=0.705$, $A=3.346\cdot 10^{-2}$ in (c) and $\varkappa=0.918$, $A=1.189$ in (d). The error decreases in reasonable agreement with (or faster than) the expected scaling $\varepsilon\sim\log{n}/\sqrt{n}$.
{\bf (e), (f)}. Error $\varepsilon$ of the MC sampling as a function of the number $m$ of the MC samples. The number of the mesh intervals is $n=300$. 
The numerical results were fitted with the function $\varepsilon=A\,\sqrt{m}$ (dashed red line), showing the expected agreement.}
\label{fig:numerical_error}
\end{figure}

Qualitatively similar results are obtained when these numerical methods are applied to the analysis of more complex shuttling scenarios, forth-and-back shuttling of a single spin and sequential shuttling of two spins (logical singlet-triplet qubit). Fig.~\ref{fig:numerical_error}(a) and (b) show the scaling of the NI error as a function of the number $n$ of the discretization  intervals. Since NI was performed with intentionally ignoring the corner-type singularity, the error scales in agreement with the expected scaling $1/n^2$ as the number $n$ of the intervals increases. For MC sampling, as shown in Fig.~\ref{fig:numerical_error}(c) and (d), the error decreases 
in agreement with (or faster than) the estimate $\log{n}/\sqrt{n}$. The statistical sampling error, shown in Fig.~\ref{fig:numerical_error}(e) and (f), decreases as expected, $\varepsilon\propto 1/\sqrt{m}$, with increasing the number $m$ of the MC samples.

Overall, we see that both numerical methods can be reliably used for numerical calculation of the dephasing factor, and the level of errors seen in several representative tests is comparable or better than expected from our discussion in Sec.~\ref{sec:method}.

Finally, note that the computational resources required for NI method and MC sampling are different. NI requires $O(n^2)$ operations for calculating the double integral for a general correlation function $K_\mathrm{B}(t_1,t_2)$. In contrast, MC sampling involves calculation of only one-dimensional integral (\ref{int_basic}), i.e.\ calculation of the sums like (\ref{eq:discr1}) and (\ref{eq:discr1}), which is to be repeated $m$ times, with the total operation count $O(mn)$. On the other hand, sampling of the noise $\mathrm{B}(t)$ via Cholesky decomposition, as discussed in Sec.~\ref{subsubsec:MC}, requires $O(n^3)$ operations for decomposition of a $(n+1)\times (n+1)$ covariance matrix $\mathbf{C}$. This is followed by multiplication of the $n+1$-dimensional column vector $w$ of i.i.d.\ normal variables  to the Cholesky matrix $L$, which involves $O(n^2)$ operations. However, we need only the value of the sums (\ref{eq:discr1}) or (\ref{eq:discr2}), not the values of individual terms in these sums. Therefore, we can pre-multiply the matrix $L$ from the left by the row vector $\xi\left((s_j + s_{j-1})/2\right)$; in this way, we can compute the relevant sums by simply multiplying the column vector $w$ by the resulting row vector with only $n+1$ multiplications and $n+1$ additions. This multiplication has to be repeated $m$ times. Thus, the MC sampling method requires in total $O(n^2+mn)$ operations. However, accuracy of the MC method scales as $1/\sqrt{m}$ and $\log{n}/\sqrt{n}$, while accuracy of the NI method scales as $1/n^2$ in the worst-case scenario. Thus, in order to achieve given accuracy in applications, it is better to use the NI method if possible, and resort to the MC sampling only if it is required by the nature of the problem (e.g.\ if the covariance function $K_\mathrm{B}(t_1,t_2)$ is not known).

\section{Summary and Outlook}
\label{sec:summary}

Realistic modeling and good understanding of decoherence during multi-qubit transport is crucial for developing, benchmarking and optimization of  distributed quantum information processing. In order to model decoherence taking place during the qubit transport, it is critical to take into account realistic correlations of the noise in both time and space. In this work we propose an approach for describing the dephasing process under such conditions, and develop a mathematical framework based on the analysis of trajectories on random sheets. The random sheets \cite{Chentsov_1956,Kitagava_1951} generalize the idea of a random process (random variables indexed with one real parameter, time $t$) to the case of a random variable indexed with multiple real parameters (for the case of a random sheet, space coordinate $x$ and time $t$). As a specific example for application of our approach, we consider electron spins shuttled in a (quasi)-one dimensional semiconductor channel, and demonstrate, how the notion of random sheet can be applied to analysis and assessment of dephasing of single- and two-spin systems under nontrivially correlated random magnetic field ${\tilde B}(x,t)$. 

When an  electron propagates as a tightly localized wavepacket centered at $x=x_c(t)$, its spin experiences random magnetic field ${\mathrm B}(t)={\tilde B}(x=x_c(t),t)$, and the impact of dephasing on the density matrix of the system is completely characterized by the characteristic functional of the field ${\mathrm B}(t)$ (or, for multiple spins, several such fields ${\mathrm B}_{{\mathbf p},{\mathbf p'}}(t)$, see Eqs.~\ref{eq:wManyQ} and \ref{eq:manyspinB}). However, analytical calculation of the characteristic functional in general is a very difficult task, and development of mathematically justified numerical methods is also a challenging problem. Thus, we need to provide a model that, on one hand, is sufficiently realistic to describe the relevant physical effects, and, on the other hand, can be mathematically tractable. 

One possible approach is to analyze the problem in its original formulation, as a motion of the wavepacket $\rho({\vec r},t; x_c(t))$ (see Eq.~\ref{eq:wavepacket}) in the field ${\tilde B}(x,t)$, similarly to the approach traditionally used in analyzing the optical waves or pulses propagating through random media \cite{RytovEtalStatRad,SobczykStochWave}. Such an approach is difficult, and usually requires a significant number of simplifications to render the problem tractable, such that most works use the model of tight wavepacket, see e.g.\  \cite{langrock_blueprint_2023,HuangHuSpinRelaxShuttl13,BoscoZouLossHighFidShuttlingSOI23}.

This model naturally leads to analyzing the effective random field ${\mathrm B}(t)$ acting on the spin. However, statistical properties of the noise ${\mathrm B}(t)$ are determined by {\em both} the properties of the random sheet ${\tilde B}(x,t)$ and the trajectory $x_c(t)$, and therefore can be very different from the commonly studied random processes. For instance, even if ${\tilde B}(x,t)$ is stationary, the derived process ${\mathrm B}(t)$ can be non-stationary. Markov property \cite{MarkovNote} is even more problematic, being not an easily defined concept for random sheets, see Appendix \ref{app:markov}. Fortunately, as we demonstrate in this work, Gaussian property is faithfully inherited by the process ${\mathrm B}(t)$ when the underlying random sheet ${\tilde B}(x,t)$ is Gaussian. We prove that this statement holds for reasonably arbitrary (even somewhat unphysical) trajectories and an arbitrary number of shuttled qubits entangled in a general way, see Sec.~\ref{sec:RSontraj}, even though in general the linear combination of correlated Gaussian variables is not necessarily Gaussian \cite{FujitaYoshidaNonGaussian23}. This is the key result for our approach, as it enables straightforward calculation of the characteristic functional analytically or numerically (in the latter case, either via numerical integration or by Monte Carlo sampling) in a rigorously justified manner. 

This approach is applied to two specific situations, forth-and-back shuttling of a single spin, and sequential shuttling of two entangled spins forming a logical singlet-triplet qubit. The first one is the simplest situation where the random sheet ${\tilde B}(x,t)$ (taken as an Ornstein-Uhlenbeck sheet with the correlation time $\tau_c=1/\kappa_t$ and length $\lambda_c=1/\kappa_x$) gives rise to nontrivially correlated noise ${\mathrm B}(t)$. The dephasing factor for this situation can be analyzed in much detail, and, in particular, we found that (i) the forth-and-back shuttling leads to stronger dephasing in comparison with an equivalent one-way shuttling, (ii) the dephasing times can differ by a factor of 2 at most, and (iii) dephasing regimes can be different in time and space, e.g.\ ballistic in time ($\chi\propto T^2$) and diffusive in space ($\chi\propto L$), see Sec.~\ref{sec:singlequbitshuttling}.

For the second situation, the explicit analytical expression for the dephasing factor has been obtained, but it appears too complex for a detailed analysis. Nevertheless, we obtained several important results. We have shown that high-fidelity shuttling, with the fidelity drop of the order of $10^{-4}$, is possible for current realistic experimental parameters. We also identified the parameter region, where shuttling of two entangled spins is more beneficial than shuttling of a single spin; this is the region of smaller shuttling velocities and shorter delay between the shuttled spins. But, perhaps most importantly, we have shown that the logical singlet-triplet qubit can be shuttled {\em arbitrarily slow} over {\em arbitrarily long distance} with {\em arbitrarily high fidelity} as long as the delay time between the two shuttled spins is small enough.

The latter conclusion can be particularly useful for experimental implementation of the high-fidelity qubit shuttling in semiconductor Si-based architectures. Besides dephasing, analyzed in this work, there is a number of other factors, not explicitly included in the model, which detrimentally affect fidelity of the shuttling. One of them is the deviation from adiabaticity during the shuttling process, and the related excitation to higher orbital states within the shuttling potential and to the excited valley states in Si/SiGe structures \cite{langrock_blueprint_2023,VolmerStruckEtalValleySplit23,JeonBenjaminFisher24RobustChargeShuttl,buonacorsi_simulated_2020}. Defects that locally reduce the valley splitting, such as atomic steps or variations in the local strain, are particularly dangerous in that respect \cite{LosertFriesenValleySplit23,TahanJoynt14,HaoRuskovTahanEtal14}, as they increase the probability of excitation at a given shuttling speed. By decreasing the shuttling velocity, one can exponentially suppress such excitation processes, and the possibility of high-fidelity low-speed shuttling, which is offered by the singlet-triplet qubit, would be of great benefit.

Another serious issue is the impact of spin-orbit coupling on the moving spin \cite{HuangHuSpinRelaxShuttl13,BoscoZouLossHighFidShuttlingSOI23}. Associated fluctuations of the transverse magnetic field lead to longitudinal relaxation of the spin, thereby reducing fidelity. Shuttling of a singlet-triplet qubit with smaller velocities and little dephasing would reduce the effect of fluctuating spin-orbit field, providing another important advantage.
Our calculations (to be presented elsewhere) with several types of exchange-only qubits \cite{EOqubit1,EOqubit2,EOqubit3,BurkardLaddPanNicholReviewQuDots23} show that these benefits are shared by other DFS-based encoding schemes, thus favoring the use of exchange-only qubits in the spin shuttling architectures.


Our approach can also be extended to a new, highly promising class of Ge-based systems, where the spatial fluctuations of the $g$-tensor, and the associated fluctuations of the {\em direction} of the effective magnetic field must be taken into account. Such an extension is conceptually straightforward, replacing the scalar random sheet ${\tilde B}(x,t)$ by a vector, i.e.\ by three correlated Gaussian random sheets ${\tilde B}_x(x,t)$, ${\tilde B}_y(x,t)$ and ${\tilde B}_z(x,t)$. The calculations presented in Sec.~\ref{sec:formulation} and Sec.~\ref{sec:model} would remain essentially the same, producing a set of nontrivially correlated but Gaussian random fields ${\mathrm B}_{a,k}(x,t)$ ($a=x,y,z$ and $k=1,\dots M$) acting on the system of $M$ entangled spins. In that case, both diagonal and off-diagonal elements of the multi-spin density matrix would change in time. Analytical solutions for such a problem would rarely be possible, so that the main tool would be numerical solution of the equations of motion for  $M$ entangled spins. Such a system of stochastic differential equations describing a Hamiltonian system with multiplicative mutually correlated Gaussian noises can be efficiently solved \cite{MilsteinTretyakovBook}, and our preliminary analysis, similar to Sec.~\ref{subsubsec:MC}, confirms feasibility and efficiency of such an approach.

Another important extension is the analysis of other noise models, such as $1/f$ noise and Johnson-Nyquist noise, which are known to be critical in semiconductor structures. In many cases, these noises are Gaussian \cite{Szankowski2017}, and the resulting magnetic noise ${\tilde B}(x,t)$ can also be modeled as Gaussian, using the framework suggested in this paper. 
More detailed assessment of the statistical properties of ${\tilde B}(x,t)$ can be achieved by calculating the electric fields created by an ensemble of individual fluctuators producing $1/f$ noise, and evaluating their influence on the electron wavepacket $\rho({\vec r},t)$. For known magnetic field gradient and/or spatial profile $g({\vec r})$, that would produce a realistic estimate for the statistics of the field ${\tilde B}(x,t)$ \cite{BoscoZouLossHighFidShuttlingSOI23,HuangHuSpinRelaxShuttl13,KepaFockeCywinskiKrzywdaChargeNoise23,KepaCywinskiKrzywdaSpinNoise23,JeonBenjaminFisher24RobustChargeShuttl}. Besides, this approach can take into account the situations where the electric noise has non-Gaussian behavior \cite{RamonNonGauss2015,YeNicholEtal24IndividualTLFs,Connors2022,Ahn2021,MMandVD_TLS24}. Such an analysis in the case of multiple qubits with nontrivial trajectories would be of much interest for experiment.

Transport of qubits through space is an important constituent of many other architectures for quantum information processing. For instance, shuttling of multiple entangled atoms is a key feature of the recently proposed fault-tolerant scheme for atom-based quantum computation \cite{ChongBernienEtalCircuitsAtomArrays23,AtomicArraysLukin2022,
XuEtalQECforAtomArrays2023,AtomicProcessorWithQECLukin2023}. The atom shuttled by the optical tweezers forms a wavepacket $\rho({\vec r};{\vec r}_c)$, tightly localized in space with the center ${\vec r}_c=\{x_c,y_c\}$ in the atomic plane. The ac Stark shift created by the optical field changes the energy distance $\hbar\omega({\vec r}_c,t)$ between the qubit energy levels, such that the effective Hamiltonian describing the atomic qubit is 
\begin{equation}
H_Z({\vec r}_c,t)/\hbar=\omega({\vec r}_c,t) S_z,
\end{equation}
cf.~Eq.~\ref{eq:tildeB}.
Due to temporal instabilities of the laser and intensity modulators which generate the trapping light, and spatial inhomogeneity (e.g. due to intermodulation within the acousto-optic deflectors shuttling the tweezers), the optical potential varies in a way that is difficult to predict or calculate in practice. Accordingly, the modulation of $\omega({\vec r}_c,t)$ acts similarly to a random field dephasing the atomic qubits. In the systems that are currently investigated, by employing active intensity stabilization, the potential is stabilized in time at the level $0.1$--$0.01$\%, such that the energy distance does not explicitly depend on time, $\omega({\vec r}_c(t),t)\equiv\omega({\vec r}_c(t))$ \cite{Conor}. In that case, the collective dynamical decoupling pulses can almost perfectly refocus the dephasing introduced during transport \cite{Conor,Dolev,AtomicProcessorWithQECLukin2023}. However, it is possible that in larger systems, when quantum computations or simulations are performed during longer periods of time, temporal fluctuations of the tweezers potential may become important, making $\omega({\vec r}_c,t)$ explicitly dependent on time. Then, the approach proposed in this work can be efficiently used for assessment of the qubit dephasing in such systems.

Finally, transport of multiple quantum single-photon pulses in space is a key element in various quantum internet applications, from quantum communication to modular quantum computing \cite{KimbleQuInternet08,AwschEtalQuInterconnects21,WehnerEtalQuInternet18}. 
Photons are used for establishing entanglement between quantum nodes, such that the phases between the states of different nodes are determined by the phases between the photon states. 
Random inhomogeneities in the medium, e.g.\ random fluctuations of the dielectric constant $\varepsilon({\vec r},t)$ of the optical fiber in space and time, produce the noise that scrambles the phase and/or polarization of the photonic qubit states, thus decohering the collective state of the quantum nodes in the network. This dephasing, together with the photon loss, is one of the most significant problems for quantum communication and distributed quantum information processing  \cite{NemannEntanglDistr248km,LiuGuoDistribQuComp23,PompiliHansonQuNetwork22,BersinLukinDixonMetroAreaQuNetwork23,PanEtalMetroAreaQuNetwork21,TutorialEntanglProtocols24HansonBorregaardEtal}. 
Correlations of the fluctuating properties of the optical fiber generally have sizeable extent in space and time   \cite{JeunhommeSingleModeFiberOpt,LeeEtalFiberOpticSensors,DongHuangLiLiuNoiseOptFibers,WanserFluctOptFibers92,GlennNoiseFluctOptFibers89}, such that the random phases accumulated by different photon states are, in general, nontrivially correlated, in the same manner as has been demonstrated in this paper. 

The photon dephasing can be modeled via direct simulation of a propagating quantum or classical light wave in a complex fluctuating medium \cite{RytovEtalStatRad,SobczykStochWave,CorneyDrummond01NoiseInOptFibersI,CorneyDrummond01NoiseInOptFibersII,PerinaTeichEtalPhotonDetectStatistics73,Tatarskii72} or scattering of photons in a medium with inhomogeneities \cite{KiilerichMolmerQuPulsesPRL19,KiilerichMolmerQuPulsesPRA20,GreenbergEtalPhotonScatOnQubit23}. However, applying this approach to complex networks with many nodes and multiple correlated propagating photons is difficult, and would require a number of radical simplifications to make the problem tractable.

Alternatively, the approach presented in this paper can be employed. When the properties of the medium fluctuate in space and time slowly in comparison with the photon wavelength and period, the medium can be described using the time- and space- varying refractive index $n({\vec r},t)=\sqrt{\varepsilon({\vec r},t)}=n_0 + {\tilde n}({\vec r},t)$, where $n_0$ and ${\tilde n}({\vec r},t)\ll n_0$ are the average (deterministic) part and the random fluctuations, respectively. 
In a locally straight single-mode optical fiber, the photon wavepacket adiabatically follows the fiber (neglecting small distortions and outward radiation caused by the fiber bending), and the photon spatiotemporal profile is well approximated as ${\mathcal E}({\vec r},t) = {\mathcal C}(y,z) u(x,t) \exp{i\Theta(x,t)}$, where the local $x$-axis is directed along the fiber; here ${\mathcal C}(y,z)$ is the fundamental transverse mode of the fiber, $u(x,t)$ is the wavepacket profile along the fiber, and $\Theta(x,t)$ is the phase. When the fluctuations ${\tilde n}({\vec r},t)$ are small and vary slowly, propagation of the wavepacket can be described using the eikonal approximation. In that case, the changes in the amplitude $u(x,t)$ are much smaller than the fluctuations of the phase \cite{RytovEtalStatRad,SobczykStochWave}, and can often be neglected; in many cases, the amplitude $u(x,t)$ is mostly affected by the photon loss (absorption, scattering to other modes, etc.) during propagation. Within this picture, the propagating photon becomes similar to a shuttled electron: the photon dephasing, like dephasing of a shuttled spin, is due to random fluctuations in the medium, while the photon loss is similar to the loss of a shuttled electron due to non-adiabaticity, excitation to other orbital or valley states, etc.

Consider a photon of the frequency $\omega$, produced by a point-like emitter at $x=0$ with the profile $u(0,t)$ and the phase $\Theta(0,t)=\omega t$. When the  photon impinges on a detector at $x=L$, its phase at time $t_0$ is a sum of the deterministic part $\alpha(t_0)=\omega (t_0-L/v)$ (where $v=c/n_0$, and $c$ is the speed of light in vacuum) and the random phase $\phi(t_0)$, cf.\ Eq.~\ref{eq:phases1spin}. The latter, within the first order in the small parameter $|{\tilde n}/n_0|\ll 1$, equals to \cite{GavrilenkoStepanov73LightSpaceTimeVarMed}
\begin{equation}
\phi(t_0)=\frac{\omega}{c} \int_{\mathcal{C}} {\tilde n}(x,t') dx
\end{equation}
where $t'=t_0-(L-x)/c_0$, $x$ is the coordinate along the fiber, and $\mathcal{C}$ is the path traveled by the photon from $x=0$ (emitter) to $x=L$ (detector), in direct analogy with Eqs.~\ref{eq:phit0onespin} and \ref{eq:phases1spin}.

Accumulation and impact of the photon phases have been considered in detail for several cases, see e.g.\ \cite{HermansBorregaardEtalSingleClick23,PompiliHansonQuNetwork22}, and constitutes an important part of analysis of the entanglement protocols \cite{TutorialEntanglProtocols24HansonBorregaardEtal}. 
However, the focus so far was on relatively simple scenarios, most often, including two nodes and four photon states, whose phases are correllated in a known manner (e.g., fully correlated or uncorrelated). 
Prospective architectures for large-scale quantum networks and modular quantum computation would include many photonic states propagating in a correlated manner. This could be photons traveling along the same fiber with some delay (e.g., in the case of the time-bin encoding or in the cluster states), or propagating forth and back along the same fiber \cite{PhotonLoopBoston24LukinLoncarEtal}. The photons may travel part of their way along the same waveguide when routed on and between photonic chips \cite{SiChipsMarchettiMinzioniEtal19,PhotonicChipsQIPReviewEnglundEtal22,ErrandoHerranzEtal20PhotIntegrCircuits}, or visit multiple quantum nodes in a coordinated fashion, 
etc \cite{ChakrabortyWehnerEtal20RoutingEntanglement,Caleffi17RoutingQuNetworks,KozlowskiWehner19LargeQuNetworks,LeeWehnerEnglundEtal22Router}. The theoretical approach proposed here would enable realistic assessment of the correlated dephasing of the photon states in such complex situations.

Summarizing, the approach proposed in this work for investigating the correlated qubit dephasing during transport through space is applicable to a wide range of experimentally important semiconductor platforms for quantum computation, providing explicit solutions for the problems that are difficult to address using other methods. Moreover, it has much potential to be a valuable tool for analyzing other architectures, and would be particularly helpful for studying photon dephasing in complex quantum networks.

\acknowledgments{
We thank D.~P.~DiVincenzo, L.~M.~K.~Vandersypen, A.~Roitershtein, M.~Russ-Rimbach, B.~Terhal, J.~Borregaard, A.~Stolk, Y.~N.~Nazarov, M.~Veldhorst, S.~Bosco, X.~Hu, C.~E.~Bradley and D.~Bluvstein (chronologically ordered) for valuable discussions.
This work is part of the research programme NWO QuTech Physics Funding (QTECH, programme 172) with project number 16QTECH02, which is (partly) financed by the Dutch Research Council (NWO).
Research was partly sponsored by the Army Research Office and was accomplished under Award Number W911NF-23-1-0110. The views and conclusions contained in this document are those of the authors and should not be interpreted as representing the official policies, either expressed or implied, of the Army Research Office or the U.S. Government. The U.S. Government is authorized to reproduce and distribute reprints for Government purposes notwithstanding any copyright notation herein.
}

\bibliography{ref-v6}

\begin{thebibliography}{163}%
\makeatletter
\providecommand \@ifxundefined [1]{%
 \@ifx{#1\undefined}
}%
\providecommand \@ifnum [1]{%
 \ifnum #1\expandafter \@firstoftwo
 \else \expandafter \@secondoftwo
 \fi
}%
\providecommand \@ifx [1]{%
 \ifx #1\expandafter \@firstoftwo
 \else \expandafter \@secondoftwo
 \fi
}%
\providecommand \natexlab [1]{#1}%
\providecommand \enquote  [1]{``#1''}%
\providecommand \bibnamefont  [1]{#1}%
\providecommand \bibfnamefont [1]{#1}%
\providecommand \citenamefont [1]{#1}%
\providecommand \href@noop [0]{\@secondoftwo}%
\providecommand \href [0]{\begingroup \@sanitize@url \@href}%
\providecommand \@href[1]{\@@startlink{#1}\@@href}%
\providecommand \@@href[1]{\endgroup#1\@@endlink}%
\providecommand \@sanitize@url [0]{\catcode `\\12\catcode `\$12\catcode
  `\&12\catcode `\#12\catcode `\^12\catcode `\_12\catcode `\%12\relax}%
\providecommand \@@startlink[1]{}%
\providecommand \@@endlink[0]{}%
\providecommand \url  [0]{\begingroup\@sanitize@url \@url }%
\providecommand \@url [1]{\endgroup\@href {#1}{\urlprefix }}%
\providecommand \urlprefix  [0]{URL }%
\providecommand \Eprint [0]{\href }%
\providecommand \doibase [0]{https://doi.org/}%
\providecommand \selectlanguage [0]{\@gobble}%
\providecommand \bibinfo  [0]{\@secondoftwo}%
\providecommand \bibfield  [0]{\@secondoftwo}%
\providecommand \translation [1]{[#1]}%
\providecommand \BibitemOpen [0]{}%
\providecommand \bibitemStop [0]{}%
\providecommand \bibitemNoStop [0]{.\EOS\space}%
\providecommand \EOS [0]{\spacefactor3000\relax}%
\providecommand \BibitemShut  [1]{\csname bibitem#1\endcsname}%
\let\auto@bib@innerbib\@empty
\bibitem [{\citenamefont {Kimble}(2018)}]{KimbleQuInternet08}%
  \BibitemOpen
  \bibfield  {author} {\bibinfo {author} {\bibfnamefont {H.~J.}\ \bibnamefont
  {Kimble}},\ }\bibfield  {title} {\bibinfo {title} {Quantum internet},\ }\href
  {https://doi.org/10.1038/nature07127} {\bibfield  {journal} {\bibinfo
  {journal} {Nature}\ }\textbf {\bibinfo {volume} {453}},\ \bibinfo {pages}
  {1023} (\bibinfo {year} {2018})}\BibitemShut {NoStop}%
\bibitem [{\citenamefont {Awschalom}\ \emph {et~al.}(2021)\citenamefont
  {Awschalom}, \citenamefont {Berggren}, \citenamefont {Bernien}, \citenamefont
  {Bhave}, \citenamefont {Carr}, \citenamefont {Davids}, \citenamefont
  {Economou}, \citenamefont {Englund}, \citenamefont {Faraon}, \citenamefont
  {Fejer}, \citenamefont {Guha}, \citenamefont {Gustafsson}, \citenamefont
  {Hu}, \citenamefont {Jiang}, \citenamefont {Kim}, \citenamefont {Korzh},
  \citenamefont {Kumar}, \citenamefont {Kwiat}, \citenamefont
  {Lon\ifmmode~\check{c}\else \v{c}\fi{}ar}, \citenamefont {Lukin},
  \citenamefont {Miller}, \citenamefont {Monroe}, \citenamefont {Nam},
  \citenamefont {Narang}, \citenamefont {Orcutt}, \citenamefont {Raymer},
  \citenamefont {Safavi-Naeini}, \citenamefont {Spiropulu}, \citenamefont
  {Srinivasan}, \citenamefont {Sun}, \citenamefont {Vu\ifmmode \check{c}\else
  \v{c}\fi{}kovi\ifmmode~\acute{c}\else \'{c}\fi{}}, \citenamefont {Waks},
  \citenamefont {Walsworth}, \citenamefont {Weiner},\ and\ \citenamefont
  {Zhang}}]{AwschEtalQuInterconnects21}%
  \BibitemOpen
  \bibfield  {author} {\bibinfo {author} {\bibfnamefont {D.}~\bibnamefont
  {Awschalom}}, \bibinfo {author} {\bibfnamefont {K.~K.}\ \bibnamefont
  {Berggren}}, \bibinfo {author} {\bibfnamefont {H.}~\bibnamefont {Bernien}},
  \bibinfo {author} {\bibfnamefont {S.}~\bibnamefont {Bhave}}, \bibinfo
  {author} {\bibfnamefont {L.~D.}\ \bibnamefont {Carr}}, \bibinfo {author}
  {\bibfnamefont {P.}~\bibnamefont {Davids}}, \bibinfo {author} {\bibfnamefont
  {S.~E.}\ \bibnamefont {Economou}}, \bibinfo {author} {\bibfnamefont
  {D.}~\bibnamefont {Englund}}, \bibinfo {author} {\bibfnamefont
  {A.}~\bibnamefont {Faraon}}, \bibinfo {author} {\bibfnamefont
  {M.}~\bibnamefont {Fejer}}, \bibinfo {author} {\bibfnamefont
  {S.}~\bibnamefont {Guha}}, \bibinfo {author} {\bibfnamefont {M.~V.}\
  \bibnamefont {Gustafsson}}, \bibinfo {author} {\bibfnamefont
  {E.}~\bibnamefont {Hu}}, \bibinfo {author} {\bibfnamefont {L.}~\bibnamefont
  {Jiang}}, \bibinfo {author} {\bibfnamefont {J.}~\bibnamefont {Kim}}, \bibinfo
  {author} {\bibfnamefont {B.}~\bibnamefont {Korzh}}, \bibinfo {author}
  {\bibfnamefont {P.}~\bibnamefont {Kumar}}, \bibinfo {author} {\bibfnamefont
  {P.~G.}\ \bibnamefont {Kwiat}}, \bibinfo {author} {\bibfnamefont
  {M.}~\bibnamefont {Lon\ifmmode~\check{c}\else \v{c}\fi{}ar}}, \bibinfo
  {author} {\bibfnamefont {M.~D.}\ \bibnamefont {Lukin}}, \bibinfo {author}
  {\bibfnamefont {D.~A.}\ \bibnamefont {Miller}}, \bibinfo {author}
  {\bibfnamefont {C.}~\bibnamefont {Monroe}}, \bibinfo {author} {\bibfnamefont
  {S.~W.}\ \bibnamefont {Nam}}, \bibinfo {author} {\bibfnamefont
  {P.}~\bibnamefont {Narang}}, \bibinfo {author} {\bibfnamefont {J.~S.}\
  \bibnamefont {Orcutt}}, \bibinfo {author} {\bibfnamefont {M.~G.}\
  \bibnamefont {Raymer}}, \bibinfo {author} {\bibfnamefont {A.~H.}\
  \bibnamefont {Safavi-Naeini}}, \bibinfo {author} {\bibfnamefont
  {M.}~\bibnamefont {Spiropulu}}, \bibinfo {author} {\bibfnamefont
  {K.}~\bibnamefont {Srinivasan}}, \bibinfo {author} {\bibfnamefont
  {S.}~\bibnamefont {Sun}}, \bibinfo {author} {\bibfnamefont {J.}~\bibnamefont
  {Vu\ifmmode \check{c}\else \v{c}\fi{}kovi\ifmmode~\acute{c}\else
  \'{c}\fi{}}}, \bibinfo {author} {\bibfnamefont {E.}~\bibnamefont {Waks}},
  \bibinfo {author} {\bibfnamefont {R.}~\bibnamefont {Walsworth}}, \bibinfo
  {author} {\bibfnamefont {A.~M.}\ \bibnamefont {Weiner}},\ and\ \bibinfo
  {author} {\bibfnamefont {Z.}~\bibnamefont {Zhang}},\ }\bibfield  {title}
  {\bibinfo {title} {Development of quantum interconnects (quics) for
  next-generation information technologies},\ }\href
  {https://doi.org/10.1103/PRXQuantum.2.017002} {\bibfield  {journal} {\bibinfo
   {journal} {PRX Quantum}\ }\textbf {\bibinfo {volume} {2}},\ \bibinfo {pages}
  {017002} (\bibinfo {year} {2021})}\BibitemShut {NoStop}%
\bibitem [{\citenamefont {Wehner}\ \emph {et~al.}(2018)\citenamefont {Wehner},
  \citenamefont {Elkouss},\ and\ \citenamefont
  {Hanson}}]{WehnerEtalQuInternet18}%
  \BibitemOpen
  \bibfield  {author} {\bibinfo {author} {\bibfnamefont {S.}~\bibnamefont
  {Wehner}}, \bibinfo {author} {\bibfnamefont {D.}~\bibnamefont {Elkouss}},\
  and\ \bibinfo {author} {\bibfnamefont {R.}~\bibnamefont {Hanson}},\
  }\bibfield  {title} {\bibinfo {title} {Quantum internet: A vision for the
  road ahead},\ }\href {https://doi.org/10.1126/science.aam9288} {\bibfield
  {journal} {\bibinfo  {journal} {Science}\ }\textbf {\bibinfo {volume}
  {362}},\ \bibinfo {pages} {eaam9288} (\bibinfo {year} {2018})}\BibitemShut
  {NoStop}%
\bibitem [{\citenamefont {Beukers}\ \emph {et~al.}(2024)\citenamefont
  {Beukers}, \citenamefont {Pasini}, \citenamefont {Choi}, \citenamefont
  {Englund}, \citenamefont {Hanson},\ and\ \citenamefont
  {Borregaard}}]{TutorialEntanglProtocols24HansonBorregaardEtal}%
  \BibitemOpen
  \bibfield  {author} {\bibinfo {author} {\bibfnamefont {H.~K.}\ \bibnamefont
  {Beukers}}, \bibinfo {author} {\bibfnamefont {M.}~\bibnamefont {Pasini}},
  \bibinfo {author} {\bibfnamefont {H.}~\bibnamefont {Choi}}, \bibinfo {author}
  {\bibfnamefont {D.}~\bibnamefont {Englund}}, \bibinfo {author} {\bibfnamefont
  {R.}~\bibnamefont {Hanson}},\ and\ \bibinfo {author} {\bibfnamefont
  {J.}~\bibnamefont {Borregaard}},\ }\bibfield  {title} {\bibinfo {title}
  {Remote-entanglement protocols for stationary qubits with photonic
  interfaces},\ }\href {https://doi.org/10.1103/PRXQuantum.5.010202} {\bibfield
   {journal} {\bibinfo  {journal} {PRX Quantum}\ }\textbf {\bibinfo {volume}
  {5}},\ \bibinfo {pages} {010202} (\bibinfo {year} {2024})}\BibitemShut
  {NoStop}%
\bibitem [{\citenamefont {Beugnon}\ \emph {et~al.}(2007)\citenamefont
  {Beugnon}, \citenamefont {Tuchendler}, \citenamefont {Marion}, \citenamefont
  {Ga{\"e}tan}, \citenamefont {Miroshnychenko}, \citenamefont {Sortais},
  \citenamefont {Lance}, \citenamefont {Jones}, \citenamefont {Messin},
  \citenamefont {Browaeys},\ and\ \citenamefont
  {Grangier}}]{BeugnonGrangierEtalOptTweezers07}%
  \BibitemOpen
  \bibfield  {author} {\bibinfo {author} {\bibfnamefont {J.}~\bibnamefont
  {Beugnon}}, \bibinfo {author} {\bibfnamefont {C.}~\bibnamefont {Tuchendler}},
  \bibinfo {author} {\bibfnamefont {H.}~\bibnamefont {Marion}}, \bibinfo
  {author} {\bibfnamefont {A.}~\bibnamefont {Ga{\"e}tan}}, \bibinfo {author}
  {\bibfnamefont {Y.}~\bibnamefont {Miroshnychenko}}, \bibinfo {author}
  {\bibfnamefont {Y.~R.~P.}\ \bibnamefont {Sortais}}, \bibinfo {author}
  {\bibfnamefont {A.~M.}\ \bibnamefont {Lance}}, \bibinfo {author}
  {\bibfnamefont {M.~P.~A.}\ \bibnamefont {Jones}}, \bibinfo {author}
  {\bibfnamefont {G.}~\bibnamefont {Messin}}, \bibinfo {author} {\bibfnamefont
  {A.}~\bibnamefont {Browaeys}},\ and\ \bibinfo {author} {\bibfnamefont
  {P.}~\bibnamefont {Grangier}},\ }\bibfield  {title} {\bibinfo {title}
  {Two-dimensional transport and transfer of a single atomic qubit in optical
  tweezers},\ }\href
  {https://doi.org/https://doi.org/10.1038/s41567-021-01357-2} {\bibfield
  {journal} {\bibinfo  {journal} {Nat. Phys.}\ }\textbf {\bibinfo {volume}
  {3}},\ \bibinfo {pages} {696} (\bibinfo {year} {2007})}\BibitemShut {NoStop}%
\bibitem [{\citenamefont {Kim}\ \emph {et~al.}(2016)\citenamefont {Kim},
  \citenamefont {Lee}, \citenamefont {Lee}, \citenamefont {Jo}, \citenamefont
  {Song},\ and\ \citenamefont {Ahn}}]{ResizeAtomicArraysAhn2016}%
  \BibitemOpen
  \bibfield  {author} {\bibinfo {author} {\bibfnamefont {H.}~\bibnamefont
  {Kim}}, \bibinfo {author} {\bibfnamefont {W.}~\bibnamefont {Lee}}, \bibinfo
  {author} {\bibfnamefont {H.-G.}\ \bibnamefont {Lee}}, \bibinfo {author}
  {\bibfnamefont {H.}~\bibnamefont {Jo}}, \bibinfo {author} {\bibfnamefont
  {Y.}~\bibnamefont {Song}},\ and\ \bibinfo {author} {\bibfnamefont
  {J.}~\bibnamefont {Ahn}},\ }\bibfield  {title} {\bibinfo {title} {In situ
  single-atom array synthesis using dynamic holographic optical tweezers},\
  }\href {https://doi.org/https://doi.org/10.1038/ncomms13317} {\bibfield
  {journal} {\bibinfo  {journal} {Nat. Commun.}\ }\textbf {\bibinfo {volume}
  {7}},\ \bibinfo {pages} {13317} (\bibinfo {year} {2016})}\BibitemShut
  {NoStop}%
\bibitem [{\citenamefont {Singh}\ \emph {et~al.}(2022)\citenamefont {Singh},
  \citenamefont {Anand}, \citenamefont {Pocklington}, \citenamefont {Kemp},\
  and\ \citenamefont {Bernien}}]{BernienEtalTwoSpeciesAtomArray22}%
  \BibitemOpen
  \bibfield  {author} {\bibinfo {author} {\bibfnamefont {K.}~\bibnamefont
  {Singh}}, \bibinfo {author} {\bibfnamefont {S.}~\bibnamefont {Anand}},
  \bibinfo {author} {\bibfnamefont {A.}~\bibnamefont {Pocklington}}, \bibinfo
  {author} {\bibfnamefont {J.~T.}\ \bibnamefont {Kemp}},\ and\ \bibinfo
  {author} {\bibfnamefont {H.}~\bibnamefont {Bernien}},\ }\bibfield  {title}
  {\bibinfo {title} {Dual-element, two-dimensional atom array with
  continuous-mode operation},\ }\href
  {https://doi.org/10.1103/PhysRevX.12.011040} {\bibfield  {journal} {\bibinfo
  {journal} {Phys. Rev. X}\ }\textbf {\bibinfo {volume} {12}},\ \bibinfo
  {pages} {011040} (\bibinfo {year} {2022})}\BibitemShut {NoStop}%
\bibitem [{\citenamefont {Kaufman}\ and\ \citenamefont
  {Ni}(2021)}]{KaufmanOptTweezerReview21}%
  \BibitemOpen
  \bibfield  {author} {\bibinfo {author} {\bibfnamefont {A.~M.}\ \bibnamefont
  {Kaufman}}\ and\ \bibinfo {author} {\bibfnamefont {K.-K.}\ \bibnamefont
  {Ni}},\ }\bibfield  {title} {\bibinfo {title} {Quantum science with optical
  tweezer arrays of ultracold atoms and molecules},\ }\href
  {https://doi.org/https://doi.org/10.1038/s41567-021-01357-2} {\bibfield
  {journal} {\bibinfo  {journal} {Nat. Phys.}\ }\textbf {\bibinfo {volume}
  {17}},\ \bibinfo {pages} {1324} (\bibinfo {year} {2021})}\BibitemShut
  {NoStop}%
\bibitem [{\citenamefont {Bluvstein}\ \emph {et~al.}(2022)\citenamefont
  {Bluvstein}, \citenamefont {Levine}, \citenamefont {Semeghini}, \citenamefont
  {Wang}, \citenamefont {Ebadi}, \citenamefont {Kalinowski}, \citenamefont
  {Keesling}, \citenamefont {Maskara}, \citenamefont {Pichler}, \citenamefont
  {Greiner}, \citenamefont {Vuletic},\ and\ \citenamefont
  {Lukin}}]{AtomicArraysLukin2022}%
  \BibitemOpen
  \bibfield  {author} {\bibinfo {author} {\bibfnamefont {D.}~\bibnamefont
  {Bluvstein}}, \bibinfo {author} {\bibfnamefont {H.}~\bibnamefont {Levine}},
  \bibinfo {author} {\bibfnamefont {G.}~\bibnamefont {Semeghini}}, \bibinfo
  {author} {\bibfnamefont {T.~T.}\ \bibnamefont {Wang}}, \bibinfo {author}
  {\bibfnamefont {S.}~\bibnamefont {Ebadi}}, \bibinfo {author} {\bibfnamefont
  {M.}~\bibnamefont {Kalinowski}}, \bibinfo {author} {\bibfnamefont
  {A.}~\bibnamefont {Keesling}}, \bibinfo {author} {\bibfnamefont
  {N.}~\bibnamefont {Maskara}}, \bibinfo {author} {\bibfnamefont
  {H.}~\bibnamefont {Pichler}}, \bibinfo {author} {\bibfnamefont
  {M.}~\bibnamefont {Greiner}}, \bibinfo {author} {\bibfnamefont
  {V.}~\bibnamefont {Vuletic}},\ and\ \bibinfo {author} {\bibfnamefont {M.~D.}\
  \bibnamefont {Lukin}},\ }\bibfield  {title} {\bibinfo {title} {A quantum
  processor based on coherent transport of entangled atom arrays},\ }\href
  {https://doi.org/https://doi.org/10.1038/s41586-022-04592-6} {\bibfield
  {journal} {\bibinfo  {journal} {Nature}\ }\textbf {\bibinfo {volume} {604}},\
  \bibinfo {pages} {451} (\bibinfo {year} {2022})}\BibitemShut {NoStop}%
\bibitem [{\citenamefont {Kielpinski}\ \emph {et~al.}(2002)\citenamefont
  {Kielpinski}, \citenamefont {Monroe},\ and\ \citenamefont
  {Wineland}}]{KielpMonroeWineland02}%
  \BibitemOpen
  \bibfield  {author} {\bibinfo {author} {\bibfnamefont {D.}~\bibnamefont
  {Kielpinski}}, \bibinfo {author} {\bibfnamefont {C.}~\bibnamefont {Monroe}},\
  and\ \bibinfo {author} {\bibfnamefont {D.~J.}\ \bibnamefont {Wineland}},\
  }\bibfield  {title} {\bibinfo {title} {Architecture for a large-scale
  ion-trap quantum computer},\ }\href
  {https://doi.org/https://doi.org/10.1038/nature00784} {\bibfield  {journal}
  {\bibinfo  {journal} {Nature}\ }\textbf {\bibinfo {volume} {417}},\ \bibinfo
  {pages} {709} (\bibinfo {year} {2002})}\BibitemShut {NoStop}%
\bibitem [{\citenamefont {Pino}\ \emph {et~al.}(2021)\citenamefont {Pino},
  \citenamefont {Dreiling}, \citenamefont {Figgatt}, \citenamefont {Gaebler},
  \citenamefont {Moses}, \citenamefont {Allman}, \citenamefont {Baldwin},
  \citenamefont {Foss-Feig}, \citenamefont {Hayes}, \citenamefont {Mayer},
  \citenamefont {Ryan-Anderson},\ and\ \citenamefont
  {Neyenhuis}}]{PinoNeyenhuisEtal21}%
  \BibitemOpen
  \bibfield  {author} {\bibinfo {author} {\bibfnamefont {J.~M.}\ \bibnamefont
  {Pino}}, \bibinfo {author} {\bibfnamefont {J.~M.}\ \bibnamefont {Dreiling}},
  \bibinfo {author} {\bibfnamefont {C.}~\bibnamefont {Figgatt}}, \bibinfo
  {author} {\bibfnamefont {J.~P.}\ \bibnamefont {Gaebler}}, \bibinfo {author}
  {\bibfnamefont {S.~A.}\ \bibnamefont {Moses}}, \bibinfo {author}
  {\bibfnamefont {M.~S.}\ \bibnamefont {Allman}}, \bibinfo {author}
  {\bibfnamefont {C.~H.}\ \bibnamefont {Baldwin}}, \bibinfo {author}
  {\bibfnamefont {M.}~\bibnamefont {Foss-Feig}}, \bibinfo {author}
  {\bibfnamefont {D.}~\bibnamefont {Hayes}}, \bibinfo {author} {\bibfnamefont
  {K.}~\bibnamefont {Mayer}}, \bibinfo {author} {\bibfnamefont
  {C.}~\bibnamefont {Ryan-Anderson}},\ and\ \bibinfo {author} {\bibfnamefont
  {B.}~\bibnamefont {Neyenhuis}},\ }\bibfield  {title} {\bibinfo {title}
  {Demonstration of the trapped-ion quantum {CCD} computer architecture},\
  }\href {https://doi.org/https://doi.org/10.1038/s41586-021-03318-4}
  {\bibfield  {journal} {\bibinfo  {journal} {Nature}\ }\textbf {\bibinfo
  {volume} {592}},\ \bibinfo {pages} {209} (\bibinfo {year}
  {2021})}\BibitemShut {NoStop}%
\bibitem [{\citenamefont {Taylor}\ \emph {et~al.}(2005)\citenamefont {Taylor},
  \citenamefont {Engel}, \citenamefont {D{\"u}r}, \citenamefont {Yacoby},
  \citenamefont {Marcus}, \citenamefont {Zoller},\ and\ \citenamefont
  {Lukin}}]{FaultTolQuDotTaylorLukin05}%
  \BibitemOpen
  \bibfield  {author} {\bibinfo {author} {\bibfnamefont {J.~M.}\ \bibnamefont
  {Taylor}}, \bibinfo {author} {\bibfnamefont {H.-A.}\ \bibnamefont {Engel}},
  \bibinfo {author} {\bibfnamefont {W.}~\bibnamefont {D{\"u}r}}, \bibinfo
  {author} {\bibfnamefont {A.}~\bibnamefont {Yacoby}}, \bibinfo {author}
  {\bibfnamefont {C.~M.}\ \bibnamefont {Marcus}}, \bibinfo {author}
  {\bibfnamefont {P.}~\bibnamefont {Zoller}},\ and\ \bibinfo {author}
  {\bibfnamefont {M.~D.}\ \bibnamefont {Lukin}},\ }\bibfield  {title} {\bibinfo
  {title} {Fault-tolerant architecture for quantum computation using
  electrically controlled semiconductor spins},\ }\href
  {https://doi.org/https://doi.org/10.1038/nphys174} {\bibfield  {journal}
  {\bibinfo  {journal} {Nature Phys.}\ }\textbf {\bibinfo {volume} {1}},\
  \bibinfo {pages} {177} (\bibinfo {year} {2005})}\BibitemShut {NoStop}%
\bibitem [{\citenamefont {Buonacorsi}\ \emph {et~al.}(2019)\citenamefont
  {Buonacorsi}, \citenamefont {Cai}, \citenamefont {Ramirez}, \citenamefont
  {Willick}, \citenamefont {Walker}, \citenamefont {Li}, \citenamefont {Shaw},
  \citenamefont {Xu}, \citenamefont {Benjamin},\ and\ \citenamefont
  {Baugh}}]{BenjaminBaughEtalTopoQCSi19}%
  \BibitemOpen
  \bibfield  {author} {\bibinfo {author} {\bibfnamefont {B.}~\bibnamefont
  {Buonacorsi}}, \bibinfo {author} {\bibfnamefont {Z.}~\bibnamefont {Cai}},
  \bibinfo {author} {\bibfnamefont {E.~B.}\ \bibnamefont {Ramirez}}, \bibinfo
  {author} {\bibfnamefont {K.~S.}\ \bibnamefont {Willick}}, \bibinfo {author}
  {\bibfnamefont {S.~M.}\ \bibnamefont {Walker}}, \bibinfo {author}
  {\bibfnamefont {J.}~\bibnamefont {Li}}, \bibinfo {author} {\bibfnamefont
  {B.~D.}\ \bibnamefont {Shaw}}, \bibinfo {author} {\bibfnamefont
  {X.}~\bibnamefont {Xu}}, \bibinfo {author} {\bibfnamefont {S.~C.}\
  \bibnamefont {Benjamin}},\ and\ \bibinfo {author} {\bibfnamefont
  {J.}~\bibnamefont {Baugh}},\ }\bibfield  {title} {\bibinfo {title} {Network
  architecture for a topological quantum computer in silicon},\ }\href
  {https://doi.org/10.1088/2058-9565/aaf3c4} {\bibfield  {journal} {\bibinfo
  {journal} {Quantum Sci. Technol.}\ }\textbf {\bibinfo {volume} {4}},\
  \bibinfo {pages} {025003} (\bibinfo {year} {2019})}\BibitemShut {NoStop}%
\bibitem [{\citenamefont {Baart}\ \emph {et~al.}(2016)\citenamefont {Baart},
  \citenamefont {Shafiei}, \citenamefont {Fujita}, \citenamefont {Reichl},
  \citenamefont {Wegscheider},\ and\ \citenamefont
  {Vandersypen}}]{VdSEtalSingleSpinCCD16}%
  \BibitemOpen
  \bibfield  {author} {\bibinfo {author} {\bibfnamefont {T.~A.}\ \bibnamefont
  {Baart}}, \bibinfo {author} {\bibfnamefont {M.}~\bibnamefont {Shafiei}},
  \bibinfo {author} {\bibfnamefont {T.}~\bibnamefont {Fujita}}, \bibinfo
  {author} {\bibfnamefont {C.}~\bibnamefont {Reichl}}, \bibinfo {author}
  {\bibfnamefont {W.}~\bibnamefont {Wegscheider}},\ and\ \bibinfo {author}
  {\bibfnamefont {L.~M.~K.}\ \bibnamefont {Vandersypen}},\ }\bibfield  {title}
  {\bibinfo {title} {Single-spin {CCD}},\ }\href
  {https://doi.org/https://doi.org/10.1038/nnano.2015.291} {\bibfield
  {journal} {\bibinfo  {journal} {Nature Nanotech.}\ }\textbf {\bibinfo
  {volume} {11}},\ \bibinfo {pages} {330} (\bibinfo {year} {2016})}\BibitemShut
  {NoStop}%
\bibitem [{\citenamefont {Vandersypen}\ \emph {et~al.}(2017)\citenamefont
  {Vandersypen}, \citenamefont {Bluhm}, \citenamefont {Clarke}, \citenamefont
  {Dzurak}, \citenamefont {Ishihara}, \citenamefont {Morello}, \citenamefont
  {Reilly}, \citenamefont {Schreiber},\ and\ \citenamefont
  {Veldhorst}}]{VdSBluhmVeldhorstEtalVision17}%
  \BibitemOpen
  \bibfield  {author} {\bibinfo {author} {\bibfnamefont {L.~M.~K.}\
  \bibnamefont {Vandersypen}}, \bibinfo {author} {\bibfnamefont
  {H.}~\bibnamefont {Bluhm}}, \bibinfo {author} {\bibfnamefont {J.~S.}\
  \bibnamefont {Clarke}}, \bibinfo {author} {\bibfnamefont {A.~S.}\
  \bibnamefont {Dzurak}}, \bibinfo {author} {\bibfnamefont {R.}~\bibnamefont
  {Ishihara}}, \bibinfo {author} {\bibfnamefont {A.}~\bibnamefont {Morello}},
  \bibinfo {author} {\bibfnamefont {D.~J.}\ \bibnamefont {Reilly}}, \bibinfo
  {author} {\bibfnamefont {L.~R.}\ \bibnamefont {Schreiber}},\ and\ \bibinfo
  {author} {\bibfnamefont {M.}~\bibnamefont {Veldhorst}},\ }\bibfield  {title}
  {\bibinfo {title} {Interfacing spin qubits in quantum dots and donors ---
  hot, dense, and coherent},\ }\href
  {https://doi.org/https://doi.org/10.1038/s41534-017-0038-y} {\bibfield
  {journal} {\bibinfo  {journal} {npj Quantum Inf.}\ }\textbf {\bibinfo
  {volume} {3}},\ \bibinfo {pages} {34} (\bibinfo {year} {2017})}\BibitemShut
  {NoStop}%
\bibitem [{\citenamefont {Langrock}\ \emph {et~al.}(2023)\citenamefont
  {Langrock}, \citenamefont {Krzywda}, \citenamefont {Focke}, \citenamefont
  {Seidler}, \citenamefont {Schreiber},\ and\ \citenamefont
  {Cywi\ifmmode~\acute{n}\else \'{n}\fi{}ski}}]{langrock_blueprint_2023}%
  \BibitemOpen
  \bibfield  {author} {\bibinfo {author} {\bibfnamefont {V.}~\bibnamefont
  {Langrock}}, \bibinfo {author} {\bibfnamefont {J.~A.}\ \bibnamefont
  {Krzywda}}, \bibinfo {author} {\bibfnamefont {N.}~\bibnamefont {Focke}},
  \bibinfo {author} {\bibfnamefont {I.}~\bibnamefont {Seidler}}, \bibinfo
  {author} {\bibfnamefont {L.~R.}\ \bibnamefont {Schreiber}},\ and\ \bibinfo
  {author} {\bibfnamefont {L.}~\bibnamefont {Cywi\ifmmode~\acute{n}\else
  \'{n}\fi{}ski}},\ }\bibfield  {title} {\bibinfo {title} {Blueprint of a
  scalable spin qubit shuttle device for coherent mid-range qubit transfer in
  disordered {Si/SiGe/SiO$_2$}},\ }\href
  {https://doi.org/10.1103/PRXQuantum.4.020305} {\bibfield  {journal} {\bibinfo
   {journal} {PRX Quantum}\ }\textbf {\bibinfo {volume} {4}},\ \bibinfo {pages}
  {020305} (\bibinfo {year} {2023})}\BibitemShut {NoStop}%
\bibitem [{\citenamefont {K{\"u}nne}\ \emph {et~al.}(2023)\citenamefont
  {K{\"u}nne}, \citenamefont {Willmes}, \citenamefont {Oberl{\"a}nder},
  \citenamefont {Gorjaew}, \citenamefont {Teske}, \citenamefont {Bhardwaj},
  \citenamefont {Beer}, \citenamefont {Kammerloher}, \citenamefont {Otten},
  \citenamefont {Seidler}, \citenamefont {Xue}, \citenamefont {Schreiber},\
  and\ \citenamefont {Bluhm}}]{KunneEtalSpinBusArch23}%
  \BibitemOpen
  \bibfield  {author} {\bibinfo {author} {\bibfnamefont {M.}~\bibnamefont
  {K{\"u}nne}}, \bibinfo {author} {\bibfnamefont {A.}~\bibnamefont {Willmes}},
  \bibinfo {author} {\bibfnamefont {M.}~\bibnamefont {Oberl{\"a}nder}},
  \bibinfo {author} {\bibfnamefont {C.}~\bibnamefont {Gorjaew}}, \bibinfo
  {author} {\bibfnamefont {J.~D.}\ \bibnamefont {Teske}}, \bibinfo {author}
  {\bibfnamefont {H.}~\bibnamefont {Bhardwaj}}, \bibinfo {author}
  {\bibfnamefont {M.}~\bibnamefont {Beer}}, \bibinfo {author} {\bibfnamefont
  {E.}~\bibnamefont {Kammerloher}}, \bibinfo {author} {\bibfnamefont
  {R.}~\bibnamefont {Otten}}, \bibinfo {author} {\bibfnamefont
  {I.}~\bibnamefont {Seidler}}, \bibinfo {author} {\bibfnamefont
  {R.}~\bibnamefont {Xue}}, \bibinfo {author} {\bibfnamefont {L.~R.}\
  \bibnamefont {Schreiber}},\ and\ \bibinfo {author} {\bibfnamefont
  {H.}~\bibnamefont {Bluhm}},\ }\href
  {https://doi.org/10.48550/arXiv.2306.16348} {\bibinfo {title} {The {SpinBus}
  architecture: Scaling spin qubits with electron shuttling}} (\bibinfo {year}
  {2023}),\ \Eprint {https://arxiv.org/abs/2306.16348} {arXiv:2306.16348}
  \BibitemShut {NoStop}%
\bibitem [{\citenamefont {Wang}\ \emph {et~al.}(2022)\citenamefont {Wang},
  \citenamefont {Ota}, \citenamefont {Edlbauer}, \citenamefont {Jadot},
  \citenamefont {Mortemousque}, \citenamefont {Richard}, \citenamefont
  {Okazaki}, \citenamefont {Nakamura}, \citenamefont {Ludwig}, \citenamefont
  {Wieck}, \citenamefont {Urdampilleta}, \citenamefont {Meunier}, \citenamefont
  {Kodera}, \citenamefont {Kaneko}, \citenamefont {Takada},\ and\ \citenamefont
  {B{\"a}uerle}}]{BauerleTakadaEtalAcousticTransportElectrons22}%
  \BibitemOpen
  \bibfield  {author} {\bibinfo {author} {\bibfnamefont {J.}~\bibnamefont
  {Wang}}, \bibinfo {author} {\bibfnamefont {S.}~\bibnamefont {Ota}}, \bibinfo
  {author} {\bibfnamefont {H.}~\bibnamefont {Edlbauer}}, \bibinfo {author}
  {\bibfnamefont {B.}~\bibnamefont {Jadot}}, \bibinfo {author} {\bibfnamefont
  {P.-A.}\ \bibnamefont {Mortemousque}}, \bibinfo {author} {\bibfnamefont
  {A.}~\bibnamefont {Richard}}, \bibinfo {author} {\bibfnamefont
  {Y.}~\bibnamefont {Okazaki}}, \bibinfo {author} {\bibfnamefont
  {S.}~\bibnamefont {Nakamura}}, \bibinfo {author} {\bibfnamefont
  {A.}~\bibnamefont {Ludwig}}, \bibinfo {author} {\bibfnamefont {A.~D.}\
  \bibnamefont {Wieck}}, \bibinfo {author} {\bibfnamefont {M.}~\bibnamefont
  {Urdampilleta}}, \bibinfo {author} {\bibfnamefont {T.}~\bibnamefont
  {Meunier}}, \bibinfo {author} {\bibfnamefont {T.}~\bibnamefont {Kodera}},
  \bibinfo {author} {\bibfnamefont {N.-H.}\ \bibnamefont {Kaneko}}, \bibinfo
  {author} {\bibfnamefont {S.}~\bibnamefont {Takada}},\ and\ \bibinfo {author}
  {\bibfnamefont {C.}~\bibnamefont {B{\"a}uerle}},\ }\bibfield  {title}
  {\bibinfo {title} {Generation of a single-cycle acoustic pulse: A scalable
  solution for transport in single-electron circuits},\ }\href
  {https://doi.org/10.1103/PhysRevX.12.031035} {\bibfield  {journal} {\bibinfo
  {journal} {Phys. Rev. X}\ }\textbf {\bibinfo {volume} {12}},\ \bibinfo
  {pages} {031035} (\bibinfo {year} {2022})}\BibitemShut {NoStop}%
\bibitem [{\citenamefont {Mills}\ \emph {et~al.}(2019)\citenamefont {Mills},
  \citenamefont {Zajac}, \citenamefont {Gullans}, \citenamefont {Schupp},
  \citenamefont {Hazard},\ and\ \citenamefont {Petta}}]{mills_shuttling_2019}%
  \BibitemOpen
  \bibfield  {author} {\bibinfo {author} {\bibfnamefont {A.~R.}\ \bibnamefont
  {Mills}}, \bibinfo {author} {\bibfnamefont {D.~M.}\ \bibnamefont {Zajac}},
  \bibinfo {author} {\bibfnamefont {M.~J.}\ \bibnamefont {Gullans}}, \bibinfo
  {author} {\bibfnamefont {F.~J.}\ \bibnamefont {Schupp}}, \bibinfo {author}
  {\bibfnamefont {T.~M.}\ \bibnamefont {Hazard}},\ and\ \bibinfo {author}
  {\bibfnamefont {J.~R.}\ \bibnamefont {Petta}},\ }\bibfield  {title} {\bibinfo
  {title} {Shuttling a single charge across a one-dimensional array of silicon
  quantum dots},\ }\href {https://doi.org/10.1038/s41467-019-08970-z}
  {\bibfield  {journal} {\bibinfo  {journal} {Nature Communications}\ }\textbf
  {\bibinfo {volume} {10}},\ \bibinfo {pages} {1063} (\bibinfo {year}
  {2019})}\BibitemShut {NoStop}%
\bibitem [{\citenamefont {Boter}\ \emph {et~al.}(2022)\citenamefont {Boter},
  \citenamefont {Dehollain}, \citenamefont {van Dijk}, \citenamefont {Xu},
  \citenamefont {Hensgens}, \citenamefont {Versluis}, \citenamefont {Naus},
  \citenamefont {Clarke}, \citenamefont {Veldhorst}, \citenamefont
  {Sebastiano},\ and\ \citenamefont
  {Vandersypen}}]{BoterVdsEtalSpiderWebArray22}%
  \BibitemOpen
  \bibfield  {author} {\bibinfo {author} {\bibfnamefont {J.~M.}\ \bibnamefont
  {Boter}}, \bibinfo {author} {\bibfnamefont {J.~P.}\ \bibnamefont
  {Dehollain}}, \bibinfo {author} {\bibfnamefont {J.~P.}\ \bibnamefont {van
  Dijk}}, \bibinfo {author} {\bibfnamefont {Y.}~\bibnamefont {Xu}}, \bibinfo
  {author} {\bibfnamefont {T.}~\bibnamefont {Hensgens}}, \bibinfo {author}
  {\bibfnamefont {R.}~\bibnamefont {Versluis}}, \bibinfo {author}
  {\bibfnamefont {H.~W.}\ \bibnamefont {Naus}}, \bibinfo {author}
  {\bibfnamefont {J.~S.}\ \bibnamefont {Clarke}}, \bibinfo {author}
  {\bibfnamefont {M.}~\bibnamefont {Veldhorst}}, \bibinfo {author}
  {\bibfnamefont {F.}~\bibnamefont {Sebastiano}},\ and\ \bibinfo {author}
  {\bibfnamefont {L.~M.}\ \bibnamefont {Vandersypen}},\ }\bibfield  {title}
  {\bibinfo {title} {Spiderweb array: A sparse spin-qubit array},\ }\href
  {https://doi.org/10.1103/PhysRevApplied.18.024053} {\bibfield  {journal}
  {\bibinfo  {journal} {Phys. Rev. Appl.}\ }\textbf {\bibinfo {volume} {18}},\
  \bibinfo {pages} {024053} (\bibinfo {year} {2022})}\BibitemShut {NoStop}%
\bibitem [{\citenamefont {Nottingham}\ \emph {et~al.}(2023)\citenamefont
  {Nottingham}, \citenamefont {Perlin}, \citenamefont {White}, \citenamefont
  {Bernien}, \citenamefont {Chong},\ and\ \citenamefont
  {Baker}}]{ChongBernienEtalCircuitsAtomArrays23}%
  \BibitemOpen
  \bibfield  {author} {\bibinfo {author} {\bibfnamefont {N.}~\bibnamefont
  {Nottingham}}, \bibinfo {author} {\bibfnamefont {M.~A.}\ \bibnamefont
  {Perlin}}, \bibinfo {author} {\bibfnamefont {R.}~\bibnamefont {White}},
  \bibinfo {author} {\bibfnamefont {H.}~\bibnamefont {Bernien}}, \bibinfo
  {author} {\bibfnamefont {F.~T.}\ \bibnamefont {Chong}},\ and\ \bibinfo
  {author} {\bibfnamefont {J.~M.}\ \bibnamefont {Baker}},\ }\href@noop {}
  {\bibinfo {title} {Decomposing and routing quantum circuits under constraints
  for neutral atom architectures}} (\bibinfo {year} {2023}),\ \Eprint
  {https://arxiv.org/abs/2307.14996} {arXiv:2307.14996 [quant-ph]} \BibitemShut
  {NoStop}%
\bibitem [{\citenamefont {Xu}\ \emph {et~al.}(2023)\citenamefont {Xu},
  \citenamefont {Ataides}, \citenamefont {Pattison}, \citenamefont
  {Raveendran}, \citenamefont {Bluvstein}, \citenamefont {Wurtz}, \citenamefont
  {Vasic}, \citenamefont {Lukin}, \citenamefont {Jiang},\ and\ \citenamefont
  {Zhou}}]{XuEtalQECforAtomArrays2023}%
  \BibitemOpen
  \bibfield  {author} {\bibinfo {author} {\bibfnamefont {Q.}~\bibnamefont
  {Xu}}, \bibinfo {author} {\bibfnamefont {J.~P.~B.}\ \bibnamefont {Ataides}},
  \bibinfo {author} {\bibfnamefont {C.~A.}\ \bibnamefont {Pattison}}, \bibinfo
  {author} {\bibfnamefont {N.}~\bibnamefont {Raveendran}}, \bibinfo {author}
  {\bibfnamefont {D.}~\bibnamefont {Bluvstein}}, \bibinfo {author}
  {\bibfnamefont {J.}~\bibnamefont {Wurtz}}, \bibinfo {author} {\bibfnamefont
  {B.}~\bibnamefont {Vasic}}, \bibinfo {author} {\bibfnamefont {M.~D.}\
  \bibnamefont {Lukin}}, \bibinfo {author} {\bibfnamefont {L.}~\bibnamefont
  {Jiang}},\ and\ \bibinfo {author} {\bibfnamefont {H.}~\bibnamefont {Zhou}},\
  }\href {https://doi.org/https://doi.org/10.48550/arXiv.2308.08648} {\bibinfo
  {title} {Constant-overhead fault-tolerant quantum computation with
  reconfigurable atom arrays}} (\bibinfo {year} {2023}),\ \Eprint
  {https://arxiv.org/abs/2308.08648} {arXiv:2308.08648 [quant-ph]} \BibitemShut
  {NoStop}%
\bibitem [{\citenamefont {Bluvstein}\ \emph {et~al.}(2024)\citenamefont
  {Bluvstein}, \citenamefont {Evered}, \citenamefont {Geim}, \citenamefont
  {Li}, \citenamefont {Zhou}, \citenamefont {Manovitz}, \citenamefont {Ebadi},
  \citenamefont {Cain}, \citenamefont {Kalinowski}, \citenamefont {Hangleiter},
  \citenamefont {Ataides}, \citenamefont {Maskara}, \citenamefont {Cong},
  \citenamefont {Gao}, \citenamefont {Rodriguez}, \citenamefont {Karolyshyn},
  \citenamefont {Semeghini}, \citenamefont {Gullans}, \citenamefont {Greiner},
  \citenamefont {Vuletic},\ and\ \citenamefont
  {Lukin}}]{AtomicProcessorWithQECLukin2023}%
  \BibitemOpen
  \bibfield  {author} {\bibinfo {author} {\bibfnamefont {D.}~\bibnamefont
  {Bluvstein}}, \bibinfo {author} {\bibfnamefont {S.~J.}\ \bibnamefont
  {Evered}}, \bibinfo {author} {\bibfnamefont {A.~A.}\ \bibnamefont {Geim}},
  \bibinfo {author} {\bibfnamefont {S.~H.}\ \bibnamefont {Li}}, \bibinfo
  {author} {\bibfnamefont {H.}~\bibnamefont {Zhou}}, \bibinfo {author}
  {\bibfnamefont {T.}~\bibnamefont {Manovitz}}, \bibinfo {author}
  {\bibfnamefont {S.}~\bibnamefont {Ebadi}}, \bibinfo {author} {\bibfnamefont
  {M.}~\bibnamefont {Cain}}, \bibinfo {author} {\bibfnamefont {M.}~\bibnamefont
  {Kalinowski}}, \bibinfo {author} {\bibfnamefont {D.}~\bibnamefont
  {Hangleiter}}, \bibinfo {author} {\bibfnamefont {J.~P.~B.}\ \bibnamefont
  {Ataides}}, \bibinfo {author} {\bibfnamefont {N.}~\bibnamefont {Maskara}},
  \bibinfo {author} {\bibfnamefont {I.}~\bibnamefont {Cong}}, \bibinfo {author}
  {\bibfnamefont {X.}~\bibnamefont {Gao}}, \bibinfo {author} {\bibfnamefont
  {P.~S.}\ \bibnamefont {Rodriguez}}, \bibinfo {author} {\bibfnamefont
  {T.}~\bibnamefont {Karolyshyn}}, \bibinfo {author} {\bibfnamefont
  {G.}~\bibnamefont {Semeghini}}, \bibinfo {author} {\bibfnamefont {M.~J.}\
  \bibnamefont {Gullans}}, \bibinfo {author} {\bibfnamefont {M.}~\bibnamefont
  {Greiner}}, \bibinfo {author} {\bibfnamefont {V.}~\bibnamefont {Vuletic}},\
  and\ \bibinfo {author} {\bibfnamefont {M.~D.}\ \bibnamefont {Lukin}},\
  }\bibfield  {title} {\bibinfo {title} {Logical quantum processor based on
  reconfigurable atom arrays},\ }\href
  {https://doi.org/https://doi.org/10.1038/s41586-023-06927-3} {\bibfield
  {journal} {\bibinfo  {journal} {Nature}\ }\textbf {\bibinfo {volume} {626}},\
  \bibinfo {pages} {58} (\bibinfo {year} {2024})}\BibitemShut {NoStop}%
\bibitem [{\citenamefont {Siegel}\ \emph {et~al.}(2024)\citenamefont {Siegel},
  \citenamefont {Strikis},\ and\ \citenamefont
  {Fogarty}}]{SiegelEtalFaultTolShuttl24}%
  \BibitemOpen
  \bibfield  {author} {\bibinfo {author} {\bibfnamefont {A.}~\bibnamefont
  {Siegel}}, \bibinfo {author} {\bibfnamefont {A.}~\bibnamefont {Strikis}},\
  and\ \bibinfo {author} {\bibfnamefont {M.}~\bibnamefont {Fogarty}},\
  }\href@noop {} {\bibinfo {title} {Towards early fault tolerance on a
  2$\times$n array of qubits equipped with shuttling}} (\bibinfo {year}
  {2024}),\ \Eprint {https://arxiv.org/abs/2402.12599} {arXiv:2402.12599
  [quant-ph]} \BibitemShut {NoStop}%
\bibitem [{\citenamefont {Borregaard}\ \emph {et~al.}(2020)\citenamefont
  {Borregaard}, \citenamefont {Pichler}, \citenamefont {Schr\"oder},
  \citenamefont {Lukin}, \citenamefont {Lodahl},\ and\ \citenamefont
  {S\o{}rensen}}]{BorregaardQuRepeater20}%
  \BibitemOpen
  \bibfield  {author} {\bibinfo {author} {\bibfnamefont {J.}~\bibnamefont
  {Borregaard}}, \bibinfo {author} {\bibfnamefont {H.}~\bibnamefont {Pichler}},
  \bibinfo {author} {\bibfnamefont {T.}~\bibnamefont {Schr\"oder}}, \bibinfo
  {author} {\bibfnamefont {M.~D.}\ \bibnamefont {Lukin}}, \bibinfo {author}
  {\bibfnamefont {P.}~\bibnamefont {Lodahl}},\ and\ \bibinfo {author}
  {\bibfnamefont {A.~S.}\ \bibnamefont {S\o{}rensen}},\ }\bibfield  {title}
  {\bibinfo {title} {One-way quantum repeater based on near-deterministic
  photon-emitter interfaces},\ }\href
  {https://doi.org/10.1103/PhysRevX.10.021071} {\bibfield  {journal} {\bibinfo
  {journal} {Phys. Rev. X}\ }\textbf {\bibinfo {volume} {10}},\ \bibinfo
  {pages} {021071} (\bibinfo {year} {2020})}\BibitemShut {NoStop}%
\bibitem [{\citenamefont {Buterakos}\ \emph {et~al.}(2017)\citenamefont
  {Buterakos}, \citenamefont {Barnes},\ and\ \citenamefont
  {Economou}}]{EconomouPhotonClusters17}%
  \BibitemOpen
  \bibfield  {author} {\bibinfo {author} {\bibfnamefont {D.}~\bibnamefont
  {Buterakos}}, \bibinfo {author} {\bibfnamefont {E.}~\bibnamefont {Barnes}},\
  and\ \bibinfo {author} {\bibfnamefont {S.~E.}\ \bibnamefont {Economou}},\
  }\bibfield  {title} {\bibinfo {title} {Deterministic generation of
  all-photonic quantum repeaters from solid-state emitters},\ }\href
  {https://doi.org/10.1103/PhysRevX.7.041023} {\bibfield  {journal} {\bibinfo
  {journal} {Phys. Rev. X}\ }\textbf {\bibinfo {volume} {7}},\ \bibinfo {pages}
  {041023} (\bibinfo {year} {2017})}\BibitemShut {NoStop}%
\bibitem [{\citenamefont {Azuma}\ \emph {et~al.}(2015)\citenamefont {Azuma},
  \citenamefont {Tamaki},\ and\ \citenamefont
  {Lo}}]{AzumaPhotonicQuRepeaters17}%
  \BibitemOpen
  \bibfield  {author} {\bibinfo {author} {\bibfnamefont {K.}~\bibnamefont
  {Azuma}}, \bibinfo {author} {\bibfnamefont {K.}~\bibnamefont {Tamaki}},\ and\
  \bibinfo {author} {\bibfnamefont {H.-K.}\ \bibnamefont {Lo}},\ }\bibfield
  {title} {\bibinfo {title} {All-photonic quantum repeaters},\ }\href
  {https://doi.org/https://doi.org/10.1038/ncomms7787} {\bibfield  {journal}
  {\bibinfo  {journal} {Nat. Commun.}\ }\textbf {\bibinfo {volume} {6}},\
  \bibinfo {pages} {6787} (\bibinfo {year} {2015})}\BibitemShut {NoStop}%
\bibitem [{\citenamefont {Cogan}\ \emph {et~al.}(2023)\citenamefont {Cogan},
  \citenamefont {Su}, \citenamefont {Kenneth},\ and\ \citenamefont
  {Gershoni}}]{CoganPhotonClusterState23}%
  \BibitemOpen
  \bibfield  {author} {\bibinfo {author} {\bibfnamefont {D.}~\bibnamefont
  {Cogan}}, \bibinfo {author} {\bibfnamefont {Z.-E.}\ \bibnamefont {Su}},
  \bibinfo {author} {\bibfnamefont {O.}~\bibnamefont {Kenneth}},\ and\ \bibinfo
  {author} {\bibfnamefont {D.}~\bibnamefont {Gershoni}},\ }\bibfield  {title}
  {\bibinfo {title} {Deterministic generation of indistinguishable photons in a
  cluster state},\ }\href {https://doi.org/10.1038/s41566-022-01152-2}
  {\bibfield  {journal} {\bibinfo  {journal} {Nat. Photon.}\ }\textbf {\bibinfo
  {volume} {17}},\ \bibinfo {pages} {324} (\bibinfo {year} {2023})}\BibitemShut
  {NoStop}%
\bibitem [{\citenamefont {Jnane}\ \emph {et~al.}(2022)\citenamefont {Jnane},
  \citenamefont {Undseth}, \citenamefont {Cai}, \citenamefont {Benjamin},\ and\
  \citenamefont {Koczor}}]{BenjaminEtalMulticoreQComp22}%
  \BibitemOpen
  \bibfield  {author} {\bibinfo {author} {\bibfnamefont {H.}~\bibnamefont
  {Jnane}}, \bibinfo {author} {\bibfnamefont {B.}~\bibnamefont {Undseth}},
  \bibinfo {author} {\bibfnamefont {Z.}~\bibnamefont {Cai}}, \bibinfo {author}
  {\bibfnamefont {S.~C.}\ \bibnamefont {Benjamin}},\ and\ \bibinfo {author}
  {\bibfnamefont {B.}~\bibnamefont {Koczor}},\ }\bibfield  {title} {\bibinfo
  {title} {Multicore quantum computing},\ }\href
  {https://doi.org/10.1103/PhysRevApplied.18.044064} {\bibfield  {journal}
  {\bibinfo  {journal} {Phys. Rev. Appl.}\ }\textbf {\bibinfo {volume} {18}},\
  \bibinfo {pages} {044064} (\bibinfo {year} {2022})}\BibitemShut {NoStop}%
\bibitem [{\citenamefont {Neumann}\ \emph {et~al.}(2022)\citenamefont
  {Neumann}, \citenamefont {Buchner}, \citenamefont {Bulla}, \citenamefont
  {Bohmann},\ and\ \citenamefont {Ursin}}]{NemannEntanglDistr248km}%
  \BibitemOpen
  \bibfield  {author} {\bibinfo {author} {\bibfnamefont {S.~P.}\ \bibnamefont
  {Neumann}}, \bibinfo {author} {\bibfnamefont {A.}~\bibnamefont {Buchner}},
  \bibinfo {author} {\bibfnamefont {L.}~\bibnamefont {Bulla}}, \bibinfo
  {author} {\bibfnamefont {M.}~\bibnamefont {Bohmann}},\ and\ \bibinfo {author}
  {\bibfnamefont {R.}~\bibnamefont {Ursin}},\ }\bibfield  {title} {\bibinfo
  {title} {Continuous entanglement distribution over a transnational 248 km
  fiber link},\ }\href
  {https://doi.org/https://doi.org/10.1038/s41467-022-33919-0} {\bibfield
  {journal} {\bibinfo  {journal} {Nat. Commun.}\ }\textbf {\bibinfo {volume}
  {13}},\ \bibinfo {pages} {6134} (\bibinfo {year} {2022})}\BibitemShut
  {NoStop}%
\bibitem [{\citenamefont {Liu}\ \emph {et~al.}(2023)\citenamefont {Liu},
  \citenamefont {Hu}, \citenamefont {Zhu}, \citenamefont {Zhang}, \citenamefont
  {Xiao}, \citenamefont {Miao}, \citenamefont {Ou}, \citenamefont {Liu},
  \citenamefont {Zhou}, \citenamefont {Li},\ and\ \citenamefont
  {Guo}}]{LiuGuoDistribQuComp23}%
  \BibitemOpen
  \bibfield  {author} {\bibinfo {author} {\bibfnamefont {X.}~\bibnamefont
  {Liu}}, \bibinfo {author} {\bibfnamefont {X.-M.}\ \bibnamefont {Hu}},
  \bibinfo {author} {\bibfnamefont {T.-X.}\ \bibnamefont {Zhu}}, \bibinfo
  {author} {\bibfnamefont {C.}~\bibnamefont {Zhang}}, \bibinfo {author}
  {\bibfnamefont {Y.-X.}\ \bibnamefont {Xiao}}, \bibinfo {author}
  {\bibfnamefont {J.-L.}\ \bibnamefont {Miao}}, \bibinfo {author}
  {\bibfnamefont {Z.-W.}\ \bibnamefont {Ou}}, \bibinfo {author} {\bibfnamefont
  {B.-H.}\ \bibnamefont {Liu}}, \bibinfo {author} {\bibfnamefont {Z.-Q.}\
  \bibnamefont {Zhou}}, \bibinfo {author} {\bibfnamefont {C.-F.}\ \bibnamefont
  {Li}},\ and\ \bibinfo {author} {\bibfnamefont {G.-C.}\ \bibnamefont {Guo}},\
  }\href@noop {} {\bibinfo {title} {Distributed quantum computing over 7.0 km}}
  (\bibinfo {year} {2023}),\ \Eprint {https://arxiv.org/abs/2307.15634}
  {arXiv:2307.15634 [quant-ph]} \BibitemShut {NoStop}%
\bibitem [{\citenamefont {Hermans}\ \emph {et~al.}(2022)\citenamefont
  {Hermans}, \citenamefont {Pompili}, \citenamefont {Beukers}, \citenamefont
  {Baier}, \citenamefont {Borregaard},\ and\ \citenamefont
  {Hanson}}]{PompiliHansonQuNetwork22}%
  \BibitemOpen
  \bibfield  {author} {\bibinfo {author} {\bibfnamefont {S.}~\bibnamefont
  {Hermans}}, \bibinfo {author} {\bibfnamefont {M.}~\bibnamefont {Pompili}},
  \bibinfo {author} {\bibfnamefont {H.}~\bibnamefont {Beukers}}, \bibinfo
  {author} {\bibfnamefont {S.}~\bibnamefont {Baier}}, \bibinfo {author}
  {\bibfnamefont {J.}~\bibnamefont {Borregaard}},\ and\ \bibinfo {author}
  {\bibfnamefont {R.}~\bibnamefont {Hanson}},\ }\bibfield  {title} {\bibinfo
  {title} {Qubit teleportation between non-neighbouring nodes in a quantum
  network},\ }\href
  {https://doi.org/https://doi.org/10.1038/s41586-022-04697-y} {\bibfield
  {journal} {\bibinfo  {journal} {Nature}\ }\textbf {\bibinfo {volume} {605}},\
  \bibinfo {pages} {663} (\bibinfo {year} {2022})}\BibitemShut {NoStop}%
\bibitem [{\citenamefont {Bersin}\ \emph {et~al.}(2023)\citenamefont {Bersin},
  \citenamefont {Grein}, \citenamefont {Sutula}, \citenamefont {Murphy},
  \citenamefont {Huan}, \citenamefont {Stevens}, \citenamefont {Suleymanzade},
  \citenamefont {Lee}, \citenamefont {Riedinger}, \citenamefont {Starling},
  \citenamefont {Stas}, \citenamefont {Knaut}, \citenamefont {Sinclair},
  \citenamefont {Assumpcao}, \citenamefont {Wei}, \citenamefont {Knall},
  \citenamefont {Machielse}, \citenamefont {Sukachev}, \citenamefont
  {Levonian}, \citenamefont {Bhaskar}, \citenamefont {Loncar}, \citenamefont
  {Hamilton}, \citenamefont {Lukin}, \citenamefont {Englund},\ and\
  \citenamefont {Dixon}}]{BersinLukinDixonMetroAreaQuNetwork23}%
  \BibitemOpen
  \bibfield  {author} {\bibinfo {author} {\bibfnamefont {E.}~\bibnamefont
  {Bersin}}, \bibinfo {author} {\bibfnamefont {M.}~\bibnamefont {Grein}},
  \bibinfo {author} {\bibfnamefont {M.}~\bibnamefont {Sutula}}, \bibinfo
  {author} {\bibfnamefont {R.}~\bibnamefont {Murphy}}, \bibinfo {author}
  {\bibfnamefont {Y.~Q.}\ \bibnamefont {Huan}}, \bibinfo {author}
  {\bibfnamefont {M.}~\bibnamefont {Stevens}}, \bibinfo {author} {\bibfnamefont
  {A.}~\bibnamefont {Suleymanzade}}, \bibinfo {author} {\bibfnamefont
  {C.}~\bibnamefont {Lee}}, \bibinfo {author} {\bibfnamefont {R.}~\bibnamefont
  {Riedinger}}, \bibinfo {author} {\bibfnamefont {D.~J.}\ \bibnamefont
  {Starling}}, \bibinfo {author} {\bibfnamefont {P.-J.}\ \bibnamefont {Stas}},
  \bibinfo {author} {\bibfnamefont {C.~M.}\ \bibnamefont {Knaut}}, \bibinfo
  {author} {\bibfnamefont {N.}~\bibnamefont {Sinclair}}, \bibinfo {author}
  {\bibfnamefont {D.~R.}\ \bibnamefont {Assumpcao}}, \bibinfo {author}
  {\bibfnamefont {Y.-C.}\ \bibnamefont {Wei}}, \bibinfo {author} {\bibfnamefont
  {E.~N.}\ \bibnamefont {Knall}}, \bibinfo {author} {\bibfnamefont
  {B.}~\bibnamefont {Machielse}}, \bibinfo {author} {\bibfnamefont {D.~D.}\
  \bibnamefont {Sukachev}}, \bibinfo {author} {\bibfnamefont {D.~S.}\
  \bibnamefont {Levonian}}, \bibinfo {author} {\bibfnamefont {M.~K.}\
  \bibnamefont {Bhaskar}}, \bibinfo {author} {\bibfnamefont {M.}~\bibnamefont
  {Loncar}}, \bibinfo {author} {\bibfnamefont {S.}~\bibnamefont {Hamilton}},
  \bibinfo {author} {\bibfnamefont {M.}~\bibnamefont {Lukin}}, \bibinfo
  {author} {\bibfnamefont {D.}~\bibnamefont {Englund}},\ and\ \bibinfo {author}
  {\bibfnamefont {P.~B.}\ \bibnamefont {Dixon}},\ }\href@noop {} {\bibinfo
  {title} {Development of a {Boston}-area 50-km fiber quantum network testbed}}
  (\bibinfo {year} {2023}),\ \Eprint {https://arxiv.org/abs/2307.15696}
  {arXiv:2307.15696 [quant-ph]} \BibitemShut {NoStop}%
\bibitem [{\citenamefont {Chen}\ \emph {et~al.}(2021)\citenamefont {Chen},
  \citenamefont {Jiang}, \citenamefont {Tang}, \citenamefont {Zhou},
  \citenamefont {Yuan}, \citenamefont {Zhou}, \citenamefont {Wang},
  \citenamefont {Liu}, \citenamefont {Chen}, \citenamefont {Liu}, \citenamefont
  {Zhang}, \citenamefont {Cui}, \citenamefont {Liang}, \citenamefont {Li},
  \citenamefont {Mao}, \citenamefont {Wang}, \citenamefont {Feng},
  \citenamefont {Chen}, \citenamefont {Zhang}, \citenamefont {Li},
  \citenamefont {Liu}, \citenamefont {Peng}, \citenamefont {Ma}, \citenamefont
  {Zhao},\ and\ \citenamefont {Pan}}]{PanEtalMetroAreaQuNetwork21}%
  \BibitemOpen
  \bibfield  {author} {\bibinfo {author} {\bibfnamefont {T.-Y.}\ \bibnamefont
  {Chen}}, \bibinfo {author} {\bibfnamefont {X.}~\bibnamefont {Jiang}},
  \bibinfo {author} {\bibfnamefont {S.-B.}\ \bibnamefont {Tang}}, \bibinfo
  {author} {\bibfnamefont {L.}~\bibnamefont {Zhou}}, \bibinfo {author}
  {\bibfnamefont {X.}~\bibnamefont {Yuan}}, \bibinfo {author} {\bibfnamefont
  {H.}~\bibnamefont {Zhou}}, \bibinfo {author} {\bibfnamefont {J.}~\bibnamefont
  {Wang}}, \bibinfo {author} {\bibfnamefont {Y.}~\bibnamefont {Liu}}, \bibinfo
  {author} {\bibfnamefont {L.-K.}\ \bibnamefont {Chen}}, \bibinfo {author}
  {\bibfnamefont {W.-Y.}\ \bibnamefont {Liu}}, \bibinfo {author} {\bibfnamefont
  {H.-F.}\ \bibnamefont {Zhang}}, \bibinfo {author} {\bibfnamefont
  {K.}~\bibnamefont {Cui}}, \bibinfo {author} {\bibfnamefont {H.}~\bibnamefont
  {Liang}}, \bibinfo {author} {\bibfnamefont {X.-G.}\ \bibnamefont {Li}},
  \bibinfo {author} {\bibfnamefont {Y.}~\bibnamefont {Mao}}, \bibinfo {author}
  {\bibfnamefont {L.-J.}\ \bibnamefont {Wang}}, \bibinfo {author}
  {\bibfnamefont {S.-B.}\ \bibnamefont {Feng}}, \bibinfo {author}
  {\bibfnamefont {Q.}~\bibnamefont {Chen}}, \bibinfo {author} {\bibfnamefont
  {Q.}~\bibnamefont {Zhang}}, \bibinfo {author} {\bibfnamefont
  {L.}~\bibnamefont {Li}}, \bibinfo {author} {\bibfnamefont {N.-L.}\
  \bibnamefont {Liu}}, \bibinfo {author} {\bibfnamefont {C.-Z.}\ \bibnamefont
  {Peng}}, \bibinfo {author} {\bibfnamefont {X.}~\bibnamefont {Ma}}, \bibinfo
  {author} {\bibfnamefont {Y.}~\bibnamefont {Zhao}},\ and\ \bibinfo {author}
  {\bibfnamefont {J.-W.}\ \bibnamefont {Pan}},\ }\bibfield  {title} {\bibinfo
  {title} {Implementation of a 46-node quantum metropolitan area network},\
  }\href {https://doi.org/https://doi.org/10.1038/s41534-021-00474-3}
  {\bibfield  {journal} {\bibinfo  {journal} {npj Quantum Inf}\ }\textbf
  {\bibinfo {volume} {7}},\ \bibinfo {pages} {134} (\bibinfo {year}
  {2021})}\BibitemShut {NoStop}%
\bibitem [{\citenamefont {Seidler}\ \emph {et~al.}(2022)\citenamefont
  {Seidler}, \citenamefont {Struck}, \citenamefont {Xue}, \citenamefont
  {Focke}, \citenamefont {Trellenkamp}, \citenamefont {Bluhm},\ and\
  \citenamefont {Schreiber}}]{seidler_conveyor-mode_2022}%
  \BibitemOpen
  \bibfield  {author} {\bibinfo {author} {\bibfnamefont {I.}~\bibnamefont
  {Seidler}}, \bibinfo {author} {\bibfnamefont {T.}~\bibnamefont {Struck}},
  \bibinfo {author} {\bibfnamefont {R.}~\bibnamefont {Xue}}, \bibinfo {author}
  {\bibfnamefont {N.}~\bibnamefont {Focke}}, \bibinfo {author} {\bibfnamefont
  {S.}~\bibnamefont {Trellenkamp}}, \bibinfo {author} {\bibfnamefont
  {H.}~\bibnamefont {Bluhm}},\ and\ \bibinfo {author} {\bibfnamefont {L.~R.}\
  \bibnamefont {Schreiber}},\ }\bibfield  {title} {\bibinfo {title}
  {Conveyor-mode single-electron shuttling in {Si}/{SiGe} for a scalable
  quantum computing architecture},\ }\href
  {https://doi.org/10.1038/s41534-022-00615-2} {\bibfield  {journal} {\bibinfo
  {journal} {npj Quantum Information}\ }\textbf {\bibinfo {volume} {8}},\
  \bibinfo {pages} {100} (\bibinfo {year} {2022})}\BibitemShut {NoStop}%
\bibitem [{\citenamefont {Yoneda}\ \emph {et~al.}(2021)\citenamefont {Yoneda},
  \citenamefont {Huang}, \citenamefont {Feng}, \citenamefont {Yang},
  \citenamefont {Chan}, \citenamefont {Tanttu}, \citenamefont {Gilbert},
  \citenamefont {Leon}, \citenamefont {Hudson}, \citenamefont {Itoh},
  \citenamefont {Morello}, \citenamefont {Bartlett}, \citenamefont {Laucht},
  \citenamefont {Saraiva},\ and\ \citenamefont
  {Dzurak}}]{yoneda_coherent_2021}%
  \BibitemOpen
  \bibfield  {author} {\bibinfo {author} {\bibfnamefont {J.}~\bibnamefont
  {Yoneda}}, \bibinfo {author} {\bibfnamefont {W.}~\bibnamefont {Huang}},
  \bibinfo {author} {\bibfnamefont {M.}~\bibnamefont {Feng}}, \bibinfo {author}
  {\bibfnamefont {C.~H.}\ \bibnamefont {Yang}}, \bibinfo {author}
  {\bibfnamefont {K.~W.}\ \bibnamefont {Chan}}, \bibinfo {author}
  {\bibfnamefont {T.}~\bibnamefont {Tanttu}}, \bibinfo {author} {\bibfnamefont
  {W.}~\bibnamefont {Gilbert}}, \bibinfo {author} {\bibfnamefont {R.~C.~C.}\
  \bibnamefont {Leon}}, \bibinfo {author} {\bibfnamefont {F.~E.}\ \bibnamefont
  {Hudson}}, \bibinfo {author} {\bibfnamefont {K.~M.}\ \bibnamefont {Itoh}},
  \bibinfo {author} {\bibfnamefont {A.}~\bibnamefont {Morello}}, \bibinfo
  {author} {\bibfnamefont {S.~D.}\ \bibnamefont {Bartlett}}, \bibinfo {author}
  {\bibfnamefont {A.}~\bibnamefont {Laucht}}, \bibinfo {author} {\bibfnamefont
  {A.}~\bibnamefont {Saraiva}},\ and\ \bibinfo {author} {\bibfnamefont {A.~S.}\
  \bibnamefont {Dzurak}},\ }\bibfield  {title} {\bibinfo {title} {Coherent spin
  qubit transport in silicon},\ }\href
  {https://doi.org/10.1038/s41467-021-24371-7} {\bibfield  {journal} {\bibinfo
  {journal} {Nature Communications}\ }\textbf {\bibinfo {volume} {12}},\
  \bibinfo {pages} {4114} (\bibinfo {year} {2021})}\BibitemShut {NoStop}%
\bibitem [{\citenamefont {Zwerver}\ \emph {et~al.}(2023)\citenamefont
  {Zwerver}, \citenamefont {Amitonov}, \citenamefont {De~Snoo}, \citenamefont
  {Madzik}, \citenamefont {Rimbach-Russ}, \citenamefont {Sammak}, \citenamefont
  {Scappucci},\ and\ \citenamefont {Vandersypen}}]{zwerver_shuttling_2023}%
  \BibitemOpen
  \bibfield  {author} {\bibinfo {author} {\bibfnamefont {A.~M.~J.}\
  \bibnamefont {Zwerver}}, \bibinfo {author} {\bibfnamefont {S.~V.}\
  \bibnamefont {Amitonov}}, \bibinfo {author} {\bibfnamefont {S.~L.}\
  \bibnamefont {De~Snoo}}, \bibinfo {author} {\bibfnamefont {M.~T.}\
  \bibnamefont {Madzik}}, \bibinfo {author} {\bibfnamefont {M.}~\bibnamefont
  {Rimbach-Russ}}, \bibinfo {author} {\bibfnamefont {A.}~\bibnamefont
  {Sammak}}, \bibinfo {author} {\bibfnamefont {G.}~\bibnamefont {Scappucci}},\
  and\ \bibinfo {author} {\bibfnamefont {L.~M.~K.}\ \bibnamefont
  {Vandersypen}},\ }\bibfield  {title} {\bibinfo {title} {Shuttling an
  {Electron} {Spin} through a {Silicon} {Quantum} {Dot} {Array}},\ }\href
  {https://doi.org/10.1103/PRXQuantum.4.030303} {\bibfield  {journal} {\bibinfo
   {journal} {PRX Quantum}\ }\textbf {\bibinfo {volume} {4}},\ \bibinfo {pages}
  {030303} (\bibinfo {year} {2023})}\BibitemShut {NoStop}%
\bibitem [{\citenamefont {van Riggelen-Doelman}\ \emph
  {et~al.}(2023)\citenamefont {van Riggelen-Doelman}, \citenamefont {Wang},
  \citenamefont {de~Snoo}, \citenamefont {Lawrie}, \citenamefont {Hendrickx},
  \citenamefont {Rimbach-Russ}, \citenamefont {Sammak}, \citenamefont
  {Scappucci}, \citenamefont {D{\'e}prez},\ and\ \citenamefont
  {Veldhorst}}]{van_riggelen-doelman_coherent_2023}%
  \BibitemOpen
  \bibfield  {author} {\bibinfo {author} {\bibfnamefont {F.}~\bibnamefont {van
  Riggelen-Doelman}}, \bibinfo {author} {\bibfnamefont {C.-A.}\ \bibnamefont
  {Wang}}, \bibinfo {author} {\bibfnamefont {S.~L.}\ \bibnamefont {de~Snoo}},
  \bibinfo {author} {\bibfnamefont {W.~I.~L.}\ \bibnamefont {Lawrie}}, \bibinfo
  {author} {\bibfnamefont {N.~W.}\ \bibnamefont {Hendrickx}}, \bibinfo {author}
  {\bibfnamefont {M.}~\bibnamefont {Rimbach-Russ}}, \bibinfo {author}
  {\bibfnamefont {A.}~\bibnamefont {Sammak}}, \bibinfo {author} {\bibfnamefont
  {G.}~\bibnamefont {Scappucci}}, \bibinfo {author} {\bibfnamefont
  {C.}~\bibnamefont {D{\'e}prez}},\ and\ \bibinfo {author} {\bibfnamefont
  {M.}~\bibnamefont {Veldhorst}},\ }\href {http://arxiv.org/abs/2308.02406}
  {\bibinfo {title} {Coherent spin qubit shuttling through germanium quantum
  dots}} (\bibinfo {year} {2023}),\ \bibinfo {note} {arXiv:2308.02406
  [cond-mat, physics:quant-ph]}\BibitemShut {NoStop}%
\bibitem [{\citenamefont {Ginzel}\ \emph {et~al.}(2020)\citenamefont {Ginzel},
  \citenamefont {Mills}, \citenamefont {Petta},\ and\ \citenamefont
  {Burkard}}]{BurkardEtalTwoDotShuttl20}%
  \BibitemOpen
  \bibfield  {author} {\bibinfo {author} {\bibfnamefont {F.}~\bibnamefont
  {Ginzel}}, \bibinfo {author} {\bibfnamefont {A.~R.}\ \bibnamefont {Mills}},
  \bibinfo {author} {\bibfnamefont {J.~R.}\ \bibnamefont {Petta}},\ and\
  \bibinfo {author} {\bibfnamefont {G.}~\bibnamefont {Burkard}},\ }\bibfield
  {title} {\bibinfo {title} {Spin shuttling in a silicon double quantum dot},\
  }\href {https://doi.org/10.1103/PhysRevB.102.195418} {\bibfield  {journal}
  {\bibinfo  {journal} {Phys. Rev. B}\ }\textbf {\bibinfo {volume} {102}},\
  \bibinfo {pages} {195418} (\bibinfo {year} {2020})}\BibitemShut {NoStop}%
\bibitem [{\citenamefont {Noiri}\ \emph {et~al.}(2022)\citenamefont {Noiri},
  \citenamefont {Takeda}, \citenamefont {Nakajima}, \citenamefont {Kobayashi},
  \citenamefont {Sammak}, \citenamefont {Scappucci},\ and\ \citenamefont
  {Tarucha}}]{noiri_shuttling-based_2022}%
  \BibitemOpen
  \bibfield  {author} {\bibinfo {author} {\bibfnamefont {A.}~\bibnamefont
  {Noiri}}, \bibinfo {author} {\bibfnamefont {K.}~\bibnamefont {Takeda}},
  \bibinfo {author} {\bibfnamefont {T.}~\bibnamefont {Nakajima}}, \bibinfo
  {author} {\bibfnamefont {T.}~\bibnamefont {Kobayashi}}, \bibinfo {author}
  {\bibfnamefont {A.}~\bibnamefont {Sammak}}, \bibinfo {author} {\bibfnamefont
  {G.}~\bibnamefont {Scappucci}},\ and\ \bibinfo {author} {\bibfnamefont
  {S.}~\bibnamefont {Tarucha}},\ }\bibfield  {title} {\bibinfo {title} {A
  shuttling-based two-qubit logic gate for linking distant silicon quantum
  processors},\ }\href {https://doi.org/10.1038/s41467-022-33453-z} {\bibfield
  {journal} {\bibinfo  {journal} {Nature Communications}\ }\textbf {\bibinfo
  {volume} {13}},\ \bibinfo {pages} {5740} (\bibinfo {year}
  {2022})}\BibitemShut {NoStop}%
\bibitem [{\citenamefont {Struck}\ \emph {et~al.}(2023)\citenamefont {Struck},
  \citenamefont {Volmer}, \citenamefont {Visser}, \citenamefont {Offermann},
  \citenamefont {Xue}, \citenamefont {Tu}, \citenamefont {Trellenkamp},
  \citenamefont {Cywi{\'n}ski}, \citenamefont {Bluhm},\ and\ \citenamefont
  {Schreiber}}]{StruckSchreiberEtalSpinPairShuttling23}%
  \BibitemOpen
  \bibfield  {author} {\bibinfo {author} {\bibfnamefont {T.}~\bibnamefont
  {Struck}}, \bibinfo {author} {\bibfnamefont {M.}~\bibnamefont {Volmer}},
  \bibinfo {author} {\bibfnamefont {L.}~\bibnamefont {Visser}}, \bibinfo
  {author} {\bibfnamefont {T.}~\bibnamefont {Offermann}}, \bibinfo {author}
  {\bibfnamefont {R.}~\bibnamefont {Xue}}, \bibinfo {author} {\bibfnamefont
  {J.-S.}\ \bibnamefont {Tu}}, \bibinfo {author} {\bibfnamefont
  {S.}~\bibnamefont {Trellenkamp}}, \bibinfo {author} {\bibfnamefont
  {{\L}.}~\bibnamefont {Cywi{\'n}ski}}, \bibinfo {author} {\bibfnamefont
  {H.}~\bibnamefont {Bluhm}},\ and\ \bibinfo {author} {\bibfnamefont {L.~R.}\
  \bibnamefont {Schreiber}},\ }\href
  {https://doi.org/10.48550/arXiv.2307.04897} {\bibinfo {title}
  {Spin-{EPR}-pair separation by conveyor-mode single electron shuttling in
  {Si}/{SiGe}}} (\bibinfo {year} {2023}),\ \Eprint
  {https://arxiv.org/abs/2307.04897} {arXiv:2307.04897 [quant-ph]} \BibitemShut
  {NoStop}%
\bibitem [{\citenamefont {Smet}\ \emph {et~al.}(2024)\citenamefont {Smet},
  \citenamefont {Matsumoto}, \citenamefont {Zwerver}, \citenamefont {Tryputen},
  \citenamefont {de~Snoo}, \citenamefont {Amitonov}, \citenamefont {Sammak},
  \citenamefont {Samkharadze}, \citenamefont {G{\"u}l}, \citenamefont
  {Wasserman}, \citenamefont {Rimbach-Russ}, \citenamefont {Scappucci},\ and\
  \citenamefont {Vandersypen}}]{SmetVandersypenEtal24SpinShuttlSilicon}%
  \BibitemOpen
  \bibfield  {author} {\bibinfo {author} {\bibfnamefont {M.~D.}\ \bibnamefont
  {Smet}}, \bibinfo {author} {\bibfnamefont {Y.}~\bibnamefont {Matsumoto}},
  \bibinfo {author} {\bibfnamefont {A.-M.~J.}\ \bibnamefont {Zwerver}},
  \bibinfo {author} {\bibfnamefont {L.}~\bibnamefont {Tryputen}}, \bibinfo
  {author} {\bibfnamefont {S.~L.}\ \bibnamefont {de~Snoo}}, \bibinfo {author}
  {\bibfnamefont {S.~V.}\ \bibnamefont {Amitonov}}, \bibinfo {author}
  {\bibfnamefont {A.}~\bibnamefont {Sammak}}, \bibinfo {author} {\bibfnamefont
  {N.}~\bibnamefont {Samkharadze}}, \bibinfo {author} {\bibfnamefont
  {{\"O}.}~\bibnamefont {G{\"u}l}}, \bibinfo {author} {\bibfnamefont
  {R.~N.~M.}\ \bibnamefont {Wasserman}}, \bibinfo {author} {\bibfnamefont
  {M.}~\bibnamefont {Rimbach-Russ}}, \bibinfo {author} {\bibfnamefont
  {G.}~\bibnamefont {Scappucci}},\ and\ \bibinfo {author} {\bibfnamefont
  {L.~M.~K.}\ \bibnamefont {Vandersypen}},\ }\href
  {https://arxiv.org/abs/2406.07267} {\bibinfo {title} {High-fidelity
  single-spin shuttling in silicon}} (\bibinfo {year} {2024}),\ \Eprint
  {https://arxiv.org/abs/2406.07267} {arXiv:2406.07267 [cond-mat.mes-hall]}
  \BibitemShut {NoStop}%
\bibitem [{\citenamefont {Jadot}\ \emph {et~al.}(2021)\citenamefont {Jadot},
  \citenamefont {Mortemousque}, \citenamefont {Chanrion}, \citenamefont
  {Thiney}, \citenamefont {Ludwig}, \citenamefont {Wieck}, \citenamefont
  {Urdampilleta}, \citenamefont {B{\"a}uerle},\ and\ \citenamefont
  {Meunier}}]{JadotMeunierEtalTwoSpinShuttling21}%
  \BibitemOpen
  \bibfield  {author} {\bibinfo {author} {\bibfnamefont {B.}~\bibnamefont
  {Jadot}}, \bibinfo {author} {\bibfnamefont {P.-A.}\ \bibnamefont
  {Mortemousque}}, \bibinfo {author} {\bibfnamefont {E.}~\bibnamefont
  {Chanrion}}, \bibinfo {author} {\bibfnamefont {V.}~\bibnamefont {Thiney}},
  \bibinfo {author} {\bibfnamefont {A.}~\bibnamefont {Ludwig}}, \bibinfo
  {author} {\bibfnamefont {A.~D.}\ \bibnamefont {Wieck}}, \bibinfo {author}
  {\bibfnamefont {M.}~\bibnamefont {Urdampilleta}}, \bibinfo {author}
  {\bibfnamefont {C.}~\bibnamefont {B{\"a}uerle}},\ and\ \bibinfo {author}
  {\bibfnamefont {T.}~\bibnamefont {Meunier}},\ }\bibfield  {title} {\bibinfo
  {title} {Distant spin entanglement via fast and coherent electron
  shuttling},\ }\href
  {https://doi.org/https://doi.org/10.1038/s41565-021-00846-y} {\bibfield
  {journal} {\bibinfo  {journal} {Nat. Nanotechnol.}\ }\textbf {\bibinfo
  {volume} {16}},\ \bibinfo {pages} {570} (\bibinfo {year} {2021})}\BibitemShut
  {NoStop}%
\bibitem [{\citenamefont {Boter}\ \emph {et~al.}(2020)\citenamefont {Boter},
  \citenamefont {Xue}, \citenamefont {Kr\"ahenmann}, \citenamefont {Watson},
  \citenamefont {Premakumar}, \citenamefont {Ward}, \citenamefont {Savage},
  \citenamefont {Lagally}, \citenamefont {Friesen}, \citenamefont
  {Coppersmith}, \citenamefont {Eriksson}, \citenamefont {Joynt},\ and\
  \citenamefont {Vandersypen}}]{BoterJoyntVdSNoiseCorrBellStates20}%
  \BibitemOpen
  \bibfield  {author} {\bibinfo {author} {\bibfnamefont {J.~M.}\ \bibnamefont
  {Boter}}, \bibinfo {author} {\bibfnamefont {X.}~\bibnamefont {Xue}}, \bibinfo
  {author} {\bibfnamefont {T.}~\bibnamefont {Kr\"ahenmann}}, \bibinfo {author}
  {\bibfnamefont {T.~F.}\ \bibnamefont {Watson}}, \bibinfo {author}
  {\bibfnamefont {V.~N.}\ \bibnamefont {Premakumar}}, \bibinfo {author}
  {\bibfnamefont {D.~R.}\ \bibnamefont {Ward}}, \bibinfo {author}
  {\bibfnamefont {D.~E.}\ \bibnamefont {Savage}}, \bibinfo {author}
  {\bibfnamefont {M.~G.}\ \bibnamefont {Lagally}}, \bibinfo {author}
  {\bibfnamefont {M.}~\bibnamefont {Friesen}}, \bibinfo {author} {\bibfnamefont
  {S.~N.}\ \bibnamefont {Coppersmith}}, \bibinfo {author} {\bibfnamefont
  {M.~A.}\ \bibnamefont {Eriksson}}, \bibinfo {author} {\bibfnamefont
  {R.}~\bibnamefont {Joynt}},\ and\ \bibinfo {author} {\bibfnamefont
  {L.~M.~K.}\ \bibnamefont {Vandersypen}},\ }\bibfield  {title} {\bibinfo
  {title} {Spatial noise correlations in a si/sige two-qubit device from bell
  state coherences},\ }\href {https://doi.org/10.1103/PhysRevB.101.235133}
  {\bibfield  {journal} {\bibinfo  {journal} {Phys. Rev. B}\ }\textbf {\bibinfo
  {volume} {101}},\ \bibinfo {pages} {235133} (\bibinfo {year}
  {2020})}\BibitemShut {NoStop}%
\bibitem [{\citenamefont {Mortemousque}\ \emph {et~al.}(2021)\citenamefont
  {Mortemousque}, \citenamefont {Jadot}, \citenamefont {Chanrion},
  \citenamefont {Thiney}, \citenamefont {B\"auerle}, \citenamefont {Ludwig},
  \citenamefont {Wieck}, \citenamefont {Urdampilleta},\ and\ \citenamefont
  {Meunier}}]{MortemousqueMeunierEtalShuttl2DArray21}%
  \BibitemOpen
  \bibfield  {author} {\bibinfo {author} {\bibfnamefont {P.-A.}\ \bibnamefont
  {Mortemousque}}, \bibinfo {author} {\bibfnamefont {B.}~\bibnamefont {Jadot}},
  \bibinfo {author} {\bibfnamefont {E.}~\bibnamefont {Chanrion}}, \bibinfo
  {author} {\bibfnamefont {V.}~\bibnamefont {Thiney}}, \bibinfo {author}
  {\bibfnamefont {C.}~\bibnamefont {B\"auerle}}, \bibinfo {author}
  {\bibfnamefont {A.}~\bibnamefont {Ludwig}}, \bibinfo {author} {\bibfnamefont
  {A.~D.}\ \bibnamefont {Wieck}}, \bibinfo {author} {\bibfnamefont
  {M.}~\bibnamefont {Urdampilleta}},\ and\ \bibinfo {author} {\bibfnamefont
  {T.}~\bibnamefont {Meunier}},\ }\bibfield  {title} {\bibinfo {title}
  {Enhanced spin coherence while displacing electron in a two-dimensional array
  of quantum dots},\ }\href {https://doi.org/10.1103/PRXQuantum.2.030331}
  {\bibfield  {journal} {\bibinfo  {journal} {PRX Quantum}\ }\textbf {\bibinfo
  {volume} {2}},\ \bibinfo {pages} {030331} (\bibinfo {year}
  {2021})}\BibitemShut {NoStop}%
\bibitem [{\citenamefont {Bosco}\ \emph {et~al.}(2024)\citenamefont {Bosco},
  \citenamefont {Zou},\ and\ \citenamefont
  {Loss}}]{BoscoZouLossHighFidShuttlingSOI23}%
  \BibitemOpen
  \bibfield  {author} {\bibinfo {author} {\bibfnamefont {S.}~\bibnamefont
  {Bosco}}, \bibinfo {author} {\bibfnamefont {J.}~\bibnamefont {Zou}},\ and\
  \bibinfo {author} {\bibfnamefont {D.}~\bibnamefont {Loss}},\ }\bibfield
  {title} {\bibinfo {title} {High-fidelity spin qubit shuttling via large
  spin-orbit interactions},\ }\href
  {https://doi.org/10.1103/PRXQuantum.5.020353} {\bibfield  {journal} {\bibinfo
   {journal} {PRX Quantum}\ }\textbf {\bibinfo {volume} {5}},\ \bibinfo {pages}
  {020353} (\bibinfo {year} {2024})}\BibitemShut {NoStop}%
\bibitem [{\citenamefont {Huang}\ and\ \citenamefont
  {Hu}(2013)}]{HuangHuSpinRelaxShuttl13}%
  \BibitemOpen
  \bibfield  {author} {\bibinfo {author} {\bibfnamefont {P.}~\bibnamefont
  {Huang}}\ and\ \bibinfo {author} {\bibfnamefont {X.}~\bibnamefont {Hu}},\
  }\bibfield  {title} {\bibinfo {title} {Spin qubit relaxation in a moving
  quantum dot},\ }\href {https://doi.org/10.1103/PhysRevB.88.075301} {\bibfield
   {journal} {\bibinfo  {journal} {Phys. Rev. B}\ }\textbf {\bibinfo {volume}
  {88}},\ \bibinfo {pages} {075301} (\bibinfo {year} {2013})}\BibitemShut
  {NoStop}%
\bibitem [{\citenamefont {Struck}\ \emph {et~al.}(2020)\citenamefont {Struck},
  \citenamefont {Hollmann}, \citenamefont {Schauer}, \citenamefont {Fedorets},
  \citenamefont {Schmidbauer}, \citenamefont {Sawano}, \citenamefont {Riemann},
  \citenamefont {Abrosimov}, \citenamefont {Cywi{\'n}ski}, \citenamefont
  {Bougeard},\ and\ \citenamefont
  {Schreiber}}]{StruckCywinskiSchreiber20QubitNoise}%
  \BibitemOpen
  \bibfield  {author} {\bibinfo {author} {\bibfnamefont {T.}~\bibnamefont
  {Struck}}, \bibinfo {author} {\bibfnamefont {A.}~\bibnamefont {Hollmann}},
  \bibinfo {author} {\bibfnamefont {F.}~\bibnamefont {Schauer}}, \bibinfo
  {author} {\bibfnamefont {O.}~\bibnamefont {Fedorets}}, \bibinfo {author}
  {\bibfnamefont {A.}~\bibnamefont {Schmidbauer}}, \bibinfo {author}
  {\bibfnamefont {K.}~\bibnamefont {Sawano}}, \bibinfo {author} {\bibfnamefont
  {H.}~\bibnamefont {Riemann}}, \bibinfo {author} {\bibfnamefont {N.~V.}\
  \bibnamefont {Abrosimov}}, \bibinfo {author} {\bibfnamefont
  {{\L}.}~\bibnamefont {Cywi{\'n}ski}}, \bibinfo {author} {\bibfnamefont
  {D.}~\bibnamefont {Bougeard}},\ and\ \bibinfo {author} {\bibfnamefont
  {L.~R.}\ \bibnamefont {Schreiber}},\ }\bibfield  {title} {\bibinfo {title}
  {Low-frequency spin qubit energy splitting noise in highly purified
  {$^{28}$Si}/{SiGe}},\ }\href {https://doi.org/10.1038/s41534-020-0276-2}
  {\bibfield  {journal} {\bibinfo  {journal} {npj Quantum Inf}\ }\textbf
  {\bibinfo {volume} {6}},\ \bibinfo {pages} {40} (\bibinfo {year}
  {2020})}\BibitemShut {NoStop}%
\bibitem [{\citenamefont {Kepa}\ \emph
  {et~al.}(2023{\natexlab{a}})\citenamefont {Kepa}, \citenamefont
  {Cywi{\'n}ski},\ and\ \citenamefont
  {Krzywda}}]{KepaCywinskiKrzywdaSpinNoise23}%
  \BibitemOpen
  \bibfield  {author} {\bibinfo {author} {\bibfnamefont {M.}~\bibnamefont
  {Kepa}}, \bibinfo {author} {\bibfnamefont {{\L}.}~\bibnamefont
  {Cywi{\'n}ski}},\ and\ \bibinfo {author} {\bibfnamefont {J.~A.}\ \bibnamefont
  {Krzywda}},\ }\bibfield  {title} {\bibinfo {title} {Correlations of spin
  splitting and orbital fluctuations due to 1/f charge noise in the {Si}/{SiGe}
  quantum dot},\ }\href {https://doi.org/10.1063/5.0156358} {\bibfield
  {journal} {\bibinfo  {journal} {Applied Physics Letters}\ }\textbf {\bibinfo
  {volume} {123}},\ \bibinfo {pages} {034003} (\bibinfo {year}
  {2023}{\natexlab{a}})}\BibitemShut {NoStop}%
\bibitem [{\citenamefont {Zou}\ \emph {et~al.}(2023)\citenamefont {Zou},
  \citenamefont {Bosco},\ and\ \citenamefont {Loss}}]{zou_spatially_2023}%
  \BibitemOpen
  \bibfield  {author} {\bibinfo {author} {\bibfnamefont {J.}~\bibnamefont
  {Zou}}, \bibinfo {author} {\bibfnamefont {S.}~\bibnamefont {Bosco}},\ and\
  \bibinfo {author} {\bibfnamefont {D.}~\bibnamefont {Loss}},\ }\href
  {http://arxiv.org/abs/2308.03054} {\bibinfo {title} {Spatially correlated
  classical and quantum noise in driven qubits: {The} good, the bad, and the
  ugly}} (\bibinfo {year} {2023})\BibitemShut {NoStop}%
\bibitem [{\citenamefont {Shalak}\ \emph {et~al.}(2023)\citenamefont {Shalak},
  \citenamefont {Delerue},\ and\ \citenamefont
  {Niquet}}]{ShalakDelerueNiquetChargeNoiseSiHole23}%
  \BibitemOpen
  \bibfield  {author} {\bibinfo {author} {\bibfnamefont {B.}~\bibnamefont
  {Shalak}}, \bibinfo {author} {\bibfnamefont {C.}~\bibnamefont {Delerue}},\
  and\ \bibinfo {author} {\bibfnamefont {Y.-M.}\ \bibnamefont {Niquet}},\
  }\bibfield  {title} {\bibinfo {title} {Modeling of spin decoherence in a {Si}
  hole qubit perturbed by a single charge fluctuator},\ }\href
  {https://doi.org/10.1103/PhysRevB.107.125415} {\bibfield  {journal} {\bibinfo
   {journal} {Phys. Rev. B}\ }\textbf {\bibinfo {volume} {107}},\ \bibinfo
  {pages} {125415} (\bibinfo {year} {2023})}\BibitemShut {NoStop}%
\bibitem [{\citenamefont {Spence}\ \emph {et~al.}(2022)\citenamefont {Spence},
  \citenamefont {Cardoso-Paz}, \citenamefont {Michal}, \citenamefont
  {Chanrion}, \citenamefont {Niegemann}, \citenamefont {Jadot}, \citenamefont
  {Mortemousque}, \citenamefont {Klemt}, \citenamefont {Thiney}, \citenamefont
  {Bertrand}, \citenamefont {Hutin}, \citenamefont {B{\"a}uerle}, \citenamefont
  {Balestro}, \citenamefont {Vinet}, \citenamefont {Niquet}, \citenamefont
  {Meunier},\ and\ \citenamefont
  {Urdampilleta}}]{SpenceNiquetMeunierEtalChargeNoise22}%
  \BibitemOpen
  \bibfield  {author} {\bibinfo {author} {\bibfnamefont {C.}~\bibnamefont
  {Spence}}, \bibinfo {author} {\bibfnamefont {B.}~\bibnamefont {Cardoso-Paz}},
  \bibinfo {author} {\bibfnamefont {V.}~\bibnamefont {Michal}}, \bibinfo
  {author} {\bibfnamefont {E.}~\bibnamefont {Chanrion}}, \bibinfo {author}
  {\bibfnamefont {D.~J.}\ \bibnamefont {Niegemann}}, \bibinfo {author}
  {\bibfnamefont {B.}~\bibnamefont {Jadot}}, \bibinfo {author} {\bibfnamefont
  {P.-A.}\ \bibnamefont {Mortemousque}}, \bibinfo {author} {\bibfnamefont
  {B.}~\bibnamefont {Klemt}}, \bibinfo {author} {\bibfnamefont
  {V.}~\bibnamefont {Thiney}}, \bibinfo {author} {\bibfnamefont
  {B.}~\bibnamefont {Bertrand}}, \bibinfo {author} {\bibfnamefont
  {L.}~\bibnamefont {Hutin}}, \bibinfo {author} {\bibfnamefont
  {C.}~\bibnamefont {B{\"a}uerle}}, \bibinfo {author} {\bibfnamefont
  {F.}~\bibnamefont {Balestro}}, \bibinfo {author} {\bibfnamefont
  {M.}~\bibnamefont {Vinet}}, \bibinfo {author} {\bibfnamefont {Y.-M.}\
  \bibnamefont {Niquet}}, \bibinfo {author} {\bibfnamefont {T.}~\bibnamefont
  {Meunier}},\ and\ \bibinfo {author} {\bibfnamefont {M.}~\bibnamefont
  {Urdampilleta}},\ }\href {https://doi.org/10.48550/arXiv.2209.01853}
  {\bibinfo {title} {Probing charge noise in few electron cmos quantum dots}}
  (\bibinfo {year} {2022}),\ \Eprint {https://arxiv.org/abs/2209.01853}
  {arXiv:2209.01853 [cond-mat.mes-hall]} \BibitemShut {NoStop}%
\bibitem [{\citenamefont {Shehata}\ \emph {et~al.}(2023)\citenamefont
  {Shehata}, \citenamefont {Simion}, \citenamefont {Li}, \citenamefont
  {Mohiyaddin}, \citenamefont {Wan}, \citenamefont {Mongillo}, \citenamefont
  {Govoreanu}, \citenamefont {Radu}, \citenamefont {De~Greve},\ and\
  \citenamefont {Van~Dorpe}}]{ShehataVanDorpeEtalChargeNoiseQuDots23}%
  \BibitemOpen
  \bibfield  {author} {\bibinfo {author} {\bibfnamefont {M.~M. E.~K.}\
  \bibnamefont {Shehata}}, \bibinfo {author} {\bibfnamefont {G.}~\bibnamefont
  {Simion}}, \bibinfo {author} {\bibfnamefont {R.}~\bibnamefont {Li}}, \bibinfo
  {author} {\bibfnamefont {F.~A.}\ \bibnamefont {Mohiyaddin}}, \bibinfo
  {author} {\bibfnamefont {D.}~\bibnamefont {Wan}}, \bibinfo {author}
  {\bibfnamefont {M.}~\bibnamefont {Mongillo}}, \bibinfo {author}
  {\bibfnamefont {B.}~\bibnamefont {Govoreanu}}, \bibinfo {author}
  {\bibfnamefont {I.}~\bibnamefont {Radu}}, \bibinfo {author} {\bibfnamefont
  {K.}~\bibnamefont {De~Greve}},\ and\ \bibinfo {author} {\bibfnamefont
  {P.}~\bibnamefont {Van~Dorpe}},\ }\bibfield  {title} {\bibinfo {title}
  {Modeling semiconductor spin qubits and their charge noise environment for
  quantum gate fidelity estimation},\ }\href
  {https://doi.org/10.1103/PhysRevB.108.045305} {\bibfield  {journal} {\bibinfo
   {journal} {Phys. Rev. B}\ }\textbf {\bibinfo {volume} {108}},\ \bibinfo
  {pages} {045305} (\bibinfo {year} {2023})}\BibitemShut {NoStop}%
\bibitem [{\citenamefont {Burkard}\ \emph {et~al.}(2023)\citenamefont
  {Burkard}, \citenamefont {Ladd}, \citenamefont {Pan}, \citenamefont
  {Nichol},\ and\ \citenamefont {Petta}}]{BurkardLaddPanNicholReviewQuDots23}%
  \BibitemOpen
  \bibfield  {author} {\bibinfo {author} {\bibfnamefont {G.}~\bibnamefont
  {Burkard}}, \bibinfo {author} {\bibfnamefont {T.~D.}\ \bibnamefont {Ladd}},
  \bibinfo {author} {\bibfnamefont {A.}~\bibnamefont {Pan}}, \bibinfo {author}
  {\bibfnamefont {J.~M.}\ \bibnamefont {Nichol}},\ and\ \bibinfo {author}
  {\bibfnamefont {J.~R.}\ \bibnamefont {Petta}},\ }\bibfield  {title} {\bibinfo
  {title} {Semiconductor spin qubits},\ }\href
  {https://doi.org/10.1103/RevModPhys.95.025003} {\bibfield  {journal}
  {\bibinfo  {journal} {Rev. Mod. Phys.}\ }\textbf {\bibinfo {volume} {95}},\
  \bibinfo {pages} {025003} (\bibinfo {year} {2023})}\BibitemShut {NoStop}%
\bibitem [{\citenamefont {Drummond}\ and\ \citenamefont
  {Corney}(2001)}]{CorneyDrummond01NoiseInOptFibersI}%
  \BibitemOpen
  \bibfield  {author} {\bibinfo {author} {\bibfnamefont {P.~D.}\ \bibnamefont
  {Drummond}}\ and\ \bibinfo {author} {\bibfnamefont {J.~F.}\ \bibnamefont
  {Corney}},\ }\bibfield  {title} {\bibinfo {title} {Quantum noise in optical
  fibers. {I}. {S}tochastic equations},\ }\href
  {https://doi.org/10.1364/JOSAB.18.000139} {\bibfield  {journal} {\bibinfo
  {journal} {J. Opt. Soc. Am. B}\ }\textbf {\bibinfo {volume} {18}},\ \bibinfo
  {pages} {139} (\bibinfo {year} {2001})}\BibitemShut {NoStop}%
\bibitem [{\citenamefont {Corney}\ and\ \citenamefont
  {Drummond}(2001)}]{CorneyDrummond01NoiseInOptFibersII}%
  \BibitemOpen
  \bibfield  {author} {\bibinfo {author} {\bibfnamefont {J.~F.}\ \bibnamefont
  {Corney}}\ and\ \bibinfo {author} {\bibfnamefont {P.~D.}\ \bibnamefont
  {Drummond}},\ }\bibfield  {title} {\bibinfo {title} {Quantum noise in optical
  fibers. {II}. {R}aman jitter in soliton communications},\ }\href
  {https://doi.org/10.1364/JOSAB.18.000153} {\bibfield  {journal} {\bibinfo
  {journal} {J. Opt. Soc. Am. B}\ }\textbf {\bibinfo {volume} {18}},\ \bibinfo
  {pages} {153} (\bibinfo {year} {2001})}\BibitemShut {NoStop}%
\bibitem [{\citenamefont {Dong}\ \emph {et~al.}(2016)\citenamefont {Dong},
  \citenamefont {Huang}, \citenamefont {Li},\ and\ \citenamefont
  {Liu}}]{DongHuangLiLiuNoiseOptFibers}%
  \BibitemOpen
  \bibfield  {author} {\bibinfo {author} {\bibfnamefont {J.}~\bibnamefont
  {Dong}}, \bibinfo {author} {\bibfnamefont {J.}~\bibnamefont {Huang}},
  \bibinfo {author} {\bibfnamefont {T.}~\bibnamefont {Li}},\ and\ \bibinfo
  {author} {\bibfnamefont {L.}~\bibnamefont {Liu}},\ }\bibfield  {title}
  {\bibinfo {title} {{Observation of fundamental thermal noise in optical
  fibers down to infrasonic frequencies}},\ }\href
  {https://doi.org/10.1063/1.4939918} {\bibfield  {journal} {\bibinfo
  {journal} {Applied Physics Letters}\ }\textbf {\bibinfo {volume} {108}},\
  \bibinfo {pages} {021108} (\bibinfo {year} {2016})}\BibitemShut {NoStop}%
\bibitem [{\citenamefont {Jeunhomme}(1990)}]{JeunhommeSingleModeFiberOpt}%
  \BibitemOpen
  \bibfield  {author} {\bibinfo {author} {\bibfnamefont {L.~B.}\ \bibnamefont
  {Jeunhomme}},\ }\href@noop {} {\emph {\bibinfo {title} {Single-mode fiber
  optics: principles and applications}}},\ \bibinfo {edition} {2nd}\ ed.\
  (\bibinfo  {publisher} {Marcel Dekker},\ \bibinfo {address} {New York},\
  \bibinfo {year} {1990})\BibitemShut {NoStop}%
\bibitem [{\citenamefont {Wanser}(1992)}]{WanserFluctOptFibers92}%
  \BibitemOpen
  \bibfield  {author} {\bibinfo {author} {\bibfnamefont {K.~H.}\ \bibnamefont
  {Wanser}},\ }\bibfield  {title} {\bibinfo {title} {Fundamental phase noise
  limit in optical fibres due to temperature fluctuations},\ }\href
  {https://digital-library.theiet.org/content/journals/10.1049/el_19920033}
  {\bibfield  {journal} {\bibinfo  {journal} {Electronics Letters}\ }\textbf
  {\bibinfo {volume} {28}},\ \bibinfo {pages} {53} (\bibinfo {year}
  {1992})}\BibitemShut {NoStop}%
\bibitem [{\citenamefont {Kubo}\ \emph {et~al.}(1985)\citenamefont {Kubo},
  \citenamefont {Toda},\ and\ \citenamefont
  {Hashitsume}}]{kubo_statistical_1985}%
  \BibitemOpen
  \bibfield  {author} {\bibinfo {author} {\bibfnamefont {R.}~\bibnamefont
  {Kubo}}, \bibinfo {author} {\bibfnamefont {M.}~\bibnamefont {Toda}},\ and\
  \bibinfo {author} {\bibfnamefont {N.}~\bibnamefont {Hashitsume}},\ }\href
  {https://doi.org/10.1007/978-3-642-96701-6} {\emph {\bibinfo {title}
  {Statistical {Physics} {II}}}},\ edited by\ \bibinfo {editor} {\bibfnamefont
  {M.}~\bibnamefont {Cardona}}, \bibinfo {editor} {\bibfnamefont
  {P.}~\bibnamefont {Fulde}},\ and\ \bibinfo {editor} {\bibfnamefont {H.-J.}\
  \bibnamefont {Queisser}},\ \bibinfo {series} {Springer {Series} in
  {Solid}-{State} {Sciences}}, Vol.~\bibinfo {volume} {31}\ (\bibinfo
  {publisher} {Springer Berlin Heidelberg},\ \bibinfo {address} {Berlin,
  Heidelberg},\ \bibinfo {year} {1985})\BibitemShut {NoStop}%
\bibitem [{\citenamefont {Gardiner}\ and\ \citenamefont
  {Zoller}(2004)}]{GardinerZoller}%
  \BibitemOpen
  \bibfield  {author} {\bibinfo {author} {\bibfnamefont {C.~W.}\ \bibnamefont
  {Gardiner}}\ and\ \bibinfo {author} {\bibfnamefont {P.}~\bibnamefont
  {Zoller}},\ }\href@noop {} {\emph {\bibinfo {title} {Quantum noise: a
  handbook of {M}arkovian and non-{M}arkovian quantum stochastic methods with
  applications to quantum optics}}},\ \bibinfo {edition} {3rd}\ ed.\ (\bibinfo
  {publisher} {Springer Verlag},\ \bibinfo {address} {Berlin; New York},\
  \bibinfo {year} {2004})\BibitemShut {NoStop}%
\bibitem [{\citenamefont {Feynman}\ and\ \citenamefont
  {Vernon~Jr}(2000)}]{Feynman2000}%
  \BibitemOpen
  \bibfield  {author} {\bibinfo {author} {\bibfnamefont {R.~P.}\ \bibnamefont
  {Feynman}}\ and\ \bibinfo {author} {\bibfnamefont {F.~L.}\ \bibnamefont
  {Vernon~Jr}},\ }\bibfield  {title} {\bibinfo {title} {The theory of a general
  quantum system interacting with a linear dissipative system},\ }\href@noop {}
  {\bibfield  {journal} {\bibinfo  {journal} {Annals of physics}\ }\textbf
  {\bibinfo {volume} {281}},\ \bibinfo {pages} {547} (\bibinfo {year}
  {2000})}\BibitemShut {NoStop}%
\bibitem [{\citenamefont {Anderson}\ and\ \citenamefont
  {Weiss}(1953)}]{AndersonWeiss53}%
  \BibitemOpen
  \bibfield  {author} {\bibinfo {author} {\bibfnamefont {P.~W.}\ \bibnamefont
  {Anderson}}\ and\ \bibinfo {author} {\bibfnamefont {P.~R.}\ \bibnamefont
  {Weiss}},\ }\bibfield  {title} {\bibinfo {title} {Exchange narrowing in
  paramagnetic resonance},\ }\href {https://doi.org/10.1103/RevModPhys.25.269}
  {\bibfield  {journal} {\bibinfo  {journal} {Rev. Mod. Phys.}\ }\textbf
  {\bibinfo {volume} {25}},\ \bibinfo {pages} {269} (\bibinfo {year}
  {1953})}\BibitemShut {NoStop}%
\bibitem [{\citenamefont {Kubo}(1954)}]{Kubo1954}%
  \BibitemOpen
  \bibfield  {author} {\bibinfo {author} {\bibfnamefont {R.}~\bibnamefont
  {Kubo}},\ }\bibfield  {title} {\bibinfo {title} {Note on the stochastic
  theory of resonance absorption},\ }\href@noop {} {\bibfield  {journal}
  {\bibinfo  {journal} {Journal of the Physical Society of Japan}\ }\textbf
  {\bibinfo {volume} {9}},\ \bibinfo {pages} {935} (\bibinfo {year}
  {1954})}\BibitemShut {NoStop}%
\bibitem [{\citenamefont {Klauder}\ and\ \citenamefont
  {Anderson}(1962)}]{KlauderAnderson62}%
  \BibitemOpen
  \bibfield  {author} {\bibinfo {author} {\bibfnamefont {J.~R.}\ \bibnamefont
  {Klauder}}\ and\ \bibinfo {author} {\bibfnamefont {P.~W.}\ \bibnamefont
  {Anderson}},\ }\bibfield  {title} {\bibinfo {title} {Spectral diffusion decay
  in spin resonance experiments},\ }\href
  {https://doi.org/10.1103/PhysRev.125.912} {\bibfield  {journal} {\bibinfo
  {journal} {Phys. Rev.}\ }\textbf {\bibinfo {volume} {125}},\ \bibinfo {pages}
  {912} (\bibinfo {year} {1962})}\BibitemShut {NoStop}%
\bibitem [{\citenamefont {Kubo}(1963)}]{Kubo1963}%
  \BibitemOpen
  \bibfield  {author} {\bibinfo {author} {\bibfnamefont {R.}~\bibnamefont
  {Kubo}},\ }\bibfield  {title} {\bibinfo {title} {Stochastic {L}iouville
  equations},\ }\href@noop {} {\bibfield  {journal} {\bibinfo  {journal}
  {Journal of Mathematical Physics}\ }\textbf {\bibinfo {volume} {4}},\
  \bibinfo {pages} {174} (\bibinfo {year} {1963})}\BibitemShut {NoStop}%
\bibitem [{\citenamefont {Chentsov}(1956)}]{Chentsov_1956}%
  \BibitemOpen
  \bibfield  {author} {\bibinfo {author} {\bibfnamefont {N.}~\bibnamefont
  {Chentsov}},\ }\bibfield  {title} {\bibinfo {title} {Wiener random fields
  depending on several parameters},\ }\href@noop {} {\bibfield  {journal}
  {\bibinfo  {journal} {Doklady Akademii Nauk SSSR}\ }\textbf {\bibinfo
  {volume} {106}},\ \bibinfo {pages} {607} (\bibinfo {year}
  {1956})}\BibitemShut {NoStop}%
\bibitem [{\citenamefont {Kitagava}(1951)}]{Kitagava_1951}%
  \BibitemOpen
  \bibfield  {author} {\bibinfo {author} {\bibfnamefont {T.}~\bibnamefont
  {Kitagava}},\ }\bibfield  {title} {\bibinfo {title} {Analysis of variance
  applied to function spaces},\ }\href@noop {} {\bibfield  {journal} {\bibinfo
  {journal} {Memoirs of the Faculty of Science, Kyushu University Series A}\
  }\textbf {\bibinfo {volume} {6}},\ \bibinfo {pages} {41} (\bibinfo {year}
  {1951})}\BibitemShut {NoStop}%
\bibitem [{\citenamefont {Volmer}\ \emph {et~al.}(2023)\citenamefont {Volmer},
  \citenamefont {Struck}, \citenamefont {Sala}, \citenamefont {Chen},
  \citenamefont {Oberl{\"a}nder}, \citenamefont {Offermann}, \citenamefont
  {Xue}, \citenamefont {Visser}, \citenamefont {Tu}, \citenamefont
  {Trellenkamp}, \citenamefont {Cywi{\'n}ski}, \citenamefont {Bluhm},\ and\
  \citenamefont {Schreiber}}]{VolmerStruckEtalValleySplit23}%
  \BibitemOpen
  \bibfield  {author} {\bibinfo {author} {\bibfnamefont {M.}~\bibnamefont
  {Volmer}}, \bibinfo {author} {\bibfnamefont {T.}~\bibnamefont {Struck}},
  \bibinfo {author} {\bibfnamefont {A.}~\bibnamefont {Sala}}, \bibinfo {author}
  {\bibfnamefont {B.}~\bibnamefont {Chen}}, \bibinfo {author} {\bibfnamefont
  {M.}~\bibnamefont {Oberl{\"a}nder}}, \bibinfo {author} {\bibfnamefont
  {T.}~\bibnamefont {Offermann}}, \bibinfo {author} {\bibfnamefont
  {R.}~\bibnamefont {Xue}}, \bibinfo {author} {\bibfnamefont {L.}~\bibnamefont
  {Visser}}, \bibinfo {author} {\bibfnamefont {J.-S.}\ \bibnamefont {Tu}},
  \bibinfo {author} {\bibfnamefont {S.}~\bibnamefont {Trellenkamp}}, \bibinfo
  {author} {\bibfnamefont {{\L}.}~\bibnamefont {Cywi{\'n}ski}}, \bibinfo
  {author} {\bibfnamefont {H.}~\bibnamefont {Bluhm}},\ and\ \bibinfo {author}
  {\bibfnamefont {L.~R.}\ \bibnamefont {Schreiber}},\ }\href@noop {} {\bibinfo
  {title} {Mapping of valley-splitting by conveyor-mode spin-coherent electron
  shuttling}} (\bibinfo {year} {2023}),\ \Eprint
  {https://arxiv.org/abs/2312.17694} {arXiv:2312.17694 [quant-ph]} \BibitemShut
  {NoStop}%
\bibitem [{\citenamefont {Lidar}\ \emph {et~al.}(1998)\citenamefont {Lidar},
  \citenamefont {Chuang},\ and\ \citenamefont
  {Whaley}}]{LidarChuangWhaleyDFS98}%
  \BibitemOpen
  \bibfield  {author} {\bibinfo {author} {\bibfnamefont {D.~A.}\ \bibnamefont
  {Lidar}}, \bibinfo {author} {\bibfnamefont {I.~L.}\ \bibnamefont {Chuang}},\
  and\ \bibinfo {author} {\bibfnamefont {K.~B.}\ \bibnamefont {Whaley}},\
  }\bibfield  {title} {\bibinfo {title} {Decoherence-free subspaces for quantum
  computation},\ }\href {https://doi.org/10.1103/PhysRevLett.81.2594}
  {\bibfield  {journal} {\bibinfo  {journal} {Phys. Rev. Lett.}\ }\textbf
  {\bibinfo {volume} {81}},\ \bibinfo {pages} {2594} (\bibinfo {year}
  {1998})}\BibitemShut {NoStop}%
\bibitem [{\citenamefont {Zanardi}\ and\ \citenamefont
  {Rasetti}(1997)}]{ZanardiRasettiDFS97}%
  \BibitemOpen
  \bibfield  {author} {\bibinfo {author} {\bibfnamefont {P.}~\bibnamefont
  {Zanardi}}\ and\ \bibinfo {author} {\bibfnamefont {M.}~\bibnamefont
  {Rasetti}},\ }\bibfield  {title} {\bibinfo {title} {Noiseless quantum
  codes},\ }\href {https://doi.org/10.1103/PhysRevLett.79.3306} {\bibfield
  {journal} {\bibinfo  {journal} {Phys. Rev. Lett.}\ }\textbf {\bibinfo
  {volume} {79}},\ \bibinfo {pages} {3306} (\bibinfo {year}
  {1997})}\BibitemShut {NoStop}%
\bibitem [{\citenamefont {Viola}\ \emph {et~al.}(2001)\citenamefont {Viola},
  \citenamefont {Fortunato}, \citenamefont {Pravia}, \citenamefont {Knill},
  \citenamefont {Laflamme},\ and\ \citenamefont
  {Cory}}]{ViolaCoryEtalDFSExp01}%
  \BibitemOpen
  \bibfield  {author} {\bibinfo {author} {\bibfnamefont {L.}~\bibnamefont
  {Viola}}, \bibinfo {author} {\bibfnamefont {E.~M.}\ \bibnamefont
  {Fortunato}}, \bibinfo {author} {\bibfnamefont {M.~A.}\ \bibnamefont
  {Pravia}}, \bibinfo {author} {\bibfnamefont {E.}~\bibnamefont {Knill}},
  \bibinfo {author} {\bibfnamefont {R.}~\bibnamefont {Laflamme}},\ and\
  \bibinfo {author} {\bibfnamefont {D.~G.}\ \bibnamefont {Cory}},\ }\bibfield
  {title} {\bibinfo {title} {Experimental realization of noiseless subsystems
  for quantum information processing},\ }\href
  {https://doi.org/10.1126/science.1064460} {\bibfield  {journal} {\bibinfo
  {journal} {Science}\ }\textbf {\bibinfo {volume} {293}},\ \bibinfo {pages}
  {2059} (\bibinfo {year} {2001})}\BibitemShut {NoStop}%
\bibitem [{\citenamefont {Jeon}\ \emph {et~al.}(2024)\citenamefont {Jeon},
  \citenamefont {Benjamin},\ and\ \citenamefont
  {Fisher}}]{JeonBenjaminFisher24RobustChargeShuttl}%
  \BibitemOpen
  \bibfield  {author} {\bibinfo {author} {\bibfnamefont {M.}~\bibnamefont
  {Jeon}}, \bibinfo {author} {\bibfnamefont {S.~C.}\ \bibnamefont {Benjamin}},\
  and\ \bibinfo {author} {\bibfnamefont {A.~J.}\ \bibnamefont {Fisher}},\
  }\href {https://arxiv.org/abs/2408.03315} {\bibinfo {title} {Robustness of
  electron charge shuttling: {A}rchitectures, pulses, charge defects and noise
  thresholds}} (\bibinfo {year} {2024}),\ \Eprint
  {https://arxiv.org/abs/2408.03315} {arXiv:2408.03315 [cond-mat.mes-hall]}
  \BibitemShut {NoStop}%
\bibitem [{\citenamefont {McNeil}\ \emph {et~al.}(2011)\citenamefont {McNeil},
  \citenamefont {Kataoka}, \citenamefont {Ford}, \citenamefont {Barnes},
  \citenamefont {Anderson}, \citenamefont {Jones}, \citenamefont {Farrer},\
  and\ \citenamefont {Ritchie}}]{ShuttlingSAWRitchie11}%
  \BibitemOpen
  \bibfield  {author} {\bibinfo {author} {\bibfnamefont {R.~P.~G.}\
  \bibnamefont {McNeil}}, \bibinfo {author} {\bibfnamefont {M.}~\bibnamefont
  {Kataoka}}, \bibinfo {author} {\bibfnamefont {C.~J.~B.}\ \bibnamefont
  {Ford}}, \bibinfo {author} {\bibfnamefont {C.~H.~W.}\ \bibnamefont {Barnes}},
  \bibinfo {author} {\bibfnamefont {D.}~\bibnamefont {Anderson}}, \bibinfo
  {author} {\bibfnamefont {G.~A.~C.}\ \bibnamefont {Jones}}, \bibinfo {author}
  {\bibfnamefont {I.}~\bibnamefont {Farrer}},\ and\ \bibinfo {author}
  {\bibfnamefont {D.~A.}\ \bibnamefont {Ritchie}},\ }\bibfield  {title}
  {\bibinfo {title} {On-demand single-electron transfer between distant quantum
  dots},\ }\href {https://doi.org/https://doi.org/10.1038/nature10444}
  {\bibfield  {journal} {\bibinfo  {journal} {Nature}\ }\textbf {\bibinfo
  {volume} {477}},\ \bibinfo {pages} {439} (\bibinfo {year}
  {2011})}\BibitemShut {NoStop}%
\bibitem [{\citenamefont {Hermelin}\ \emph {et~al.}(2011)\citenamefont
  {Hermelin}, \citenamefont {Takada}, \citenamefont {Yamamoto}, \citenamefont
  {Tarucha}, \citenamefont {Wieck}, \citenamefont {Saminadayar}, \citenamefont
  {B{\"a}uerle},\ and\ \citenamefont {Meunier}}]{ShuttlingSAWMeunier11}%
  \BibitemOpen
  \bibfield  {author} {\bibinfo {author} {\bibfnamefont {S.}~\bibnamefont
  {Hermelin}}, \bibinfo {author} {\bibfnamefont {S.}~\bibnamefont {Takada}},
  \bibinfo {author} {\bibfnamefont {M.}~\bibnamefont {Yamamoto}}, \bibinfo
  {author} {\bibfnamefont {S.}~\bibnamefont {Tarucha}}, \bibinfo {author}
  {\bibfnamefont {A.~D.}\ \bibnamefont {Wieck}}, \bibinfo {author}
  {\bibfnamefont {L.}~\bibnamefont {Saminadayar}}, \bibinfo {author}
  {\bibfnamefont {C.}~\bibnamefont {B{\"a}uerle}},\ and\ \bibinfo {author}
  {\bibfnamefont {T.}~\bibnamefont {Meunier}},\ }\bibfield  {title} {\bibinfo
  {title} {Electrons surfing on a sound wave as a platform for quantum optics
  with flying electrons},\ }\href
  {https://doi.org/https://doi.org/10.1038/nature10416} {\bibfield  {journal}
  {\bibinfo  {journal} {Nature}\ }\textbf {\bibinfo {volume} {477}},\ \bibinfo
  {pages} {435} (\bibinfo {year} {2011})}\BibitemShut {NoStop}%
\bibitem [{\citenamefont {Schuetz}\ \emph {et~al.}(2017)\citenamefont
  {Schuetz}, \citenamefont {Kn\"orzer}, \citenamefont {Giedke}, \citenamefont
  {Vandersypen}, \citenamefont {Lukin},\ and\ \citenamefont
  {Cirac}}]{SchuetzLukinVandersypenEtal17SAW}%
  \BibitemOpen
  \bibfield  {author} {\bibinfo {author} {\bibfnamefont {M.~J.~A.}\
  \bibnamefont {Schuetz}}, \bibinfo {author} {\bibfnamefont {J.}~\bibnamefont
  {Kn\"orzer}}, \bibinfo {author} {\bibfnamefont {G.}~\bibnamefont {Giedke}},
  \bibinfo {author} {\bibfnamefont {L.~M.~K.}\ \bibnamefont {Vandersypen}},
  \bibinfo {author} {\bibfnamefont {M.~D.}\ \bibnamefont {Lukin}},\ and\
  \bibinfo {author} {\bibfnamefont {J.~I.}\ \bibnamefont {Cirac}},\ }\bibfield
  {title} {\bibinfo {title} {Acoustic traps and lattices for electrons in
  semiconductors},\ }\href {https://doi.org/10.1103/PhysRevX.7.041019}
  {\bibfield  {journal} {\bibinfo  {journal} {Phys. Rev. X}\ }\textbf {\bibinfo
  {volume} {7}},\ \bibinfo {pages} {041019} (\bibinfo {year}
  {2017})}\BibitemShut {NoStop}%
\bibitem [{\citenamefont {Wang}\ \emph {et~al.}(2024)\citenamefont {Wang},
  \citenamefont {John}, \citenamefont {Tidjani}, \citenamefont {Yu},
  \citenamefont {Ivlev}, \citenamefont {D{\'e}prez}, \citenamefont {van
  Riggelen-Doelman}, \citenamefont {Woods}, \citenamefont {Hendrickx},
  \citenamefont {Lawrie}, \citenamefont {Stehouwer}, \citenamefont
  {Oosterhout}, \citenamefont {Sammak}, \citenamefont {Friesen}, \citenamefont
  {Scappucci}, \citenamefont {de~Snoo}, \citenamefont {Rimbach-Russ},
  \citenamefont {Borsoi},\ and\ \citenamefont
  {Veldhorst}}]{WangVeldhorstEtalGeQDs24}%
  \BibitemOpen
  \bibfield  {author} {\bibinfo {author} {\bibfnamefont {C.-A.}\ \bibnamefont
  {Wang}}, \bibinfo {author} {\bibfnamefont {V.}~\bibnamefont {John}}, \bibinfo
  {author} {\bibfnamefont {H.}~\bibnamefont {Tidjani}}, \bibinfo {author}
  {\bibfnamefont {C.~X.}\ \bibnamefont {Yu}}, \bibinfo {author} {\bibfnamefont
  {A.~S.}\ \bibnamefont {Ivlev}}, \bibinfo {author} {\bibfnamefont
  {C.}~\bibnamefont {D{\'e}prez}}, \bibinfo {author} {\bibfnamefont
  {F.}~\bibnamefont {van Riggelen-Doelman}}, \bibinfo {author} {\bibfnamefont
  {B.~D.}\ \bibnamefont {Woods}}, \bibinfo {author} {\bibfnamefont {N.~W.}\
  \bibnamefont {Hendrickx}}, \bibinfo {author} {\bibfnamefont {W.~I.~L.}\
  \bibnamefont {Lawrie}}, \bibinfo {author} {\bibfnamefont {L.~E.~A.}\
  \bibnamefont {Stehouwer}}, \bibinfo {author} {\bibfnamefont {S.~D.}\
  \bibnamefont {Oosterhout}}, \bibinfo {author} {\bibfnamefont
  {A.}~\bibnamefont {Sammak}}, \bibinfo {author} {\bibfnamefont
  {M.}~\bibnamefont {Friesen}}, \bibinfo {author} {\bibfnamefont
  {G.}~\bibnamefont {Scappucci}}, \bibinfo {author} {\bibfnamefont {S.~L.}\
  \bibnamefont {de~Snoo}}, \bibinfo {author} {\bibfnamefont {M.}~\bibnamefont
  {Rimbach-Russ}}, \bibinfo {author} {\bibfnamefont {F.}~\bibnamefont
  {Borsoi}},\ and\ \bibinfo {author} {\bibfnamefont {M.}~\bibnamefont
  {Veldhorst}},\ }\bibfield  {title} {\bibinfo {title} {Operating semiconductor
  quantum processors with hopping spins},\ }\href
  {https://doi.org/10.1126/science.ado5915} {\bibfield  {journal} {\bibinfo
  {journal} {Science}\ }\textbf {\bibinfo {volume} {385}},\ \bibinfo {pages}
  {447} (\bibinfo {year} {2024})}\BibitemShut {NoStop}%
\bibitem [{Note1()}]{Note1}%
  \BibitemOpen
  \bibinfo {note} {The confining potential created by the clavier gates
  slightly differs from an ideal traveling wave due to the finite gate pitch:
  the potential periodically changes its shape, and so does the wavepacket of
  the shuttled electron. If the phase shift between the adjacent pairs of the
  gates is small enough then these changes are small~\cite
  *{langrock_blueprint_2023,JeonBenjaminFisher24RobustChargeShuttl} and can be
  neglected. Otherwise, the periodic changes in the shape of the wavepacket,
  and the associated time variation of $B_0$ in Eq.~\ref {eq:tildeB} can be
  taken into account within the outlined approach.}\BibitemShut {Stop}%
\bibitem [{\citenamefont {Rytov}\ \emph {et~al.}(1989)\citenamefont {Rytov},
  \citenamefont {Kravtsov},\ and\ \citenamefont
  {Tatarskii}}]{RytovEtalStatRad}%
  \BibitemOpen
  \bibfield  {author} {\bibinfo {author} {\bibfnamefont {S.~M.}\ \bibnamefont
  {Rytov}}, \bibinfo {author} {\bibfnamefont {Y.~A.}\ \bibnamefont
  {Kravtsov}},\ and\ \bibinfo {author} {\bibfnamefont {V.~I.}\ \bibnamefont
  {Tatarskii}},\ }\href@noop {} {\emph {\bibinfo {title} {Principles of
  Statistical Radiophysics}}},\ \bibinfo {edition} {2nd}\ ed.\ (\bibinfo
  {publisher} {Springer Verlag},\ \bibinfo {address} {Berlin, Heidelberg, New
  York},\ \bibinfo {year} {1989})\BibitemShut {NoStop}%
\bibitem [{\citenamefont {Sobczyk}(1985)}]{SobczykStochWave}%
  \BibitemOpen
  \bibfield  {author} {\bibinfo {author} {\bibfnamefont {K.}~\bibnamefont
  {Sobczyk}},\ }\href@noop {} {\emph {\bibinfo {title} {Stochastic wave
  propagation}}}\ (\bibinfo  {publisher} {Elsevier, PWN --- Polish Scientific
  Publishers},\ \bibinfo {address} {Amsterdam, Warszawa},\ \bibinfo {year}
  {1985})\BibitemShut {NoStop}%
\bibitem [{\citenamefont {Smith}\ and\ \citenamefont
  {Raymer}(2006)}]{SmithRaynerTwoPhotonWaveMechanics06}%
  \BibitemOpen
  \bibfield  {author} {\bibinfo {author} {\bibfnamefont {B.~J.}\ \bibnamefont
  {Smith}}\ and\ \bibinfo {author} {\bibfnamefont {M.~G.}\ \bibnamefont
  {Raymer}},\ }\bibfield  {title} {\bibinfo {title} {Two-photon wave
  mechanics},\ }\href {https://doi.org/10.1103/PhysRevA.74.062104} {\bibfield
  {journal} {\bibinfo  {journal} {Phys. Rev. A}\ }\textbf {\bibinfo {volume}
  {74}},\ \bibinfo {pages} {062104} (\bibinfo {year} {2006})}\BibitemShut
  {NoStop}%
\bibitem [{\citenamefont {Paterson}(2005)}]{PatersonOrbAngMomPhotons05}%
  \BibitemOpen
  \bibfield  {author} {\bibinfo {author} {\bibfnamefont {C.}~\bibnamefont
  {Paterson}},\ }\bibfield  {title} {\bibinfo {title} {Atmospheric turbulence
  and orbital angular momentum of single photons for optical communication},\
  }\href {https://doi.org/10.1103/PhysRevLett.94.153901} {\bibfield  {journal}
  {\bibinfo  {journal} {Phys. Rev. Lett.}\ }\textbf {\bibinfo {volume} {94}},\
  \bibinfo {pages} {153901} (\bibinfo {year} {2005})}\BibitemShut {NoStop}%
\bibitem [{\citenamefont {Semenov}\ and\ \citenamefont
  {Vogel}(2010)}]{SemenovVogelEntanglPhotonsTurbAtmosph10}%
  \BibitemOpen
  \bibfield  {author} {\bibinfo {author} {\bibfnamefont {A.~A.}\ \bibnamefont
  {Semenov}}\ and\ \bibinfo {author} {\bibfnamefont {W.}~\bibnamefont
  {Vogel}},\ }\bibfield  {title} {\bibinfo {title} {Entanglement transfer
  through the turbulent atmosphere},\ }\href
  {https://doi.org/10.1103/PhysRevA.81.023835} {\bibfield  {journal} {\bibinfo
  {journal} {Phys. Rev. A}\ }\textbf {\bibinfo {volume} {81}},\ \bibinfo
  {pages} {023835} (\bibinfo {year} {2010})}\BibitemShut {NoStop}%
\bibitem [{\citenamefont {Vasylyev}\ \emph {et~al.}(2016)\citenamefont
  {Vasylyev}, \citenamefont {Semenov},\ and\ \citenamefont
  {Vogel}}]{VasylyevSemenovVogelAtmosphTransmittance16}%
  \BibitemOpen
  \bibfield  {author} {\bibinfo {author} {\bibfnamefont {D.}~\bibnamefont
  {Vasylyev}}, \bibinfo {author} {\bibfnamefont {A.~A.}\ \bibnamefont
  {Semenov}},\ and\ \bibinfo {author} {\bibfnamefont {W.}~\bibnamefont
  {Vogel}},\ }\bibfield  {title} {\bibinfo {title} {Atmospheric quantum
  channels with weak and strong turbulence},\ }\href
  {https://doi.org/10.1103/PhysRevLett.117.090501} {\bibfield  {journal}
  {\bibinfo  {journal} {Phys. Rev. Lett.}\ }\textbf {\bibinfo {volume} {117}},\
  \bibinfo {pages} {090501} (\bibinfo {year} {2016})}\BibitemShut {NoStop}%
\bibitem [{\citenamefont {Pe\ifmmode~\check{r}\else \v{r}\fi{}ina}\ \emph
  {et~al.}(1973)\citenamefont {Pe\ifmmode~\check{r}\else \v{r}\fi{}ina},
  \citenamefont {Pe\ifmmode~\check{r}\else \v{r}\fi{}inov\'a}, \citenamefont
  {Teich},\ and\ \citenamefont
  {Diament}}]{PerinaTeichEtalPhotonDetectStatistics73}%
  \BibitemOpen
  \bibfield  {author} {\bibinfo {author} {\bibfnamefont {J.}~\bibnamefont
  {Pe\ifmmode~\check{r}\else \v{r}\fi{}ina}}, \bibinfo {author} {\bibfnamefont
  {V.}~\bibnamefont {Pe\ifmmode~\check{r}\else \v{r}\fi{}inov\'a}}, \bibinfo
  {author} {\bibfnamefont {M.~C.}\ \bibnamefont {Teich}},\ and\ \bibinfo
  {author} {\bibfnamefont {P.}~\bibnamefont {Diament}},\ }\bibfield  {title}
  {\bibinfo {title} {Two descriptions for the photocounting detection of
  radiation passed through a random medium: A comparison for the turbulent
  atmosphere},\ }\href {https://doi.org/10.1103/PhysRevA.7.1732} {\bibfield
  {journal} {\bibinfo  {journal} {Phys. Rev. A}\ }\textbf {\bibinfo {volume}
  {7}},\ \bibinfo {pages} {1732} (\bibinfo {year} {1973})}\BibitemShut
  {NoStop}%
\bibitem [{\citenamefont {Tatarskii}(1972)}]{Tatarskii72}%
  \BibitemOpen
  \bibfield  {author} {\bibinfo {author} {\bibfnamefont {V.~I.}\ \bibnamefont
  {Tatarskii}},\ }\bibfield  {title} {\bibinfo {title} {Fluctuations of photon
  flux in a medium with random inhomogeneities of the dielectric constant},\
  }\href@noop {} {\bibfield  {journal} {\bibinfo  {journal} {Sov. Phys. JETP}\
  }\textbf {\bibinfo {volume} {34}},\ \bibinfo {pages} {969} (\bibinfo {year}
  {1972})}\BibitemShut {NoStop}%
\bibitem [{\citenamefont {Bloembergen}\ \emph {et~al.}(1948)\citenamefont
  {Bloembergen}, \citenamefont {Purcell},\ and\ \citenamefont
  {Pound}}]{BloembPurcellPound48}%
  \BibitemOpen
  \bibfield  {author} {\bibinfo {author} {\bibfnamefont {N.}~\bibnamefont
  {Bloembergen}}, \bibinfo {author} {\bibfnamefont {E.~M.}\ \bibnamefont
  {Purcell}},\ and\ \bibinfo {author} {\bibfnamefont {R.~V.}\ \bibnamefont
  {Pound}},\ }\bibfield  {title} {\bibinfo {title} {Relaxation effects in
  nuclear magnetic resonance absorption},\ }\href
  {https://doi.org/10.1103/PhysRev.73.679} {\bibfield  {journal} {\bibinfo
  {journal} {Phys. Rev.}\ }\textbf {\bibinfo {volume} {73}},\ \bibinfo {pages}
  {679} (\bibinfo {year} {1948})}\BibitemShut {NoStop}%
\bibitem [{\citenamefont {Uhlenbeck}\ and\ \citenamefont
  {Ornstein}(1930)}]{UhlenbeckOrnsteinRandomProc}%
  \BibitemOpen
  \bibfield  {author} {\bibinfo {author} {\bibfnamefont {G.~E.}\ \bibnamefont
  {Uhlenbeck}}\ and\ \bibinfo {author} {\bibfnamefont {L.~S.}\ \bibnamefont
  {Ornstein}},\ }\bibfield  {title} {\bibinfo {title} {On the {T}heory of the
  {B}rownian {M}otion},\ }\href {https://doi.org/10.1103/PhysRev.36.823}
  {\bibfield  {journal} {\bibinfo  {journal} {Phys. Rev.}\ }\textbf {\bibinfo
  {volume} {36}},\ \bibinfo {pages} {823} (\bibinfo {year} {1930})}\BibitemShut
  {NoStop}%
\bibitem [{\citenamefont {Chandrasekhar}(1943)}]{ChandrasekharRandomProc}%
  \BibitemOpen
  \bibfield  {author} {\bibinfo {author} {\bibfnamefont {S.}~\bibnamefont
  {Chandrasekhar}},\ }\bibfield  {title} {\bibinfo {title} {Stochastic problems
  in physics and astronomy},\ }\href {https://doi.org/10.1103/RevModPhys.15.1}
  {\bibfield  {journal} {\bibinfo  {journal} {Rev. Mod. Phys.}\ }\textbf
  {\bibinfo {volume} {15}},\ \bibinfo {pages} {1} (\bibinfo {year}
  {1943})}\BibitemShut {NoStop}%
\bibitem [{\citenamefont {Wang}\ and\ \citenamefont
  {Uhlenbeck}(1945)}]{WangUhlenbeckRandomProc}%
  \BibitemOpen
  \bibfield  {author} {\bibinfo {author} {\bibfnamefont {M.~C.}\ \bibnamefont
  {Wang}}\ and\ \bibinfo {author} {\bibfnamefont {G.~E.}\ \bibnamefont
  {Uhlenbeck}},\ }\bibfield  {title} {\bibinfo {title} {On the {T}heory of the
  {B}rownian {M}otion {II}},\ }\href
  {https://doi.org/10.1103/RevModPhys.17.323} {\bibfield  {journal} {\bibinfo
  {journal} {Rev. Mod. Phys.}\ }\textbf {\bibinfo {volume} {17}},\ \bibinfo
  {pages} {323} (\bibinfo {year} {1945})}\BibitemShut {NoStop}%
\bibitem [{Mar()}]{MarkovNote}%
  \BibitemOpen
  \href@noop {} {}\bibinfo {note} {Markov property of the random process is
  often confused with infinitely short correlation time. Note that even a
  static random process, not varying in time and having infinite correlation
  time, is Markovian, see e.g.~\cite*{GardinerRandomProc,VanKampenRandomProc}.
  Following the established terminology, we call the process with infinitely
  short correlation time ``uncorrelated noise'' or ``delta-correlated
  process''.}\BibitemShut {Stop}%
\bibitem [{\citenamefont {Jeske}\ and\ \citenamefont
  {Cole}(2013)}]{JeskeColeSpatiallyCorrDecoh13}%
  \BibitemOpen
  \bibfield  {author} {\bibinfo {author} {\bibfnamefont {J.}~\bibnamefont
  {Jeske}}\ and\ \bibinfo {author} {\bibfnamefont {J.~H.}\ \bibnamefont
  {Cole}},\ }\bibfield  {title} {\bibinfo {title} {Derivation of markovian
  master equations for spatially correlated decoherence},\ }\href
  {https://doi.org/10.1103/PhysRevA.87.052138} {\bibfield  {journal} {\bibinfo
  {journal} {Phys. Rev. A}\ }\textbf {\bibinfo {volume} {87}},\ \bibinfo
  {pages} {052138} (\bibinfo {year} {2013})}\BibitemShut {NoStop}%
\bibitem [{\citenamefont {Jeske}\ \emph {et~al.}(2013)\citenamefont {Jeske},
  \citenamefont {Vogt},\ and\ \citenamefont
  {Cole}}]{JeskeVogtColeSpatiallyCorrNoise13}%
  \BibitemOpen
  \bibfield  {author} {\bibinfo {author} {\bibfnamefont {J.}~\bibnamefont
  {Jeske}}, \bibinfo {author} {\bibfnamefont {N.}~\bibnamefont {Vogt}},\ and\
  \bibinfo {author} {\bibfnamefont {J.~H.}\ \bibnamefont {Cole}},\ }\bibfield
  {title} {\bibinfo {title} {Excitation and state transfer through spin chains
  in the presence of spatially correlated noise},\ }\href
  {https://doi.org/10.1103/PhysRevA.88.062333} {\bibfield  {journal} {\bibinfo
  {journal} {Phys. Rev. A}\ }\textbf {\bibinfo {volume} {88}},\ \bibinfo
  {pages} {062333} (\bibinfo {year} {2013})}\BibitemShut {NoStop}%
\bibitem [{Note2()}]{Note2}%
  \BibitemOpen
  \bibinfo {note} {Data and codes used in this work could be accessed at
  \protect \url
  {https://data.4tu.nl/datasets/d0d1007f-c27d-491d-b7e1-cc60e38047b4}, DOI
  10.4121/d0d1007f-c27d-491d-b7e1-cc60e38047b4. The codes for modeling of the
  multi-qubit shuttling are available on Github repository \protect \url
  {https://github.com/EigenSolver/SpinShuttling.jl} and \protect \url
  {https://eigensolver.github.io/SpinShuttling.jl/dev/}.}\BibitemShut {Stop}%
\bibitem [{\citenamefont {Gardiner}(1997)}]{GardinerRandomProc}%
  \BibitemOpen
  \bibfield  {author} {\bibinfo {author} {\bibfnamefont {C.~W.}\ \bibnamefont
  {Gardiner}},\ }\href@noop {} {\emph {\bibinfo {title} {Handbook of
  {S}tochastic {M}ethods for {P}hysics, {C}hemistry and the {N}atural
  {S}ciences}}},\ \bibinfo {edition} {2nd}\ ed.\ (\bibinfo  {publisher}
  {Springer Verlag},\ \bibinfo {address} {Berlin, Heidelberg},\ \bibinfo {year}
  {1997})\BibitemShut {NoStop}%
\bibitem [{\citenamefont {Van~Kampen}(2007)}]{VanKampenRandomProc}%
  \BibitemOpen
  \bibfield  {author} {\bibinfo {author} {\bibfnamefont {N.~G.}\ \bibnamefont
  {Van~Kampen}},\ }\href
  {https://doi.org/https://doi.org/10.1016/B978-0-444-52965-7.50022-2} {\emph
  {\bibinfo {title} {Stochastic Processes in Physics and Chemistry}}},\
  \bibinfo {edition} {3rd}\ ed.\ (\bibinfo  {publisher} {Elsevier},\ \bibinfo
  {address} {Amsterdam},\ \bibinfo {year} {2007})\BibitemShut {NoStop}%
\bibitem [{\citenamefont {Dynkin}(1980)}]{DynkinMarkovProcAndRandFields}%
  \BibitemOpen
  \bibfield  {author} {\bibinfo {author} {\bibfnamefont {E.~B.}\ \bibnamefont
  {Dynkin}},\ }\bibfield  {title} {\bibinfo {title} {{Markov processes and
  random fields}},\ }\href@noop {} {\bibfield  {journal} {\bibinfo  {journal}
  {Bulletin (New Series) of the American Mathematical Society}\ }\textbf
  {\bibinfo {volume} {3}},\ \bibinfo {pages} {975 } (\bibinfo {year}
  {1980})}\BibitemShut {NoStop}%
\bibitem [{\citenamefont {Walsh}(1986)}]{Walsh1986}%
  \BibitemOpen
  \bibfield  {author} {\bibinfo {author} {\bibfnamefont {J.~B.}\ \bibnamefont
  {Walsh}},\ }\href@noop {} {\emph {\bibinfo {title} {An Introduction to
  Stochastic Partial Differential Equations}}}\ (\bibinfo  {publisher}
  {Springer Berlin Heidelberg},\ \bibinfo {year} {1986})\BibitemShut {NoStop}%
\bibitem [{\citenamefont {Werner}\ and\ \citenamefont
  {Powell}(2021)}]{WernerPowellLectNotesGFF21}%
  \BibitemOpen
  \bibfield  {author} {\bibinfo {author} {\bibfnamefont {W.}~\bibnamefont
  {Werner}}\ and\ \bibinfo {author} {\bibfnamefont {E.}~\bibnamefont
  {Powell}},\ }\href@noop {} {\bibinfo {title} {Lecture notes on the {G}aussian
  free field}} (\bibinfo {year} {2021}),\ \Eprint
  {https://arxiv.org/abs/2004.04720} {arXiv:2004.04720 [math.PR]} \BibitemShut
  {NoStop}%
\bibitem [{\citenamefont {Milstein}\ and\ \citenamefont
  {Tretyakov}(2021)}]{MilsteinTretyakovBook}%
  \BibitemOpen
  \bibfield  {author} {\bibinfo {author} {\bibfnamefont {G.~N.}\ \bibnamefont
  {Milstein}}\ and\ \bibinfo {author} {\bibfnamefont {M.~V.}\ \bibnamefont
  {Tretyakov}},\ }\href
  {https://doi.org/https://doi.org/10.1007/978-3-030-82040-4} {\emph {\bibinfo
  {title} {Stochastic numerics for mathematical physics}}},\ \bibinfo {edition}
  {2nd}\ ed.\ (\bibinfo  {publisher} {Springer},\ \bibinfo {address} {Cham,
  Switzerland},\ \bibinfo {year} {2021})\BibitemShut {NoStop}%
\bibitem [{\citenamefont {Dodson}\ \emph {et~al.}(2022)\citenamefont {Dodson},
  \citenamefont {Ercan}, \citenamefont {Corrigan}, \citenamefont {Losert},
  \citenamefont {Holman}, \citenamefont {McJunkin}, \citenamefont {Edge},
  \citenamefont {Friesen}, \citenamefont {Coppersmith},\ and\ \citenamefont
  {Eriksson}}]{DodsonFriesenValleyAndInterface22}%
  \BibitemOpen
  \bibfield  {author} {\bibinfo {author} {\bibfnamefont {J.~P.}\ \bibnamefont
  {Dodson}}, \bibinfo {author} {\bibfnamefont {H.~E.}\ \bibnamefont {Ercan}},
  \bibinfo {author} {\bibfnamefont {J.}~\bibnamefont {Corrigan}}, \bibinfo
  {author} {\bibfnamefont {M.~P.}\ \bibnamefont {Losert}}, \bibinfo {author}
  {\bibfnamefont {N.}~\bibnamefont {Holman}}, \bibinfo {author} {\bibfnamefont
  {T.}~\bibnamefont {McJunkin}}, \bibinfo {author} {\bibfnamefont {L.~F.}\
  \bibnamefont {Edge}}, \bibinfo {author} {\bibfnamefont {M.}~\bibnamefont
  {Friesen}}, \bibinfo {author} {\bibfnamefont {S.~N.}\ \bibnamefont
  {Coppersmith}},\ and\ \bibinfo {author} {\bibfnamefont {M.~A.}\ \bibnamefont
  {Eriksson}},\ }\bibfield  {title} {\bibinfo {title} {How valley-orbit states
  in silicon quantum dots probe quantum well interfaces},\ }\href
  {https://doi.org/10.1103/PhysRevLett.128.146802} {\bibfield  {journal}
  {\bibinfo  {journal} {Phys. Rev. Lett.}\ }\textbf {\bibinfo {volume} {128}},\
  \bibinfo {pages} {146802} (\bibinfo {year} {2022})}\BibitemShut {NoStop}%
\bibitem [{\citenamefont {Losert}\ \emph {et~al.}(2023)\citenamefont {Losert},
  \citenamefont {Eriksson}, \citenamefont {Joynt}, \citenamefont {Rahman},
  \citenamefont {Scappucci}, \citenamefont {Coppersmith},\ and\ \citenamefont
  {Friesen}}]{LosertFriesenValleySplit23}%
  \BibitemOpen
  \bibfield  {author} {\bibinfo {author} {\bibfnamefont {M.~P.}\ \bibnamefont
  {Losert}}, \bibinfo {author} {\bibfnamefont {M.~A.}\ \bibnamefont
  {Eriksson}}, \bibinfo {author} {\bibfnamefont {R.}~\bibnamefont {Joynt}},
  \bibinfo {author} {\bibfnamefont {R.}~\bibnamefont {Rahman}}, \bibinfo
  {author} {\bibfnamefont {G.}~\bibnamefont {Scappucci}}, \bibinfo {author}
  {\bibfnamefont {S.~N.}\ \bibnamefont {Coppersmith}},\ and\ \bibinfo {author}
  {\bibfnamefont {M.}~\bibnamefont {Friesen}},\ }\bibfield  {title} {\bibinfo
  {title} {Practical strategies for enhancing the valley splitting in {Si/SiGe}
  quantum wells},\ }\href {https://doi.org/10.1103/PhysRevB.108.125405}
  {\bibfield  {journal} {\bibinfo  {journal} {Phys. Rev. B}\ }\textbf {\bibinfo
  {volume} {108}},\ \bibinfo {pages} {125405} (\bibinfo {year}
  {2023})}\BibitemShut {NoStop}%
\bibitem [{Note3()}]{Note3}%
  \BibitemOpen
  \bibinfo {note} {We assume the convention that $\protect \mathbb {R}_+$
  includes zero.}\BibitemShut {Stop}%
\bibitem [{Note4()}]{Note4}%
  \BibitemOpen
  \bibinfo {note} {Note that the term ``random field'' is used in literature to
  describe a number of similar but different objects with different properties,
  such that its concrete meaning depends on the context. In particular, it is
  important not to confuse an arbitrary Gaussian field, considered in this
  section, with one specific case, the standard object in statistical physics
  known as Gaussian Free Fields (GFF)~\cite
  *{friedli2017statistical,sheffield2007gaussian}. GFF is a specific type of
  Gaussian field on a graph (discrete case) or on a Sobolev space of functions
  on some continuous domain $D \subset \protect \mathbb {R}^{n}$ (continuous
  case). The discrete GFF is a random function $h(x)$ on the graph with the
  probability density $ \exp \left (-\DOTSB \sum@ \slimits@ _{x \sim y} p_{x,y}
  (h(x) - h(y))^2 \right ), $ where the summation goes over all pairs of
  adjacent vertices, and $p_{x,y}$ are the Markov chain transition
  probabilities. In the continuum limit, when the domain $D$ is two- or
  higher-dimensional, the variance of a GFF becomes infinite, and therefore
  $h(x)$ has to be understood as a random element (random variable with
  non-numerical values), whose values belong to the space of bounded linear
  functionals on the Sobolev space $H^1(D)$, i.e., it is a distribution over
  distributions or a stochastic field indexed by functions. In contrast, the
  Gaussian fields considered in this paper have no such singularities, and
  admit simpler mathematical description.}\BibitemShut {Stop}%
\bibitem [{\citenamefont {Baran}\ \emph {et~al.}(2003)\citenamefont {Baran},
  \citenamefont {Pap},\ and\ \citenamefont {Van~Zuijlen}}]{Baran2003}%
  \BibitemOpen
  \bibfield  {author} {\bibinfo {author} {\bibfnamefont {S.}~\bibnamefont
  {Baran}}, \bibinfo {author} {\bibfnamefont {G.}~\bibnamefont {Pap}},\ and\
  \bibinfo {author} {\bibfnamefont {M.~C.}\ \bibnamefont {Van~Zuijlen}},\
  }\bibfield  {title} {\bibinfo {title} {Estimation of the mean of stationary
  and nonstationary {Ornstein}--{U}hlenbeck processes and sheets},\ }\href@noop
  {} {\bibfield  {journal} {\bibinfo  {journal} {Computers \& Mathematics with
  Applications}\ }\textbf {\bibinfo {volume} {45}},\ \bibinfo {pages} {563}
  (\bibinfo {year} {2003})}\BibitemShut {NoStop}%
\bibitem [{\citenamefont {Baran}\ and\ \citenamefont
  {Sikolya}(2012)}]{Baran2012}%
  \BibitemOpen
  \bibfield  {author} {\bibinfo {author} {\bibfnamefont {S.}~\bibnamefont
  {Baran}}\ and\ \bibinfo {author} {\bibfnamefont {K.}~\bibnamefont
  {Sikolya}},\ }\bibfield  {title} {\bibinfo {title} {Parameter estimation in
  linear regression driven by a {Gaussian} sheet},\ }\href@noop {} {\bibfield
  {journal} {\bibinfo  {journal} {Acta Scientiarum Mathematicarum}\ }\textbf
  {\bibinfo {volume} {78}},\ \bibinfo {pages} {689} (\bibinfo {year}
  {2012})}\BibitemShut {NoStop}%
\bibitem [{\citenamefont {Adler}(1981)}]{AdlerBookGeomRandF}%
  \BibitemOpen
  \bibfield  {author} {\bibinfo {author} {\bibfnamefont {R.~J.}\ \bibnamefont
  {Adler}},\ }\href@noop {} {\emph {\bibinfo {title} {The geometry of random
  fields}}}\ (\bibinfo  {publisher} {John Wiley \& Sons Inc},\ \bibinfo
  {address} {Chichester},\ \bibinfo {year} {1981})\BibitemShut {NoStop}%
\bibitem [{\citenamefont {Sun}(2005)}]{Sun2005MercerThm}%
  \BibitemOpen
  \bibfield  {author} {\bibinfo {author} {\bibfnamefont {H.}~\bibnamefont
  {Sun}},\ }\bibfield  {title} {\bibinfo {title} {{M}ercer theorem for {RKHS}
  on noncompact sets},\ }\href
  {https://doi.org/https://doi.org/10.1016/j.jco.2004.09.002} {\bibfield
  {journal} {\bibinfo  {journal} {Journal of Complexity}\ }\textbf {\bibinfo
  {volume} {21}},\ \bibinfo {pages} {337} (\bibinfo {year} {2005})}\BibitemShut
  {NoStop}%
\bibitem [{\citenamefont {Fujita}\ and\ \citenamefont
  {Yoshida}(2023)}]{FujitaYoshidaNonGaussian23}%
  \BibitemOpen
  \bibfield  {author} {\bibinfo {author} {\bibfnamefont {T.}~\bibnamefont
  {Fujita}}\ and\ \bibinfo {author} {\bibfnamefont {N.}~\bibnamefont
  {Yoshida}},\ }\bibfield  {title} {\bibinfo {title} {An example showing that
  the sum of two normal random variables may not be normal},\ }\href
  {https://doi.org/10.1080/0020739X.2023.2248992} {\bibfield  {journal}
  {\bibinfo  {journal} {International Journal of Mathematical Education in
  Science and Technology}\ ,\ \bibinfo {pages} {1}} (\bibinfo {year}
  {2023})}\BibitemShut {NoStop}%
\bibitem [{\citenamefont {Skorokhod}(1965)}]{Skorokhod1965}%
  \BibitemOpen
  \bibfield  {author} {\bibinfo {author} {\bibfnamefont {A.~V.}\ \bibnamefont
  {Skorokhod}},\ }\bibfield  {title} {\bibinfo {title} {Constructive methods of
  specifying stochastic processes},\ }\href@noop {} {\bibfield  {journal}
  {\bibinfo  {journal} {Russian Mathematical Surveys}\ }\textbf {\bibinfo
  {volume} {20}},\ \bibinfo {pages} {63} (\bibinfo {year} {1965})}\BibitemShut
  {NoStop}%
\bibitem [{Note5()}]{Note5}%
  \BibitemOpen
  \bibinfo {note} {Although in almost all real physical situations the function
  $F(s)$ is continuous, we include the case of discontinuities mostly for
  generality of treatment.}\BibitemShut {Stop}%
\bibitem [{\citenamefont {S{\"u}li}\ and\ \citenamefont
  {Mayers}(2003)}]{suli2003introduction}%
  \BibitemOpen
  \bibfield  {author} {\bibinfo {author} {\bibfnamefont {E.}~\bibnamefont
  {S{\"u}li}}\ and\ \bibinfo {author} {\bibfnamefont {D.~F.}\ \bibnamefont
  {Mayers}},\ }\href@noop {} {\emph {\bibinfo {title} {An introduction to
  numerical analysis}}}\ (\bibinfo  {publisher} {Cambridge University Press},\
  \bibinfo {address} {Cambridge, UK},\ \bibinfo {year} {2003})\BibitemShut
  {NoStop}%
\bibitem [{\citenamefont {Press}\ \emph {et~al.}(1992)\citenamefont {Press},
  \citenamefont {Teukolsky}, \citenamefont {Vettering},\ and\ \citenamefont
  {Flannery}}]{NumRecipes}%
  \BibitemOpen
  \bibfield  {author} {\bibinfo {author} {\bibfnamefont {W.~H.}\ \bibnamefont
  {Press}}, \bibinfo {author} {\bibfnamefont {S.~A.}\ \bibnamefont
  {Teukolsky}}, \bibinfo {author} {\bibfnamefont {W.~T.}\ \bibnamefont
  {Vettering}},\ and\ \bibinfo {author} {\bibfnamefont {B.~P.}\ \bibnamefont
  {Flannery}},\ }\href@noop {} {\emph {\bibinfo {title} {Numerical Recipes in
  {C}: The Art of Scientific Computing}}},\ \bibinfo {edition} {2nd}\ ed.\
  (\bibinfo  {publisher} {Cambridge University Press},\ \bibinfo {address}
  {Cambridge, UK},\ \bibinfo {year} {1992})\BibitemShut {NoStop}%
\bibitem [{\citenamefont {Yoneda}\ \emph {et~al.}(2023)\citenamefont {Yoneda},
  \citenamefont {Rojas-Arias}, \citenamefont {Stano}, \citenamefont {Takeda},
  \citenamefont {Noiri}, \citenamefont {Nakajima}, \citenamefont {Loss},\ and\
  \citenamefont {Tarucha}}]{yoneda_noise-correlation_2023}%
  \BibitemOpen
  \bibfield  {author} {\bibinfo {author} {\bibfnamefont {J.}~\bibnamefont
  {Yoneda}}, \bibinfo {author} {\bibfnamefont {J.~S.}\ \bibnamefont
  {Rojas-Arias}}, \bibinfo {author} {\bibfnamefont {P.}~\bibnamefont {Stano}},
  \bibinfo {author} {\bibfnamefont {K.}~\bibnamefont {Takeda}}, \bibinfo
  {author} {\bibfnamefont {A.}~\bibnamefont {Noiri}}, \bibinfo {author}
  {\bibfnamefont {T.}~\bibnamefont {Nakajima}}, \bibinfo {author}
  {\bibfnamefont {D.}~\bibnamefont {Loss}},\ and\ \bibinfo {author}
  {\bibfnamefont {S.}~\bibnamefont {Tarucha}},\ }\bibfield  {title} {\bibinfo
  {title} {Noise-correlation spectrum for a pair of spin qubits in silicon},\
  }\href {https://doi.org/10.1038/s41567-023-02238-6} {\bibfield  {journal}
  {\bibinfo  {journal} {Nat. Phys.}\ }\textbf {\bibinfo {volume} {19}},\
  \bibinfo {pages} {1793} (\bibinfo {year} {2023})}\BibitemShut {NoStop}%
\bibitem [{\citenamefont {Rojas-Arias}\ \emph {et~al.}(2023)\citenamefont
  {Rojas-Arias}, \citenamefont {Noiri}, \citenamefont {Stano}, \citenamefont
  {Nakajima}, \citenamefont {Yoneda}, \citenamefont {Takeda}, \citenamefont
  {Kobayashi}, \citenamefont {Sammak}, \citenamefont {Scappucci}, \citenamefont
  {Loss},\ and\ \citenamefont {Tarucha}}]{rojas-arias_spatial_2023}%
  \BibitemOpen
  \bibfield  {author} {\bibinfo {author} {\bibfnamefont {J.}~\bibnamefont
  {Rojas-Arias}}, \bibinfo {author} {\bibfnamefont {A.}~\bibnamefont {Noiri}},
  \bibinfo {author} {\bibfnamefont {P.}~\bibnamefont {Stano}}, \bibinfo
  {author} {\bibfnamefont {T.}~\bibnamefont {Nakajima}}, \bibinfo {author}
  {\bibfnamefont {J.}~\bibnamefont {Yoneda}}, \bibinfo {author} {\bibfnamefont
  {K.}~\bibnamefont {Takeda}}, \bibinfo {author} {\bibfnamefont
  {T.}~\bibnamefont {Kobayashi}}, \bibinfo {author} {\bibfnamefont
  {A.}~\bibnamefont {Sammak}}, \bibinfo {author} {\bibfnamefont
  {G.}~\bibnamefont {Scappucci}}, \bibinfo {author} {\bibfnamefont
  {D.}~\bibnamefont {Loss}},\ and\ \bibinfo {author} {\bibfnamefont
  {S.}~\bibnamefont {Tarucha}},\ }\bibfield  {title} {\bibinfo {title} {Spatial
  noise correlations beyond nearest neighbors in ${}^{28}\mathrm{Si}/${Si}-{Ge}
  spin qubits},\ }\href {https://doi.org/10.1103/PhysRevApplied.20.054024}
  {\bibfield  {journal} {\bibinfo  {journal} {Phys. Rev. Appl.}\ }\textbf
  {\bibinfo {volume} {20}},\ \bibinfo {pages} {054024} (\bibinfo {year}
  {2023})}\BibitemShut {NoStop}%
\bibitem [{\citenamefont {Riberi}\ \emph {et~al.}(2022)\citenamefont {Riberi},
  \citenamefont {Norris}, \citenamefont {Beaudoin},\ and\ \citenamefont
  {Viola}}]{ViolaSpaceTimeNoise22}%
  \BibitemOpen
  \bibfield  {author} {\bibinfo {author} {\bibfnamefont {F.}~\bibnamefont
  {Riberi}}, \bibinfo {author} {\bibfnamefont {L.~M.}\ \bibnamefont {Norris}},
  \bibinfo {author} {\bibfnamefont {F.}~\bibnamefont {Beaudoin}},\ and\
  \bibinfo {author} {\bibfnamefont {L.}~\bibnamefont {Viola}},\ }\bibfield
  {title} {\bibinfo {title} {Frequency estimation under non-{M}arkovian
  spatially correlated quantum noise},\ }\href
  {https://doi.org/10.1088/1367-2630/ac92a2} {\bibfield  {journal} {\bibinfo
  {journal} {New Journal of Physics}\ }\textbf {\bibinfo {volume} {24}},\
  \bibinfo {pages} {103011} (\bibinfo {year} {2022})}\BibitemShut {NoStop}%
\bibitem [{\citenamefont {Kepa}\ \emph
  {et~al.}(2023{\natexlab{b}})\citenamefont {Kepa}, \citenamefont {Focke},
  \citenamefont {Cywi{\'n}ski},\ and\ \citenamefont
  {Krzywda}}]{KepaFockeCywinskiKrzywdaChargeNoise23}%
  \BibitemOpen
  \bibfield  {author} {\bibinfo {author} {\bibfnamefont {M.}~\bibnamefont
  {Kepa}}, \bibinfo {author} {\bibfnamefont {N.}~\bibnamefont {Focke}},
  \bibinfo {author} {\bibfnamefont {{\L}.}~\bibnamefont {Cywi{\'n}ski}},\ and\
  \bibinfo {author} {\bibfnamefont {J.~A.}\ \bibnamefont {Krzywda}},\
  }\bibfield  {title} {\bibinfo {title} {Simulation of $1/f$ charge noise
  affecting a quantum dot in a {Si}/{SiGe} structure},\ }\href
  {https://doi.org/10.1063/5.0151029} {\bibfield  {journal} {\bibinfo
  {journal} {Applied Physics Letters}\ }\textbf {\bibinfo {volume} {123}},\
  \bibinfo {pages} {034005} (\bibinfo {year} {2023}{\natexlab{b}})}\BibitemShut
  {NoStop}%
\bibitem [{\citenamefont {Kalos}\ and\ \citenamefont
  {Whitlock}(2008)}]{KalosWhitlock}%
  \BibitemOpen
  \bibfield  {author} {\bibinfo {author} {\bibfnamefont {M.~H.~K.}\
  \bibnamefont {Kalos}}\ and\ \bibinfo {author} {\bibfnamefont {P.~A.}\
  \bibnamefont {Whitlock}},\ }\href@noop {} {\emph {\bibinfo {title} {{M}onte
  {C}arlo methods}}},\ \bibinfo {edition} {2nd}\ ed.\ (\bibinfo  {publisher}
  {Wiley-VCH},\ \bibinfo {address} {Weinheim},\ \bibinfo {year}
  {2008})\BibitemShut {NoStop}%
\bibitem [{\citenamefont {Azmoodeh}\ \emph {et~al.}(2014)\citenamefont
  {Azmoodeh}, \citenamefont {Sottinen}, \citenamefont {Viitasaari},\ and\
  \citenamefont {Yazigi}}]{Azmoodeh2014}%
  \BibitemOpen
  \bibfield  {author} {\bibinfo {author} {\bibfnamefont {E.}~\bibnamefont
  {Azmoodeh}}, \bibinfo {author} {\bibfnamefont {T.}~\bibnamefont {Sottinen}},
  \bibinfo {author} {\bibfnamefont {L.}~\bibnamefont {Viitasaari}},\ and\
  \bibinfo {author} {\bibfnamefont {A.}~\bibnamefont {Yazigi}},\ }\bibfield
  {title} {\bibinfo {title} {Necessary and sufficient conditions for
  {H{\"o}}lder continuity of {G}aussian processes},\ }\href
  {https://doi.org/https://doi.org/10.1016/j.spl.2014.07.030} {\bibfield
  {journal} {\bibinfo  {journal} {Statistics Probability Letters}\ }\textbf
  {\bibinfo {volume} {94}},\ \bibinfo {pages} {230} (\bibinfo {year}
  {2014})}\BibitemShut {NoStop}%
\bibitem [{\citenamefont {Alomari}(2014)}]{Alomari2014}%
  \BibitemOpen
  \bibfield  {author} {\bibinfo {author} {\bibfnamefont {M.~W.}\ \bibnamefont
  {Alomari}},\ }\href@noop {} {\bibinfo {title} {New sharp inequalities of
  {O}strowski and generalized trapezoid type for the {R}iemann--{S}tieltjes
  integrals and applications}} (\bibinfo {year} {2014}),\ \bibinfo {note}
  {arXiv preprint arXiv:1408.1497}\BibitemShut {NoStop}%
\bibitem [{\citenamefont {Dragomir}(2011)}]{Dragomir2015}%
  \BibitemOpen
  \bibfield  {author} {\bibinfo {author} {\bibfnamefont {S.~S.}\ \bibnamefont
  {Dragomir}},\ }\bibfield  {title} {\bibinfo {title} {Approximating the
  {R}iemann--{S}tieltjes integral by a trapezoidal quadrature rule with
  applications},\ }\href@noop {} {\bibfield  {journal} {\bibinfo  {journal}
  {Mathematical and Computer Modelling}\ }\textbf {\bibinfo {volume} {54}},\
  \bibinfo {pages} {243} (\bibinfo {year} {2011})}\BibitemShut {NoStop}%
\bibitem [{\citenamefont {Deheuvels}\ \emph {et~al.}(2006)\citenamefont
  {Deheuvels}, \citenamefont {Peccati},\ and\ \citenamefont
  {Yor}}]{Deheuvels2006}%
  \BibitemOpen
  \bibfield  {author} {\bibinfo {author} {\bibfnamefont {P.}~\bibnamefont
  {Deheuvels}}, \bibinfo {author} {\bibfnamefont {G.}~\bibnamefont {Peccati}},\
  and\ \bibinfo {author} {\bibfnamefont {M.}~\bibnamefont {Yor}},\ }\bibfield
  {title} {\bibinfo {title} {On quadratic functionals of the {Brownian} sheet
  and related processes},\ }\href@noop {} {\bibfield  {journal} {\bibinfo
  {journal} {Stochastic Processes and their Applications}\ }\textbf {\bibinfo
  {volume} {116}},\ \bibinfo {pages} {493} (\bibinfo {year}
  {2006})}\BibitemShut {NoStop}%
\bibitem [{\citenamefont {Skordos}\ and\ \citenamefont
  {Sutcliffe}(2008)}]{Skordos2008}%
  \BibitemOpen
  \bibfield  {author} {\bibinfo {author} {\bibfnamefont {A.~A.}\ \bibnamefont
  {Skordos}}\ and\ \bibinfo {author} {\bibfnamefont {M.~P.}\ \bibnamefont
  {Sutcliffe}},\ }\bibfield  {title} {\bibinfo {title} {Stochastic simulation
  of woven composites forming},\ }\href@noop {} {\bibfield  {journal} {\bibinfo
   {journal} {Composites Science and Technology}\ }\textbf {\bibinfo {volume}
  {68}},\ \bibinfo {pages} {283} (\bibinfo {year} {2008})}\BibitemShut
  {NoStop}%
\bibitem [{\citenamefont {Cywi\ifmmode~\acute{n}\else \'{n}\fi{}ski}\ \emph
  {et~al.}(2008)\citenamefont {Cywi\ifmmode~\acute{n}\else \'{n}\fi{}ski},
  \citenamefont {Lutchyn}, \citenamefont {Nave},\ and\ \citenamefont
  {Das~Sarma}}]{CywinskiLutchynDasSarmaDD08}%
  \BibitemOpen
  \bibfield  {author} {\bibinfo {author} {\bibfnamefont {L.}~\bibnamefont
  {Cywi\ifmmode~\acute{n}\else \'{n}\fi{}ski}}, \bibinfo {author}
  {\bibfnamefont {R.~M.}\ \bibnamefont {Lutchyn}}, \bibinfo {author}
  {\bibfnamefont {C.~P.}\ \bibnamefont {Nave}},\ and\ \bibinfo {author}
  {\bibfnamefont {S.}~\bibnamefont {Das~Sarma}},\ }\bibfield  {title} {\bibinfo
  {title} {How to enhance dephasing time in superconducting qubits},\ }\href
  {https://doi.org/10.1103/PhysRevB.77.174509} {\bibfield  {journal} {\bibinfo
  {journal} {Phys. Rev. B}\ }\textbf {\bibinfo {volume} {77}},\ \bibinfo
  {pages} {174509} (\bibinfo {year} {2008})}\BibitemShut {NoStop}%
\bibitem [{\citenamefont {Viola}\ and\ \citenamefont
  {Lloyd}(1998)}]{ViolaLloydDD}%
  \BibitemOpen
  \bibfield  {author} {\bibinfo {author} {\bibfnamefont {L.}~\bibnamefont
  {Viola}}\ and\ \bibinfo {author} {\bibfnamefont {S.}~\bibnamefont {Lloyd}},\
  }\bibfield  {title} {\bibinfo {title} {Dynamical suppression of decoherence
  in two-state quantum systems},\ }\href
  {https://doi.org/10.1103/PhysRevA.58.2733} {\bibfield  {journal} {\bibinfo
  {journal} {Phys. Rev. A}\ }\textbf {\bibinfo {volume} {58}},\ \bibinfo
  {pages} {2733} (\bibinfo {year} {1998})}\BibitemShut {NoStop}%
\bibitem [{Note6()}]{Note6}%
  \BibitemOpen
  \bibinfo {note} {Due to the symmetry of $K_\protect \mathrm {B}(t_1,t_2)$ in
  that case, it is sufficient to perform integration over only one of those
  regions.}\BibitemShut {Stop}%
\bibitem [{\citenamefont {Buonacorsi}\ \emph {et~al.}(2020)\citenamefont
  {Buonacorsi}, \citenamefont {Shaw},\ and\ \citenamefont
  {Baugh}}]{buonacorsi_simulated_2020}%
  \BibitemOpen
  \bibfield  {author} {\bibinfo {author} {\bibfnamefont {B.}~\bibnamefont
  {Buonacorsi}}, \bibinfo {author} {\bibfnamefont {B.}~\bibnamefont {Shaw}},\
  and\ \bibinfo {author} {\bibfnamefont {J.}~\bibnamefont {Baugh}},\ }\bibfield
   {title} {\bibinfo {title} {Simulated coherent electron shuttling in silicon
  quantum dots},\ }\href {https://doi.org/10.1103/PhysRevB.102.125406}
  {\bibfield  {journal} {\bibinfo  {journal} {Physical Review B}\ }\textbf
  {\bibinfo {volume} {102}},\ \bibinfo {pages} {125406} (\bibinfo {year}
  {2020})}\BibitemShut {NoStop}%
\bibitem [{\citenamefont {Tahan}\ and\ \citenamefont
  {Joynt}(2014)}]{TahanJoynt14}%
  \BibitemOpen
  \bibfield  {author} {\bibinfo {author} {\bibfnamefont {C.}~\bibnamefont
  {Tahan}}\ and\ \bibinfo {author} {\bibfnamefont {R.}~\bibnamefont {Joynt}},\
  }\bibfield  {title} {\bibinfo {title} {Relaxation of excited spin, orbital,
  and valley qubit states in ideal silicon quantum dots},\ }\href
  {https://doi.org/10.1103/PhysRevB.89.075302} {\bibfield  {journal} {\bibinfo
  {journal} {Phys. Rev. B}\ }\textbf {\bibinfo {volume} {89}},\ \bibinfo
  {pages} {075302} (\bibinfo {year} {2014})}\BibitemShut {NoStop}%
\bibitem [{\citenamefont {Hao}\ \emph {et~al.}(2014)\citenamefont {Hao},
  \citenamefont {Ruskov}, \citenamefont {Xiao}, \citenamefont {Tahan},\ and\
  \citenamefont {Jiang}}]{HaoRuskovTahanEtal14}%
  \BibitemOpen
  \bibfield  {author} {\bibinfo {author} {\bibfnamefont {X.}~\bibnamefont
  {Hao}}, \bibinfo {author} {\bibfnamefont {R.}~\bibnamefont {Ruskov}},
  \bibinfo {author} {\bibfnamefont {M.}~\bibnamefont {Xiao}}, \bibinfo {author}
  {\bibfnamefont {C.}~\bibnamefont {Tahan}},\ and\ \bibinfo {author}
  {\bibfnamefont {H.}~\bibnamefont {Jiang}},\ }\bibfield  {title} {\bibinfo
  {title} {Electron spin resonance and spin-valley physics in a silicon double
  quantum dot},\ }\href {https://doi.org/10.1038/ncomms4860} {\bibfield
  {journal} {\bibinfo  {journal} {Nat. Commun.}\ }\textbf {\bibinfo {volume}
  {5}},\ \bibinfo {pages} {3860} (\bibinfo {year} {2014})}\BibitemShut
  {NoStop}%
\bibitem [{\citenamefont {Bacon}\ \emph {et~al.}(2000)\citenamefont {Bacon},
  \citenamefont {Kempe}, \citenamefont {Lidar},\ and\ \citenamefont
  {Whaley}}]{EOqubit1}%
  \BibitemOpen
  \bibfield  {author} {\bibinfo {author} {\bibfnamefont {D.}~\bibnamefont
  {Bacon}}, \bibinfo {author} {\bibfnamefont {J.}~\bibnamefont {Kempe}},
  \bibinfo {author} {\bibfnamefont {D.~A.}\ \bibnamefont {Lidar}},\ and\
  \bibinfo {author} {\bibfnamefont {K.~B.}\ \bibnamefont {Whaley}},\ }\bibfield
   {title} {\bibinfo {title} {Universal fault-tolerant quantum computation on
  decoherence-free subspaces},\ }\href
  {https://doi.org/10.1103/PhysRevLett.85.1758} {\bibfield  {journal} {\bibinfo
   {journal} {Phys. Rev. Lett.}\ }\textbf {\bibinfo {volume} {85}},\ \bibinfo
  {pages} {1758} (\bibinfo {year} {2000})}\BibitemShut {NoStop}%
\bibitem [{\citenamefont {DiVincenzo}\ \emph {et~al.}(2000)\citenamefont
  {DiVincenzo}, \citenamefont {Bacon}, \citenamefont {Kempe}, \citenamefont
  {Burkard},\ and\ \citenamefont {Whaley}}]{EOqubit2}%
  \BibitemOpen
  \bibfield  {author} {\bibinfo {author} {\bibfnamefont {D.~P.}\ \bibnamefont
  {DiVincenzo}}, \bibinfo {author} {\bibfnamefont {D.}~\bibnamefont {Bacon}},
  \bibinfo {author} {\bibfnamefont {J.}~\bibnamefont {Kempe}}, \bibinfo
  {author} {\bibfnamefont {G.}~\bibnamefont {Burkard}},\ and\ \bibinfo {author}
  {\bibfnamefont {K.~B.}\ \bibnamefont {Whaley}},\ }\bibfield  {title}
  {\bibinfo {title} {Universal quantum computation with the exchange
  interaction},\ }\href {https://www.nature.com/articles/35042541} {\bibfield
  {journal} {\bibinfo  {journal} {Nature}\ }\textbf {\bibinfo {volume} {408}},\
  \bibinfo {pages} {339} (\bibinfo {year} {2000})}\BibitemShut {NoStop}%
\bibitem [{\citenamefont {Russ}\ and\ \citenamefont
  {Burkard}(2017)}]{EOqubit3}%
  \BibitemOpen
  \bibfield  {author} {\bibinfo {author} {\bibfnamefont {M.}~\bibnamefont
  {Russ}}\ and\ \bibinfo {author} {\bibfnamefont {G.}~\bibnamefont {Burkard}},\
  }\bibfield  {title} {\bibinfo {title} {Three-electron spin qubits},\ }\href
  {https://doi.org/10.1088/1361-648X/aa761f} {\bibfield  {journal} {\bibinfo
  {journal} {J. Phys.: Condens. Matter}\ }\textbf {\bibinfo {volume} {29}},\
  \bibinfo {pages} {393001} (\bibinfo {year} {2017})}\BibitemShut {NoStop}%
\bibitem [{\citenamefont {Sza{\'{n}}kowski}\ \emph {et~al.}(2017)\citenamefont
  {Sza{\'{n}}kowski}, \citenamefont {Ramon}, \citenamefont {Krzywda},
  \citenamefont {Kwiatkowski},\ and\ \citenamefont
  {Cywi{\'{n}}ski}}]{Szankowski2017}%
  \BibitemOpen
  \bibfield  {author} {\bibinfo {author} {\bibfnamefont {P.}~\bibnamefont
  {Sza{\'{n}}kowski}}, \bibinfo {author} {\bibfnamefont {G.}~\bibnamefont
  {Ramon}}, \bibinfo {author} {\bibfnamefont {J.}~\bibnamefont {Krzywda}},
  \bibinfo {author} {\bibfnamefont {D.}~\bibnamefont {Kwiatkowski}},\ and\
  \bibinfo {author} {\bibfnamefont {{\L}.}~\bibnamefont {Cywi{\'{n}}ski}},\
  }\bibfield  {title} {\bibinfo {title} {Environmental noise spectroscopy with
  qubits subjected to dynamical decoupling},\ }\href
  {https://doi.org/10.1088/1361-648x/aa7648} {\bibfield  {journal} {\bibinfo
  {journal} {Journal of Physics: Condensed Matter}\ }\textbf {\bibinfo {volume}
  {29}},\ \bibinfo {pages} {333001} (\bibinfo {year} {2017})}\BibitemShut
  {NoStop}%
\bibitem [{\citenamefont {Ramon}(2015)}]{RamonNonGauss2015}%
  \BibitemOpen
  \bibfield  {author} {\bibinfo {author} {\bibfnamefont {G.}~\bibnamefont
  {Ramon}},\ }\bibfield  {title} {\bibinfo {title} {Non-{G}aussian signatures
  and collective effects in charge noise affecting a dynamically decoupled
  qubit},\ }\href {https://doi.org/10.1103/PhysRevB.92.155422} {\bibfield
  {journal} {\bibinfo  {journal} {Phys. Rev. B}\ }\textbf {\bibinfo {volume}
  {92}},\ \bibinfo {pages} {155422} (\bibinfo {year} {2015})}\BibitemShut
  {NoStop}%
\bibitem [{\citenamefont {Ye}\ \emph {et~al.}(2024)\citenamefont {Ye},
  \citenamefont {Ellaboudy}, \citenamefont {Albrecht}, \citenamefont {Vudatha},
  \citenamefont {Jacobson},\ and\ \citenamefont
  {Nichol}}]{YeNicholEtal24IndividualTLFs}%
  \BibitemOpen
  \bibfield  {author} {\bibinfo {author} {\bibfnamefont {F.}~\bibnamefont
  {Ye}}, \bibinfo {author} {\bibfnamefont {A.}~\bibnamefont {Ellaboudy}},
  \bibinfo {author} {\bibfnamefont {D.}~\bibnamefont {Albrecht}}, \bibinfo
  {author} {\bibfnamefont {R.}~\bibnamefont {Vudatha}}, \bibinfo {author}
  {\bibfnamefont {N.~T.}\ \bibnamefont {Jacobson}},\ and\ \bibinfo {author}
  {\bibfnamefont {J.~M.}\ \bibnamefont {Nichol}},\ }\href
  {https://arxiv.org/abs/2401.14541} {\bibinfo {title} {Characterization of
  individual charge fluctuators in si/sige quantum dots}} (\bibinfo {year}
  {2024}),\ \Eprint {https://arxiv.org/abs/2401.14541} {arXiv:2401.14541
  [cond-mat.mes-hall]} \BibitemShut {NoStop}%
\bibitem [{\citenamefont {Connors}\ \emph {et~al.}(2022)\citenamefont
  {Connors}, \citenamefont {Nelson}, \citenamefont {Edge},\ and\ \citenamefont
  {Nichol}}]{Connors2022}%
  \BibitemOpen
  \bibfield  {author} {\bibinfo {author} {\bibfnamefont {E.~J.}\ \bibnamefont
  {Connors}}, \bibinfo {author} {\bibfnamefont {J.~J.}\ \bibnamefont {Nelson}},
  \bibinfo {author} {\bibfnamefont {L.~F.}\ \bibnamefont {Edge}},\ and\
  \bibinfo {author} {\bibfnamefont {J.~M.}\ \bibnamefont {Nichol}},\ }\bibfield
   {title} {\bibinfo {title} {Charge-noise spectroscopy of {Si/SiGe} quantum
  dots via dynamically-decoupled exchange oscillations},\ }\href
  {https://doi.org/10.1038/s41467-022-28519-x} {\bibfield  {journal} {\bibinfo
  {journal} {Nat. Commun.}\ }\textbf {\bibinfo {volume} {13}},\ \bibinfo
  {pages} {940} (\bibinfo {year} {2022})}\BibitemShut {NoStop}%
\bibitem [{\citenamefont {Ahn}\ \emph {et~al.}(2021)\citenamefont {Ahn},
  \citenamefont {Das~Sarma},\ and\ \citenamefont {Kestner}}]{Ahn2021}%
  \BibitemOpen
  \bibfield  {author} {\bibinfo {author} {\bibfnamefont {S.}~\bibnamefont
  {Ahn}}, \bibinfo {author} {\bibfnamefont {S.}~\bibnamefont {Das~Sarma}},\
  and\ \bibinfo {author} {\bibfnamefont {J.~P.}\ \bibnamefont {Kestner}},\
  }\bibfield  {title} {\bibinfo {title} {Microscopic bath effects on noise
  spectra in semiconductor quantum dot qubits},\ }\href
  {https://link.aps.org/doi/10.1103/PhysRevB.103.L041304} {\bibfield  {journal}
  {\bibinfo  {journal} {Phys. Rev. B}\ }\textbf {\bibinfo {volume} {103}},\
  \bibinfo {pages} {L041304} (\bibinfo {year} {2021})}\BibitemShut {NoStop}%
\bibitem [{\citenamefont {Mehmandoost}\ and\ \citenamefont
  {Dobrovitski}(2024)}]{MMandVD_TLS24}%
  \BibitemOpen
  \bibfield  {author} {\bibinfo {author} {\bibfnamefont {M.}~\bibnamefont
  {Mehmandoost}}\ and\ \bibinfo {author} {\bibfnamefont {V.~V.}\ \bibnamefont
  {Dobrovitski}},\ }\bibfield  {title} {\bibinfo {title} {Decoherence induced
  by a sparse bath of two-level fluctuators: {P}eculiar features of $1/f$ noise
  in high-quality qubits},\ }\href
  {https://doi.org/10.1103/PhysRevResearch.6.033175} {\bibfield  {journal}
  {\bibinfo  {journal} {Phys. Rev. Res.}\ }\textbf {\bibinfo {volume} {6}},\
  \bibinfo {pages} {033175} (\bibinfo {year} {2024})}\BibitemShut {NoStop}%
\bibitem [{Con()}]{Conor}%
  \BibitemOpen
  \href@noop {} {}\bibinfo {note} {C.~E.~Bradley (private
  communication)}\BibitemShut {NoStop}%
\bibitem [{Dol()}]{Dolev}%
  \BibitemOpen
  \href@noop {} {}\bibinfo {note} {D.~Bluvstein (private
  communication)}\BibitemShut {NoStop}%
\bibitem [{\citenamefont {Lee}\ \emph {et~al.}(2012)\citenamefont {Lee},
  \citenamefont {Kim}, \citenamefont {Park}, \citenamefont {Eom}, \citenamefont
  {Kim}, \citenamefont {Rho},\ and\ \citenamefont
  {Choi}}]{LeeEtalFiberOpticSensors}%
  \BibitemOpen
  \bibfield  {author} {\bibinfo {author} {\bibfnamefont {B.~H.}\ \bibnamefont
  {Lee}}, \bibinfo {author} {\bibfnamefont {Y.~H.}\ \bibnamefont {Kim}},
  \bibinfo {author} {\bibfnamefont {K.~S.}\ \bibnamefont {Park}}, \bibinfo
  {author} {\bibfnamefont {J.~B.}\ \bibnamefont {Eom}}, \bibinfo {author}
  {\bibfnamefont {M.~J.}\ \bibnamefont {Kim}}, \bibinfo {author} {\bibfnamefont
  {B.~S.}\ \bibnamefont {Rho}},\ and\ \bibinfo {author} {\bibfnamefont {H.~Y.}\
  \bibnamefont {Choi}},\ }\bibfield  {title} {\bibinfo {title} {Interferometric
  fiber optic sensors},\ }\href {https://doi.org/10.3390/s120302467} {\bibfield
   {journal} {\bibinfo  {journal} {Sensors}\ }\textbf {\bibinfo {volume}
  {12}},\ \bibinfo {pages} {2467} (\bibinfo {year} {2012})}\BibitemShut
  {NoStop}%
\bibitem [{\citenamefont {Glenn}(1989)}]{GlennNoiseFluctOptFibers89}%
  \BibitemOpen
  \bibfield  {author} {\bibinfo {author} {\bibfnamefont {W.}~\bibnamefont
  {Glenn}},\ }\bibfield  {title} {\bibinfo {title} {Noise in interferometric
  optical systems: an optical {N}yquist theorem},\ }\href
  {https://doi.org/10.1109/3.29251} {\bibfield  {journal} {\bibinfo  {journal}
  {IEEE Journal of Quantum Electronics}\ }\textbf {\bibinfo {volume} {25}},\
  \bibinfo {pages} {1218} (\bibinfo {year} {1989})}\BibitemShut {NoStop}%
\bibitem [{\citenamefont {Kiilerich}\ and\ \citenamefont
  {M\o{}lmer}(2019)}]{KiilerichMolmerQuPulsesPRL19}%
  \BibitemOpen
  \bibfield  {author} {\bibinfo {author} {\bibfnamefont {A.~H.}\ \bibnamefont
  {Kiilerich}}\ and\ \bibinfo {author} {\bibfnamefont {K.}~\bibnamefont
  {M\o{}lmer}},\ }\bibfield  {title} {\bibinfo {title} {Input-output theory
  with quantum pulses},\ }\href
  {https://doi.org/10.1103/PhysRevLett.123.123604} {\bibfield  {journal}
  {\bibinfo  {journal} {Phys. Rev. Lett.}\ }\textbf {\bibinfo {volume} {123}},\
  \bibinfo {pages} {123604} (\bibinfo {year} {2019})}\BibitemShut {NoStop}%
\bibitem [{\citenamefont {Kiilerich}\ and\ \citenamefont
  {M\o{}lmer}(2020)}]{KiilerichMolmerQuPulsesPRA20}%
  \BibitemOpen
  \bibfield  {author} {\bibinfo {author} {\bibfnamefont {A.~H.}\ \bibnamefont
  {Kiilerich}}\ and\ \bibinfo {author} {\bibfnamefont {K.}~\bibnamefont
  {M\o{}lmer}},\ }\bibfield  {title} {\bibinfo {title} {Quantum interactions
  with pulses of radiation},\ }\href
  {https://doi.org/10.1103/PhysRevA.102.023717} {\bibfield  {journal} {\bibinfo
   {journal} {Phys. Rev. A}\ }\textbf {\bibinfo {volume} {102}},\ \bibinfo
  {pages} {023717} (\bibinfo {year} {2020})}\BibitemShut {NoStop}%
\bibitem [{\citenamefont {Greenberg}\ \emph {et~al.}(2023)\citenamefont
  {Greenberg}, \citenamefont {Moiseev},\ and\ \citenamefont
  {Shtygashev}}]{GreenbergEtalPhotonScatOnQubit23}%
  \BibitemOpen
  \bibfield  {author} {\bibinfo {author} {\bibfnamefont {Y.~S.}\ \bibnamefont
  {Greenberg}}, \bibinfo {author} {\bibfnamefont {A.~G.}\ \bibnamefont
  {Moiseev}},\ and\ \bibinfo {author} {\bibfnamefont {A.~A.}\ \bibnamefont
  {Shtygashev}},\ }\bibfield  {title} {\bibinfo {title} {Single-photon
  scattering on a qubit: Space-time structure of the scattered field},\ }\href
  {https://doi.org/10.1103/PhysRevA.107.013519} {\bibfield  {journal} {\bibinfo
   {journal} {Phys. Rev. A}\ }\textbf {\bibinfo {volume} {107}},\ \bibinfo
  {pages} {013519} (\bibinfo {year} {2023})}\BibitemShut {NoStop}%
\bibitem [{\citenamefont {Gavrilenko}\ and\ \citenamefont
  {Stepanov}(1973)}]{GavrilenkoStepanov73LightSpaceTimeVarMed}%
  \BibitemOpen
  \bibfield  {author} {\bibinfo {author} {\bibfnamefont {V.~G.}\ \bibnamefont
  {Gavrilenko}}\ and\ \bibinfo {author} {\bibfnamefont {N.~S.}\ \bibnamefont
  {Stepanov}},\ }\bibfield  {title} {\bibinfo {title} {Transformation of the
  wave spectrum in a medium having smooth space-time fluctuations},\ }\href
  {https://doi.org/https://doi.org/10.1007/BF01080794} {\bibfield  {journal}
  {\bibinfo  {journal} {Radiophysics and Quantum Electronics}\ }\textbf
  {\bibinfo {volume} {16}},\ \bibinfo {pages} {50} (\bibinfo {year}
  {1973})}\BibitemShut {NoStop}%
\bibitem [{\citenamefont {Hermans}\ \emph {et~al.}(2023)\citenamefont
  {Hermans}, \citenamefont {Pompili}, \citenamefont {Dos Santos~Martins},
  \citenamefont {Montblanch}, \citenamefont {Beukers}, \citenamefont {Baier},
  \citenamefont {Borregaard},\ and\ \citenamefont
  {Hanson}}]{HermansBorregaardEtalSingleClick23}%
  \BibitemOpen
  \bibfield  {author} {\bibinfo {author} {\bibfnamefont {S.~L.~N.}\
  \bibnamefont {Hermans}}, \bibinfo {author} {\bibfnamefont {M.}~\bibnamefont
  {Pompili}}, \bibinfo {author} {\bibfnamefont {L.}~\bibnamefont {Dos
  Santos~Martins}}, \bibinfo {author} {\bibfnamefont {A.~R.-P.}\ \bibnamefont
  {Montblanch}}, \bibinfo {author} {\bibfnamefont {H.~K.~C.}\ \bibnamefont
  {Beukers}}, \bibinfo {author} {\bibfnamefont {S.}~\bibnamefont {Baier}},
  \bibinfo {author} {\bibfnamefont {J.}~\bibnamefont {Borregaard}},\ and\
  \bibinfo {author} {\bibfnamefont {R.}~\bibnamefont {Hanson}},\ }\bibfield
  {title} {\bibinfo {title} {Entangling remote qubits using the single-photon
  protocol: an in-depth theoretical and experimental study},\ }\href
  {https://doi.org/DOI 10.1088/1367-2630/acb004} {\bibfield  {journal}
  {\bibinfo  {journal} {New Journal of Physics}\ }\textbf {\bibinfo {volume}
  {25}},\ \bibinfo {pages} {013011} (\bibinfo {year} {2023})}\BibitemShut
  {NoStop}%
\bibitem [{\citenamefont {Knaut}\ \emph {et~al.}(2024)\citenamefont {Knaut},
  \citenamefont {Suleymanzade}, \citenamefont {Wei}, \citenamefont {Assumpcao},
  \citenamefont {Stas}, \citenamefont {Huan}, \citenamefont {Machielse},
  \citenamefont {Knall}, \citenamefont {Sutula}, \citenamefont {Baranes},
  \citenamefont {Sinclair}, \citenamefont {De-Eknamkul}, \citenamefont
  {Levonian}, \citenamefont {Bhaskar}, \citenamefont {Park}, \citenamefont
  {Lon{\v c}ar},\ and\ \citenamefont
  {Lukin}}]{PhotonLoopBoston24LukinLoncarEtal}%
  \BibitemOpen
  \bibfield  {author} {\bibinfo {author} {\bibfnamefont {C.~M.}\ \bibnamefont
  {Knaut}}, \bibinfo {author} {\bibfnamefont {A.}~\bibnamefont {Suleymanzade}},
  \bibinfo {author} {\bibfnamefont {Y.-C.}\ \bibnamefont {Wei}}, \bibinfo
  {author} {\bibfnamefont {D.~R.}\ \bibnamefont {Assumpcao}}, \bibinfo {author}
  {\bibfnamefont {P.-J.}\ \bibnamefont {Stas}}, \bibinfo {author}
  {\bibfnamefont {Y.~Q.}\ \bibnamefont {Huan}}, \bibinfo {author}
  {\bibfnamefont {B.}~\bibnamefont {Machielse}}, \bibinfo {author}
  {\bibfnamefont {E.~N.}\ \bibnamefont {Knall}}, \bibinfo {author}
  {\bibfnamefont {M.}~\bibnamefont {Sutula}}, \bibinfo {author} {\bibfnamefont
  {G.}~\bibnamefont {Baranes}}, \bibinfo {author} {\bibfnamefont
  {N.}~\bibnamefont {Sinclair}}, \bibinfo {author} {\bibfnamefont
  {C.}~\bibnamefont {De-Eknamkul}}, \bibinfo {author} {\bibfnamefont {D.~S.}\
  \bibnamefont {Levonian}}, \bibinfo {author} {\bibfnamefont {M.~K.}\
  \bibnamefont {Bhaskar}}, \bibinfo {author} {\bibfnamefont {H.}~\bibnamefont
  {Park}}, \bibinfo {author} {\bibfnamefont {M.}~\bibnamefont {Lon{\v c}ar}},\
  and\ \bibinfo {author} {\bibfnamefont {M.~D.}\ \bibnamefont {Lukin}},\
  }\bibfield  {title} {\bibinfo {title} {Entanglement of nanophotonic quantum
  memory nodes in a telecom network},\ }\href
  {https://doi.org/10.1038/s41586-024-07252-z} {\bibfield  {journal} {\bibinfo
  {journal} {Nature}\ }\textbf {\bibinfo {volume} {629}},\ \bibinfo {pages}
  {573} (\bibinfo {year} {2024})}\BibitemShut {NoStop}%
\bibitem [{\citenamefont {Marchetti}\ \emph {et~al.}(2019)\citenamefont
  {Marchetti}, \citenamefont {Lacava}, \citenamefont {Carroll}, \citenamefont
  {Gradkowski},\ and\ \citenamefont
  {Minzioni}}]{SiChipsMarchettiMinzioniEtal19}%
  \BibitemOpen
  \bibfield  {author} {\bibinfo {author} {\bibfnamefont {R.}~\bibnamefont
  {Marchetti}}, \bibinfo {author} {\bibfnamefont {C.}~\bibnamefont {Lacava}},
  \bibinfo {author} {\bibfnamefont {L.}~\bibnamefont {Carroll}}, \bibinfo
  {author} {\bibfnamefont {K.}~\bibnamefont {Gradkowski}},\ and\ \bibinfo
  {author} {\bibfnamefont {P.}~\bibnamefont {Minzioni}},\ }\bibfield  {title}
  {\bibinfo {title} {Coupling strategies for silicon photonics integrated
  chips},\ }\href {https://doi.org/10.1364/PRJ.7.000201} {\bibfield  {journal}
  {\bibinfo  {journal} {Photon. Res.}\ }\textbf {\bibinfo {volume} {7}},\
  \bibinfo {pages} {201} (\bibinfo {year} {2019})}\BibitemShut {NoStop}%
\bibitem [{\citenamefont {Pelucchi}\ \emph {et~al.}(2022)\citenamefont
  {Pelucchi}, \citenamefont {Fagas}, \citenamefont {Aharonovich}, \citenamefont
  {Englund}, \citenamefont {Figueroa}, \citenamefont {Gong}, \citenamefont
  {Hannes}, \citenamefont {Liu}, \citenamefont {Lu}, \citenamefont {Matsuda},
  \citenamefont {Pan}, \citenamefont {Schreck}, \citenamefont {Sciarrino},
  \citenamefont {Silberhorn}, \citenamefont {Wang},\ and\ \citenamefont
  {J{\"o}ns}}]{PhotonicChipsQIPReviewEnglundEtal22}%
  \BibitemOpen
  \bibfield  {author} {\bibinfo {author} {\bibfnamefont {E.}~\bibnamefont
  {Pelucchi}}, \bibinfo {author} {\bibfnamefont {G.}~\bibnamefont {Fagas}},
  \bibinfo {author} {\bibfnamefont {I.}~\bibnamefont {Aharonovich}}, \bibinfo
  {author} {\bibfnamefont {D.}~\bibnamefont {Englund}}, \bibinfo {author}
  {\bibfnamefont {E.}~\bibnamefont {Figueroa}}, \bibinfo {author}
  {\bibfnamefont {Q.}~\bibnamefont {Gong}}, \bibinfo {author} {\bibfnamefont
  {H.}~\bibnamefont {Hannes}}, \bibinfo {author} {\bibfnamefont
  {J.}~\bibnamefont {Liu}}, \bibinfo {author} {\bibfnamefont {C.-Y.}\
  \bibnamefont {Lu}}, \bibinfo {author} {\bibfnamefont {N.}~\bibnamefont
  {Matsuda}}, \bibinfo {author} {\bibfnamefont {J.-W.}\ \bibnamefont {Pan}},
  \bibinfo {author} {\bibfnamefont {F.}~\bibnamefont {Schreck}}, \bibinfo
  {author} {\bibfnamefont {F.}~\bibnamefont {Sciarrino}}, \bibinfo {author}
  {\bibfnamefont {C.}~\bibnamefont {Silberhorn}}, \bibinfo {author}
  {\bibfnamefont {J.}~\bibnamefont {Wang}},\ and\ \bibinfo {author}
  {\bibfnamefont {K.~D.}\ \bibnamefont {J{\"o}ns}},\ }\bibfield  {title}
  {\bibinfo {title} {The potential and global outlook of integrated photonics
  for quantum technologies},\ }\href
  {https://doi.org/10.1038/s42254-021-00398-z} {\bibfield  {journal} {\bibinfo
  {journal} {Nature Reviews Physics}\ }\textbf {\bibinfo {volume} {4}},\
  \bibinfo {pages} {194} (\bibinfo {year} {2022})}\BibitemShut {NoStop}%
\bibitem [{\citenamefont {Errando-Herranz}\ \emph {et~al.}(2020)\citenamefont
  {Errando-Herranz}, \citenamefont {Takabayashi}, \citenamefont {Edinger},
  \citenamefont {Sattari}, \citenamefont {Gylfason},\ and\ \citenamefont
  {Quack}}]{ErrandoHerranzEtal20PhotIntegrCircuits}%
  \BibitemOpen
  \bibfield  {author} {\bibinfo {author} {\bibfnamefont {C.}~\bibnamefont
  {Errando-Herranz}}, \bibinfo {author} {\bibfnamefont {A.~Y.}\ \bibnamefont
  {Takabayashi}}, \bibinfo {author} {\bibfnamefont {P.}~\bibnamefont
  {Edinger}}, \bibinfo {author} {\bibfnamefont {H.}~\bibnamefont {Sattari}},
  \bibinfo {author} {\bibfnamefont {K.~B.}\ \bibnamefont {Gylfason}},\ and\
  \bibinfo {author} {\bibfnamefont {N.}~\bibnamefont {Quack}},\ }\bibfield
  {title} {\bibinfo {title} {Mems for photonic integrated circuits},\ }\href
  {https://doi.org/10.1109/JSTQE.2019.2943384} {\bibfield  {journal} {\bibinfo
  {journal} {IEEE Journal of Selected Topics in Quantum Electronics}\ }\textbf
  {\bibinfo {volume} {26}},\ \bibinfo {pages} {1} (\bibinfo {year}
  {2020})}\BibitemShut {NoStop}%
\bibitem [{\citenamefont {Chakraborty}\ \emph {et~al.}(2020)\citenamefont
  {Chakraborty}, \citenamefont {Elkouss}, \citenamefont {Rijsman},\ and\
  \citenamefont {Wehner}}]{ChakrabortyWehnerEtal20RoutingEntanglement}%
  \BibitemOpen
  \bibfield  {author} {\bibinfo {author} {\bibfnamefont {K.}~\bibnamefont
  {Chakraborty}}, \bibinfo {author} {\bibfnamefont {D.}~\bibnamefont
  {Elkouss}}, \bibinfo {author} {\bibfnamefont {B.}~\bibnamefont {Rijsman}},\
  and\ \bibinfo {author} {\bibfnamefont {S.}~\bibnamefont {Wehner}},\
  }\bibfield  {title} {\bibinfo {title} {Entanglement distribution in a quantum
  network: A multicommodity flow-based approach},\ }\href
  {https://doi.org/10.1109/TQE.2020.3028172} {\bibfield  {journal} {\bibinfo
  {journal} {IEEE Transactions on Quantum Engineering}\ }\textbf {\bibinfo
  {volume} {1}},\ \bibinfo {pages} {1} (\bibinfo {year} {2020})}\BibitemShut
  {NoStop}%
\bibitem [{\citenamefont {Caleffi}(2017)}]{Caleffi17RoutingQuNetworks}%
  \BibitemOpen
  \bibfield  {author} {\bibinfo {author} {\bibfnamefont {M.}~\bibnamefont
  {Caleffi}},\ }\bibfield  {title} {\bibinfo {title} {Optimal routing for
  quantum networks},\ }\href {https://doi.org/10.1109/ACCESS.2017.2763325}
  {\bibfield  {journal} {\bibinfo  {journal} {IEEE Access}\ }\textbf {\bibinfo
  {volume} {5}},\ \bibinfo {pages} {22299} (\bibinfo {year}
  {2017})}\BibitemShut {NoStop}%
\bibitem [{\citenamefont {Kozlowski}\ and\ \citenamefont
  {Wehner}(2019)}]{KozlowskiWehner19LargeQuNetworks}%
  \BibitemOpen
  \bibfield  {author} {\bibinfo {author} {\bibfnamefont {W.}~\bibnamefont
  {Kozlowski}}\ and\ \bibinfo {author} {\bibfnamefont {S.}~\bibnamefont
  {Wehner}},\ }\bibfield  {title} {\bibinfo {title} {Towards large-scale
  quantum networks},\ }in\ \href {https://doi.org/10.1145/3345312.3345497}
  {\emph {\bibinfo {booktitle} {Proceedings of the Sixth Annual ACM
  International Conference on Nanoscale Computing and Communication}}},\
  \bibinfo {series and number} {NANOCOM '19}\ (\bibinfo  {publisher}
  {Association for Computing Machinery},\ \bibinfo {address} {New York, NY,
  USA},\ \bibinfo {year} {2019})\BibitemShut {NoStop}%
\bibitem [{\citenamefont {Lee}\ \emph {et~al.}(2022)\citenamefont {Lee},
  \citenamefont {Bersin}, \citenamefont {Dahlberg}, \citenamefont {Wehner},\
  and\ \citenamefont {Englund}}]{LeeWehnerEnglundEtal22Router}%
  \BibitemOpen
  \bibfield  {author} {\bibinfo {author} {\bibfnamefont {Y.}~\bibnamefont
  {Lee}}, \bibinfo {author} {\bibfnamefont {E.}~\bibnamefont {Bersin}},
  \bibinfo {author} {\bibfnamefont {A.}~\bibnamefont {Dahlberg}}, \bibinfo
  {author} {\bibfnamefont {S.}~\bibnamefont {Wehner}},\ and\ \bibinfo {author}
  {\bibfnamefont {D.}~\bibnamefont {Englund}},\ }\bibfield  {title} {\bibinfo
  {title} {A quantum router architecture for high-fidelity entanglement flows
  in quantum networks},\ }\href {https://doi.org/10.1038/s41534-022-00582-8}
  {\bibfield  {journal} {\bibinfo  {journal} {npj Qu. Inf.}\ }\textbf {\bibinfo
  {volume} {8}},\ \bibinfo {pages} {75} (\bibinfo {year} {2022})}\BibitemShut
  {NoStop}%
\bibitem [{\citenamefont {Friedli}\ and\ \citenamefont
  {Velenik}(2017)}]{friedli2017statistical}%
  \BibitemOpen
  \bibfield  {author} {\bibinfo {author} {\bibfnamefont {S.}~\bibnamefont
  {Friedli}}\ and\ \bibinfo {author} {\bibfnamefont {Y.}~\bibnamefont
  {Velenik}},\ }\href@noop {} {\emph {\bibinfo {title} {Statistical mechanics
  of lattice systems: a concrete mathematical introduction}}}\ (\bibinfo
  {publisher} {Cambridge University Press},\ \bibinfo {year}
  {2017})\BibitemShut {NoStop}%
\bibitem [{\citenamefont {Sheffield}(2007)}]{sheffield2007gaussian}%
  \BibitemOpen
  \bibfield  {author} {\bibinfo {author} {\bibfnamefont {S.}~\bibnamefont
  {Sheffield}},\ }\bibfield  {title} {\bibinfo {title} {Gaussian free fields
  for mathematicians},\ }\href@noop {} {\bibfield  {journal} {\bibinfo
  {journal} {Probability theory and related fields}\ }\textbf {\bibinfo
  {volume} {139}},\ \bibinfo {pages} {521} (\bibinfo {year}
  {2007})}\BibitemShut {NoStop}%
\bibitem [{\citenamefont {Pitt}(1971)}]{Pitt1971}%
  \BibitemOpen
  \bibfield  {author} {\bibinfo {author} {\bibfnamefont {L.~D.}\ \bibnamefont
  {Pitt}},\ }\bibfield  {title} {\bibinfo {title} {A {M}arkov property for
  {G}aussian processes with a multidimensional parameter},\ }\href
  {https://doi.org/https://doi.org/10.1007/BF00252003} {\bibfield  {journal}
  {\bibinfo  {journal} {Archive for Rational Mechanics and Analysis}\ }\textbf
  {\bibinfo {volume} {43}},\ \bibinfo {pages} {367} (\bibinfo {year}
  {1971})}\BibitemShut {NoStop}%
\bibitem [{\citenamefont {Balan{\c{c}}a}\ and\ \citenamefont
  {Herbin}(2012)}]{Balanca2012}%
  \BibitemOpen
  \bibfield  {author} {\bibinfo {author} {\bibfnamefont {P.}~\bibnamefont
  {Balan{\c{c}}a}}\ and\ \bibinfo {author} {\bibfnamefont {E.}~\bibnamefont
  {Herbin}},\ }\bibfield  {title} {\bibinfo {title} {{A Set-Indexed
  {O}rnstein--{U}hlenbeck Process}},\ }\href
  {https://doi.org/10.1214/ECP.v17-1903} {\bibfield  {journal} {\bibinfo
  {journal} {Electronic Communications in Probability}\ }\textbf {\bibinfo
  {volume} {17}},\ \bibinfo {pages} {1 } (\bibinfo {year} {2012})}\BibitemShut
  {NoStop}%
\bibitem [{\citenamefont {Balan{\c{c}}a}(2015)}]{Balanca2015}%
  \BibitemOpen
  \bibfield  {author} {\bibinfo {author} {\bibfnamefont {P.}~\bibnamefont
  {Balan{\c{c}}a}},\ }\bibfield  {title} {\bibinfo {title} {An increment-type
  set-indexed {M}arkov property},\ }\href@noop {} {\bibfield  {journal}
  {\bibinfo  {journal} {Journal of Theoretical Probability}\ }\textbf {\bibinfo
  {volume} {28}},\ \bibinfo {pages} {1271} (\bibinfo {year}
  {2015})}\BibitemShut {NoStop}%
\bibitem [{\citenamefont {Noack}\ \emph {et~al.}(2023)\citenamefont {Noack},
  \citenamefont {Krishnan}, \citenamefont {Risser},\ and\ \citenamefont
  {Reyes}}]{Noack2023}%
  \BibitemOpen
  \bibfield  {author} {\bibinfo {author} {\bibfnamefont {M.~M.}\ \bibnamefont
  {Noack}}, \bibinfo {author} {\bibfnamefont {H.}~\bibnamefont {Krishnan}},
  \bibinfo {author} {\bibfnamefont {M.~D.}\ \bibnamefont {Risser}},\ and\
  \bibinfo {author} {\bibfnamefont {K.~G.}\ \bibnamefont {Reyes}},\ }\bibfield
  {title} {\bibinfo {title} {Exact {G}aussian processes for massive datasets
  via non-stationary sparsity-discovering kernels},\ }\href@noop {} {\bibfield
  {journal} {\bibinfo  {journal} {Scientific Reports}\ }\textbf {\bibinfo
  {volume} {13}},\ \bibinfo {pages} {3155} (\bibinfo {year}
  {2023})}\BibitemShut {NoStop}%
\bibitem [{\citenamefont {Matthias}(2005)}]{Matthias2005}%
  \BibitemOpen
  \bibfield  {author} {\bibinfo {author} {\bibfnamefont {S.}~\bibnamefont
  {Matthias}},\ }\href@noop {} {\bibinfo {title} {Low rank updates for the
  {C}holesky decomposition}} (\bibinfo {year} {2005}),\ \bibinfo {note}
  {{T{\"u}bingen}: Max Planck Society}\BibitemShut {NoStop}%
\bibitem [{\citenamefont {{Wolfram Research{,} Inc.}}()}]{Mathematica13}%
  \BibitemOpen
  \bibfield  {author} {\bibinfo {author} {\bibnamefont {{Wolfram Research{,}
  Inc.}}},\ }\href {https://www.wolfram.com/mathematica} {\bibinfo {title}
  {Mathematica, {V}ersion 13.3}},\ \bibinfo {note} {{Champaign}, {IL},
  2024}\BibitemShut {NoStop}%
\end{thebibliography}%

\appendix

\section{Markov property of random sheets \label{app:markov}}

The standard way to define the transition probabilities and the Markov property for ordinary random processes (indexed with a single parameter) is via the natural linear time ordering, which is clearly defined only for a single parameter. Extending this definition to the multi-parameter case is non-trivial, and naive attempts may lead to disappointing results, see \cite{DynkinMarkovProcAndRandFields,AdlerBookGeomRandF}. For instance, if  Markovianity is defined by conditioning on the values of the random process at the boundary of a many-dimensional region (so-called quasi-Markov property), then the process ceases to be random outside of this region \cite{AdlerBookGeomRandF}.

It is possible to define a version of the Markov property for random sheets. Recall that we are working on a probability space $(\Omega, \mathcal{B}, \mathbb{P})$ and $\mathcal{B}$ is a $\sigma$-algebra of events. For a ``standard'' random process $X(t,\omega) = X(t)$, indexed by a single parameter $t\in\mathbb{R}_{+}$, the Markov property can be understood as follows: if we aim to make predictions about the future of the process $X(t)$ for $t>s$ (for instance if we want to estimate average of some observable in the future), it is enough to take in account only the current state $X(s)$, ignoring past. This is expressed as
\begin{equation}\label{ordinary_markov_prop}
\begin{split}
&\mathbb{E}\left( f(X(r+s)) | \mathcal{F}_{s}  \right) =\\
&\mathbb{E}\left( f(X(r+s)) | X(s)  \right), \forall s\in \mathbb{R}_{+}, \forall r \ge 0,
\end{split}
\end{equation}
where $\mathcal{F}_{s},\ s\in \mathbb{R}_{+}$ is the natural filtration of $\sigma$-algebras for the process $X(s)$, and $f$ is a Borel function. Each $\mathcal{F}_s$ is a sub-algebra of the initial $\sigma$-algebra $\mathcal{B}$ of all events. The $\sigma$-algebra $\mathcal{F}_{s}$ for the moment $s$ is the minimal $\sigma$-algebra such that all $X(t),\ t \le s$ are measurable with respect to $\mathcal{F}_{s}$. We want to have the simplest probabilistic construction, which, at the moment $s$, captures all information about what has already happened to the random variables $X(t)$ for $t \le s$, i.e.\ up to the current moment. Thus, we include in $\mathcal{F}_{s}$ all past events necessary for each random variable $X(t)$ with $t \le s$ to be measurable with respect to $\mathcal{F}_{s}$. Further we go in time, the broader our $\sigma$-algebra should be, since more random variables $X(t)$ at $t \le s$ should be taken into account. The property (\ref{ordinary_markov_prop}) essentially means that the past imposes no additional requirements beyond those imposed by the current moment $s$, i.e.\ the past is irrelevant for predicting the possible future values. 

Therefore, if we are able to stop the process at the moment $s$, fix the current state $X(s)$ (e.g.\ store it in memory), and later start the process again, then, in order to predict all probabilistic properties of the process in the future, we don't need anything else but the state $X(s)$. This requirement generalized to the random stopping time known as the strong Markov property. Stopping time is a random variable ${\mathrm s}: \Omega \rightarrow \mathbb{R}_{+}$, such that the event $S_t = \{\omega \in \Omega: {\mathrm s}(\omega)<t \} \in \mathcal{F}_{t}$. We can shift the process $X(t)$ with a stopping time, defining the new process 
\begin{equation}
X_{{\mathrm s}}(t) = X(t+{\mathrm s}(\omega),  \omega).
\label{eq:xtil}
\end{equation}
The process is called strongly Markovian if, for each stopping time ${\mathrm s}$, we have 
\begin{equation}\label{strongly_markov_prop}
\mathbb{E}\left( f(X_{{\mathrm s}}(t)) | \mathcal{F}_{{\mathrm s}}  \right) = \mathbb{E}\left( f(X_{{\mathrm s}}(t)) | X({\mathrm s})  \right),
\end{equation}
where $\mathcal{F}_{{\mathrm s}} = \{B\in \mathcal{B}: B\cap S_t \in \mathcal{F}_t\ \forall t \in \mathbb{R}_{+} \}$ is the stopped $\sigma$-algebra.

There are several approaches to define the Markov property for sheets. To introduce an intuitive approach, which echoes the strong Markov property, we need to define the stopping point \cite{Walsh1986}. Suppose that we have the Brownian sheet ${\mathscr W}$, parametrized by $\theta=(x,t)$ on the probability space $(\Omega, \mathcal{B}, \mathbb{P})$. It generates a collection of $\sigma$-algebras
\begin{equation}
\begin{split}
&\mathcal{F}_{(x,t)} = \sigma \{ {\mathscr W}(u,s), \; \text{where}\ u\le x \ \text{or}\ s\le t \},\\ 
&(x,t) \in \mathbb{R}_+^2.
\end{split}
\end{equation}
This is the smallest $\sigma$-algebra, with respect to which all random variables are measurable.
\begin{definition}[Weak stopping point]
A two-dimensional random vector $({\mathrm u}, {\mathrm s})$ is called a weak stopping point if
\begin{equation}
\{ {\mathrm u} < x, {\mathrm s} <t \} \in \mathcal{F}_{(x,t)},\quad \forall (x,t) \in \mathbb{R}_+^2.
\end{equation}
\end{definition}
Similar to the case of one parameter, the stopping point is the random starting position for a random sheet. 
Stopped $\sigma$-algebra in that case is
\begin{equation}
\mathcal{F}_{({\mathrm u},{\mathrm s})} = \{ A \in \mathcal{B}: A \cap \{ {\mathrm u} < x, {\mathrm s} <t \} \in \mathcal{F}_{(x,t)} \}.
\end{equation}

Starting from the Brownian sheet on $\mathbb{R}_+^2$, it is possible to introduce its analog defined on the set $\mathfrak{A}$ of closed rectangles lying in $\mathbb{R}_+^2$. Let us characterize a rectangle by two points, $\theta_0=(u_0,s_0)$ and $\theta_1=(u_1,s_1)$, such that $u_0<u_1$ and $s_0<s_1$, i.e.\ the vertices of the rectangle which are, respectively, the closest to, and the farthest from the origin. We define a random function ${\widetilde {\mathscr W}}$ on a closed rectangle as 
\begin{equation}
\begin{split}
&{\widetilde {\mathscr W}}\left( [\theta_0,\theta_1] \right) = \\
&{\mathscr W}(u_1,s_1) - {\mathscr W}(u_0,s_1)- {\mathscr W}(u_1,s_0)+{\mathscr W}(u_0,s_0),   
\end{split}
\end{equation}
which produces a random sheet evolving from the ``moment'' $\theta_s=(u,s)$
\begin{equation}
\begin{split}
&{\mathscr W}_{{\mathrm u}, {\mathrm s}}(x,t) = \\ 
&{\widetilde {\mathscr W}}( [({\mathrm u},{\mathrm s}),(x+{\mathrm u},t+{\mathrm s})] ), \quad (x,t)\in \mathbb{R}^2_{+}.
\end{split}
\end{equation}
The Brownian sheet has a following version of the strong Markov property~\cite{Pitt1971,Walsh1986}.
\begin{theorem}
The random sheet ${\mathscr W}_{{\mathrm u}, {\mathrm s}}(x,t)$ is a Brownian sheet independent of $\mathcal{F}_{({\mathrm u},{\mathrm s})}$.
\end{theorem}
If we assume a constant stopping point $({\mathrm u},{\mathrm s}) \equiv (u_0,s_0)$, then we obtain a Brownian sheet evolving from the point $(u_0,s_0)$, independent from the ``past''. If we apply the Lamperti transform to this sheet we will get the OU sheet independent from the ``past''. This construction, while quite intuitive, has not yet been developed, to our knowledge, at sufficient level of rigor.

Instead, in order to show Markovianity, we can utilize the set-indexed construction for the OU sheet $Y(A),\ A\in \mathfrak{A}$ \cite{Balanca2012}. Explicit construction is quite involved and could be defined for different index sets $\mathfrak{A}$, but in our case it suffices that $\mathfrak{A}$ is the set of all closed rectangles. The standard OU sheet will have the following description in this manner,
\begin{equation}
Y(x,t) = Y([(0,0),(x,t)]).
\end{equation}
The process $Y(A)$ is known to be $\mathcal{C}$-Markov \cite{Balanca2012}, this property also has a rather complex definition. We also need the definition of the flow of sets.
\begin{definition}[Flow] \label{def:flow}
Let $[a,b]\subset \mathbb{R}_{+}$ be an interval. A set-valued mapping
\begin{equation}
f(t): [a,b] \rightarrow \mathfrak{A}
\end{equation}
is called flow if the following conditions hold
\begin{itemize}
\item Increasing $a\le s\le t \le b \Rightarrow f(s)\subseteq f(t)$
\item Inner-continuity $a\le s < b: f(s) = \cap_{v>s} f(v)$
\item Outer-continuity $a < s < b: f(s) = \mathrm{cl}\left(\cup_{v<s} f(v)\right)^{}$
\end{itemize}
\end{definition}
For any flow $f(t): [0,T] \rightarrow \mathfrak{A}$, $\mathcal{C}$-Markovianity implies that the process $Y(f(t))$ will be Markovian, that is a consequence of Proposition~2.10 of \cite{Balanca2015}.

\begin{figure}[tbp!]
\centering
\includegraphics[width=\linewidth]{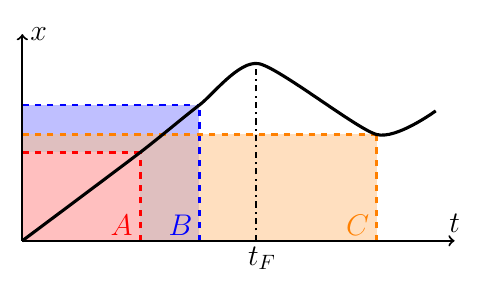}
\caption{The path $\Gamma(t)\equiv (x(t),t)$ (shown as a thick black curve) generates a flow until the moment of time $t_F$: at times $0\le t\le t_F$, the set of rectangles generated by this path are nested (two such rectangles are shown, colored with red and blue, denoted $A$ and $B$, respectively). At times $t>t_F$, this path ceases to generate a flow, since the generated rectangles are not nested anymore (the rectangle $C$, colored with orange, is not nested with respect to $B$). Correspondingly, the random process induced by the path $\Gamma(t)$ is a Markov process until $t=t_F$, and ceases to be Markovian after that.}
\label{fig:markov}
\end{figure}

The notion of flow can be informally visualized as follows. On the $(x,t)$ plane, we take a path $\Gamma(t) = (x(t),t)$ parametrized by the variable $t$,  and consider a set of rectangles, whose lower-left corner is at the origin $(0,0)$ and the upper-right corner coincides with the point on the path $\Gamma(t)$ with the coordinates $(x(t),t)$. If all such rectangles generated by the path $\Gamma(t)$ are nested, then this path generates a flow, see Fig.~\ref{fig:markov}.
Therefore, it is very natural that the following can be proven: the path $\Gamma(t) = (x(t),t)$ defines the flow iff $x(t)$ is continuous and non-decreasing. 

In the case of the one-way shuttling, the path on the plane $(x_c,t)$ is defined  
as $\Gamma(t)=(x_c(t),t)$ with $x_c(t)=vt$, such that $x_c(t)$ is  continuous and non-decreasing, and all rectangles generated by this path are nested, i.e.\ this path defines flow and the process $\mathrm{B}(t)$ is Markovian.
For the case of forth-and-back shuttling, the path does not define flow at times $t>T$, see Fig.~\ref{fig:traj1and2}a, and the probability to obtain certain values of the process at the time $t>T$ depends on the values the process had on the time interval $[0,T)$. 
Similarly, in the situation when two qubits are shuttled sequentially, see Fig.~\ref{fig:traj1and2}b, we have two separate paths $\Gamma_1(t)$ and $\Gamma_2(t)$, one for each qubit. We can combine them in a single path $\Gamma(t)$, having a shape of parallelogram, but this path does not generate a flow. Correspondingly, the values assumed by the process $\mathrm{B}_1(t)$, which corresponds to the part $\Gamma_1(t)$ of the path $\Gamma(t)$, are correlated with the values taken by the process $\mathrm{B}_2(t)$, which corresponds to the part $\Gamma_2(t)$ of the path $\Gamma(t)$.

\section{Problem of the dephasing hotspots \label{app:hotspot}}

Dephasing hotspots are tightly localized areas of the sample where dephasing is greatly enhanced, e.g.\ due to an atomic-sized defect in a Si/SiGe nanostructure, so that the variance of the noise in the vicinity of a hotspot is much larger than in other areas of the sample. While this work is focused mostly on stationary random sheets, where the variance is the same in all regions, it is useful to demonstrate that the hotspots can also be efficiently treated within the approach suggested here, and show, how the corresponding kernels of the Gaussian sheets are to be modified by the hotspots.

Mathematically, such areas on the random sheets could be modeled with the bump function kernels \cite{Noack2023}. 
We consider the bump functions only in the coordinate domain, while in the time domain the kernel of the Gaussian sheet is assumed constant; this means that the positions of the hotspots in the shuttling channel are fixed and their properties do not vary in time. We also assume that the shuttling time is finite. 
Let $B_{r}(x_0)$ be an open interval of length $2r$ centered at the point $x_0$, and $\mathds{1}_{B_{r}(x_0)}(x)$ denotes the indicator function of the interval. The bump function is defined as 
\begin{equation}\label{bumpfun}
\Psi_{x_0;r,a,c}(x) =
c\,\exp\left(-\frac{a}{1- \frac{|x-x_0|^2}{r^2}} \right) \mathds{1}_{B_{r}(x_0)}(x),
\end{equation}
with parameters $a, c>0$. The bump function has the compact support, the interval $B_{r}(x_0)$, and has derivatives of all orders.

Let us consider a certain number $P$ of hotspots, which are centered at $x_k, 1\le k \le P$, have the radii $r_k$, and are described by the parameters $a_k,c_k$ of the corresponding bump functions. Denoting $\Psi_k(x) = \Psi_{x_k;r_k,a_k,c_k}(x)$, we modify the kernel of the Gaussian sheet as follows:
\begin{equation}\label{ker_full}
\begin{split}
&K_{\mathrm{full}} (x,t,x',t') = \\
&K_{\mathrm{OU}} (x,t,x',t') + \sum_{k=1}^{N} \Psi_k(x)\Psi_k(x'),
\end{split}
\end{equation}
thus obtaining the covariance function for a stationary Gaussian OU sheet with added hotspots. 

This technique could be useful if the characteristic functional is calculated numerically by sampling the random process $\mathrm{V}(s)$, see Sec.~\ref{sec:method}. 

To justify applicability of the procedure described there to the case of added hotspots, let us consider the process $Z(s,\omega)$, which is obtained from, for instance, an OU process $X(s,\omega)$, by adding a hotspot with the bump function~(\ref{bumpfun}) $\Psi$. The resulting process remains centered, and 
\begin{widetext}
\begin{equation}
\begin{split}    
&\sqrt{\mathbb{E}\left[\left(Z(r,\omega) - Z(s,\omega)\right)^2\right]} = \sqrt{K_{Z}(r,r) + K_{Z}(s,s) -2 K_{Z}(r,s)} =\\
&\sqrt{2\left[1 - K_{OU}(x(r),t(r);x(s),t(s)) \right] + \Psi^2(x(r)) + \Psi^2(x(s)) - 2 \Psi(x(r))\Psi(x(s))},
\end{split}
\end{equation}
\end{widetext}
since bump functions are smooth and uniformly bounded, adding the hotspots does not change the estimate for $\alpha_0$ and the method remains valid.

Similarly, numerical sampling of the process $Z(s,\omega)$, described in Sec~\ref{sec:method}, is easily performed. 
Assume that we have already calculated the Cholesky decomposition of the matrix $\mathbf{C}$ in Eq.~\ref{eq:discrcov}. If we want to update it by adding a hotspot, we add to $\mathbf{C}$ the rank-one matrix 
\begin{equation}
\left( \Psi(x_j)\Psi(x_p) \right)_{p,j = 1}^{n} = v^{T} v,
\end{equation}
which is constructed from the vector $v = \left( \Psi(x_p) \right)_{p = 1}^{n}$, 
where $n$ is the number of the nodes $x_j$, $j=1,\dots n$, and perform a computationally easy rank-one update of the Cholesky decomposition \cite{Matthias2005}.

\section{Evaluation of the optimal parameter $\varepsilon$ for Monte Carlo sampling \label{app:mceval}}

The overall accuracy of numerical evaluation of the characteristic functional is a sum of two terms, one being the quantity $\Delta_M = M_0 \exp(-n^{\kappa(\alpha -\varepsilon)})$ in Eq.~\ref{eq:rvar_esteem} and the other being the discretization error $\Delta_d = C_d n^{-\varepsilon}$, where $C_d$ is a constant depending on the integrator $F(s)$, see Eqs.~\ref{quad_fl} and \ref{eq:q_quad}. We can choose the parameter $\varepsilon$ in an optimal way to obtain a good bound on the accuracy. This can be done in a straightforward manner, by minimizing the sum $\Delta_\mathrm{tot}=\Delta_M+\Delta_d$ over the quantity $\delta=\alpha-\varepsilon$. 

We aim to minimize the sum 
$\Delta_\mathrm{tot}=M_0\exp(-n^{\kappa\delta})+C_d n^{-\alpha}\,n^\delta$ as a function of $\delta$. Introducing the quantity $\nu=\ln{n}$, we write
$$
 \Delta_\mathrm{tot}(\delta) = \Delta_M(\delta) + \Delta_d(\delta) 
 = M_0 \exp(-{\mathrm e}^{\kappa\nu\delta}) + C_d {\mathrm e}^{-\alpha\nu}{\mathrm e}^{\nu\delta},
$$
and, as $\delta$ increases, the term $\Delta_M(\delta)$ rapidly decreases, while  $\Delta_d(\delta)$ grows. 
So, in order for $\Delta_\mathrm{tot}$ to have minimum at a finite value of $\delta>0$, we must have $\Delta_M(0)>\Delta_d(0)$: otherwise, the minimum of $\Delta_\mathrm{tot}$ is reached at $\delta=0$, and the bound on the total accuracy is $\Delta_\mathrm{tot}(0)=M_0+C_d \exp{-\alpha\nu}$, i.e.\ is limited by a constant $M_0$, and cannot be improved by increasing the number $n$ of the integration nodes.

Thus, we assume that $n$ is sufficiently large to ensure that $\Delta_M>\Delta_d$ at $\delta=0$, which means that $M_0>C_d\exp{-\alpha\nu}$. Taking the derivative of $\Delta_\mathrm{tot}(\delta)$ with respect to $\delta$ and setting it to zero, we obtain the equation
$$
C_d {\mathrm e}^{-\alpha\nu} {\mathrm e}^{\nu\delta} = {\mathrm e}^{\kappa\nu\delta} {\mathrm e}^{-\exp{\kappa\nu\delta}},
$$
and, taking the logarithm of both sides, the equation determining the minimum of $\Delta_\mathrm{tot}(\delta)$ becomes
$$
\nu \delta - A_\nu = \kappa\nu \delta - {\mathrm e}^{\kappa\nu \delta},
$$
where $A_\nu = \alpha\nu - \ln{[C_d/(\kappa M_0)]}$. Again, we assume that $\nu=\ln{n}$ is large enough to ensure that $A_\nu<-1$, such that the equation above has an admissible solution for $\delta$, satisfying the requirements  $0<\delta<\alpha$. Denoting $y=\kappa\nu\delta$ and rewriting this equation as
$$
y - {\mathrm e}^y = \frac{y}{\kappa} - A_\nu,
$$
we see that it has admissible solutions for $0<\kappa<2$ (these bounds are imposed by Theorem~\ref{thm:MCalpha}) as long as $n$ is large enough. For our purposes, it is convenient to choose $\kappa=1$; this may not give the tightest possible bound on accuracy, but makes it easy to obtain the explicit solution for $\delta$:
$$
\delta = \frac{1}{\nu} \ln{A_\nu} = \frac{1}{\nu} \ln{\left[-\alpha\nu  + \ln(C_d/M_0)\right]}.
$$
With such a choice of $\kappa$ and $\delta$, we have $n^{\kappa\delta} = \exp{\kappa\nu\delta} = A_\nu$, 
such that 
$$
\Delta_M = M_0 {\mathrm e}^{-A_\nu},\quad \Delta_d = C_d A_\nu {\mathrm e}^{-\alpha\nu},
$$
and therefore
$$
\Delta_\mathrm{tot} = C_d n^{-\alpha} \left[1 + \alpha \ln{n} - \ln(C_d/M_0)\right],
$$
which is $O(n^{-\alpha}\ln{n})$ for large $n$. Finally, the parameter $\alpha$ can assume any value as long as $\alpha<\alpha_0$, and for practical purposes can be just replaced by $\alpha_0$, giving the estimate $O(n^{-\alpha_0}\log{n})$ for the accuracy of our approximation.

\section{Calculation of the decoherence factor for the forth-and-back shuttling of a single spin qubit \label{app:forthback}}

Here we present the details of the calculation of the decoherence factor $W(t_0)$ for the scenario where a single electron is first shuttled forth, from $x_c=0$ at $t=0$ to $x_c=L$ at $t=T$ with the velocity $v=L/T$, and then back with the same velocity, from $x_c=L$ at $t=T$ to $x_c=0$ at $t=t_0=2T$. The qubit trajectory $x_c(t)$ consists of two branches, $x_c(t)=vt$ for $t\in [0,T]$ and $x_c(t)=2L-vt$ for $t\in [T,2T]$, as shown in Fig.~\ref{fig:traj1and2}. This trajectory defines the random process ${\mathrm B}(t)={\tilde B}(x_c(t),t)$, which is derived from the underlying OU random sheet ${\tilde B}(x_c(t),t)$. The process ${\mathrm B}(t)$ inherits Gaussian property from the random sheet ${\tilde B}(x_c(t),t)$, and therefore it is completely specified by its covariance function 
\begin{equation}
K_{\mathrm B}(t_1,t_2)=K_{\tilde B}(x_c(t_1),t_1; x_c(t_2),t_2),
\label{eq:KBvsKOU}
\end{equation}
where $K_{\tilde B}(x_1,t_1;x_2,t_2)$ is the covariance of the OU sheet given in Eq.~\ref{eq:random_sheet_corr}. Explicitly, 
\begin{equation}
K_{\mathrm B}(t_1,t_2)=\sigma^2_{\mathrm B}\,\exp{-Q(t_1,t_2)},
\end{equation}
with 
\begin{align*}
Q(t_1,t_2) = 
\begin{cases}
(\kappa_t+\kappa_x v) |t_1-t_2|\;\ {\mathrm{in}}\ R_1, \\
\kappa_t (t_2-t_1) + \kappa_x v|(t_1+t_2)-2T|\;\ {\mathrm{in}}\ R_2, \\
\kappa_t (t_1-t_2) + \kappa_x v|(t_1+t_2)-2T|\;\ {\mathrm{in}}\ R_3, \\
(\kappa_t+\kappa_x v) |t_1-t_2|\;\ {\mathrm{in}}\ R_4, 
\end{cases}
\end{align*}
where we divided the relevant domain $[0,2T]\times [0,2T]$ of the variables $(t_1,t_2)$ into four parts:
\begin{eqnarray}
\nonumber
R_1,&\ &{\mathrm{where}}\ t_1\in[0,T],\, t_2\in [0,T],\\ \nonumber
R_2,&\ &{\mathrm{where}}\ t_1\in[0,T],\, t_2\in [T,2T],\\ \nonumber
R_3,&\ &{\mathrm{where}}\ t_1\in[T,2T],\, t_2\in [0,T],\\ \nonumber
R_4,&\ &{\mathrm{where}}\ t_1\in[T,2T],\, t_2\in [T,2T].
\end{eqnarray}
As explained in the main text, the decoherence factor $W(t_0)$ is given by a double integral of the covariance function $K_{\mathrm B}(t_1,t_2)$
\begin{eqnarray}
W(t_0)&=&\exp{-\chi(t_0)}\\ \nonumber
\chi(t_0)&=&\frac{1}{2}\int_0^{t_0} dt_1 \int_0^{t_0} dt_2 \; K_{\mathrm B}(t_1,t_2).
\end{eqnarray}
The double integral $\chi(t_0)$ naturally splits into a sum of four integrals $I_{1}$, $I_{2}$, $I_{3}$, and $I_{4}$, which  correspond to integration over the regions $R_1$--$R_4$, respectively. Due to the symmetry of the covariance function $K_{\mathrm B}(t_1,t_2)$, we immediately see that $I_{1}=I_{4}$ and $I_{2}=I_{3}$, so that $\chi(t_0)=I_{1}+I_{2}$. Explicit calculation of the integrals gives the answers:  
\begin{eqnarray}
\nonumber
I_{1} &=& \int_0^{T} dt_1 \int_0^{T} dt_2 \, K_{\mathrm B}(t_1,t_2)  
  = (\sigma_{\mathrm B}/\kappa_t)^2\,\Lambda_1,\\ 
\Lambda_1 &=& \frac{2(\beta+\gamma+{\mathrm e}^{-\beta-\gamma}-1)} {(1+\gamma/\beta)^2}, 
\end{eqnarray}
where we introduced the dimensionless shuttling length $\gamma=\kappa_x L$ and time $\beta=\kappa_t T$, and 
\begin{eqnarray}
\nonumber
I_{2} &=& \int_0^{T} dt_1 \int_T^{2T} dt_2 \, K_{\mathrm B}(t_1,t_2) 
  = (\sigma_{\mathrm B}/\kappa_t)^2\,\Lambda_2,\\ 
\Lambda_2 &=& \frac{\left({\mathrm e}^{-2\beta}-1\right) (\gamma /\beta) - 2 {\mathrm e}^{-(\beta+\gamma)}+{\mathrm e}^{-2\beta}+1}{1-\gamma ^2/\beta ^2},
\end{eqnarray}
such that 
\begin{equation}
W(t_0)={\mathrm e}^{-\chi(t_0)},\quad \chi(t_0)=(\sigma_{\mathrm B}/\kappa_t)^2 \Lambda(\beta,\gamma),
\end{equation}
where $\Lambda(\beta,\gamma)=\Lambda_1+\Lambda_2$.
Notice that the term $\Lambda_1$ has the same form as the integral appearing in the calculation of dephasing by the standard OU process, cf.\ Eq.~\ref{eq:wT}. This coincidence is natural, because the regions $R_1$ and $R_4$ correspond to the separate branches 1 and 2 of the shuttling trajectory, and ${\mathrm B}(t)$ is a usual OU process there. The non-trivial contribution comes from the correlations between the branches, and is taken into account during integration over the regions $R_2$ and $R_3$, which produces the term $\Lambda_2$ that has a very different form.

\section{Calculation of the decoherence factor for sequential shuttling of two spin qubits \label{apdx:derivation}}

The calculations were performed using the covariance function of the process $\delta\mathrm{B}(t)$ given by Eq.~\ref{eq:cov2}. The calculation are quite lengthy and tedious, and were performed mostly using the package Mathematica 13.3 \cite{Mathematica13}.

\subsection{Autocorrelation terms}

First, we calculate the autocorrelation terms, which describe decoherence of individual spins.

\begin{figure}[h]
\centering
\includegraphics[width=0.49\linewidth]{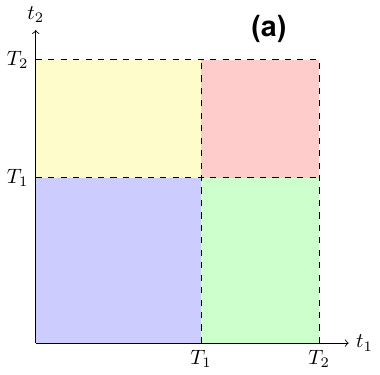}
\includegraphics[width=0.49\linewidth]{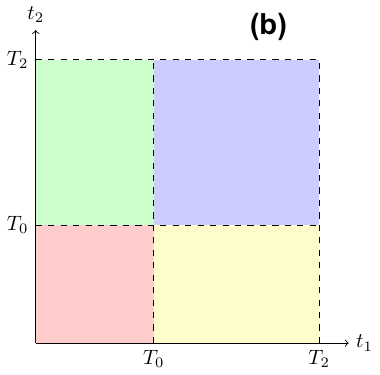}
\caption{Division of the integration area into different regions when calculating the contributions of the autocorrelation terms. {\bf(a)}: Integration of $K_{\mathrm{B}1}(t_1,t_2)=\mathbb{E}[B_1(t_1)B_1(t_2)]$. {\bf(b)}: Integration of $K_{\mathrm{B}2}(t_1,t_2)=\mathbb{E}[B_2(t_1)B_2(t_2)]$.}
\label{fig:integral_area_autocorr}
\end{figure}

We start by evaluating the contribution produced by the autocorrelation of the field $\mathrm{B}_1(t)$ acting on the first spin, calculating the integral
\begin{equation}
I_{11}=
\exp{-\int_0^T \dd t_1 \int_0^T \dd t_2 \, K_{\mathrm{B}1}(t_1,t_2)},
\end{equation}
where $K_{\mathrm{B}1}(t_1,t_2)=\mathbb{E}\left[\mathrm{B}_1(t_1) \mathrm{B}_1(t_2)\right]$ is the covariance function of the field $\mathrm{B}_1(t)$  
(the first term in Eq.~\ref{eq:cov2}), and therefore $K_{\mathrm{B}1}(t_1,t_2)=K_{OU}(x_{c1}(t_1),t_1;x_{c1}(t_2),t_2)$. This function depends on the difference $x_{c1}(t_1)-x_{c1}(t_2)$, which is equal to
\begin{equation}
\label{eq:Kxc1}
x_{c1}(t_1)-x_{c1}(t_2)=
\begin{cases}
v (t_1-t_2) & \;\ {\mathrm{in}}\ \textbf{B}, \\
L-v t_2 & \;\ {\mathrm{in}}\ \textbf{G}, \\
v t_1 -L& \;\ {\mathrm{in}}\ \textbf{Y}, \\
0 & \;\ {\mathrm{in}}\ \textbf{R},
\end{cases}
\end{equation}
where we divided the whole integration area into four regions as shown in Fig.~\ref{fig:integral_area_autocorr}a, shading each region with the corresponding color:
\begin{eqnarray}
\textbf{B}\ \mathrm{(blue)}&:&\ \ t_1 \in [0,T_1],\ t_2 \in [0,T_1],\\ \nonumber
\textbf{G}\ \mathrm{(green)}&:&\ \ t_1 \in [T_1,T_2],\ t_2 \in [0,T_1],\\ \nonumber
\textbf{Y}\ \mathrm{(yellow)}&:&\ \ t_1 \in [0,T_1],\ t_2 \in [T_1,T_2],\\ \nonumber
\textbf{R}\ \mathrm{(red)}&:&\ \ t_1 \in [T_1,T_2], \ t_2 \in [T_1,T_2].
\end{eqnarray}
Evaluating the integral $I_{11}$ in each region, we obtain:
\begin{eqnarray}
\textbf{B:}\;\nonumber\hfill P_1&&=\int_0^{T_1} \dd t_1 \int_0^{T_1} \dd t_2 \, \exp{ -(\kappa _t+\kappa _x L/T_1)| t_1-t_2| }\nonumber \\&&
=2 T_1^2 \cdot \frac{e^{-L \kappa _x-T_1 \kappa _t}+L \kappa _x+T_1 \kappa _t-1}{\left(L \kappa _x+T_1 \kappa _t\right)^2}
\\
\textbf{R:}\;\nonumber\hfill P_2&&=\int_{T_1}^{T_2} \dd t_1 \int_{T_1}^{T_2} \dd t_2 \,    \exp{ -\kappa _t| t_1-t_2| }
\nonumber \\&&
=\frac{2}{\kappa _t^2} \left[\kappa _tT_0+e^{-\kappa _t T_0 }-1\right]
\\
\textbf{G:}\;\nonumber\hfill P_3&&=\\ \nonumber
&&\!\!\!\!\!\! \int_{T_1}^{T_2} \dd t_1 \int_0^{T_1} \dd t_2 \,\exp{ -\kappa _t (t_1-t_2)-\kappa_x L(1-t_2/T_1) }
\nonumber \\&&
=\frac{T_1 \left(1-e^{-T_0 \kappa _t}\right) \left(1-e^{-L \kappa _x-T_1 \kappa _t}\right)}{\kappa _t \left(L \kappa _x+T_1 \kappa _t\right)}\\
\textbf{Y:}\;\nonumber\hfill P_4&&=\\ \nonumber
&&\!\!\!\!\!\! \int_0^{T_1} \dd t_1 \int_{T_1}^{T_2} \dd t_2 \,\exp{ -\kappa _t (t_2-t_1)-\kappa_x L(1-t_1/T_1) }
\nonumber \\&&
=\frac{T_1 \left(1-e^{-T_0 \kappa _t}\right) \left(1-e^{-L \kappa _x-T_1 \kappa _t}\right)}{\kappa _t \left(L \kappa _x+T_1 \kappa _t\right)}
\end{eqnarray}
Note that the correlation function $K_{\mathrm{B}1}(t_1,t_2)$ is symmetric with respect to interchange of the variables $t_1$ and $t_2$, such that the integrals $P_3$ and $P_4$ are equal.

In the same way, we calculate the terms involving the autocorrelations of the field $\mathrm{B}_2(t)$ acting on the second spin, described by the covariance function $K_{\mathrm{B}2}(t_1,t_2)$:
\begin{equation}
I_{22}=\exp{-\int_0^T \dd t_1 \int_0^T \dd t_2 \, K_{\mathrm{B}2}(t_1,t_2)}.
\end{equation}
The function $K_{\mathrm{B}2}(t_1,t_2)$ is determined by the trajectory $x_{c2}(t)$ of the second spin, and involves the difference $x_{c2}(t_1)-x_{c2}(t_2)$. Again, we divide the integration area into four regions as shown in Fig.~\ref{fig:integral_area_autocorr}b, but now the color coding is different:
\begin{eqnarray}
\textbf{B}\ \mathrm{(blue)}&:&\ \ t_1 \in [T_0,T_2],\ t_2 \in [T_0,T_2],\\ \nonumber
\textbf{G}\ \mathrm{(green)}&:&\ \ t_1 \in [0,T_0],\ t_2 \in [T_0,T_2],\\ \nonumber
\textbf{Y}\ \mathrm{(yellow)}&:&\ \ t_1 \in [T_0,T_2],\ t_2 \in [0,T_0],\\ \nonumber
\textbf{R}\ \mathrm{(red)}&:&\ \ t_1 \in [0,T_0], \ t_2 \in [0,T_0].
\end{eqnarray}
With this color coding, the quantity $x_{c2}(t_1)-x_{c2}(t_2)$ is equal to 
\begin{equation}
\label{eq:Kxc2}
x_{c2}(t_1)-x_{c2}(t_2)=
\begin{cases}
v (t_1-t_2) & \;\ {\mathrm{in}}\ \textbf{B}, \\
v (T_0-t_2) & \;\ {\mathrm{in}}\ \textbf{G}, \\
v (t_1-T_0)& \;\ {\mathrm{in}}\ \textbf{Y}, \\
0 & \;\ {\mathrm{in}}\ \textbf{R}.
\end{cases}
\end{equation}
The reason for different encoding becomes clear when we compare Eqs.~\ref{eq:Kxc2} and \ref{eq:Kxc1}, take into account that, firstly, $T_2-T_0=T_1$ (both spins are shuttled with the same velocity, so the shuttling times are equal) and secondly, that the covariance function of the underlying OU sheet $K_{OU}(x,t;x',t')$ is translationally invariant in time and space. As a result, it is not difficult to see that the integrals $I_{22}$ and $I_{11}$ over the regions denoted by the same letters/colors are equal, and we have 
\begin{equation}
I_{11}=I_{22}=P_1+P_2+P_3+P_4.
\end{equation}

\subsection{Cross-correlation terms}

Next, we evaluate the contribution from the cross-correlations between the fields $\mathrm{B}_1(t)$ and $\mathrm{B}_2(t)$ (the term $\mathbb{E}\left[\mathrm{B}_1(t_1) \mathrm{B}_2(t_2)\right]$ in Eq.~\ref{eq:cov2}), calculating the integral
\begin{equation}
I_{12}=
\exp{-\int_0^T \dd t_1 \int_0^T \dd t_2 \, K_{{\mathrm{B}12}}(t_1,t_2)},
\end{equation}
where the cross-covariance function $K_{{\mathrm{B}12}}(t_1,t_2)=K_{OU}[t_1,x_{c1}(t_1);t_2, x_{c2}(t_2)]$. Since both spins are shuttled with the same velocity, and the underlying OU sheet is stationary, the contribution coming from the other cross-correlation term, $\mathbb{E}\left[\mathrm{B}_2(t_1) \mathrm{B}_1(t_2)\right]$ in Eq.~\ref{eq:cov2}, is also equal to $I_{12}$.

\begin{figure}[bph!]
\centering
\includegraphics[width=0.49\linewidth]{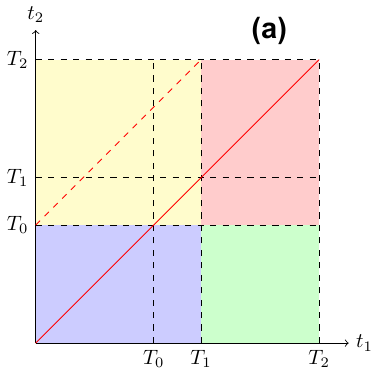}
\includegraphics[width=0.49\linewidth]{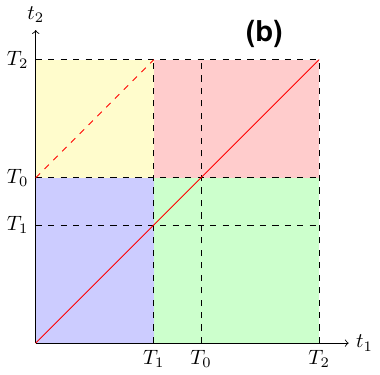}
\caption{Division of the integration area into different regions when calculating the contributions of the cross-correlation terms. {\bf(a)}: The case when $T_1>T_0$. {\bf(b)}: The case when $T_1<T_0$.}
\label{fig:integral_area_correlated}
\end{figure}

As before, we subdivide the whole integration area in four regions, see Fig.~\ref{fig:integral_area_correlated},
\begin{eqnarray}
\textbf{B}\ \mathrm{(blue)}&:&\ \ t_1 \in [0,T_1],\ t_2 \in [0,T_0]\\
\nonumber
\textbf{G}\ \mathrm{(green)}&:&\ \ t_1 \in [T_1,T_2],\ t_2 \in [0,T_0]\\
\nonumber
\textbf{Y}\ \mathrm{(yellow)}&:&\ \ t_1 \in [0,T_1],\ t_2 \in [T_0,T_2]\\
\nonumber
\textbf{R}\ \mathrm{(red)}&:&\ \ t_1 \in [T_1,T_2],\ t_2 \in [T_0,T_2]
\end{eqnarray}
in accordance with the quantity $x_1(t_1)-x_2(t_2)$, which is equal to 
\begin{equation}
x_{c1}(t_1)-x_{c2}(t_2)=
\begin{cases}
v t_1& \;\ {\mathrm{in}}\ \textbf{B}, \\
L & \;\ {\mathrm{in}}\ \textbf{G}, \\
v(t_1-t_2+T_0)& \;\ {\mathrm{in}}\ \textbf{Y}, \\
L -v(t_2-T_0)& \;\ {\mathrm{in}}\ \textbf{R}.
\end{cases}
\end{equation}
This quantity is positive in the regions \textbf{B}, \textbf{G}, and \textbf{R}, but in the region \textbf{Y}, the sign of $x_1(t_1)-x_2(t_2)$ changes as the line $t_2=t_1 +T_0$ is crossed (shown as dashed red line in Fig.~\ref{fig:integral_area_correlated}). Besides, the quantity $t_1-t_2$ changes the sign when crossing the line $t_1=t_2$ (solid red line in Fig.~\ref{fig:integral_area_correlated}). Thus, the whole integration area is naturally divided into several parts, schematically shown in Fig.~\ref{fig:integral_area_correlated}, according to the form of the covariance function $K_{OU}[t_1,x_{c1}(t_1);t_2, x_{c2}(t_2)]$, and these parts have different shape, depending on whether the delay time $T_0$ is larger or smaller than the single-spin shuttling time $T_1$.

Before performing integration, it is worthwhile to notice that in the case $T_0>T_1$ the integral over the region \textbf{R} has the form 
\begin{eqnarray}
\int _{T_1}^{T_2}\int _{T_1}^{0}\exp{-\kappa _t \left|T_2- t_2'-t_1\right| -\frac{\kappa _x L t_2'}{T_1}} \dd t_2' \dd t_1,
    \nonumber     \\
=\int _{T_0}^{0}\int _{T_1}^{0}\exp{-\kappa _t \left|t_1'- t_2'\right| -\frac{\kappa _x L t_2'}{T_1}} \dd t_2' \dd t_1, \nonumber
\end{eqnarray}
and is equal to the integral over the region \textbf{B} (the second line of the equation above); here we used the change of variables $t_2'=T_1+T_0-t_2=T_2-t_2$ and $t_1'=T_2-t_1$. In a similar manner, one can see that the integrals over \textbf{B} and \textbf{R} are also equal in the case $T_0<T_1$.

Performing the calculations for the case $T_0>T_1$, we obtain

\begin{widetext}
\begin{eqnarray}
&&\textbf{G:}\ \ \ 
C_1' =\int_{T_1}^{T_2} \int_{0}^{T_0}  \exp{-\kappa_t|t_1-t_2|-\kappa_x L} \dd t_2 \dd t_1\\  \nonumber
&&=\frac{e^{-L \kappa _x}}{\kappa _t^2} \left[2 (T_0-T_1) \kappa _t-e^{-T_1 \kappa _t}+e^{\left(T_1-T_0\right) \kappa _t}-e^{\left(T_0-T_2\right) \kappa _t}+e^{-T_2 \kappa _t}\right]\\
&&\textbf{B:}\ \ \ 
C_2' =\int _0^{T_1}\int _0^{T_0}\exp{-\kappa _t \left| t_1-t_2\right| -\frac{\kappa _x L t_1}{T_1}} \dd t_2 \dd t_1\\ \nonumber
&&=\frac{T_1 \left(e^{T_1 \kappa _t}-e^{T_0 \kappa _t}\right) e^{-L \kappa _x-\left(T_0+T_1\right) \kappa _t} \left(e^{L \kappa _x}-e^{T_1 \kappa _t}\right)}{\kappa _t \left(T_1 \kappa _t-L \kappa _x\right)}+\frac{T_1}{\kappa _t} \left(\frac{e^{-T_1 \kappa _t}-e^{-L \kappa _x}}{T_1 \kappa _t-L \kappa _x}+\frac{e^{-L \kappa _x-T_1 \kappa _t}-1}{L \kappa _x+T_1 \kappa _t}+\frac{2-2 e^{-L \kappa _x}}{L \kappa _x}\right)\\
&&\textbf{Y:}\ \ \ 
C_3' =\int _0^{T_1}\int _{T_0}^{T_2}\exp{-\kappa _t \left(t_2-t_1\right)+\frac{\kappa _x L}{T_1} |t_1-t_2+T_0|}\dd t_2 \dd t_1\\ \nonumber
&&=\frac{T_1^2 e^{-T_0 \kappa _t} \left(e^{-L \kappa _x-T_1 \kappa _t}+L \kappa _x+T_1 \kappa _t-1\right)}{\left(L \kappa _x+T_1 \kappa _t\right)^2}
+\frac{T_1^2 e^{-T_0 \kappa _t} \left(e^{T_1 \kappa _t-L \kappa _x}+L \kappa _x-T_1 \kappa _t-1\right)}{\left(T_1 \kappa _t-L \kappa _x\right)^2}\\ 
&&\textbf{R:}\ \ \ 
C_4'=C_2'.\\ \nonumber
\end{eqnarray}

In the opposite case, when $T_0<T_1$, we have

%
\begin{eqnarray}
&&\textbf{G:}\ \ \ 
C_1'' =\frac{1}{\kappa _t^2}\left(e^{T_0 \kappa _t}-1\right)^2 e^{-L \kappa _x-\left(\left(T_0+T_1\right) \kappa _t\right)}\\
&&\textbf{B:}\\ \nonumber
&&C_2'' =\frac{T_1}{\kappa _t} 
\biggl\{\frac{-e^{\left(T_0-T_1\right) \kappa _t-L \kappa _x}+e^{-L \kappa _x-T_1 \kappa _t}+e^{-\frac{L T_0 \kappa _x}{T_1}}-1}{L \kappa _x+T_1 \kappa _t}
+\frac{1-e^{-\frac{L T_0 \kappa _x}{T_1}}}{L \kappa _x}
+\frac{T_1 \kappa _t \left(1-e^{-\frac{L T_0 \kappa _x}{T_1}}\right)+L \kappa _x \left(e^{-T_0 \kappa _t}-1\right)}{L \kappa _x \left(T_1 \kappa _t-L \kappa _x\right)}
\biggr\}\\
&&\textbf{Y:}\ \ \ 
C_3'' =
{T_1^2}\biggl\{\frac{e^{-T_0 \left(\frac{L \kappa _x}{T_1}+\kappa _t\right)} \left(e^{T_0 \kappa _t} \left[L\kappa _x (T_0-T_1)/T_1 +\left(T_1-T_0\right) \kappa _t+1\right]-e^{\frac{L T_0 \kappa _x}{T_1}} \left(-L \kappa _x+T_1 \kappa _t+1\right)\right)}{\left(T_1 \kappa _t-L \kappa _x\right)^2}\\ \nonumber
+&&\frac{e^{-T_0 \kappa _t} \left(e^{-L \kappa _x-T_1 \kappa _t}+L \kappa _x+T_1 \kappa _t-1\right)+e^{\left(T_0-T_1\right) \kappa _t-L \kappa _x}+e^{-\frac{L T_0 \kappa _x}{T_1}} \left[L \kappa _x (1- T_0/T_1)+(T_1-T_0) \kappa _t-1\right]}{\left(L \kappa _x+T_1 \kappa _t\right)^2}\biggr\}\\
&&\textbf{R:} \ \ \ 
C_4''=C_2''
\end{eqnarray}
\end{widetext}

\end{document}